\documentclass[aps,prd,superscriptaddress,11pt,showpacs,notitlepage]{revtex4}

	\usepackage{graphicx}
	\usepackage{amsmath,amssymb,mathrsfs}
	\usepackage[colorlinks=true, a4paper=true, pdfstartview=FitV,
linkcolor=blue, citecolor=blue, urlcolor=blue]{hyperref}

\usepackage{subfigure}	
\usepackage{epsf}
\usepackage{epsfig}
\usepackage[usenames,dvipsnames]{xcolor}
\usepackage{bbm}
\usepackage{color}

\newcommand{\be}{\begin{equation}}
\newcommand{\ee}{\end{equation}}
\newcommand{\bea}{\begin{eqnarray}}
\newcommand{\eea}{\end{eqnarray}}

\newcommand{\bb}{\bibitem}
 
\newcommand{\eqn}{\begin{eqnarray}}
\newcommand{\eqnx}{\end{eqnarray}}

\numberwithin{equation}{section}

\begin{document}

\title{Oscillons in a perturbed signum-Gordon model}
\author{P. Klimas}
\author{J. S. Streibel}
\affiliation{Departamento de F\'isica, Universidade Federal de Santa Catarina, Campus Trindade, 88040-900, Florian\'o—polis-SC, Brazil}
\author{A. Wereszczynski}
\affiliation{Institute of Physics,  Jagiellonian University,
Lojasiewicza 11, Krak\'{o}w, Poland}
\author{W. J.  Zakrzewski}
\affiliation{Department of Mathematical Sciences, Durham University, Durham DH1 3LE, U.K.}

\begin{abstract}
We study various properties of a perturbed signum-Gordon model, which has been obtained through the dimensional reduction 
of the called `first BPS submodel of the Skyrme model'. This study is motivated by the observation that the first BPS submodel of the Skyrme model may
be partially responsible for the good qualities of the rational map ansatz approximation to the solutions of the Skyrme model.
We investigate the existence, stability and various properties of oscillons and other time-dependent states in this perturbed signum-Gordon model.
\end{abstract}

\maketitle 
\section{Introduction}

The Skyrme model \cite{skyrme} is a field theoretical framework offering an approach for a consistent and unified description of baryonic matter 
on all scales - from nucleons and atomic nuclei to neutron stars - where baryons are realized as topological solitons in a purely mesonic 
(pionic) theory. Recently two important steps have been made allowing us to treat the model as an efficient tool in the {\it quantitative} 
description of nuclear matter. First of all, some attemps have been made \cite{ASW-BPS}, \cite{mar}, \cite{Sut-BPS}, \cite{Leeds-BPS}, \cite{bjarke},
to improve the long standing issue of unphysically large binding energies of nucleons in nuclei as predicted by the standard Skyrme model.
 This has involved modifying the original Skyrme model by bringing it closer to a near BPS model. Secondly, an improvement of the 
standard rigid rotor quantization \cite{nappi}, known as the vibrational modes quantization, has led to a more complete description of the excitation 
spectra of light nuclei which has brought them towards a much better agreement with experimental data \cite{vib}. 

Although the structure of static Skyrmions is fairly well understood both numerically and analytically (especially in the minimal Skyrme model 
\cite{rat-map}, the loosely bound Skyrme models \cite{Leeds-BPS}, \cite{bjarke}, \cite{stefano} and the BPS Skyrme model \cite{ASW-BPS}), the time dependent
 solutions are much more difficult to obtain and to analyse. On the other hand, time-dependent configurations are needed in many 
physical applications of Skyrmions - especially in the (vibrational) quantization procedure. For example, small amplitude radial 
perturbations of static solitons contribute to some vibrational states, which have been interpreted as the Roper resonances \cite{roper-old}, 
\cite{bizon}, \cite{roper-BPS}, \cite{roper-new}. Furthermore, (iso)-rotations influence shapes of Skyrmions and and so can change their moments of 
inertia which in turn can modify the energies of quantum states \cite{iso-rot}. Finally, scattering of Skyrmions \cite{scatter} can be 
interpreted as scattering of atomic nuclei (heavy ions) and teach us some physics of fusion/fission processes.

In the present paper we study something, which at first sight may appear very different, namely,
the existence, stability and various properties of a perturbed signum-Gordon model in (1+1) dimensions. 
However, the model which we study is related to dimensionally reduced submodel of the Skyrme model, namely, the so-called
first BPS submodel of the Skyrme model \cite{new-BPS1}. And this submodel, in turn, has been claimed to partially explain the successes
 of the rational map approximation, or in other words, an almost rational map structure of Skyrmions in some Skyrme type models.
 Since, some geometric properties of static solutions of the full theory are encoded in the first BPS submodel one may ask whether other
 solutions of this submodel may also teach us something useful about the full Skyrme theory. In particular, one can look at non-trivial time-dependent solutions,
 especially oscillons, of the submodel.
 As this submodel is a BPS theory there is a chance that we may answer this question in a (partially)
 analytical manner. In fact, it has recently been reported that the first BPS submodel has the signum-Gordon breathers (on $\mathbb{R}_+$) 
as its approximate solutions \cite{new-BPS2}. The true solutions of this submodel get some corrections from a subleading term in the potential
 resulting in a very special deformation of the signum-Gordon breather.

 For these reasons, analysis of oscillons in such a modified signum-Gordon model is the main aim of this work. Of course, we hope that this work,
in addition to being an important study of oscillons by themselves, is also the first step in gaining some (analytical and numerical) understanding 
of the time dependent solutions of the Skyrme model, in its topologically trivial sector, which physically is related to the pionic sector,
 and which therefore, has some influence on all possible interactions between Skyrmions. 

Independently of all of this and looking from a wider field theoretic perspective, we want to gain a better understanding of transitions from a breather ({\it i.e.}, an infinitely long-lived time periodic solution of a model) to an oscillon, once a small perturbation term has been added to the signum-Gordon Lagrangian. 
Thus, we begin our analysis by considering the signum-Gordon model as well as its modifications, on the full infinite line $\mathbb{R}$
and only later restrict it to a half infinite line $\mathbb{R}_+$, relevant in the Skyrme considerations in which $\mathbb{R}_+$ corresponds to the radial
distance from the centre of a given Skyrmion.

\vspace*{0.2cm}

The paper is organized as follows. In Sec. II we present the first BPS submodel of the Skyrme model and discuss its relation to a modified signum-Gordon
 model on the semi-real line $\mathbb{R}_+$. In Sec. III we recall some known results on breathers as well as on self-similar 
solutions of the signum-Gordon model on the real line $\mathbb{R}$. In addition, in this section we also discuss perturbed breathers. In Sec. IV we describe
 modifications of the previous discussion when the signum-Gordon model is defined on a semi-infinite line $\mathbb{R}_+$. Sec. V is devoted to 
the investigations of oscillons in the modified signum-Gordon model on $\mathbb{R}$. In this section we also present an analytical construction 
of approximate oscillons in this model and we compare them with the numerical solutions of the model.
 In Sec. VI we present similar results for the modified signum-Gordon model on $\mathbb{R}_+$ and we finish this paper with a section describing our 
conclusions and indicating our plans for the future work.

\section{The first BPS submodel of the Skyrme model}
The first BPS submodel of the Skyrme model \cite{new-BPS1} is defined by the following Lagrangian density:
\be
\mathcal{L}_{24}^{(1)} \;\; = \; \; 4\sin^2\xi \frac{u_\mu \bar{u}^\mu}{(1+|u|^2)^2}  - 
 4 \sin^2 \xi \left( \xi_\mu \xi^\mu \frac{u_\mu \bar{u}^\mu}{(1+|u|^2)^2} -\frac{\xi_\mu \bar{u}^\mu \;\xi_\mu u^\nu}{(1+|u|^2)^2} \right), \label{BPS-1}
\ee
where $\xi$ and $u$ are, respectively,  real and complex fields and $u_{\mu}\equiv\partial_{\mu}u$, $\xi_{\mu}\equiv\partial_{\mu}\xi$. These field variables parametrize an
 $SU(2)$  matrix field $U=e^{i\xi \vec{\tau}\cdot \vec{n}}$, which is the commonly used field of the Skyrme model. 
The unit vector field $\vec{n}$ is related to the complex field $u$ by  the standard stereographic projection. Finally, the components of
 vector $\vec{\tau}$ are the Pauli matrices. The importance of the submodel stems from the observation that this model, together with the
 so-called `second BPS submodel of the Skyrme submodel', combine into the massless Skyrme model \cite{skyrme}. Both submodels
 are separately BPS theories, {\it i.e.}  their relevant topological bounds are saturated for solutions of  the corresponding Bogomolny equations.
 Furthermore, the solutions of (\ref{BPS-1}) are of the form of rational maps which, as it was mentioned in \cite{new-BPS2}, suggests
 a possible explanation of the success of the rational map ansatz (RMA) approximation to the solutions of the full Skyrme model. 

In fact, the first BPS submodel (\ref{BPS-1}) possesses the following BPS solutions 
\be
\xi=\xi(r)=\left\{
\begin{array}{ll}
\pi-r & r \leq \pi, \\
0 & r \geq \pi
\end{array}
\right. \label{sol xi}
\ee
and 
\be
u=u(z)=\frac{p(z)}{q(z)},
\ee
where $z=\tan \frac{\theta}{2} e^{i\varphi}$ is the stereographic coordinate on the unit sphere parametrized by the standard spherical
 angles $\theta, \varphi$ and $r$ is the radial coordinate. Here $p,q$ are polynomials of a (maximal) finite degree $B \in \mathbb{Z}$
 (no common divisors). In other words, $u$ is {\it an arbitrary} rational map of a given degree.
 The baryon charge of the corresponding BPS Skyrmion is inherited from the degree of this underlying rational
 map provided that the profile function $\xi$ obeys the boundary conditions:
\be
\xi(0)=\pi, \;\;\; \xi(R)=0,
\ee 
where $R$ is the geometric size of Skyrmion and can be infinite for usual infinitely extended solitons or finite for compactons. For the static solution (\ref{sol xi}) we have $R=\pi$. So, as discussed in (\ref{BPS-1}) the size of Skyrmion of the first BPS submodel is fixed at $\pi$.

The large arbitrariness of the angular part of such a Skyrmion, related to the arbitrariness of the rational map, follows from 
an observation that the static energy functional of the model (\ref{BPS-1}) can be written as a product (if the spherical ansatz is assumed, {\it i.e.} $\xi=\xi(r), u=u(z)$) of two terms, corresponding, respectively, to the static energy function of the $CP^1$ model, $E_u$, on the unit two sphere described by the spherical 
angles and the radial term $E_\xi$ given by
\be
E_{24}^{(1)}=2 E_\xi E_u, 
\ee
\be
 E_u =\int d\Omega_{\mathbb{S}^2} \frac{(1+z\bar{z})^2}{(1+u\bar{u})^2} (u_z\bar{u}_{\bar{z}}+u_{\bar{z}}\bar{u}_z),
 \;\;\;\; E_\xi=\int dr \sin^2 \xi (1+\xi^2_r).
\ee

\vspace*{0.2cm}

It is quite remarkable that the same product form decomposition of the first BPS submodel Lagrangian also holds if the profile function $\xi$ is additionally a function of time; {\it i.e.}
 $\xi=\xi(r,t)$.
 The angular part is still described by a rational map, exactly as in the static case, and we arrive at 
the following reduced Lagrangian describing the time dependence of the profile function $\xi(r,t)$ (effectively a field theory in (1+1) dimensions):
\be
L^{(1)}=4\pi \int dr  \left[ 2B \sin^2 \xi (\xi_\mu^2-1) \right] = 16\pi B \int dr  \left[\frac{1}{2}  \bar{\eta}_\mu^2 -\bar{\eta} +\frac{1}{2}\bar{\eta}^2\right],\label{radialMSG}
\ee
where $\mu=(0,1)$ and $x^0=t, x^1=r$ and where we have introduced a new target space coordinate $\bar{\eta}=1-\cos \xi$.
Looking at $L^{(1)}$ we note that the
 restriction of the target space variable, $\bar{\eta}$ to $\bar{\eta} \in [0,2]$, can be incorporated into the form of
 the potential 
for $L^{(1)}$ to have infinite barriers 
 at $\bar{\eta}=0$ and $\bar{\eta}=2$, namely:
\begin{eqnarray}
V(\bar{\eta})=
\left\{\begin{array}{ccl}
\infty&{\rm for}&\bar{\eta}<0,\\
\bar\eta-\frac{1}{2}\bar{\eta}^2&{\rm for}&0\le \bar{\eta}\le 2,\\
\infty&{\rm for}&\bar{\eta}>2.
\end{array}\right.\label{Vb}
\end{eqnarray}
The  potential \eqref{Vb} has two minima at $\bar{\eta}=0$ and $\bar{\eta}=2$ and it belongs to the class of so-called V-shaped potentials.
 
A formally identical Lagrangian (although in a different physical context) was studied in \cite{arodzklimas}. 
Following the ideas of \cite{arodzklimas}, we introduce a new scalar field $\eta$ and extend its target space to the 
full $\mathbb{R}$ and make the potential \eqref{Vb} (now  considered as a function of $\eta$) periodic. 
 \begin{figure}[h!]
\centering
\subfigure{\includegraphics[width=0.55\textwidth,height=0.2\textwidth, angle =0]{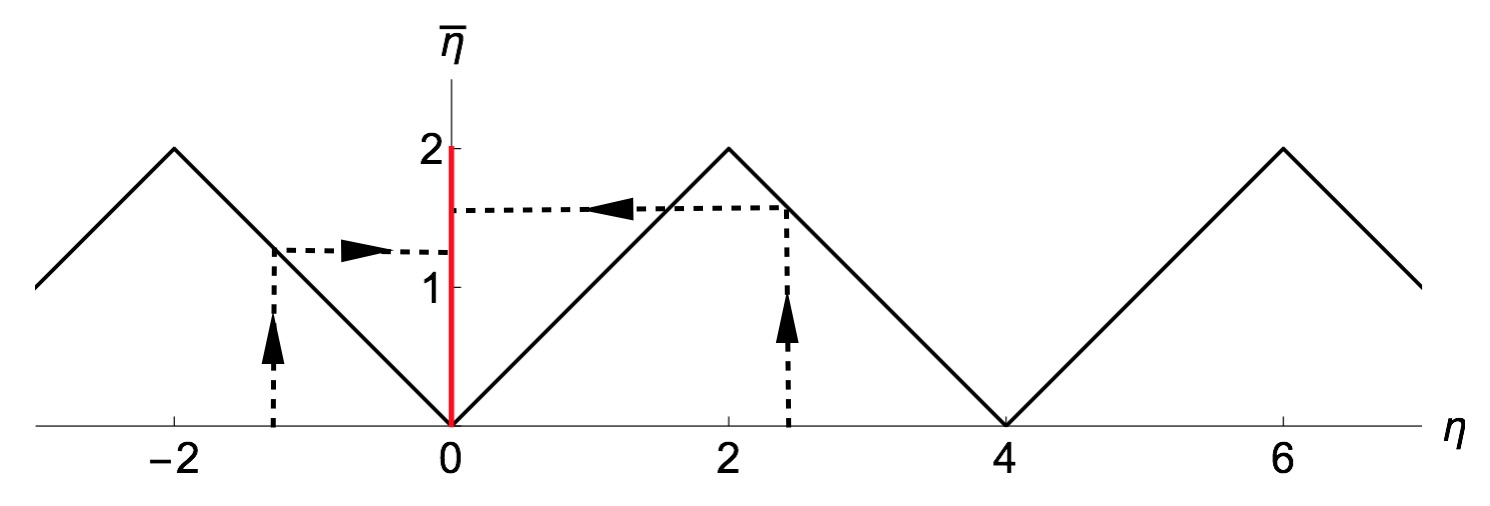}}
\caption{Folding transformation ${\mathbb R}\ni \eta\mapsto\bar \eta\in[0,2]$.}\label{folding}
\end{figure}
The change of the target space field ${\mathbb R}\ni \eta\mapsto\bar \eta\in[0,2]$, called the folding transformation,
 can then be cast in the form
\be
\bar\eta=\sum_{n=-\infty}^{\infty}|\eta-4n|H_n(\eta),\label{bareta}
\ee
where we have introduced a double step function
\be
H_n(\eta):=\theta(\eta-4n+2)-\theta(\eta-4n-2)
\ee
defined as a combination of the Heaviside's step functions such that $H_n(\eta)=1$ for $\eta\in(-2+2n,2+2n)$ and $H_n(\eta)=0$ outside this 
segment. The folding transformation is plotted in Fig.\ref{folding}.

\begin{figure}[h!]
\centering
\subfigure{\includegraphics[width=0.65\textwidth,height=0.17\textwidth, angle =0]{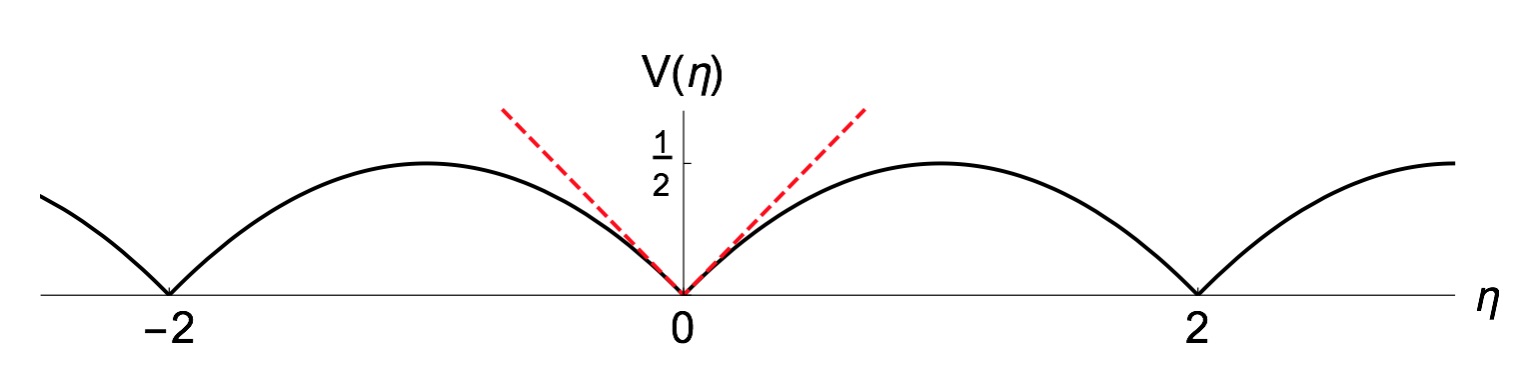}}
\caption{Periodic extension of the potential \eqref{Vb}. For $|\eta|\ll 1$ the potential behaves as $V(\eta)\approx|\eta|$.}\label{unfold}
\end{figure}

The periodic potential is defined, as a function of $\eta$ {\it i.e.} by
\be
V(\eta):= \bar \eta -\frac{1}{2}\bar \eta^2,\label{Vunfold}
 \ee
where $\bar \eta$ is given by \eqref{bareta}. Alternatively, one can take the periodic potential in the form
\be
V(\eta):=\sum_{n=-\infty}^{\infty}\left(|\eta-4n|-\frac{1}{2}(\eta-4n)^2\right)H_n(\eta).\label{Vunfold2}
\ee
The potential \eqref{Vunfold2} is sketched in Fig.\ref{unfold}. 

The folding procedure is very useful because it allows us to avoid the inconvenience of having to deal with infinite barriers.
 In such an approach, the folding transformation maps the evolution of the auxiliary system onto the evolution of the original model.

\vspace{0.2cm}

In the next sections we will analyze the time-dependent solutions of the resultant model, effectively, the (1+1) dimensional
 model
\be
L^{(1)}=16\pi B \int dr  \left(  \frac{1}{2} \eta_\mu^2 -V(\eta) \right),
\label{model}
\ee
where $V(\eta)$ is given by \eqref{Vunfold}. This Lagrangian, for reasons that we will explain below, we call the modified 
signum-Gordon model. 

Some further comments should bem added here.  First of all we would like to stress that the resulting Lagrangian 
 ($L^{(1)}$) describes 
effectively  a (1+1) dimensional model with the spatial variable being the radial coordinate $r$. Although 
the problem looks like an usual scalar field theory in (1+1) dimension with a certain potential, one has to remember that the solutions
 of this model describe objects living in (3+1) dimensional space-time with the angular part defined by an appropriate rational map.
 Thus, we are dealing with the time-dependent configurations in three spatial dimensions.

Secondly, the reduced model (\ref{model}) is a theory on a semi-infinite line as $r \in \mathbb{R}_+$. However,
 we will begin our analysis by replacing the radial coordinate by $x \in \mathbb{R}$. As we will see, due to the compact nature
 of the solutions, many features are independent of this replacement. On the other hand, as we will discuss later, some
 properties hold only for the model defined on $\mathbb{R}_+$.

Thirdly, the model (\ref{model}) is a generalisation (specific perturbation) of the (1+1) dimensional signum-Gordon model
\be
L_{sG}= \int
 dx \left(  \frac{1}{2} \phi_{\mu}^2 -|\phi| \right).\label{SG}
\ee 
Indeed, in the small amplitude limit $|\eta(t,r)|  \ll 1$ this model (\ref{model}) simplifies to 
 the signum-Gordon model on $\mathbb{R}_+$
\be
L^{(1)}\approx 16\pi B \int dr  \left(  \frac{1}{2} \eta_\mu^2 - \left| \eta \right| \right).\label{radialSG}
\ee

This shows that small amplitude solutions of (\ref{model}), relevant for the zero-charge, {\it i.e.} non-topological sector
 of the first BPS submodel of the Skyrme theory, are approximately given by exact solutions of the signum-Gordon model
 on $\mathbb{R}_+$. Thus, we begin our discussion here by recalling the analytical time dependent solutions of the
 signum-Gordon \cite{arodz1}, \cite{arodz2}.

\section{Time dependent solutions of the signum-Gordon model}
\subsection{The signum-Gordon breather}\label{exbreather}
The original signum-Gordon model \cite{arodz1} is a field theory defined by the following Lagrangian density (here we 
use the notation $x^1=x\in{\mathbb R}$ as this model was considered in the usual (1+1) dimensional space-time)
\be
\mathcal{L}_{SG}=\frac{1}{2}(\partial_t\phi)^2-\frac{1}{2}(\partial_x\phi)^2-|\phi|,\label{mod1}
\ee 
where $\phi(t,x)$ is a real scalar field in (1+1) dimensions. Its classical solution field $\phi$ obeys the second order
 differential equation:
\be
\partial^2_t\phi-\partial^2_x\phi+{\rm sgn}(\phi)=0.
\ee
The model \eqref{mod1} possesses also a periodic compact solution (called `breather') which satisfies the 
conditions
\be
\phi(0,x)=0,\qquad\partial_t\phi(0,x)=v(x).\label{initial}
\ee
The function $v(x)$ is determined from the boundary conditions:
\be
\phi(t,0)=0=\phi(t,1),\qquad \partial_x\phi(t,0)=0=\partial_x\phi(t,1),\label{bc}
\ee
which lead to the following expression
\begin{figure}[h!]
\centering
\includegraphics[width=0.5\textwidth,height=0.5\textwidth, angle =0]{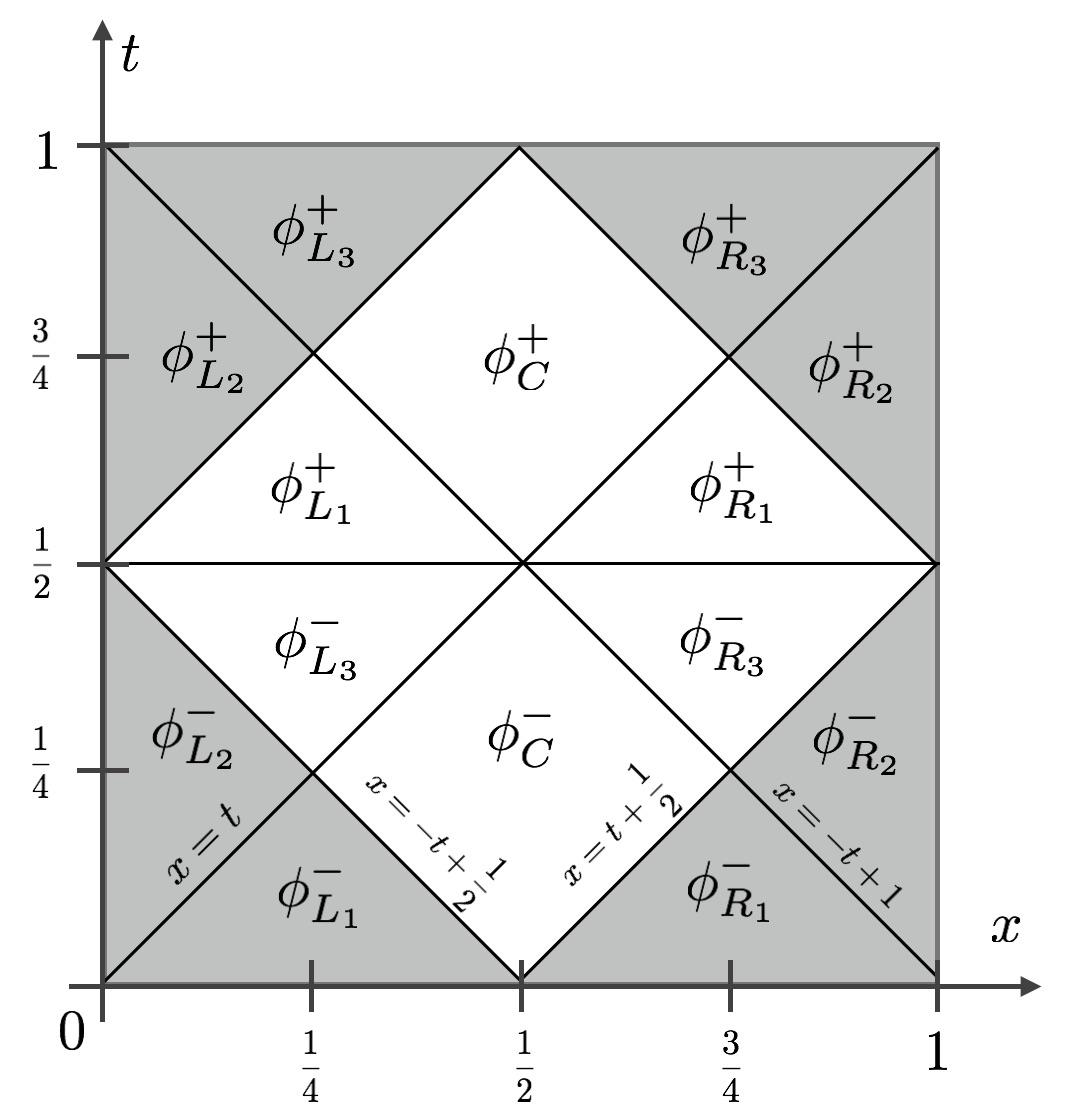}
\caption{Partial solutions that, together, describe the exact breather of the signum-Gordon model.}\label{fig1}
\end{figure}
\begin{eqnarray}
v(x)=\left\{\begin{array}{ccl}
0&{\rm for}&x\le 0,\\
|x-{\textstyle\frac{1}{2}}|-{\textstyle\frac{1}{2}}&{\rm for}&0\le x\le 1,\\
0&{\rm for}&x\ge 1.
\end{array}\right.\label{rightvelocity}
\end{eqnarray}

 This resulting exact breather, {\it i.e.} indefinitely long living oscillon which does not emit any radiation, can be
 constructed as follows: (see different regions in Fig. \ref{fig1}). (We prefer to call this solution as breather rather than oscillon. Note 
that here we differ from the original terminology used in \cite{arodz1}.) The analytical form of this solution consists of 
the following partial solutions:
\begin{align}
\phi^-_{L_1}(t,x)&=\frac{t^2}{2}-tx,\label{ps1}
\\
\phi^-_{L_2}(t,x)&=-\frac{x^2}{2},\label{ps2}
\\
\phi^-_{L_3}(t,x)&=\frac{t^2}{2}+tx-\frac{t}{2}-\frac{x}{2}+\frac{1}{8},\label{ps3}
\\
\phi^-_{C}(t,x)&=t^2+\frac{x^2}{2}-\frac{x}{2}-\frac{t}{2}+\frac{1}{8}\label{ps4}
\end{align}
and the further solutions that can be obtained from \eqref{ps1}-\eqref{ps4} by certain transformations.
 Here $L$ stands for left-hand-side partial solutions and $C$ for central partial solutions. The right-hand-side partial 
solutions are given in terms of the left-hand-side partial solutions by
\be
\phi^-_{R_k}(t,x)=\phi^-_{L_k}(t,1-x)\quad{\rm where}\quad k=1,2,3.\label{sym1}
\ee
Similarly, further partial solutions valid for $t\in [\frac{1}{2},1]$ can be obtained from the ones mentioned above
through the transformations
\be
\phi^+_{\alpha}(t,x)=-\phi^-_\alpha(t-{\textstyle \frac{1}{2}},x),\quad {\rm where}\quad \alpha=\{L_k,R_k\}.\label{sym2}
\ee

The solutions $\phi_{L_1}$ and $\phi_{R_1}$ can be determined  directly from the initial conditions. They are given by the formula 
\be
\phi_{L_1/R_1}(t,x)=\frac{t^2}{2}+\frac{1}{2}\int_{x-t}^{x+t} dsv(s),
\ee
 where $v(x)=-x$ for $L_1$ and $v(x)=1-x$ for $R_1$. The remaining $^-$  solutions have the general form 
\be
\phi(t,x)=\frac{t^2}{2}+F(x+t)+G(x-t),\label{gensol}
\ee
 where the functions $F(x+t)$ and $G(x-t)$ are obtained from the matching conditions at the surfaces of the light cones.  For instance, the solution $\phi_C^-(t,x)$ must satisfy the conditions 
\be
\phi_C^-(t,-t+{\textstyle\frac{1}{2}})=\phi_{L_1}^-(t,-t+{\textstyle\frac{1}{2}}),
\ee
and
\be
\phi_C^-(t,t+{\textstyle\frac{1}{2}})=\phi_{R_1}^-(t,t+{\textstyle\frac{1}{2}}).
\ee
Similarly, we require that 
\be
\phi^-_{L_2}(t,0)=0\qquad{\rm and}\qquad  \phi^-_{L_2}(t,t)=\phi_{L_1}^-(t,t).
\ee

The solution $\phi^-_{L_3}(t,x)$ is obtained after imposing the conditions 
\be
\phi^-_{L_3}(t,-t+{\textstyle\frac{1}{2}})=\phi^-_{L_2}(t,-t+{\textstyle\frac{1}{2}})
\ee
and
\be
\phi^-_{L_3}(t,t)=\phi^-_{C}(t,t).
\ee
All the remaining partial solutions can be obtained by applying transformations \eqref{sym1} and \eqref{sym2}.

Let us note that this breather solution is not unique. One can use a symmetry of the signum-Gordon equation and construct an infinitely 
large family of breathers parametrised by a scale parameter $l$
\be
\phi_l = l^2 \phi \left( \frac{t}{l}, \frac{x}{l}\right). \label{scale}
\ee

It can be easily checked that the amplitude of the $\phi_1$ solution is $l^2/16$ while the period and the 
size (support) of the breather are given, respectively, by:
\be
T=l, \;\;\; R=l.
\ee
Finally, their energy is
\be
E=\frac{1}{24} l^3.\label{breatherenergy}
\ee

It is interesting to note that as one decreases $l$, ({\it i.e.} considers small amplitude
 Breathers), they oscillate faster but carry smaller values of energy. Thus such breathers probably dominate the
 interaction and radiation regimes and so play a crucial role in kink anti-kink annihilation processes.
In this behaviour the breather solutions  of the signum-Gordon model resemble breathers of the well known sine-Gordon model.

\subsection{Self-similar components of oscillon solutions in the signum-Gordon model}

\subsubsection{The breather case}

According to the previous section, the basic signum-Gordon breather, $l=1$,  exists on the support
 $x\in [0,1]$. Its strictly fixed size is a direct consequence of a very special form of the partial 
solutions $\phi_{L_2/R_2}$ that constitute borders of the breather, {\it i.e.} they  match the vacuum 
solution $\phi=0$ at $x=0$ and $x=1$. Let us discuss some properties of this solution and its origin.

 First of all, we note that the static character of the partial solutions,
which together generate the breather solution,  permits them to preserve the time independence of
 the points where the breathers  match the vacuum solution. Secondly, solutions $\phi_{L_2/R_2}(t,x)$ together 
with solutions $\phi_{L_1/R_1}(t,x)$ belong to the class of {\it self-similar} solutions of the signum-Gordon model.
 This implies that such solutions can be obtained from the self-similar initial data which are determined in terms of two constant parameters. This family of self-similar solutions of the signum-Gordon model was studied in \cite{arodz3}. In this section we will only discuss those of them that have a direct relation to the breathers.

The existence of self-similar solutions of the signum -Gordon equations is closely related to the scaling symmetry (\ref{scale}). All such solutions have the following form
\be
\phi(t,x)=x^2 S\left( \frac{t}{x} \right),\label{SSansatz}
\ee
where $S(y)$ is a scalar function of $y=t/x$. The use of the  ansatz \eqref{SSansatz} describing their form allows us to reduce the signum-Gordon equation to an ordinary differential equation
\be
(1-y^2)S''+2yS'-2S+{\rm sgn}(S)=0.\label{eqS}
\ee
The self-similar initial data have the form
\begin{eqnarray}
\phi(0,x)=
\left\{\begin{array}{ccc}
R_0x^2&{\rm dla}&x\le0\\
S_0x^2&{\rm dla}&x\ge0
\end{array}\right.,\qquad
\partial_t\phi(t,x)|_{t=0}=
\left\{\begin{array}{ccc}
\dot R_0x&{\rm dla}&x\le0\\
\dot S_0x&{\rm dla}&x\ge0
\end{array}\right..
\end{eqnarray}

Such initial data, with constant $R_0$, $\dot R_0$ and $S_0$, $\dot S_0$ chosen independently, are self-similar because the point $x=0$ remains unchanged under the rescaling transformation $x\rightarrow x/\lambda$.

 In what follows we will restrict our considerations to the case $R_0=0$ and $\dot R_0=0$. Constants $S_0$ and $\dot S_0$ determine the value of the function $S(y)$ and its first derivative at $y=0$. Namely, $S_0=S(0)$ and $\dot{S}_0=\partial_y S |_{y=0}$. 

Note that the scale invariance of the self-similar solutions leads to the   potential infiniteness of their total energy. This may seem to be bad but, in fact, it does not make these solutions unphysical. In our case, only fragments of such solutions (partial solutions restricted to a finite support) appear in physical configurations and their  total energy is finite. This is exactly what happens for the breather in the signum-Gordon model.

Solutions of \eqref{eqS} have the general form
\begin{equation}
S_k(y)=\frac{(-1)^k}{2}[\beta_k(y^2+1)-\alpha_ky-1],\label{Sk}
\end{equation}
where coefficients $\alpha_k$ and $\beta_k$ are constant parameters which must be determined by their dependence on $S_0$, $\dot S_0$ and by matching conditions at points $a_k$ such that $S(a_k)=0$. Since the initial data possess the symmetry $x\rightarrow 1-x$, then it is enough to consider its partial solutions in the segment $x\in[0,\frac{1}{2}]$.  Initial conditions \eqref{initial} and \eqref{rightvelocity} imply that  $S_0=0$ and $\dot{S}_0=-1$. This leads to the partial solutions  \eqref{ps1} and \eqref{ps2}. These partial solutions can be cast in the form that makes their relation to the function $S(y)$ explicit:
\begin{align}
\phi^-_{L_1}(t,x)&=x^2\left[\frac{1}{2}\left(\frac{t}{x}\right)^2-\frac{t}{x}\right],\\
\phi^-_{L_2}(t,x)&=x^2\left[-\frac{1}{2}\right].
\end{align}
Shaded regions in Fig.  \ref{fig1} represent all partial solutions of the 
breather which originate in self-similar solutions or can be obtained from such solutions
 by the transformations $\phi(t,x)\rightarrow \phi(t,1-x)$ and $\phi(t,x)\rightarrow -\phi(t-\frac{1}{2},x)$. 
Thus, the self-similar solutions govern the behaviour of the breathers close to the boundary.

\subsubsection{Perturbed breathers}

In this section we look at the slightly modified initial conditions, namely 
\be
\phi(0,x)=0,\qquad \partial_t\phi(t,x)|_{t=0}=\epsilon v(x),
\ee   
where $v(x)$ is given by \eqref{rightvelocity} and $\epsilon$ is a free parameter, $\epsilon>0$.
 Such initial conditions, when the discussion is restricted to the segment $x\le \frac{1}{2}$ still correspond to the self-similar initial data, $S_0=0$ and $\dot S_0=-\epsilon$. Initial data on the remaining sector $x\ge \frac{1}{2}$ 
are given by the transformation $x\rightarrow 1-x$ of the initial data for $x\le \frac{1}{2}$.
 According to \cite{arodz3}, the self-similar solutions determined by $S_0=0$ and $\dot S_0\neq 1$ are
 not static. This implies that in the very initial phase of its evolution the support of the oscillon is 
expected to expand or shrink.  The value of  $\epsilon$ determines the class of self-similar solutions
 that appear shortly after the initial instant. Such initial conditions are not suitable (except for the $\epsilon=1$ case)
 for the studies of constant support breathers. Physically this implies that we have the unit size breather with a perturbation on
 top of it. Thus, our investigations of such initial conditions make a contribution to the stability analysis of exact
 signum-Gordon breathers. For that we have to distinguish few qualitatively different cases. 

\begin{enumerate}
\item Case $\epsilon \ge \frac{1}{2}$.\newline
The solution is given by a piece of a self-similar solution, restricted to the region $x\le \frac{1}{2}-t$ 
\begin{eqnarray}
\phi_L(t,x)=
\left\{\begin{array}{ccl}
0&{\rm for}&x\le v_0t,\\
x^2\left[-\frac{1}{2(1-v_0^2)}\left(1-v_0\frac{t}{x}\right)^2\right]&{\rm for}&v_0t\le x\le t,\\
x^2\left[\frac{1}{2}\frac{t}{x}\left(\frac{t}{x}-2\epsilon \right)\right]&{\rm for}&t\le x\le \frac{1}{2}-t,
\end{array}\right.\label{solfinite}
\end{eqnarray}
where $v_0=\frac{1}{\epsilon}-1$ is the velocity  of the point at which a nontrivial partial solution and
 the vacuum solution $\phi=0$ are sewn together. Let us note that there are further subcases within the class of solutions
 described by \eqref{solfinite}. 

\begin{enumerate}
\item For $\epsilon>1$, the velocity $v_0$ is negative, $v_0<0,$ and it has the limit $v_0=-1$
 as $\varepsilon\rightarrow \infty$.

\begin{figure*}[h!]
\centering
\subfigure[]{\includegraphics[width=0.3\textwidth,height=0.15\textwidth, angle =0]{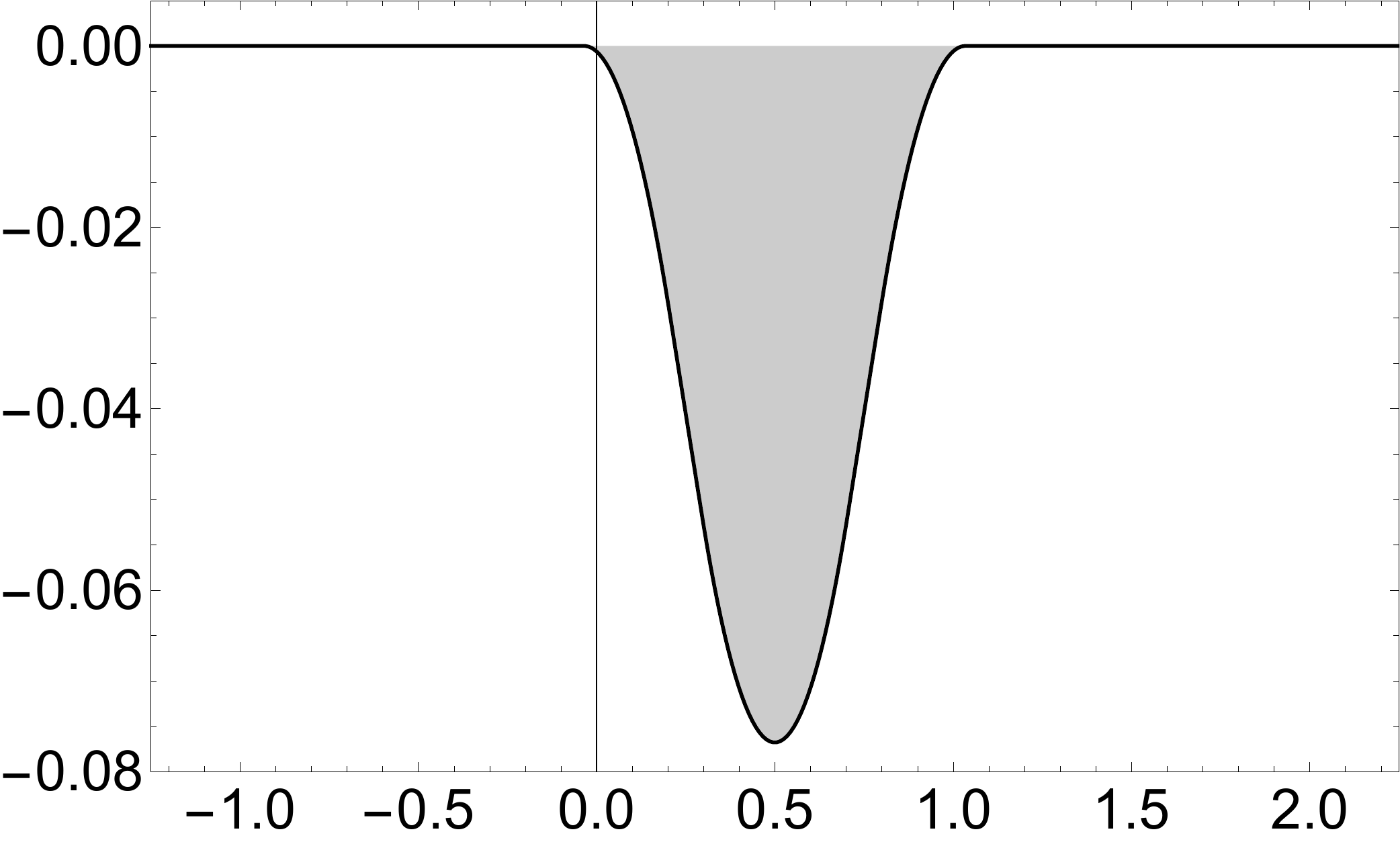}}
\subfigure[]{\includegraphics[width=0.3\textwidth,height=0.15\textwidth, angle =0]{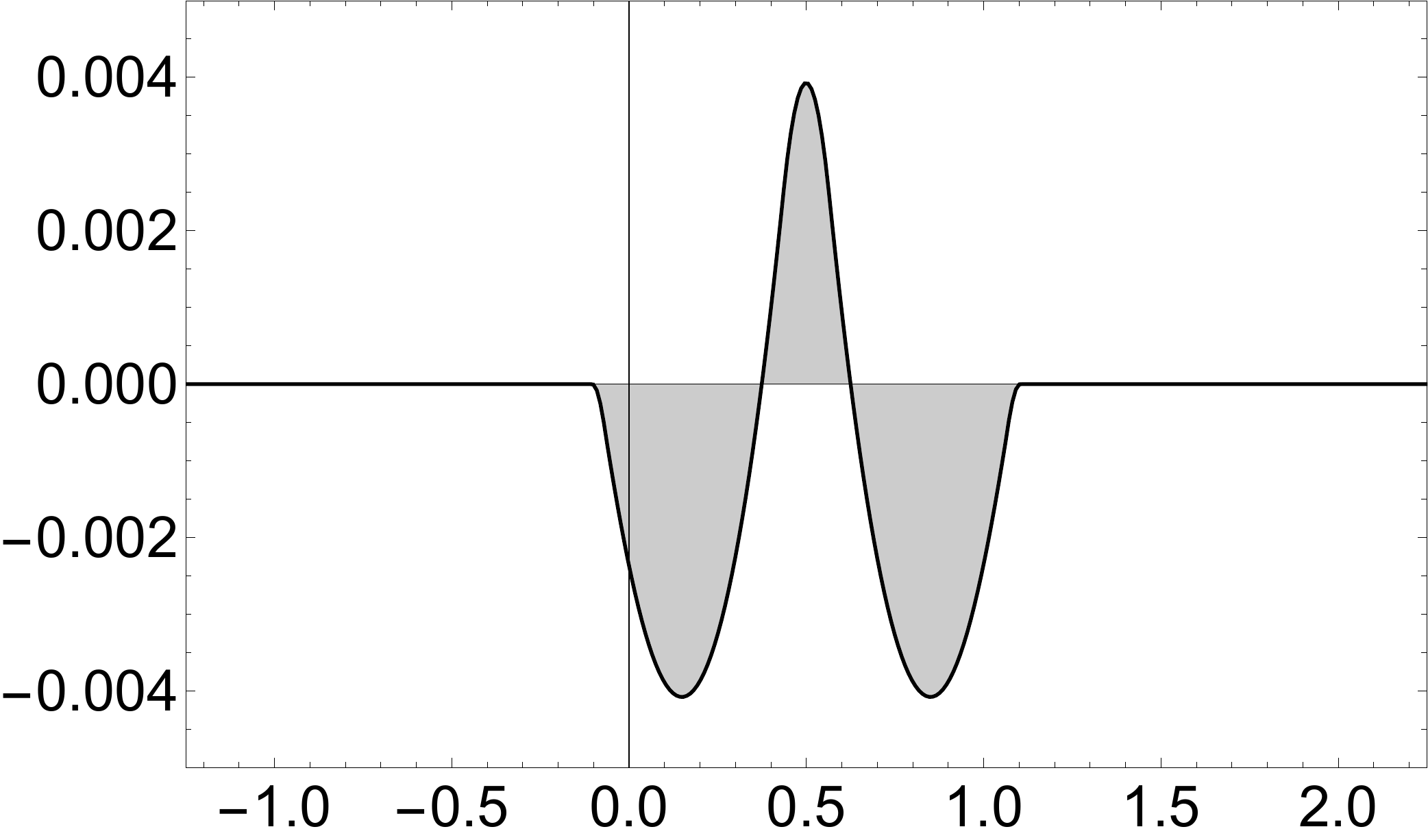}}
\subfigure[]{\includegraphics[width=0.3\textwidth,height=0.15\textwidth, angle =0]{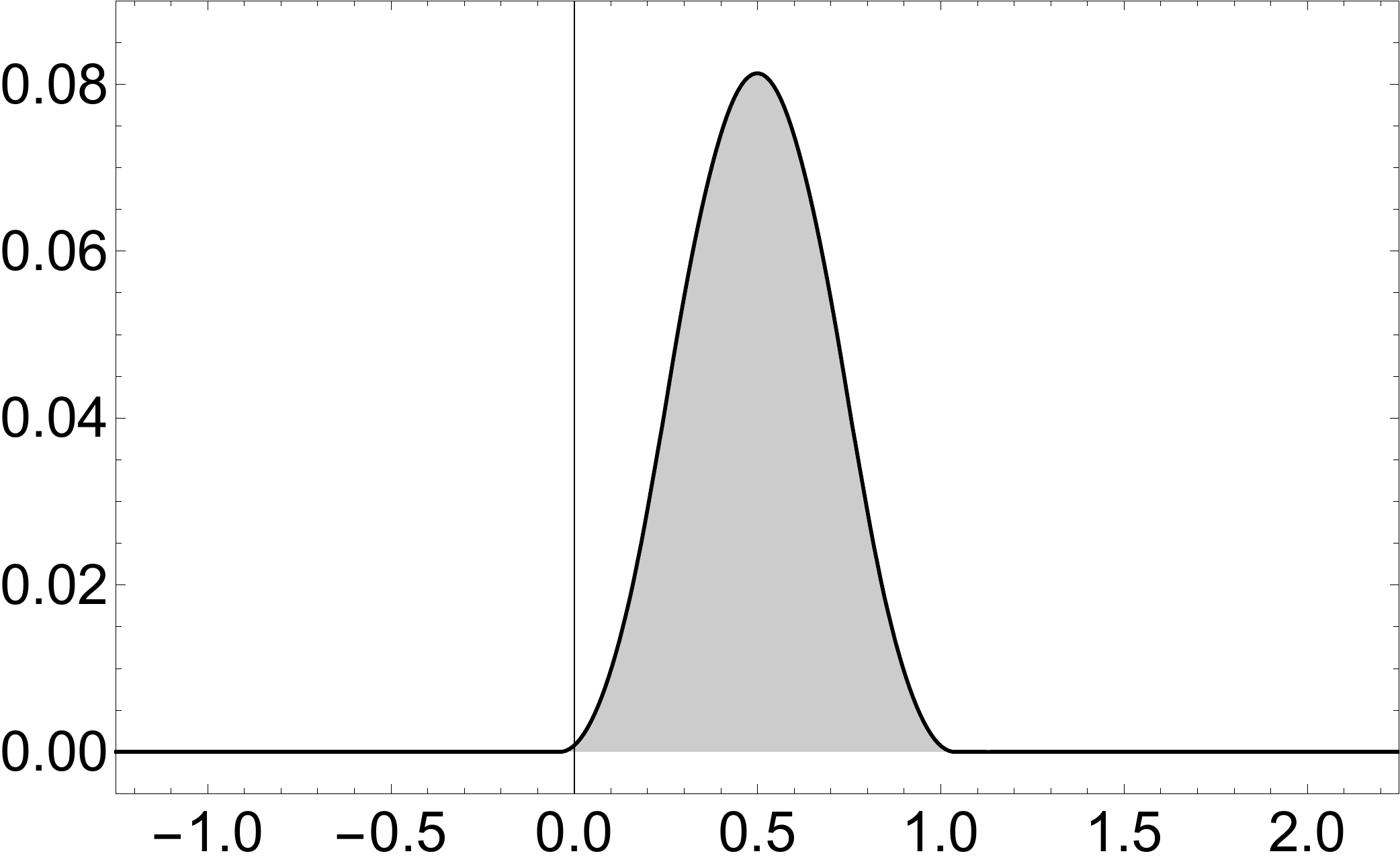}}
\subfigure[]{\includegraphics[width=0.3\textwidth,height=0.15\textwidth, angle =0]{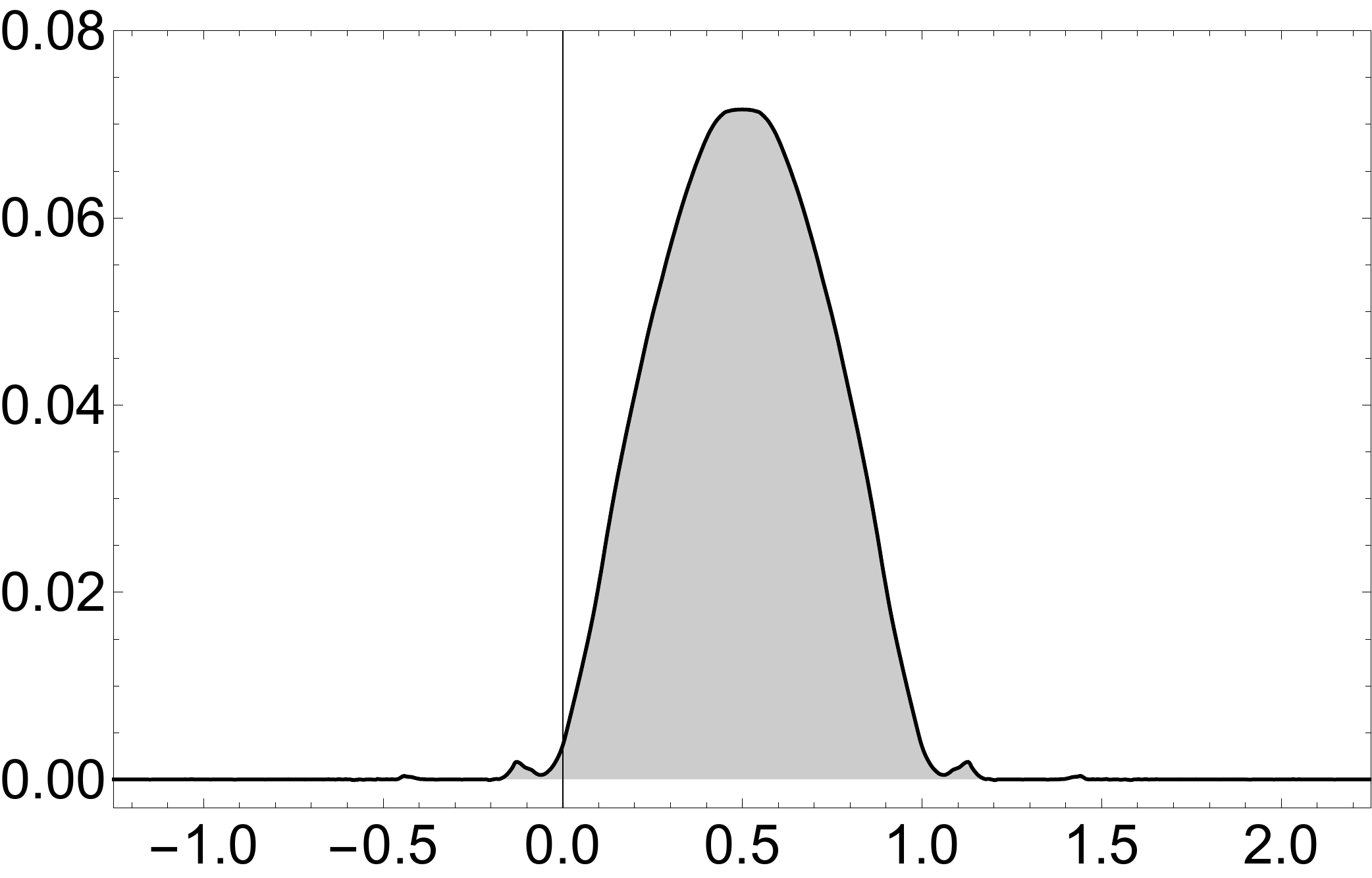}}
\subfigure[]{\includegraphics[width=0.3\textwidth,height=0.15\textwidth, angle =0]{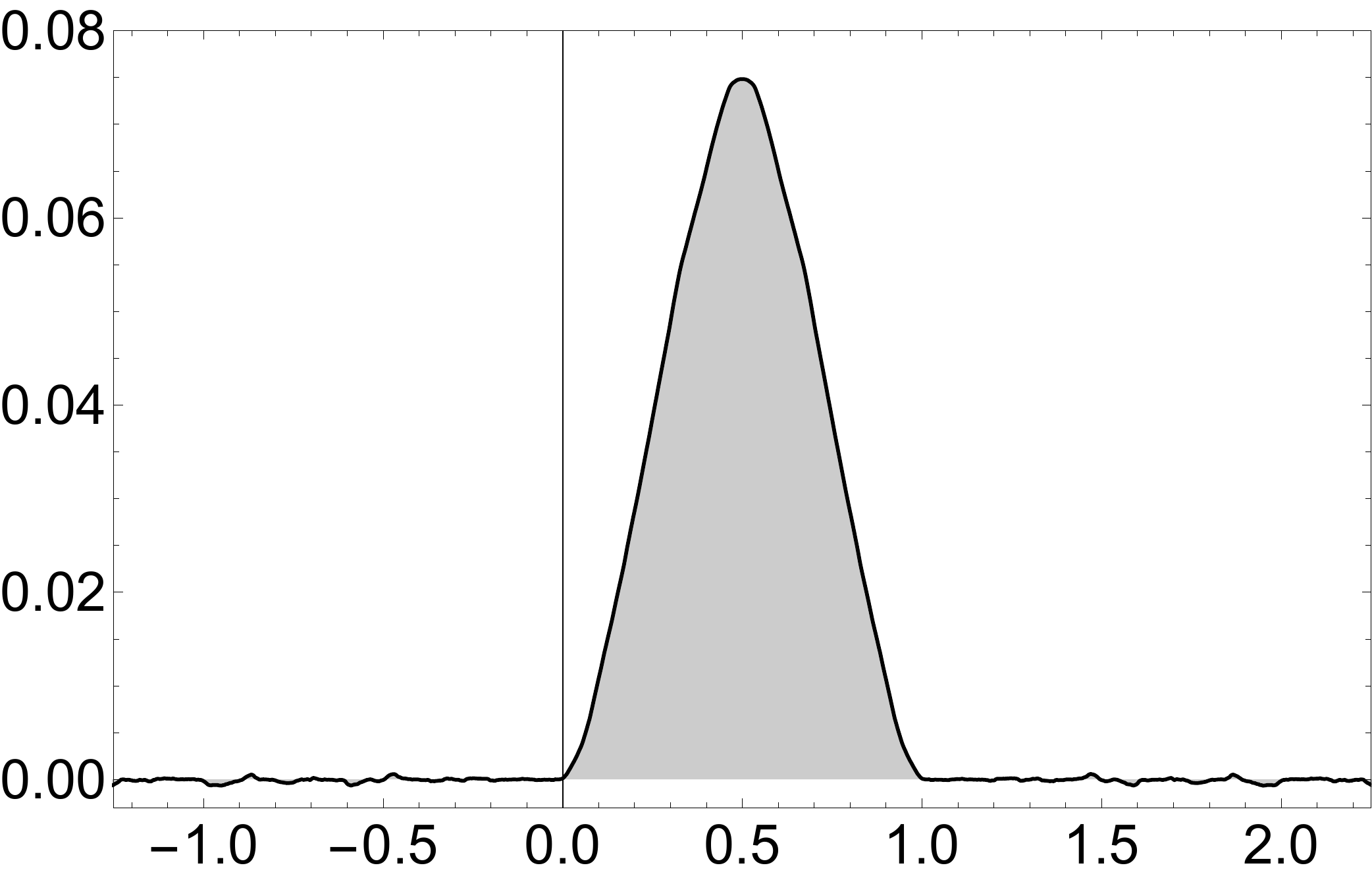}}
\caption{The signum-Gordon field for $\epsilon=1.2$ at (a) $t=0.2052$, (b) $t=1.1350$, (c) $t=5.4025$, (d) $t=5.6383$, (e) $t=13.5579$, (f) $t=27.5035$.}\label{SG-e-1.2}
\end{figure*}
\begin{figure}[h!]
\centering 
\subfigure[]{\includegraphics[width=0.4\textwidth,height=0.15\textwidth, angle =0]{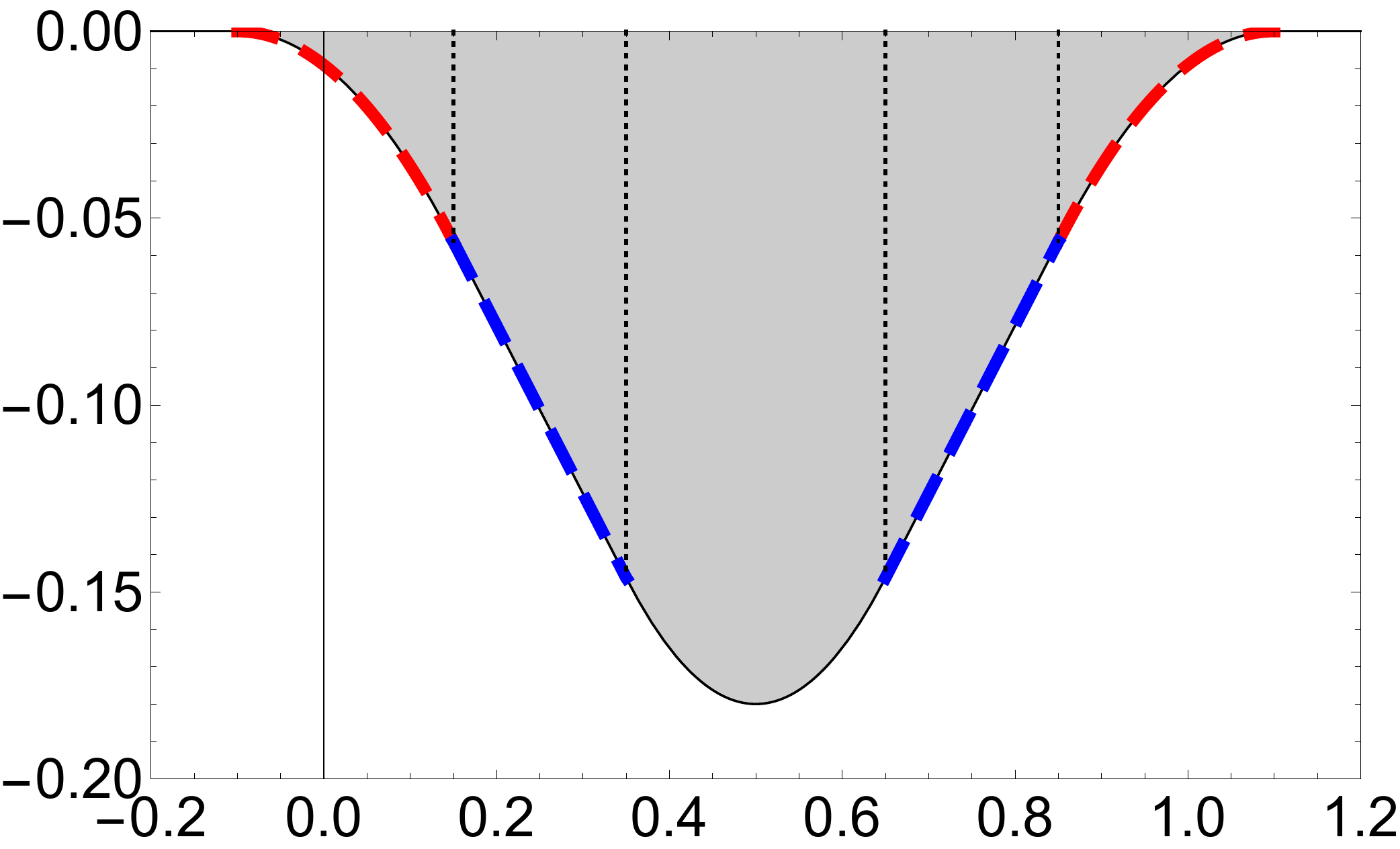}}
\subfigure[]{\includegraphics[width=0.4\textwidth,height=0.15\textwidth, angle =0]{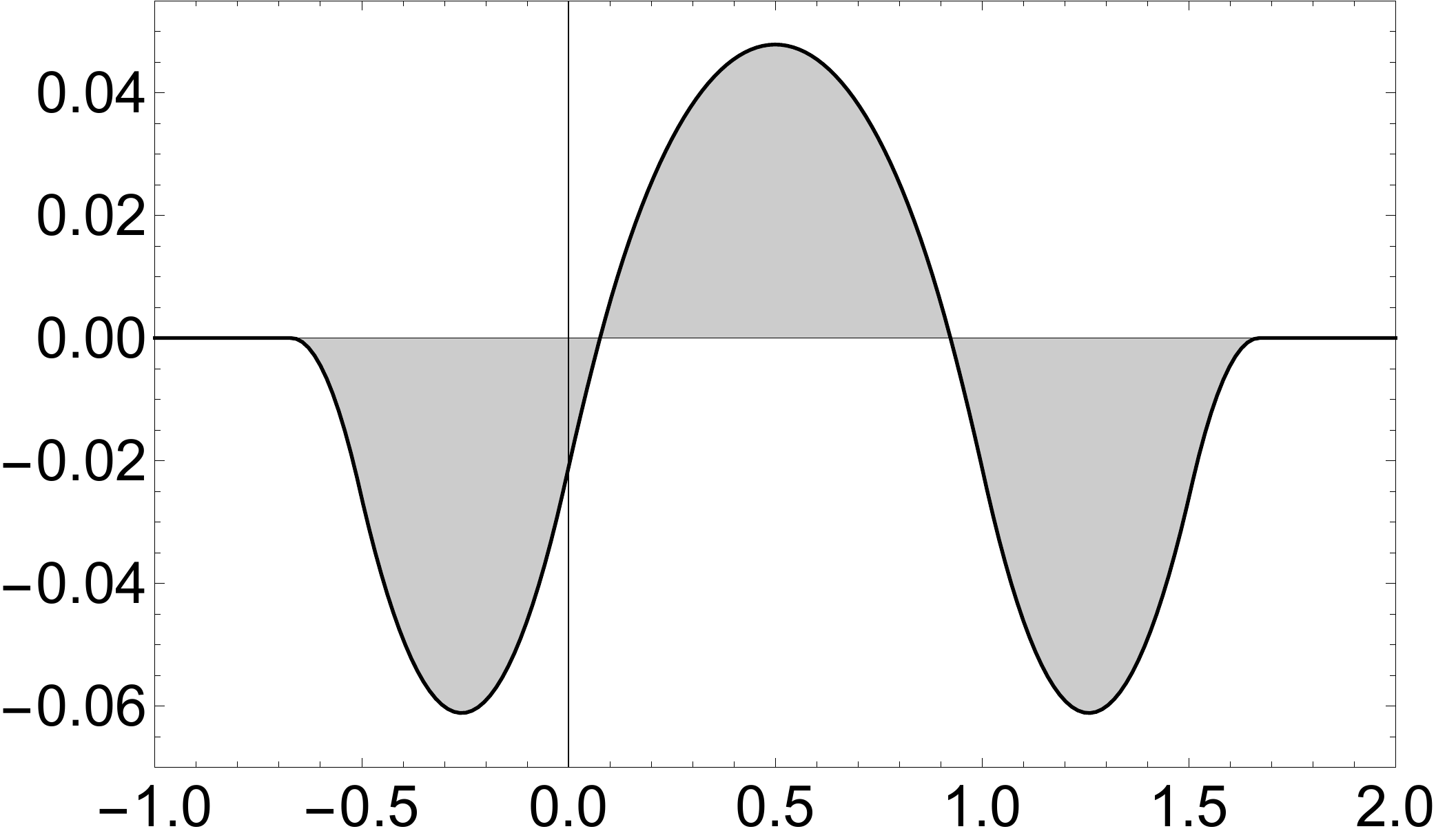}}
\subfigure[]{\includegraphics[width=0.4\textwidth,height=0.15\textwidth, angle =0]{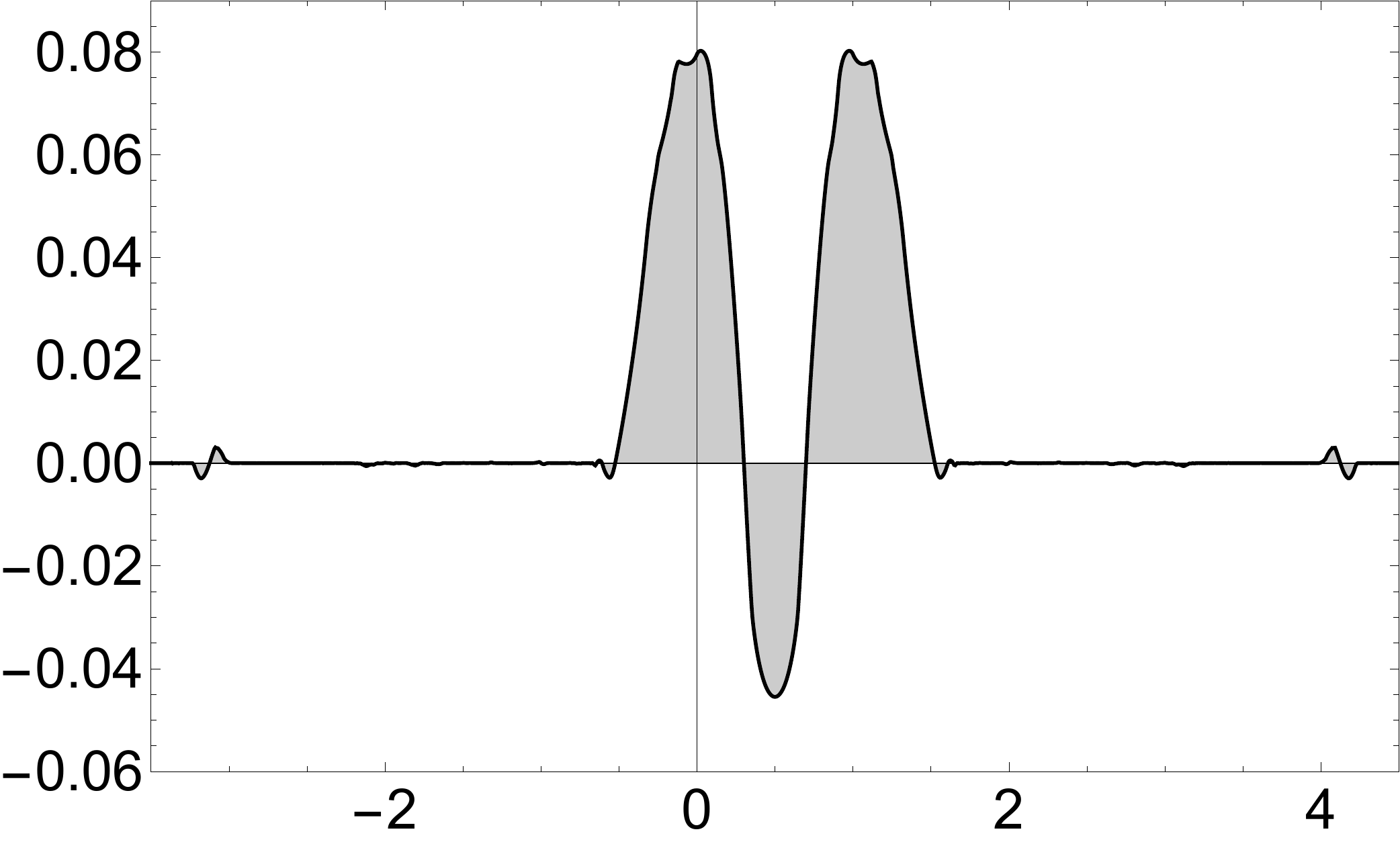}}
\subfigure[]{\includegraphics[width=0.4\textwidth,height=0.15\textwidth, angle =0]{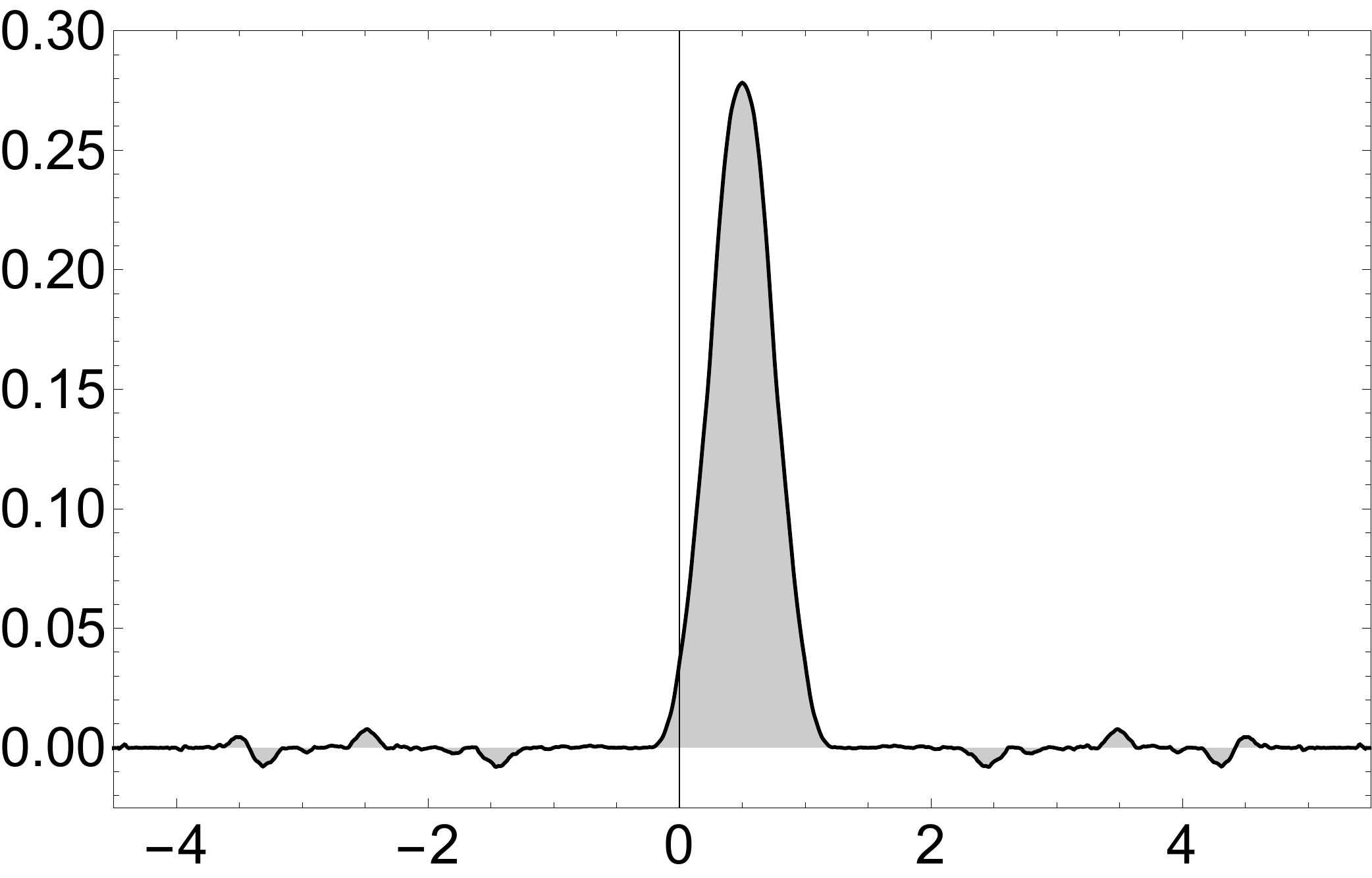}}
\caption[]{Numerical perturbed breather solution (shadowed region under solid line) and  self-similar  exact solution (dashed lines) for $\epsilon=3.0$. Snapshots  correspond to (a) $t=0.15$,  (b) $t=1.01$, (c) $t=8.96$,  (d) $t=21.23$. }\label{SG-e-3}
\end{figure}
\begin{figure}
\centering 
\subfigure[]{\includegraphics[width=0.4\textwidth,height=0.15\textwidth, angle =0]{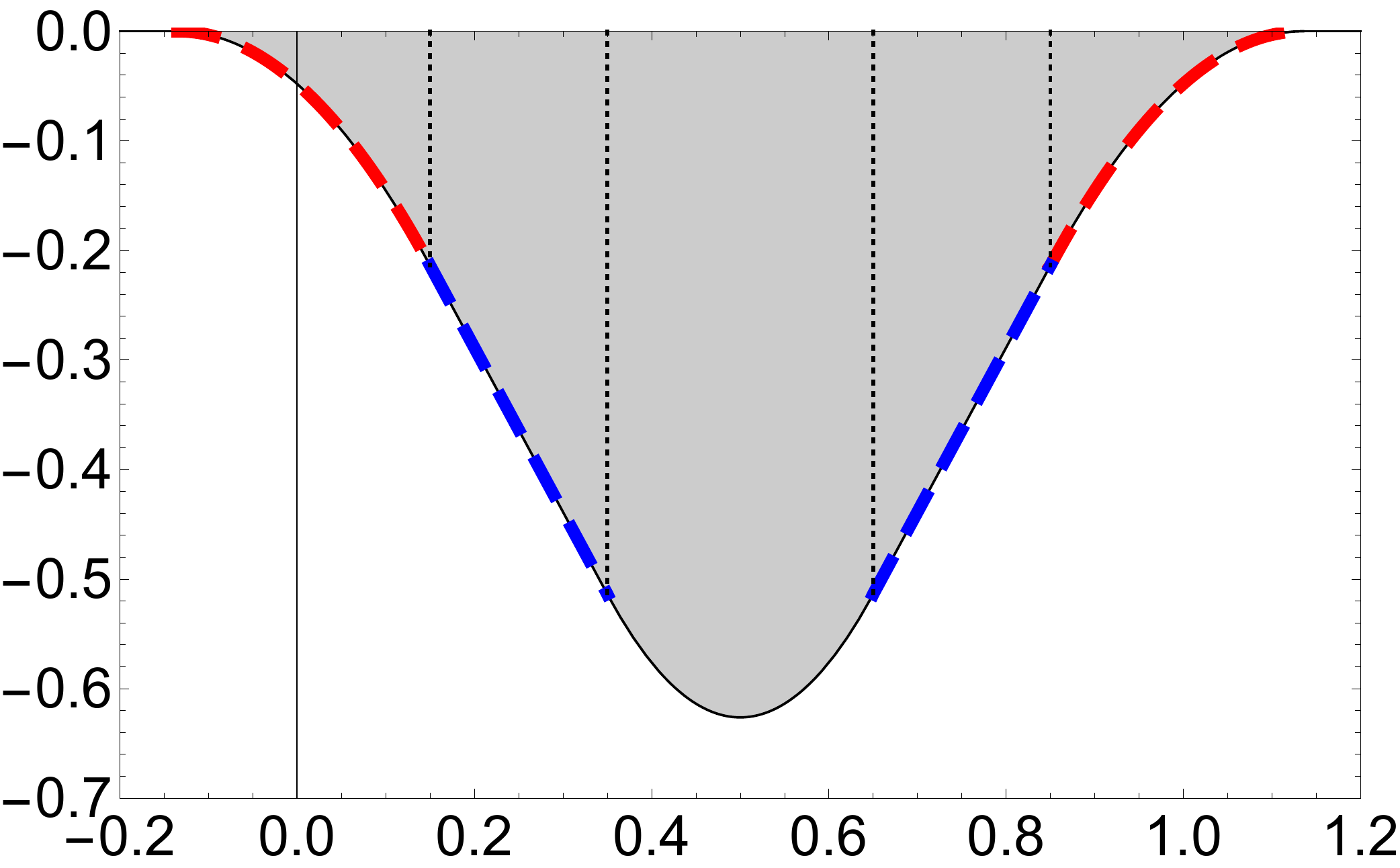}}
\subfigure[]{\includegraphics[width=0.4\textwidth,height=0.15\textwidth, angle =0]{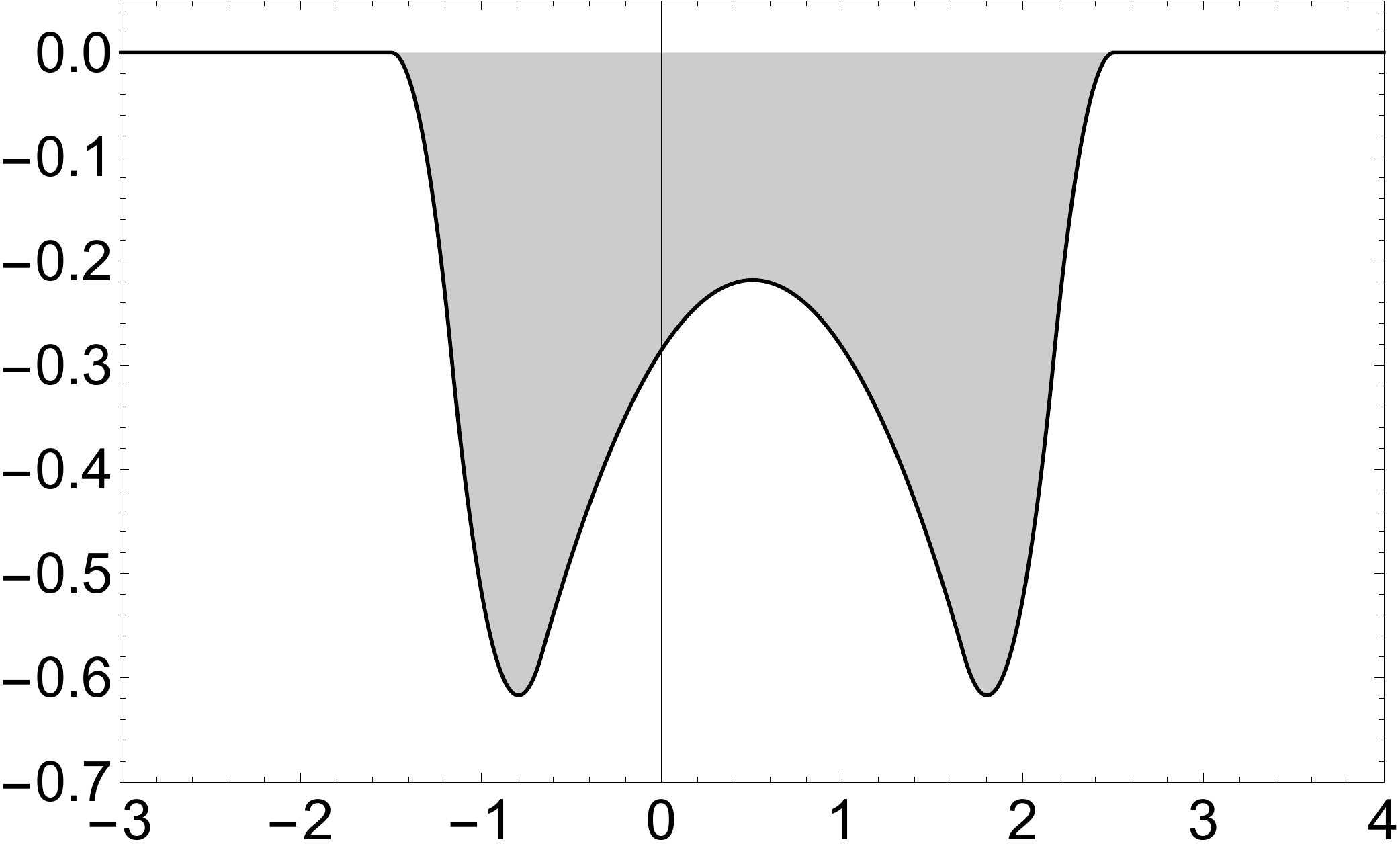}}
\subfigure[]{\includegraphics[width=0.4\textwidth,height=0.15\textwidth, angle =0]{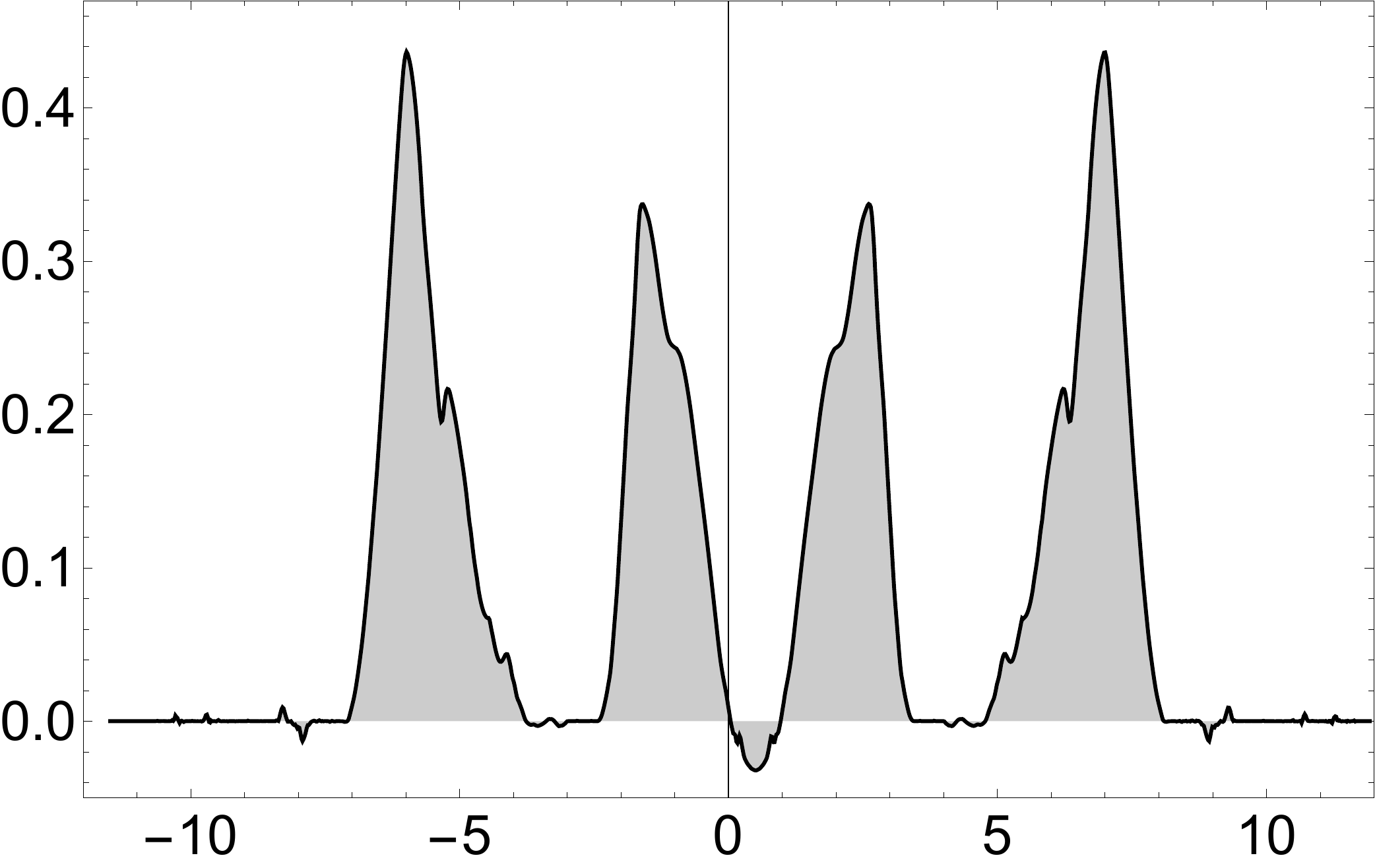}}
\subfigure[]{\includegraphics[width=0.4\textwidth,height=0.15\textwidth, angle =0]{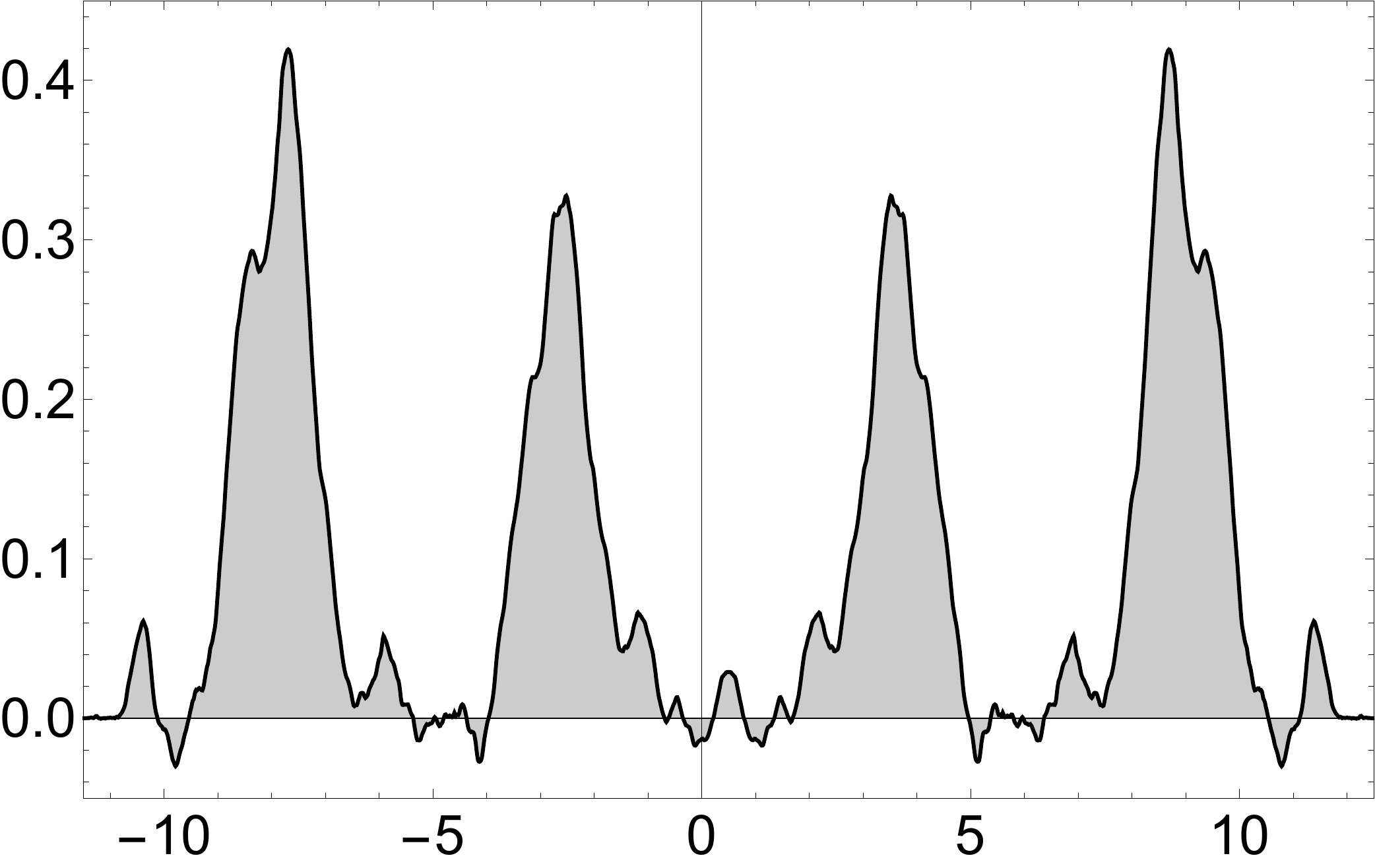}}
\caption[]{Numerical perturbed breather solution (shadowed region under solid line) and  self-similar exact solution (dashed lines) for $\epsilon=10.0$. Snapshots  correspond to (a) $t=0.15$,  (b) $t=1.67$, (c) $t=20.79$,  (d) $t=41.03$.}\label{SG-e-10}
\end{figure}
In the very initial phase of the evolution, the vacuum solution and the oscillating solution 
are matched at the point $x_L(t)=v_0t$ which moves to the left with the speed $|v_0|$.
 The corresponding symmetric matching point $x_R(t)=1-x_L(t)$ moves to the right with the speed
 $|v_0|$. This means that the support of the solution expands. Such behaviour is sketched 
in Figures \ref{SG-e-3}(a) and \ref{SG-e-10}(a) which present numerical and analytic solutions for $\epsilon=3.0$,
 $\epsilon=10.0$.  During the initial part of the evolution the solution has self-similar components (dashed lines).
 The initial configuration of the field on $x\in[0,1]$ has some extra kinetic energy when compared to the energy of 
the exact breather and as one can expect that this surplus of the energy would lead to the expansion of the oscillating region.
 Moreover, there is also some radiation which has the form of some small localized packages (small breathers or oscillons)
 that propagate outside of the region occupied by the oscillon. 

Speaking qualitatively, we see that the initially perturbed breathers are rather surprisingly stable objects
 although there are no topological obstacles to prevent their decay nor there is any  mass gap in the energy spectrum. 
\\
For small perturbations represented here by $\epsilon=1.2$ the initial configuration oscillates as a single smooth object
 performing 25 oscillations. Then, at $t \approx 13$ small packages of energy (small oscillons) are emitted; this
 is the main mechanism by which the system gets rid of its excess of energy. This emission is not a continuous process.
 On the contrary, small oscillons are emitted only at a few instances of time. We observe that the emission of oscillons with smaller
 energy takes place more often than the emission of more massive structures. All the time central oscillon does 
not lose its identity and dominates over the emitted little oscillons Fig.\ref{SG-e-1.2}(d), (e). It would be 
interesting to evolve the system longer and verify whether the final state of the evolution results in a signum-Gordon breather 
with the unit support or if the emission of small oscillons leads to a slow but unavoidable annihilation of the
 initial oscillon. We have found that even for a quite large deformation of the initial condition $(\epsilon=3.0)$ the same qualitative picture is valid - Fig. \ref{SG-e-3}. 

For very large deformations the dynamics is drastically different. In this case the final state decomposes into 
large substructures (large finite support oscillons) which seem to stay together for some time. However, they slightly
 repel each other. As they are compactons they do not interact directly but via the emission of small oscillons which 
results in a week repulsion. For example, for $\epsilon=10$ one can see four such substructures, Fig. \ref{SG-e-10}. 
The issue of the appearance of a given number of oscillons in the final state is another very interesting problem to
 investigate,
 which however, is beyond the scope of the present paper.

\item For $\epsilon=1$, the velocity $v_0$ vanishes, {\it i.e.} $v_0=0$. This is exactly the case of 
the pure breather. In this case the exact solution is known. The exact breather exists for infinitely long 
time and does not radiate. We have made use of this solution to check how far we can trust our numerical simulations.
 After 60 oscillations the discrepancy between analytical calculations and the results of our numerics was still below $0.5\%$.

\item For $\frac{1}{2}<\epsilon<1$, the velocity $v_0$ is positive
 and is still less than unity, $0<v_0<1$. The matching points $x_L(t)$ and $x_R(t)$ move towards the
 centre of the solution.  This leads to the shrinking of the support in the very initial phase of the evolution.
 An example of such a solution, that corresponds to $\epsilon=0.8$, is shown in Fig.\ref{F6}. The energy of initial 
field configuration is smaller than the energy  of the exact breather with unit support. On the other hand, we can 
treat this initial configuration as a perturbation on the top of an exact breather with smaller support and  whose 
amplitude  has been obtained from the $l=1$ breather by the scaling transformation. Then, again, such a solution seems 
to relax to a sort
 of a breather state, from which some smaller oscillons are emitted during the relaxation time. 

\begin{figure}[h!]
\centering 
\subfigure[]{\includegraphics[width=0.3\textwidth,height=0.15\textwidth, angle =0]{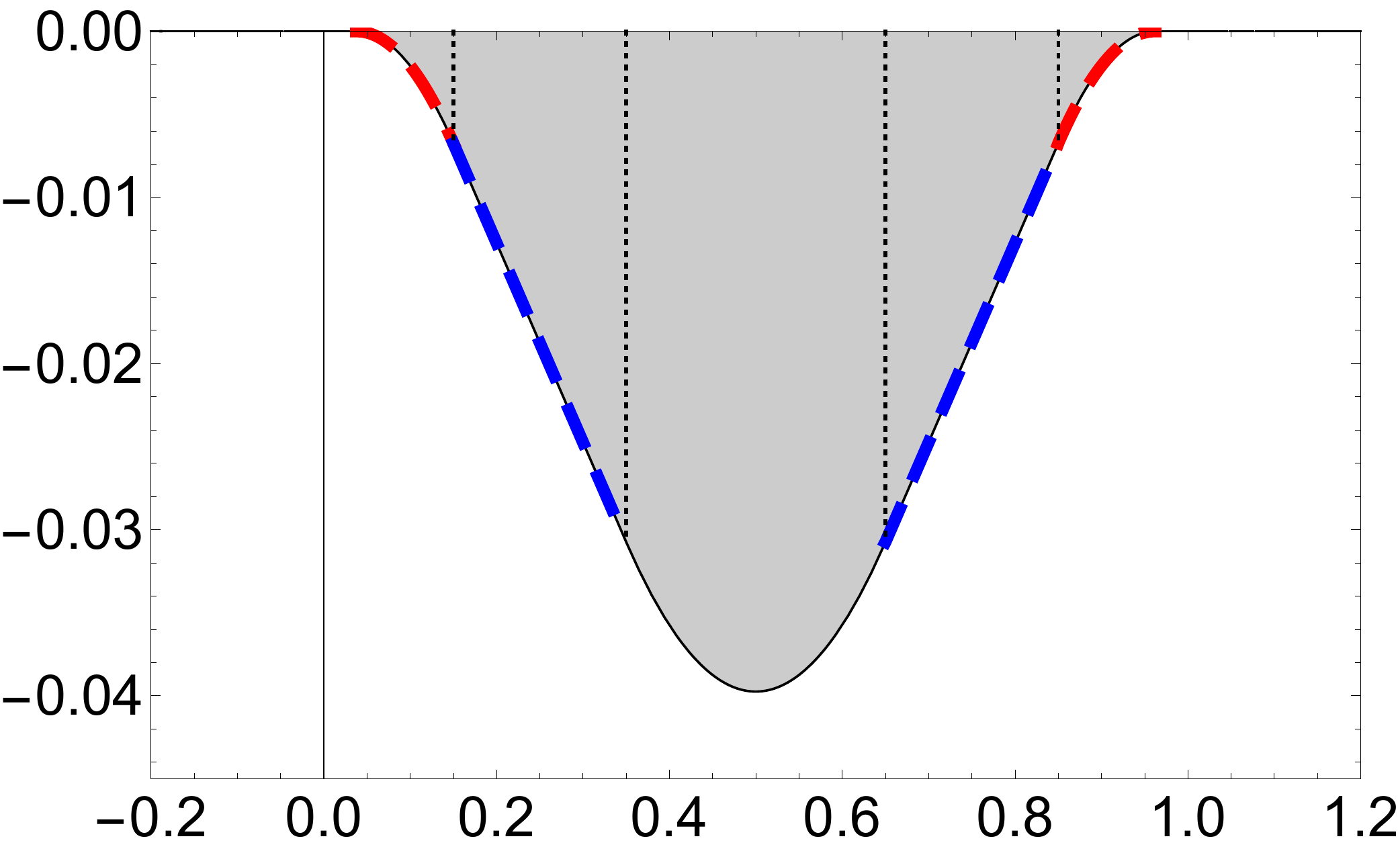}}
\subfigure[]{\includegraphics[width=0.3\textwidth,height=0.15\textwidth, angle =0]{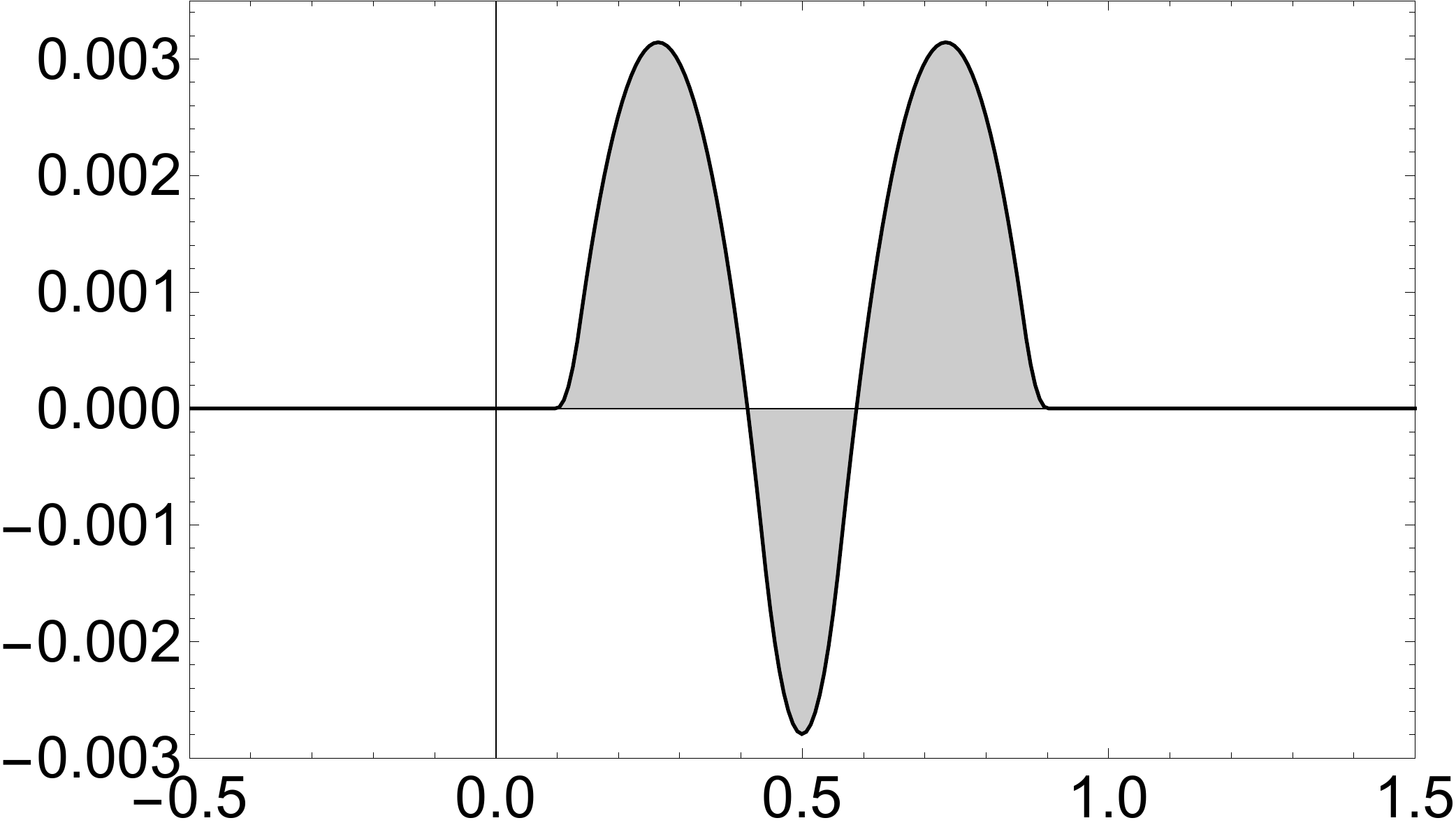}}
\subfigure[]{\includegraphics[width=0.3\textwidth,height=0.15\textwidth, angle =0]{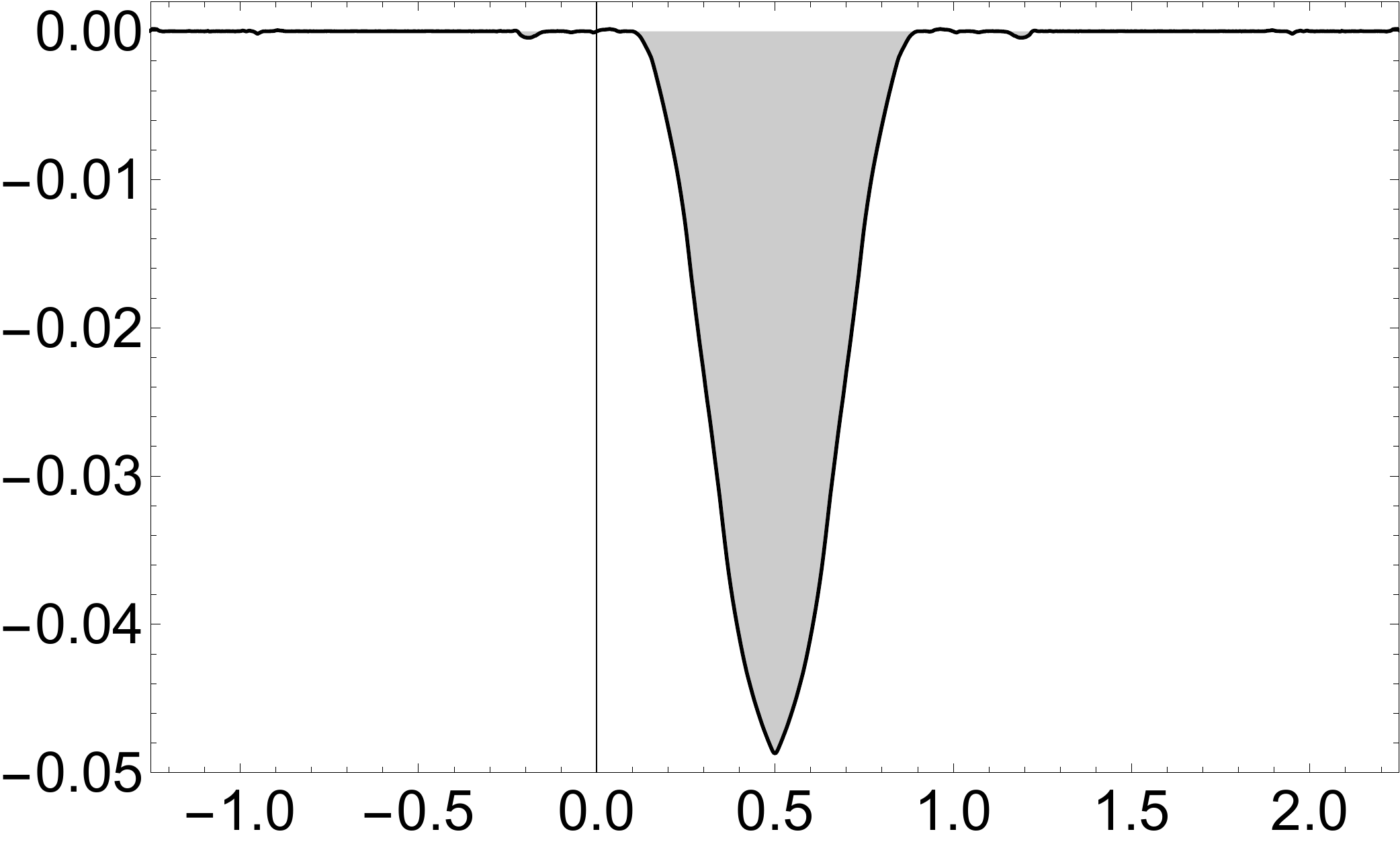}}
\caption[]{Numerical perturbed breather solution (shadowed region under solid line) and  self-similar exact  solution (dashed lines) for $\epsilon=0.8$. Snapshots  correspond to (a) $t=0.15$,  (b) $t=0.44$ and (c) $t=10.33$.}\label{F6}
\end{figure}

\item For $\epsilon=\frac{1}{2}$, the velocity $v_0=1$.
In this case the support of the parabolic partial solution vanishes. The self-similar  partial solution consists of a single component which is linear in variable $x$. The spatial derivative $\partial_x\phi$ is discontinuous at the matching points $x=t$ and $x=1-t$. Note, that discontinuity of the derivative is admissible at light cones. The numerical solution for $\epsilon=0.5$ is shown in Fig.\ref{F7}.

\begin{figure}[h!]
\centering 
\subfigure{\includegraphics[width=0.3\textwidth,height=0.15\textwidth, angle =0]{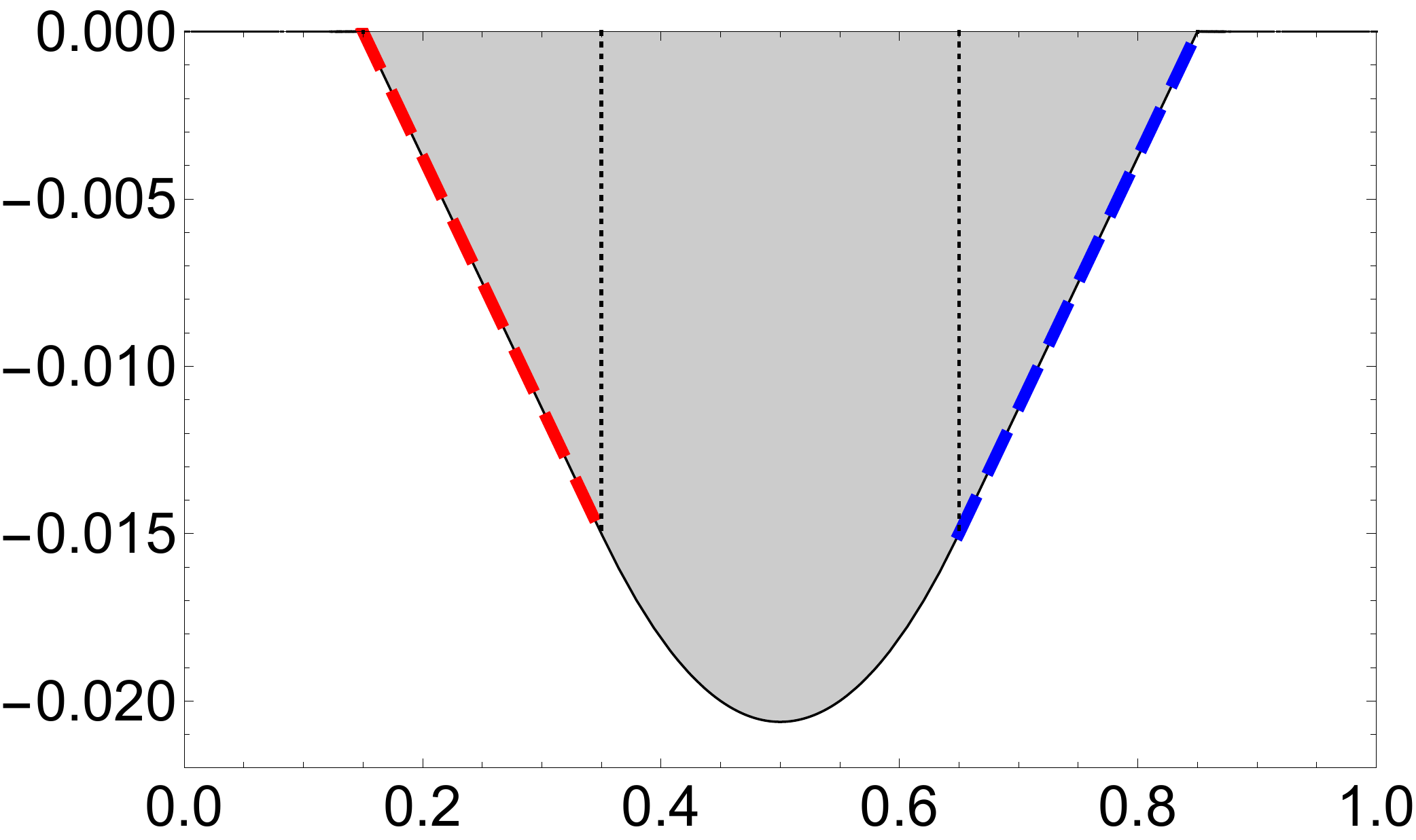}}
\subfigure{\includegraphics[width=0.3\textwidth,height=0.15\textwidth, angle =0]{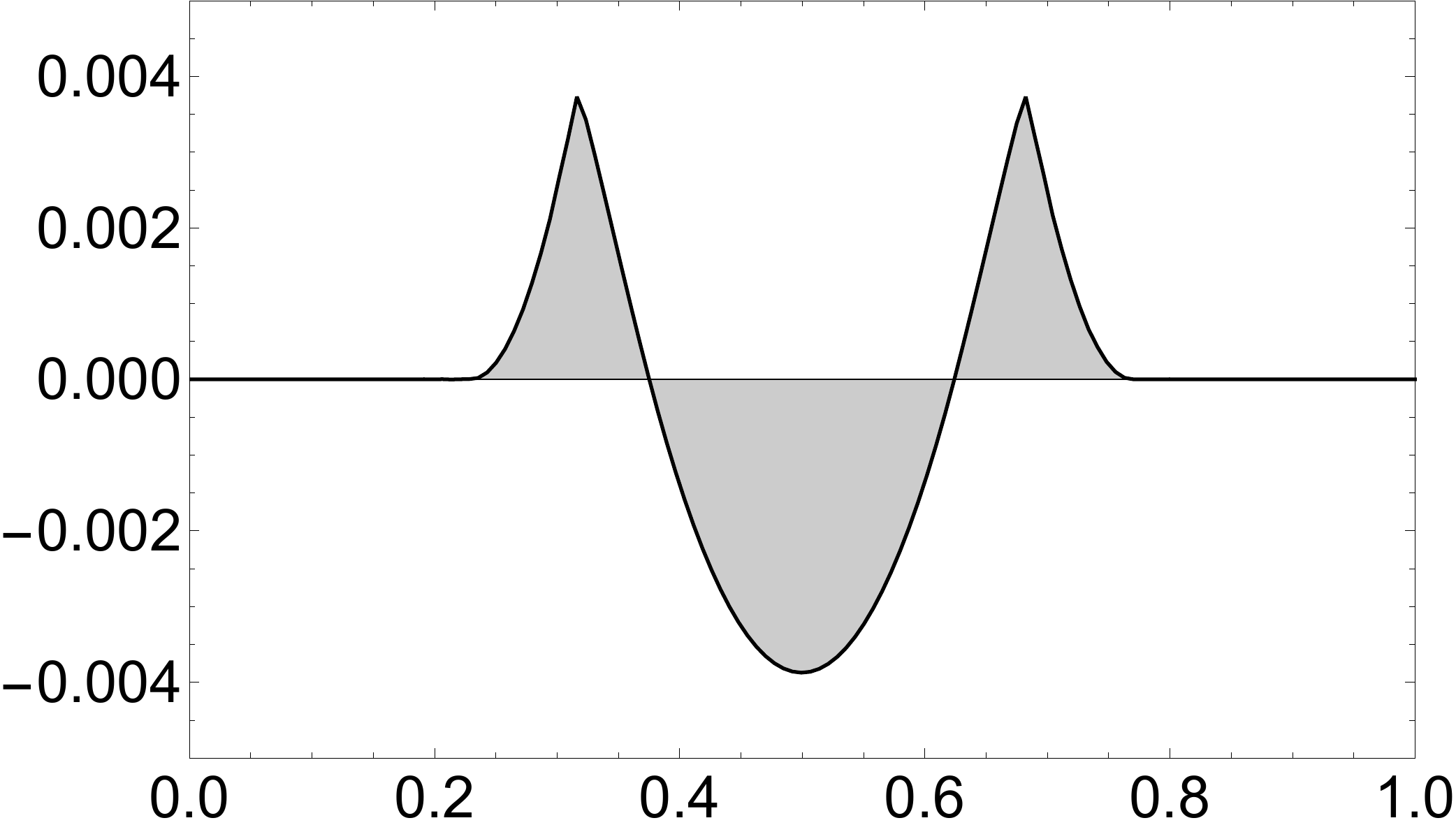}}
\subfigure{\includegraphics[width=0.3\textwidth,height=0.15\textwidth, angle =0]{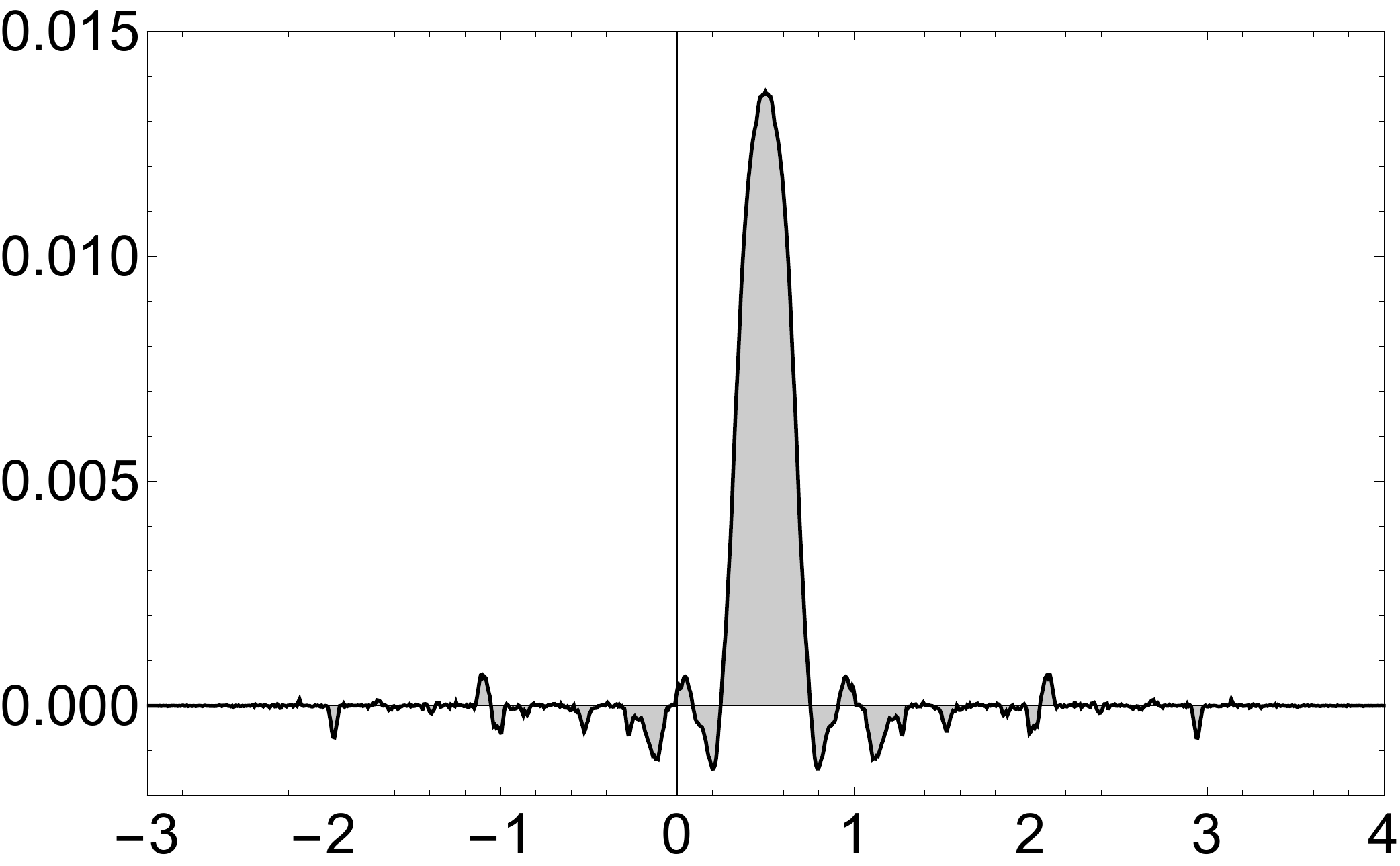}}
\caption[]{Numerical perturbed breather solution (shadowed region under the solid curve) and  self-similar  exact  solution (dashed lines) for $\epsilon=0.5$. Snapshots  correspond to (a) $t=0.15$,  (b) $t=0.32$, (c) $t=7.76$.}\label{F7}
\end{figure}
\end{enumerate}

\item Case $0<\epsilon<\frac{1}{2}$. In this case the solution is technically more complicated than the solutions discussed 
before. This is due to the fact that
 for $\epsilon <\frac{1}{2}$, the relevant self-similar solution consists of infinitely many smoothly joined
 parabolic solutions $S_k$, $k \in \mathbb{N}$, given by \eqref{Sk}.

The coefficients $\alpha_k$ and $\beta_k$ and matching points $a_k$ are given in terms of parameters of the first 
 parabola $S_1(y)$. Here, $S_0=0$ and $\dot S_0=-\epsilon$, so the first parabola is parametrized by
 $\alpha_1=2\dot S_0=-2\epsilon$ and $\beta_1=1-2S_0=1$.  Since the solution $S_1(y)=-\frac{y}{2}(y+2\epsilon)\ge 0$
 has the support $y\in[-2\epsilon,0]$,  it does not arise in the self-similar solution which is considered
 for $x\ge 0$. However, the partial solutions for $k=2,3,\cdots$ give rise to the self-similar solutions because 
supports belong to the interval $y\ge 0$. 
The parameters of the partial solutions $\alpha_k$, $\beta_k$ and the matching points $a_k$ are determined
 by the conditions $S_k(a_k)=0=S_{k+1}(a_k)$ and $S'_k(a_k)=S'_{k+1}(a_k)$. 

\begin{figure}[h!]
\centering 
\subfigure{\includegraphics[width=0.6\textwidth,height=0.2\textwidth, angle =0]{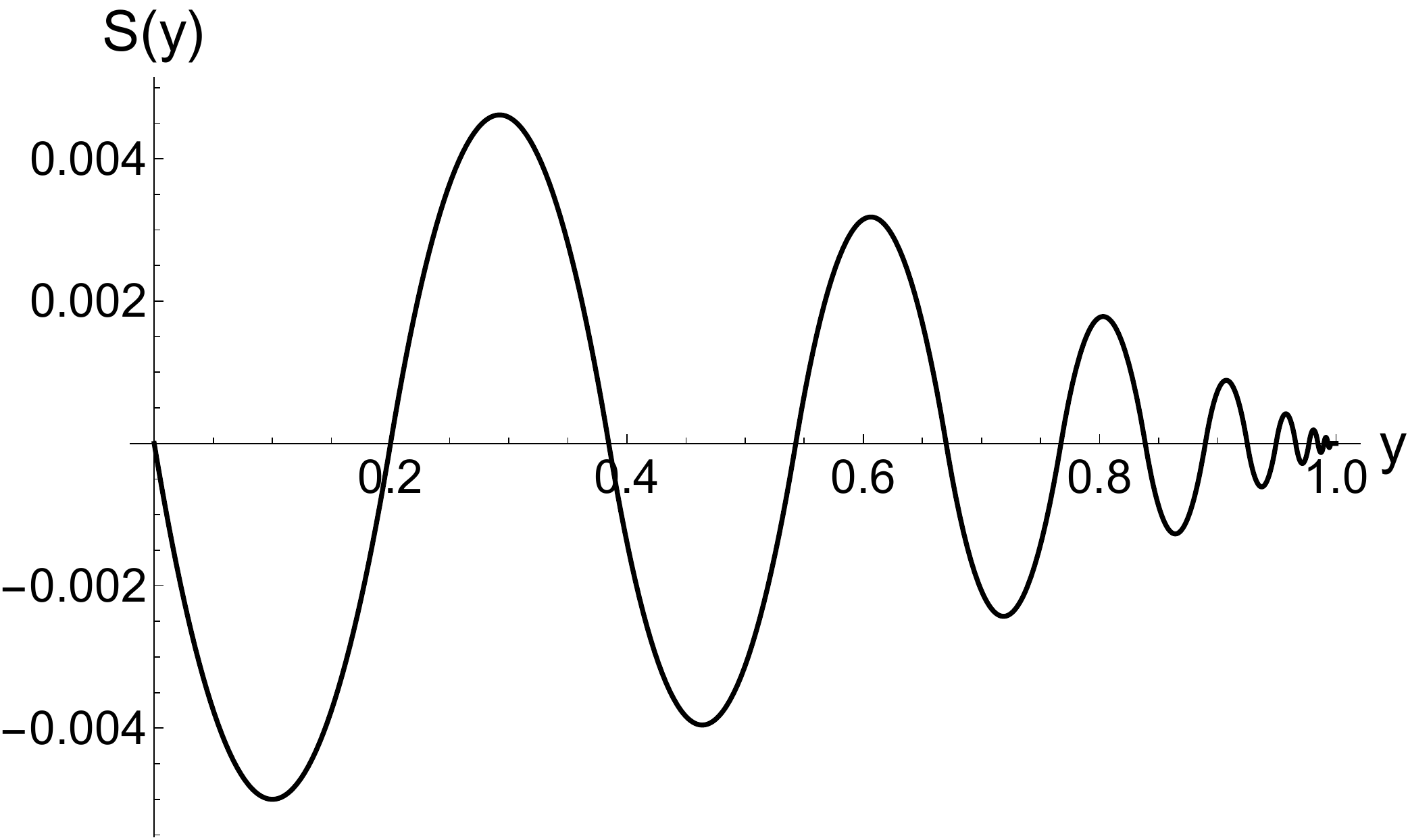}}
\caption{Solutions $S_k(y)$, $k=2,3,\cdots$ for $\epsilon=0.1$.}\label{F8}
\end{figure}
\begin{figure}[h!]
\centering 
\subfigure[]{\includegraphics[width=0.45\textwidth,height=0.2\textwidth, angle =0]{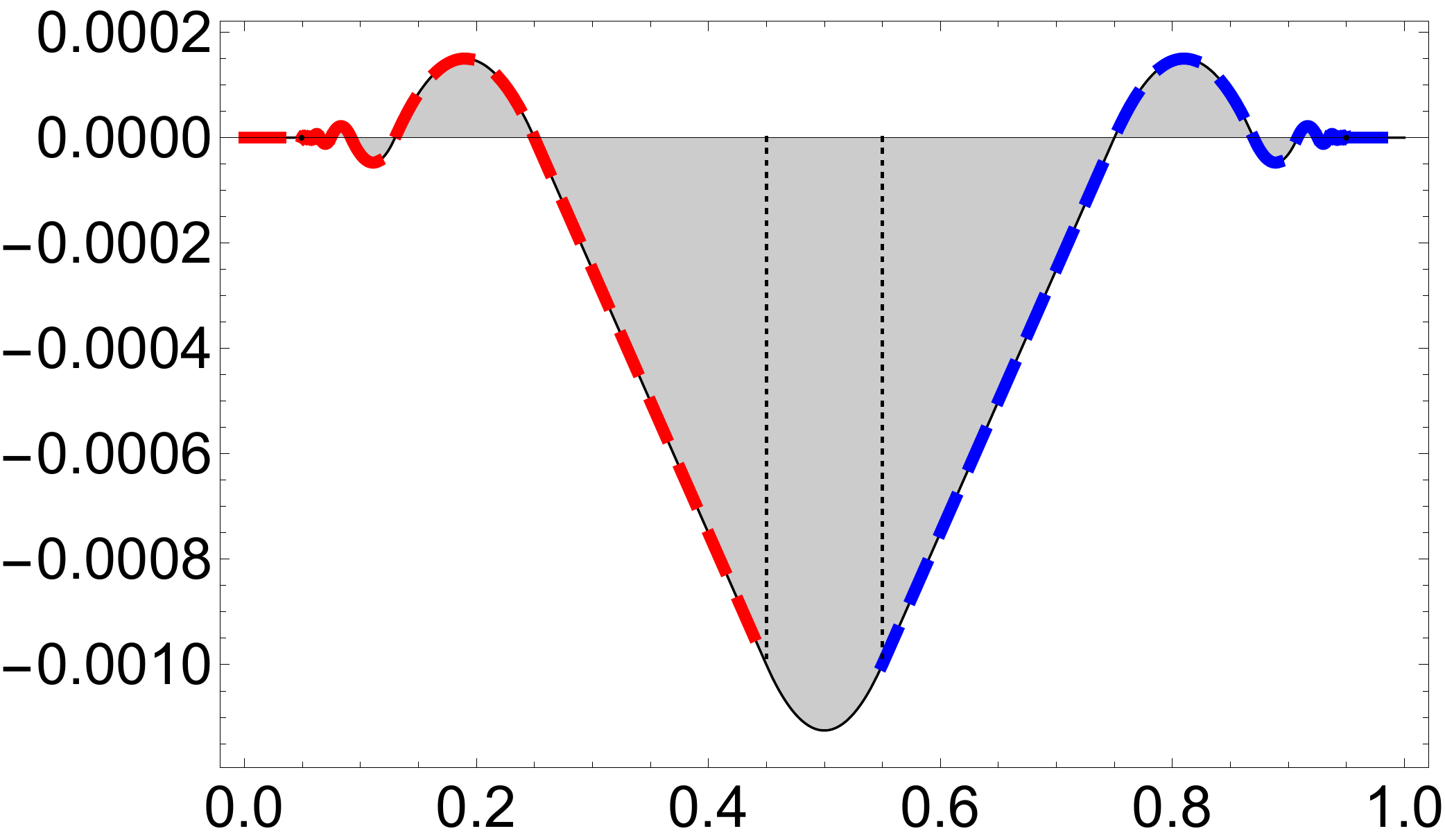}}
\subfigure[]{\includegraphics[width=0.45\textwidth,height=0.2\textwidth, angle =0]{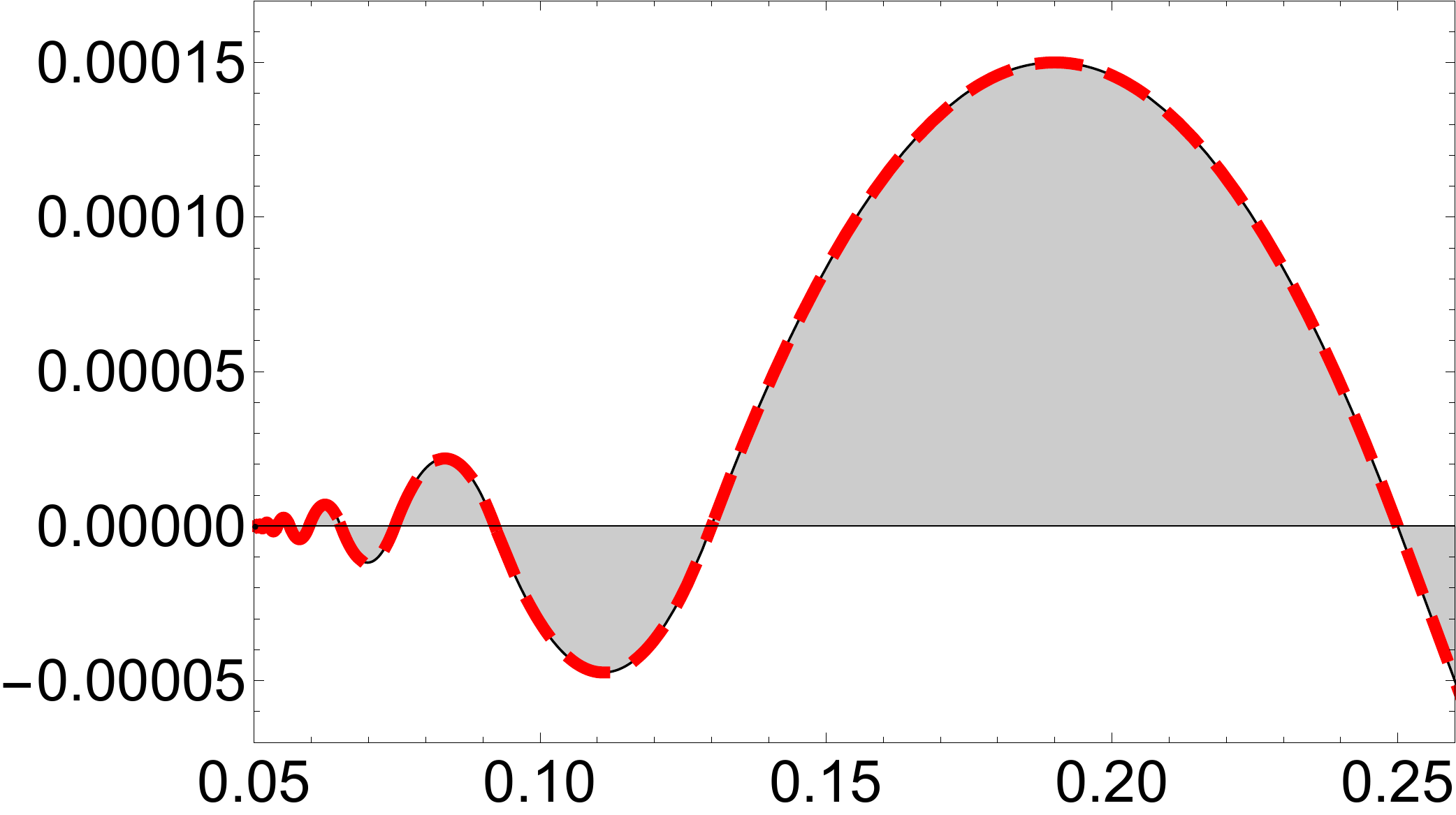}}
\subfigure[]{\includegraphics[width=0.45\textwidth,height=0.2\textwidth, angle =0]{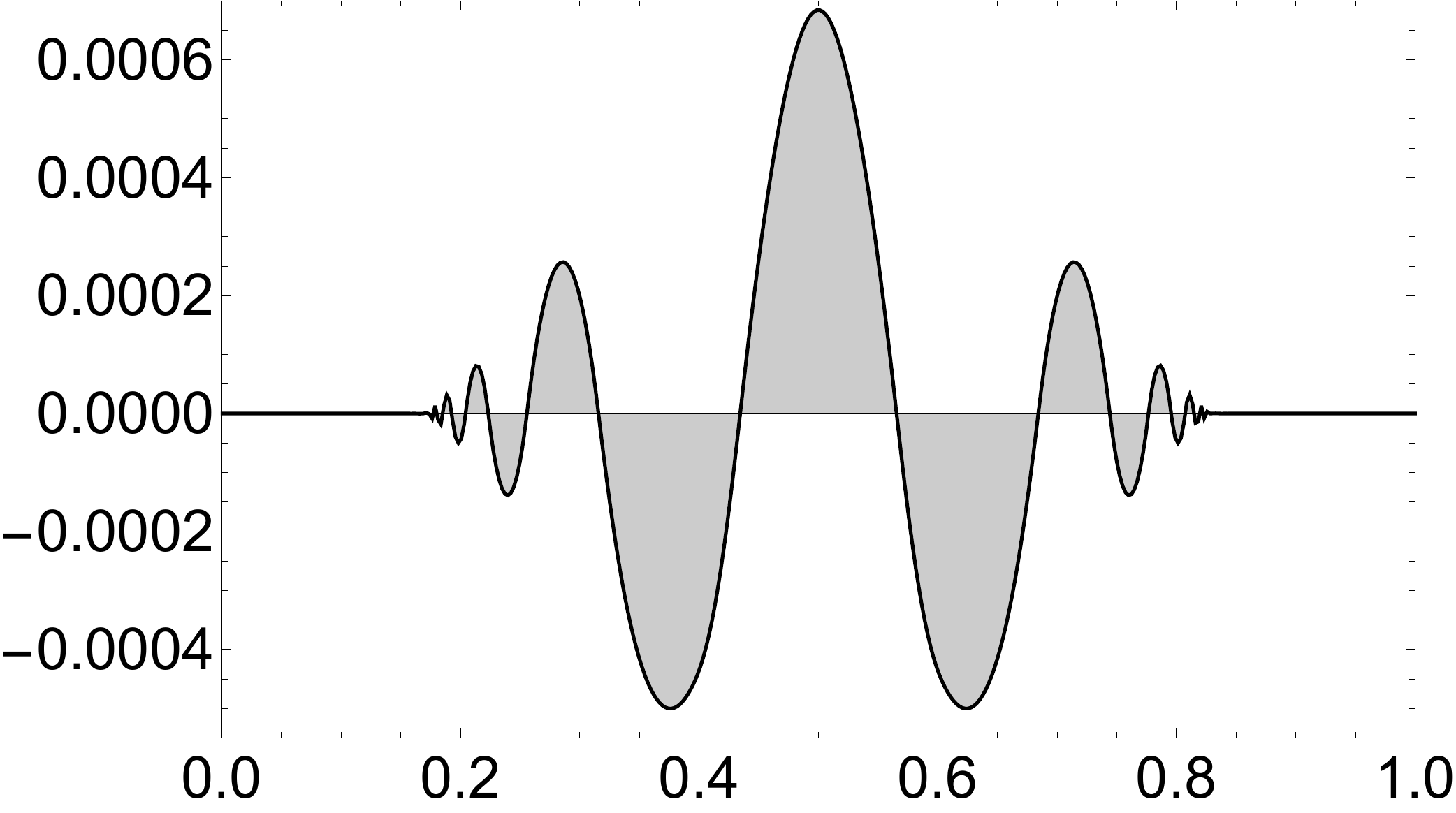}}
\subfigure[]{\includegraphics[width=0.45\textwidth,height=0.2\textwidth, angle =0]{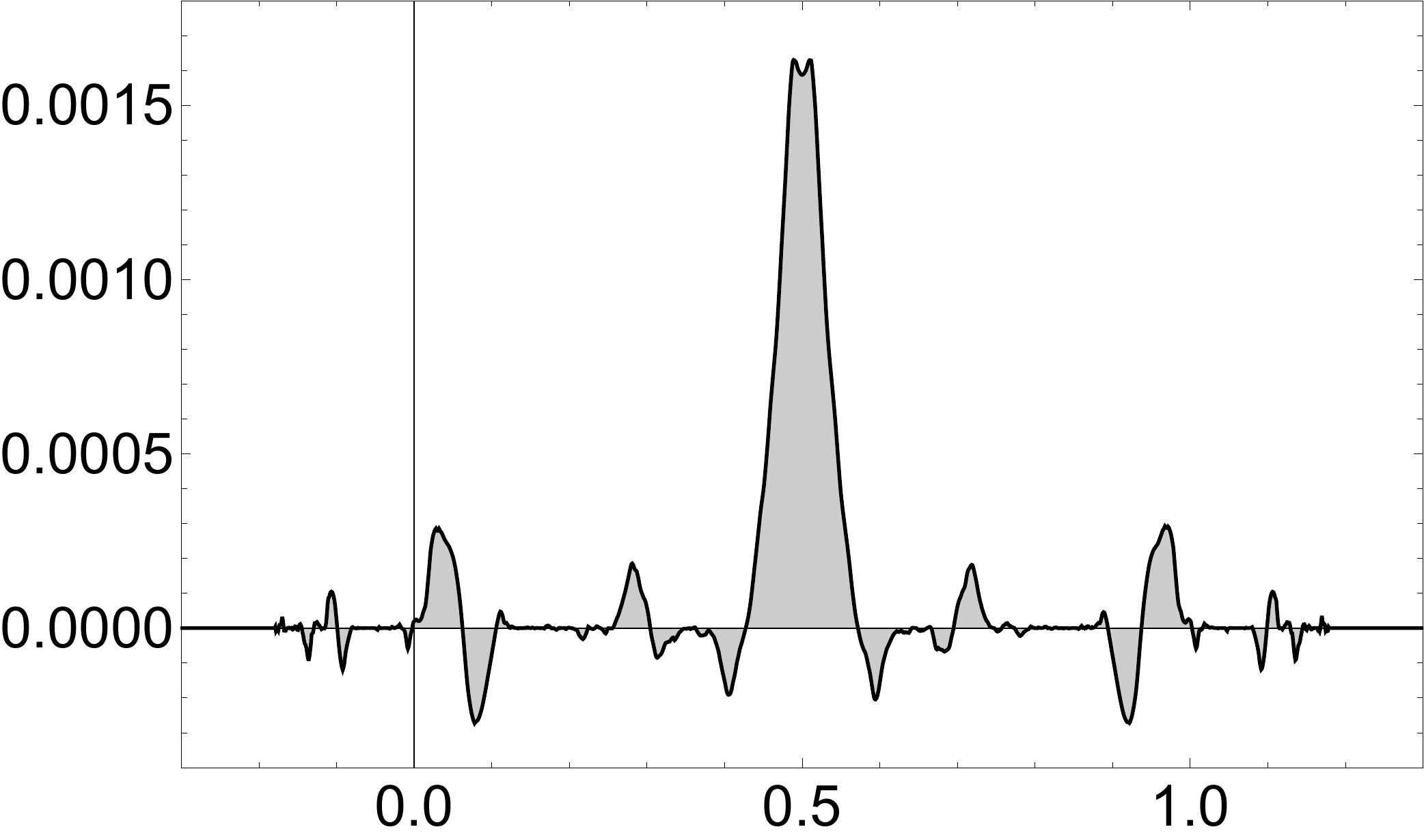}}
\caption{(a,b) Numerical solution  (shadowed region under solid line) and  self-similar  exact 
 components (dashed line) for $\epsilon=0.1$. (a,b) Field $\phi$ at $t=0.05$, (c) field $\phi$ at $t=0.17$,
  (d) field $\phi$ at $t=1.18$.}\label{F9}
\end{figure}

In terms of the auxiliary quantities:
\[
p:=\frac{1-a_0}{1+a_0},\qquad q:=\frac{1-a_1}{1+a_1}, \qquad r:=\frac{q}{p},
\]
where $a_0=-2\epsilon$ and $a_1=0$ are zeros of $S_1(y)$,  the parameters of the solutions take the form
\begin{eqnarray}
\alpha_k&=&\frac{1}{1+r}\left[\frac{1}{p r^{k-1}}-pr^k+(-1)^k\left(\frac{1}{p}-q\right)\right]-(-1)^k\alpha_1,\\
\beta_k&=&\frac{1}{2}+\frac{1}{2(1+r)}\left[\frac{1}{p r^{k-1}}+pr^k+(-1)^k\left(\frac{1}{p}+1\right)(1+q)\right]-(-1)^k\beta_1,\\
a_k&=&\frac{p^{k-1}-q^k}{p^{k-1}+q^k}.
\end{eqnarray}
In Fig.\ref{F8} we plot the parabolas $S_k(y)$ for $k=2,3,\cdots$ and $\epsilon=0.1$. They give rise to
 solutions $\phi(t,x)$. Such solutions are shown in figure Fig.\ref{F9}. In this case the emitted radiation
 becomes visible very soon. The numerical simulations have shows that even only after the collision of two fronts of infinite waves 
the system exhibits a large number of short-length oscillations. Such oscillations propagate outside of the region where the oscillon
 is localized.

\end{enumerate}

We conclude that self-similar solutions, restricted to compact supports,  can appear as partial solutions of some 
finite energy configurations of the signum-Gordon model. In particular, they are components of the exact signum-Gordon
 breather and they also arise in more complicated perturbed breather-like configurations. 
 All such perturbed breathers radiate, for which the main mechanism, at least for $\epsilon>1$, is the emission of smaller
 oscillons. We have also observed that this radiation is more intense for solutions with $\epsilon<1$ than for solutions
 with $\epsilon>1$.  In the sector  $\epsilon<0.5$  the whole solution gets very quickly converted into radiation.
 This is certainly related to the infinite number of oscillations which are present in the self-similar components of the 
whole system. A collision of two such wave trains leads to the transfer of energy into a large number of small-scale
 oscillations. We see  that a change of the initial data from $\epsilon>\frac{1}{2}$ to $\epsilon<\frac{1}{2}$ 
leads to a very significant difference in the 
behaviour of the signum-Gordon fields.

\section{The signum-Gordon model on a semi-infinite line}

As we have mentioned before the Skyrme theory motivated model is defined on a semi-infinite line as it involves the
 radial variable $r \in \mathbb{R}_+$. For small amplitude solutions, the modified Lagrangian (\ref{model}) coincides
 with the signum-Gordon model, again defined on $\mathbb{R}_+$. So here we analyse properties of such
 a signum-Gordon theory
\be
\mathcal{L}_{sG_+}=\frac{1}{2}(\partial_t\phi)^2-\frac{1}{2}(\partial_r\phi)^2-|\phi|, \;\;\;\; r\geq 0.\label{mod1+} 
\ee

First of all, due to the compact nature of the breather solutions in the model on $\mathbb{R}$ one can also easily 
construct
the corresponding breathers in the $\mathbb{R}_+$ case.  The field equation 
\be
\partial^2_t\phi-\partial^2_r\phi+{\rm sgn}(\phi)=0
\ee
has a compact breather solution with the usual initial 
\be
\phi(0,r)=0,\qquad\partial_t\phi(0,r)=v(r)
\ee
and boundary conditions 
\be
\phi(t,0)=0=\phi(t,1),\qquad \partial_r\phi(t,0)=0=\partial_r\phi(t,1)
\ee
where
\begin{eqnarray}
v(r)=\left\{\begin{array}{ccl}
 |r-{\textstyle\frac{1}{2}}|-{\textstyle\frac{1}{2}}&{\rm for}&0\le r\le 1,\\
0&{\rm for}&r\ge 1.
\end{array}\right. 
\end{eqnarray}

The resulting exact breather shell solution on the unit segment $r \in [0,1]$ can be constructed using the same partial
 solutions as for the model on $\mathbb{R}$ (\ref{ps1})-(\ref{ps4}) and following the same prescription summarised
 in Fig. \ref{fig1}. 
\\

Then, applying the translation $r\rightarrow r+R$ we can shift the breather shell to any position. Now, we get our breather 
shell solutions  with the support $[R,R+1]$, where the vacuum is in the inner ball $r \in [0,R]$ and in the outer 
space $r \in [R+1,\infty)$.
 Note that although the volume of such a solution grows quadratically with $R$ its energy remain unchanged. 
This situation (energy being independent of the volume) is completely opposite to what happens in topologically
 non-trivial BPS sectors, where energy of BPS solutions grows linearly with the  topological baryon charge $B$ while
 the volume remains fixed (volume is independent of the energy).

Let us note that we can still apply the scaling symmetry (\ref{scale}) with the positive scaling
 parameter $l$. The period, size and the energy scale identically as in the $\mathbb{R}$ case. 

\vspace*{0.2cm}
\begin{figure*}[h!]
\centering
\subfigure[]{\includegraphics[width=0.45\textwidth,height=0.15\textwidth, angle =0]{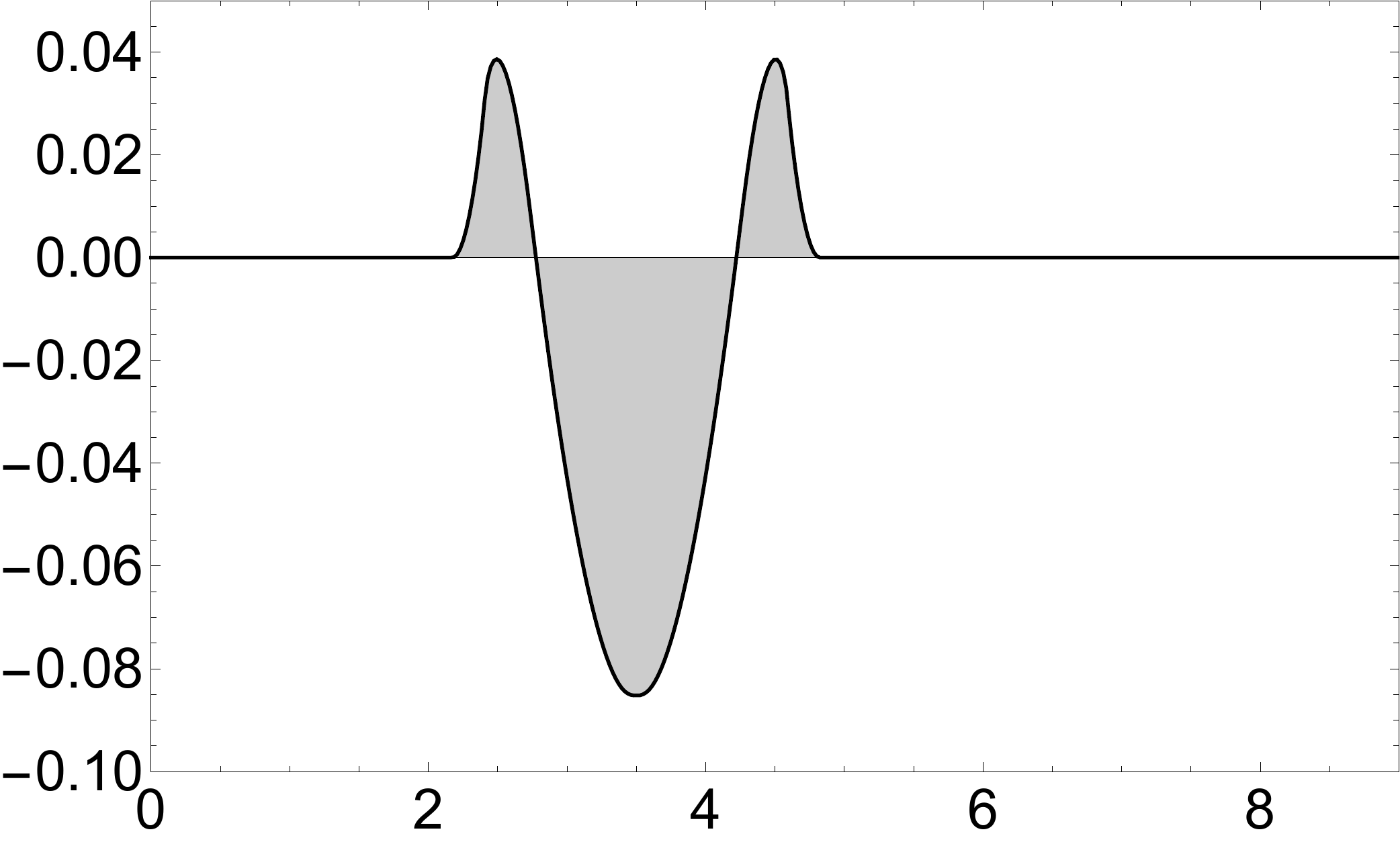}}
\subfigure[]{\includegraphics[width=0.45\textwidth,height=0.15\textwidth, angle =0]{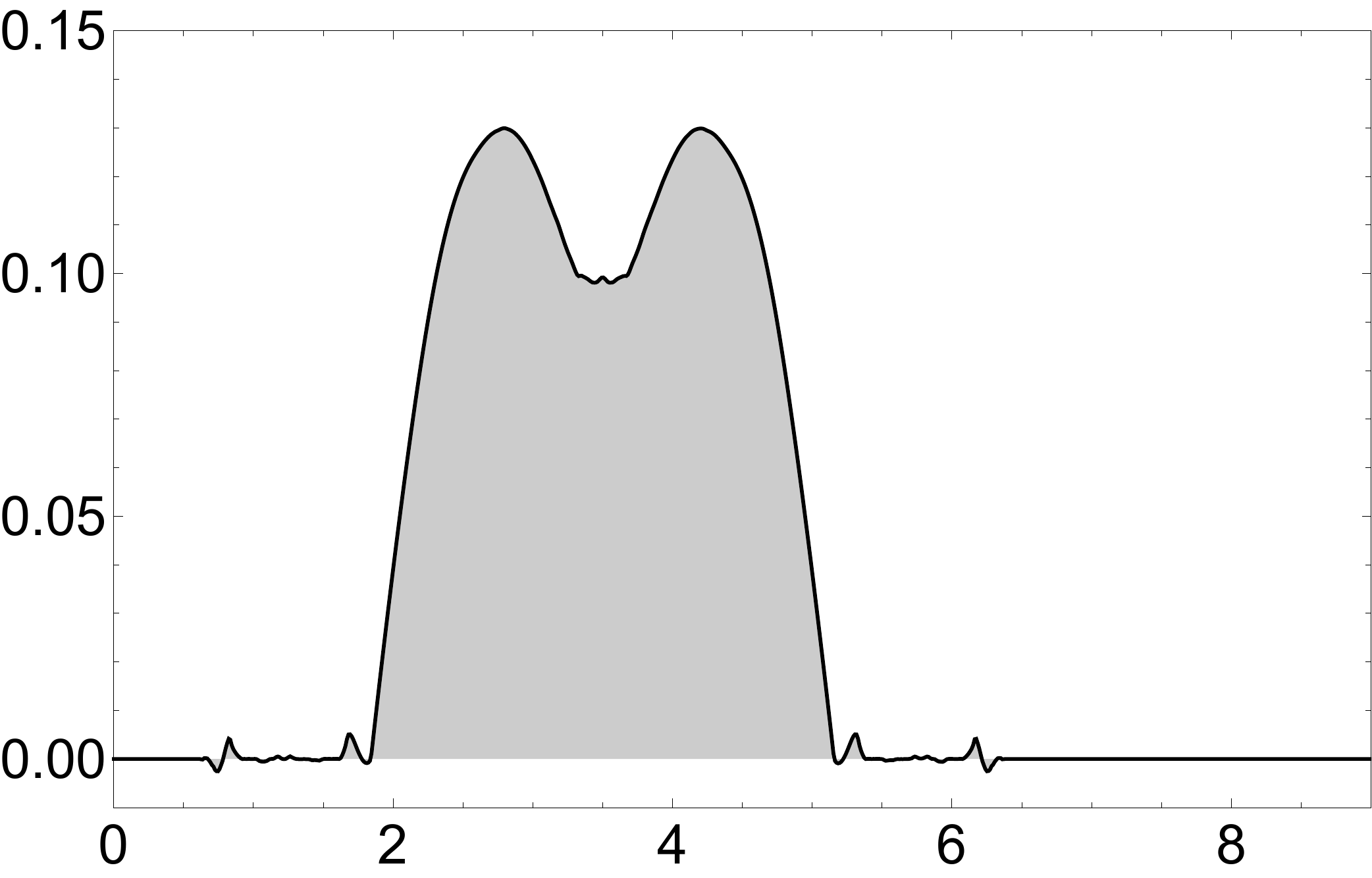}}
\subfigure[]{\includegraphics[width=0.45\textwidth,height=0.15\textwidth, angle =0]{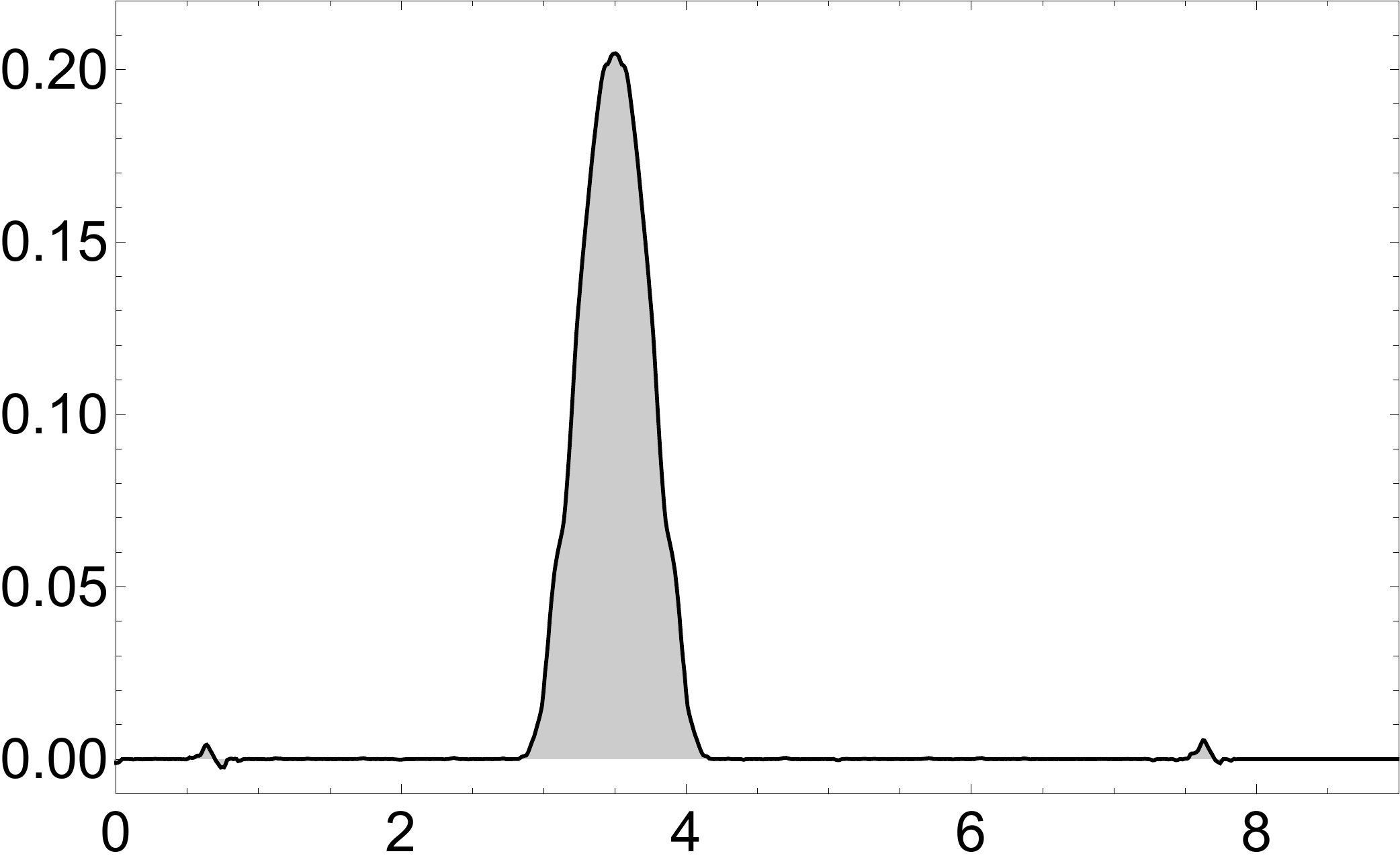}}
\subfigure[]{\includegraphics[width=0.45\textwidth,height=0.15\textwidth, angle =0]{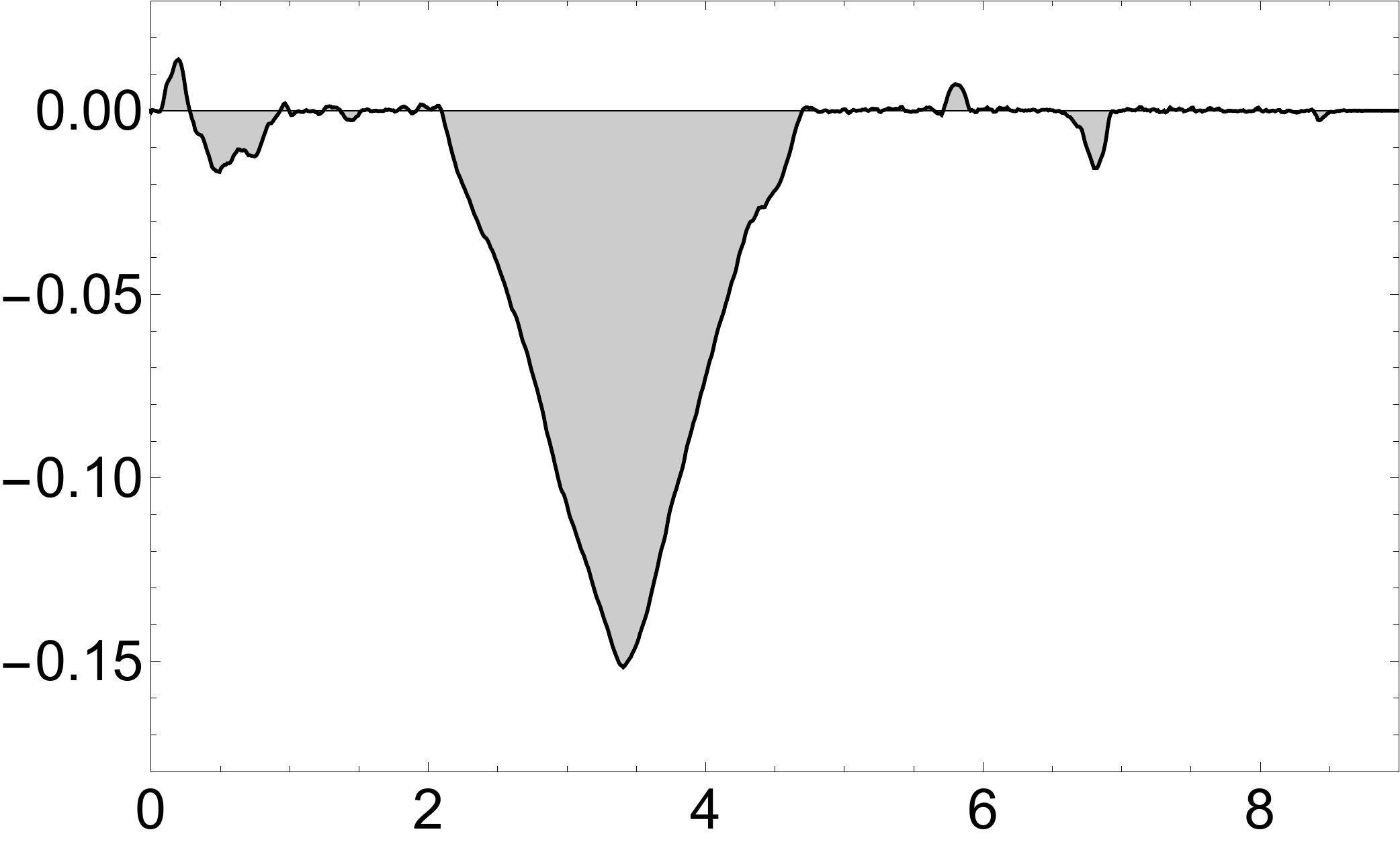}}
\caption{Perturbed breather in the signum-Gordon model on ${\mathbb R}_+$ for $\epsilon=3.0$ at  (a) $t=2.97$, (b) $t=7.436$, (c) $t=9.61$, (d) $t=29.79$.}\label{F19}
\end{figure*}

A novel feature arises when the field does not approach the vacuum value at the origin. This corresponds to a ball-like
 solution with support $r\in [0,r_0]$. Specifically, the boundary conditions are
 \be
\phi(t,r_0)=0,\qquad \partial_r\phi(t,r_0)=0 
\ee
and the behaviour of $\phi$ at the origin is arbitrary. Resultant solutions are just partial solutions of the breather of
the  $\mathbb{R}$ theory. For example, 
if $r_0=\frac{1}{2}$, the relevant breather solution can be obtained from solutions $\phi^{\pm}_{\alpha}(t,x)$ which 
describe the right half of the breather \eqref{sym1}. Namely, the partial solutions have the form
\be
\phi^{\pm}_\alpha(t,r):=\phi^{\pm}_{\alpha}(t,x=r+\textstyle{\frac{1}{2}}),\qquad \alpha=\{C,R_k\},\label{breathercentrum}
\ee
where $\phi^+_{\alpha}(t,x)=-\phi^-_{\alpha}(t-\textstyle{\frac{1}{2}},x)$ and $\phi^-_{R_k}(t,x)=\phi^-_{L_k}(t,1-x)$. 

\vspace*{0.2cm}

Finally, let us say a few words about perturbed breathers. For a shell breather with any perturbation, 
the initial behaviour of the evolution is the same as in the $\mathbb{R}$ model. 
The only qualitative difference arises when the left moving radiation gets to the origin where it bounces off
 and returns to the breather. This leads to a more involved and faster relaxation of the initial state. 

\section{Oscillons in the modified signum-Gordon model}
\subsection{The model}
Let us now consider the modified signum-Gordon model
\be
L=\frac{1}{2}(\partial_t \eta)^2-\frac{1}{2}(\partial_x \eta)^2-V(\eta),\label{BPS-L}
\ee 
where $V(\eta)$ is given by \eqref{Vunfold} or equivalently by \eqref{Vunfold2}. This time we go beyond the (infinitesimal)
 small amplitude approximation and analyze the fate of the breathers of the original signum-Gordon model in which 
 a quadratic term has been added to the potential. Again, we begin our consideration with the model defined on the full
 real line $x \in \mathbb{R}$. Modifications to the half-line model will be discussed later. 

\begin{figure}[h!]
\centering 
\subfigure{\includegraphics[width=0.6\textwidth,height=0.2\textwidth, angle =0]{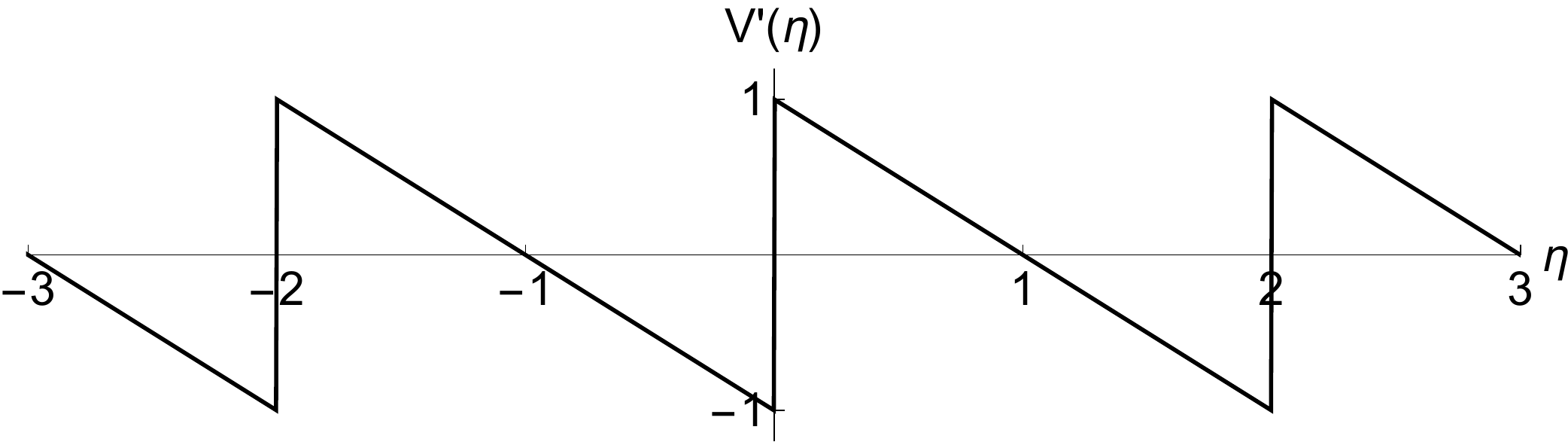}}
\caption{Function $V'(\eta).$}\label{F10}
\end{figure}

The field equation for the unfolded variable $\eta$ is now of the form
\be
(\partial^2_t-\partial^2_x)\eta+V'(\eta)=0,
\ee
where the derivative of the potential is described by a saw-shape function
\be
V'(\eta)=\sum_{n=-\infty}^{\infty}\left[{\rm sgn}(\eta-4n)-(\eta-4n)\right]H_n(\eta).\label{diffV}
\ee
A plot of this function is presented in Fig.\ref{F10}

As long as the amplitude of the field is smaller than the second vacuum $|\eta|<2$,  we can restrict considerations 
to terms $n=0$. This implies that \eqref{bareta} simplifies to $\bar \eta=|\eta|$, 
\eqref{Vunfold} to $V(\eta)=|\eta|-\frac{1}{2}\eta^2$ and \eqref{diffV} to $V'(\eta)={\rm sgn}(\eta)-\eta$. 
The model can then be rewritten as
\be
L=\frac{1}{2}(\partial_t \eta)^2-\frac{1}{2}(\partial_x \eta)^2-\left| \eta \right| +  \frac{1}{2} \eta ^2. \label{L-gen}
\ee 

Note that one can easily generalize this model and introduce a coupling constant in the quadratic term.
 However, such a model 
can be always transformed to the form (\ref{L-gen}). Indeed,  in this case 
\be
L=\frac{1}{2}(\partial_t \eta)^2-\frac{1}{2}(\partial_x \eta)^2-\left| \eta \right| +  \frac{\lambda}{2} \eta^2 
\ee 
and we get a $\lambda$-dependent generalization of the signum-Gordon equation
\be
(\partial^2_t -\partial^2_x) \eta-\lambda \eta+{\rm sgn}(\eta)=0. \label{eqpert1}
\ee
However, this equation can be brought to the $\lambda=1$ form by the following simple change of variables
\be
\eta=\frac{1}{\lambda} \eta' (t',  x'), \;\;\; t'=\sqrt{\lambda} t, \;\; x'=\sqrt{\lambda} x.
\ee
After this transformation has been performed (\ref{eqpert1}) becomes
\be
(\partial^2_t -\partial^2_x) \eta- \eta+{\rm sgn}(\eta)=0, \label{L-gen+}
\ee
in which the primes have been removed. Hence, this extra coupling constant can always be scaled away.
 So from now on we will use $\lambda=1$.

 Let us note  that our transformation has exactly the same form as the symmetry transformation of the signum-Gordon
 equation. Here, it does not produce any new solutions but it allows us to eliminate the coupling constant.
 Of course, the original symmetry of the signum-Gordon model is broken by the quadratic term in the potential. 
This also results in the lack of self-similar solutions. However, once the amplitude of the solution goes to zero,
 the symmetry is effectively restored. Since this happens close to the boundary of the oscillons one can conclude that
the self-similar solutions of the signum-Gordon model should  provide us with a good approximation to oscillons at
 least close to the compacton boundary.

\subsection{Approximate oscillon solutions}
In order to construct an approximate solution which would be the counterpart of the exact breather of the 
signum-Gordon model we assume that the field $\eta(t,x)$ satisfies the identical initial conditions 
as the field $\phi(t,x)$ which describes the exact breather in the model \eqref{mod1}.  From now on we shall use $\eta$ to denote
 the field of the modified signum-Gordon model while $\phi$ will refer to its undeformed version.

We are interested in a solution with support $x\in[0,l]$ where $l$ describes the characteristic size of the oscillon.
 The signum-Gordon breathers characterized by $l$ can be obtained from the basic breather by the scaling
 transformation \eqref{scale}. Since the perturbed model has no scaling symmetry, its solutions cannot be obtained 
from the solution with $l=1$. Thus the solution of the perturbed model must depend on  $l$ from the very beginning.
 For this reason we consider the initial condition
\be
\eta(0,x)=0 \qquad {\rm and}\qquad\partial_t \eta(t,x)|_{t=0}=v_l(x),\label{ic2}
\ee 
where $v_l(x)=|x-\frac{l}{2}|-\frac{l}{2}$ for $x\in[0,l]$ and $v_l(x)=0$ outside this segment.

The fundamental difficulty which we have to face is the fact that we do not know the general form  of the solution 
of \eqref{eqpert1} (as general as \eqref{gensol}). To make any progress with this problem we make an approximation 
which is similar in nature to what was made in \cite{Klimas}.  This approximation led to the study of the evolution of some self-similar
 initial data in a model with a broken scaling symmetry \cite{Klimas}. We have to stress, however, that, in general, it is
 not possible to construct any analytical solution which is valid for all times. This origin of this difficulty lies in problems with 
the determination of trajectories of some zeros of the partial solutions and resides in problems with analytical calculation
 of some integrals. The analytical solutions presented in this section contain all partial solutions which can be obtained 
without having to determine their zeros. 

\subsubsection{Solution for $t\in[0,\frac{l}{2}]$}
The partial solutions, which follow directly from the initial conditions, can be obtained by the reduction of the partial 
differential equation \eqref{eqpert1} to a system containing three ordinary differential equations.
 As it was pointed out in \cite{Klimas}, the perturbative method applied to a self-similar initial data leads to an exact 
solution. However, it also requires some extremely lengthy computations.

Fortunately, a solution can be obtained almost immediately with the help of the following ansatz:
\be
\eta(t,x)=a(t)x^2+b(t)x+c(t),\label{ansatz}
\ee
where the coefficients $a(t)$, $b(t)$ and $c(t)$ obey, respectively, the equations: 
\begin{align}
a''(t)-a(t)&=0,\label{ho1}\\
b''(t)-b(t)&=0,\label{ho2}\\
c''(t)-c(t)&=2a(t)+1,\label{ho3}
\end{align}
and where ${\rm sgn}(\eta)=-1$ at the beginning of the evolution due to the initial condition
 $\partial_t\eta(0,x)=v_l(x)<0$.

The equations \eqref{ho1}-\eqref{ho3} have solutions
\begin{align}
a(t)&=a_1\sinh(t)+a_2\cosh(t),\\
b(t)&=b_1\sinh(t)+b_2\cosh(t),\\
c(t)&=c_1\sinh(t)+c_2\cosh(t)+\int_0^{\infty}dt'D(t-t')(2a(t')+1),
\end{align}
where $D(t)=\theta(t)\sinh(t)$ is the fundamental solution of the equation $c''(t)-c(t)=\delta(t)$.
The integration constants $a_1,\cdots, c_2$ must be chosen so that \eqref{ic2} holds.

Such partial solutions are the counterparts of $l^2\phi^{-}_{L_1/R_1}(\frac{t}{l},\frac{x}{l})$ of section \eqref{exbreather},
 and so we denote them as $\eta^{-}_{L_1/R_1}(t,x)$.
 Note that the solutions $\phi^{-}_{L_1/R_1}(t,x)$ are self-similar and so expressions
 $l^2\phi^{-}_{L_1/R_1}(\frac{t}{l},\frac{x}{l})$ and $\phi^{-}_{L_1/R_1}(t,x)$ are equal. 
 The initial conditions \eqref{ic2} fix the free constants and we get:
\begin{align}
\eta^{-}_{L_1}(t,x)&=-x\sinh(t)+\cosh(t)-1,\label{fL1}\\
\eta^{-}_{R_1}(t,x)&=\eta^{-}_{L_1}(t,l-x).\label{fR1}
\end{align}

The partial solutions \eqref{fL1} and \eqref{fR1} have  supports $x\in[t,-t+\frac{l}{2}]$ and  $x\in[\frac{l}{2}+t,-t+l]$. 
The supports shrink to single points at $t=\frac{l}{4}$.

Note that, these are the {\it exact} partial solutions of the perturbed model.
 Solutions \eqref{fL1} and \eqref{fR1} tend to $\phi^{-}_{L_1/R_1}(t,x)$ in the limit of small amplitudes
 ($t\ll1$ for $l\ll 1$).

The counterpart of the solution $l^2\phi_C^-(\frac{t}{l},\frac{x}{l})$, which we will call $\eta_{C}^-(t,x)$,
 where $x\in[-t+\frac{l}{2}, t+\frac{l}{2}]$, must have a more general form than the solution given by \eqref{ansatz}.
 As the general solution of the perturbed equation is not known we approximate it by a solution of
 the non-homogeneous wave equation replacing term proportional to $\eta^-_{C}(t,x)$ by
 $ l^2\phi^-_C(\frac{t}{l},\frac{x}{l})$. A similar approximation can be made for other partial solutions 
$\eta^-_{L_2/R_2}(t,x)$ and  $\eta^-_{L_3/R_3}(t,x)$.

 For this reason we consider the following {\it approximate} equation
\be
(\partial^2_t-\partial^2_x)\eta^-_\alpha(t,x)={\textstyle1+l^2\phi^-_\alpha(\frac{t}{l},\frac{x}{l})},\label{eqapprox}
\ee
where $\alpha=\{C,L_2,R_2,L_3,R_3\}$. It is easy to see that equation \eqref{eqapprox} possesses a solution of the form
\be
\eta^-_\alpha(t,x)=F_{\alpha}(x+t)+G_{\alpha}(x-t)+h_{\alpha}(t,x),
\ee
where $h_{\alpha}(t,x)$ is a particular solution of the nonhomogeneous equation \eqref{eqapprox} and has the form: 
\be
h_{\alpha}(t,x)=-{\textstyle\frac{1}{4}}(x^2-t^2)+I_{\alpha}(t,x),
\ee
where
\begin{align}
I_{\alpha}(t,x)&=-
\frac{l^2}{4}\int_0^{x+t}du\int_0^{x-t}dw\, \phi^-_\alpha{\textstyle\left(\frac{u-w}{2l},\frac{u+w}{2l}\right)}.\label{int}
\end{align}

The functions $F_{\alpha}(x+t)$ and $G_{\alpha}(x-t)$ are determined by the matching conditions at surfaces 
of the light cones $x=\pm t+\frac{l}{2}$, $x=t$ and $x=-t+l$. We need to compute explicitly
 only the solutions $\eta^-_C$,  $\eta^-_{L_2}$ and $\eta^-_{L_3}$. The remaining partial solutions can be obtained
 by performing the transformation $x\rightarrow l-x$ applied to the left-hand-side partial solutions.
The integrals \eqref{int}  take the form
\begin{align}
I_C(t,x)&=-{\textstyle\frac{1}{64}(x^2-t^2)\left(2l^2-4l(x+t)+3x^2+5t^2\right)},\\
I_{L_2}(t,x)&={\textstyle\frac{1}{192} (x^2-t^2)(t^2+7x^2)},\\
 I_{L_3}(t,x)&=-{\textstyle\frac{1}{192} (x^2-t^2)} (6l^2 - 12l(t+x)+ 7 t^2 + 16 t x + x^2).
\end{align}

The solution $\eta^-_C(t,x)$ satisfies the boundary conditions 
$${\textstyle\eta^-_C(t,-t+\frac{l}{2})=\eta^-_{L_1}(t,-t+\frac{l}{2}),\qquad\eta^-_C(t,t+\frac{l}{2})=\eta^-_{R_1}(t,t+\frac{l}{2})},$$
and so takes the form 
\begin{align}
\eta^-_C(t,x)&= \cosh\left(\frac{x+t}{2}-\frac{l}{4}\right)+\cosh\left(\frac{x-t}{2}-\frac{l}{4}\right)-2\nonumber
\\
&+\left(\frac{x+t}{2}-\frac{3l}{4}\right)\sinh\left(\frac{x+t}{2}-\frac{l}{4}\right)+\left(\frac{x-t}{2}+\frac{l}{4}\right)\sinh\left(\frac{x-t}{2}-\frac{l}{4}\right)\nonumber\\
&-\frac{1}{4}\left(x+t-\frac{l}{2}\right)\left(x-t-\frac{l}{2}\right)\left[1+\frac{1}{64}(3l^2-16lt+20t^2-12lx+12x^2)\right].
\end{align}


The solution $\eta^-_{L_2}(t,x)$ has to match the partial solution $\eta_{L_1}(t,x)$ on the light cone $x=t$
 so we require that $\eta^-_{L_2}(t,t)=\eta^-_{L_1}(t,t)$. At the other end it must match the vacuum solution $\eta=0$. 
If the matching point $x_0(t)$ belongs to the light cone then there is no condition on $\partial_x\eta$
 of the partial solution at  $x_0(t)$. Otherwise, the derivative with respect to $x$ must vanish. In such a case 
we have two conditions at $x_0(t)$ and one condition at $x=t$. Such a problem can be solved if another partial solution is taken into account. In general $x_0$ is a function of time. In order to avoid such difficulties we choose boundary condition $\eta^-_{L_2}(t,0)=0$.  This condition is very simple and consistent with our small amplitude approximation. According to our numerical analysis such a condition is well satisfied for $l$ up to $l\approx 1.3$.    The partial solution which satisfies matching conditions at $x=0$ and $x=t$ is of the form
\begin{align}
\eta^-_{L_2}(t,x)&=\cosh\left(\frac{x+t}{2}\right)-\cosh\left(\frac{x-t}{2}\right)-\frac{x+t}{2}\sinh\left(\frac{x+t}{2}\right)+\frac{x-t}{2}\sinh\left(\frac{x-t}{2}\right) \nonumber
\\& +\frac{1}{48}x(x-t)\left(2x^2+tx+t^2-24\right).
\end{align}
The right hand side partial solution is given by  $\eta^-_{R_2}(t,x)=\eta^-_{L_2}(t,l-x)$.

The solution $\eta^-_{L_3}(t,x)$ has to satisfy the boundary conditions $\eta^-_{L_3}(t,-t+\frac{1}{2})=\eta^-_{L_2}(t,-t+\frac{1}{2})$ and $\eta^-_{L_3}(t,t)=\eta^-_C(t,t)$. A partial solution which satisfies these conditions is of the form
\begin{align}
\eta^-_{L_3}(t,x)&=\cosh\left(\frac{x+t}{2}-\frac{l}{4}\right)-\cosh\left(\frac{x-t}{2}\right)+\cosh\left(\frac{l}{4}\right)-1\nonumber\\
&+\left(\frac{x+t}{2}-\frac{3l}{4}\right)\sinh\left(\frac{x+t}{2}-\frac{l}{4}\right)+\frac{x-t}{2}\sinh\left(\frac{x-t}{2}\right)-\frac{l}{4}\sinh\left(\frac{l}{4}\right)\nonumber\\
&+\frac{1}{3072}(2t-l)[9l^3-26l^2t+64t^3+8l(24-5t^2)]\nonumber\\
&+\frac{x}{768}[48t(t^2+8)-12l(t^2-8)-54l^2t+19l^3]\nonumber\\
&+\frac{x^2}{256}[28lt-17l^2-128]+\frac{5x^3}{192}(3l-4t).
\end{align}
\begin{figure*}[h]
\centering
\subfigure[]{\includegraphics[width=0.4\textwidth,height=0.15\textwidth, angle =0]{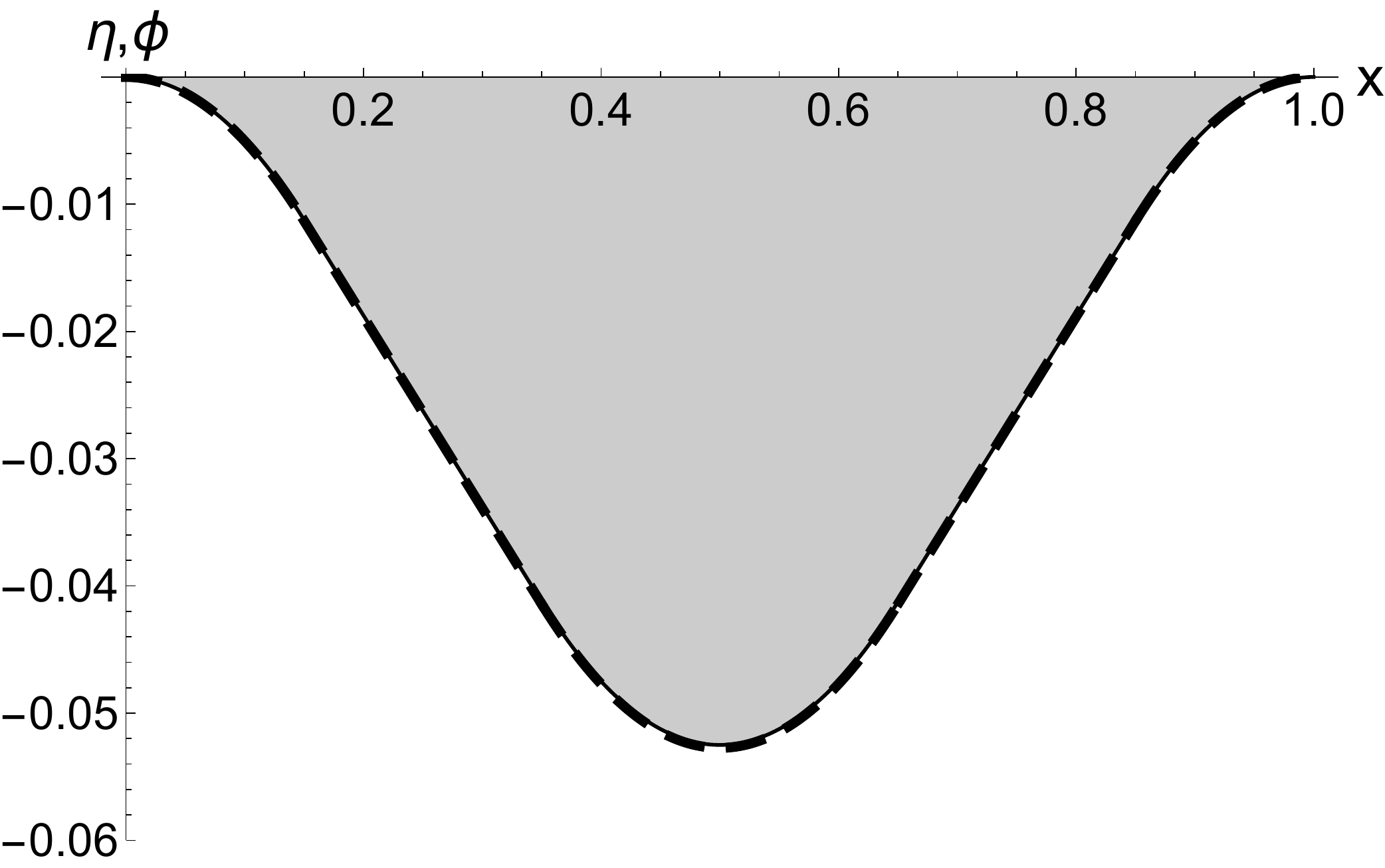}}
\subfigure[]{\includegraphics[width=0.4\textwidth,height=0.15\textwidth, angle =0]{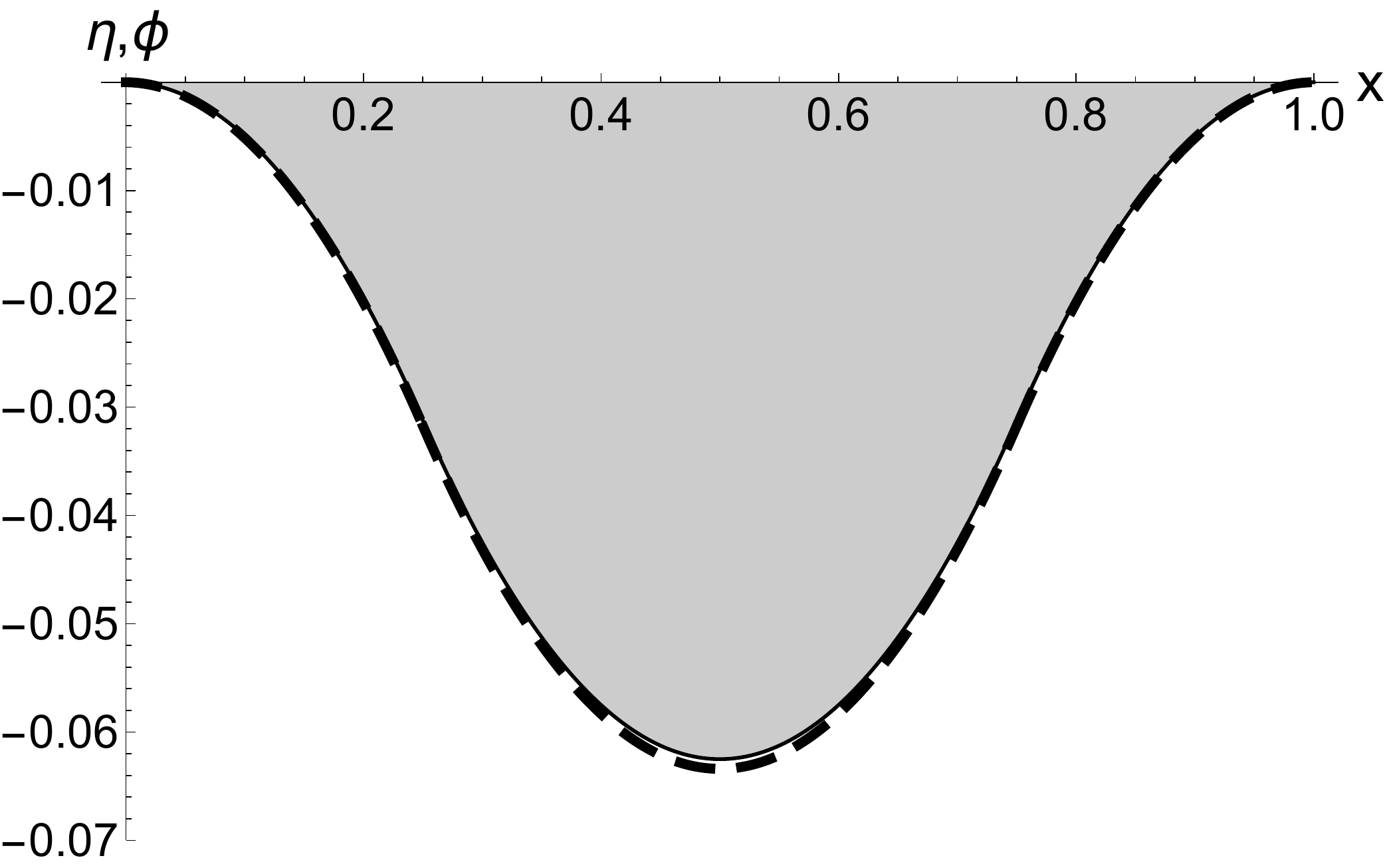}}
\subfigure[]{\includegraphics[width=0.4\textwidth,height=0.15\textwidth, angle =0]{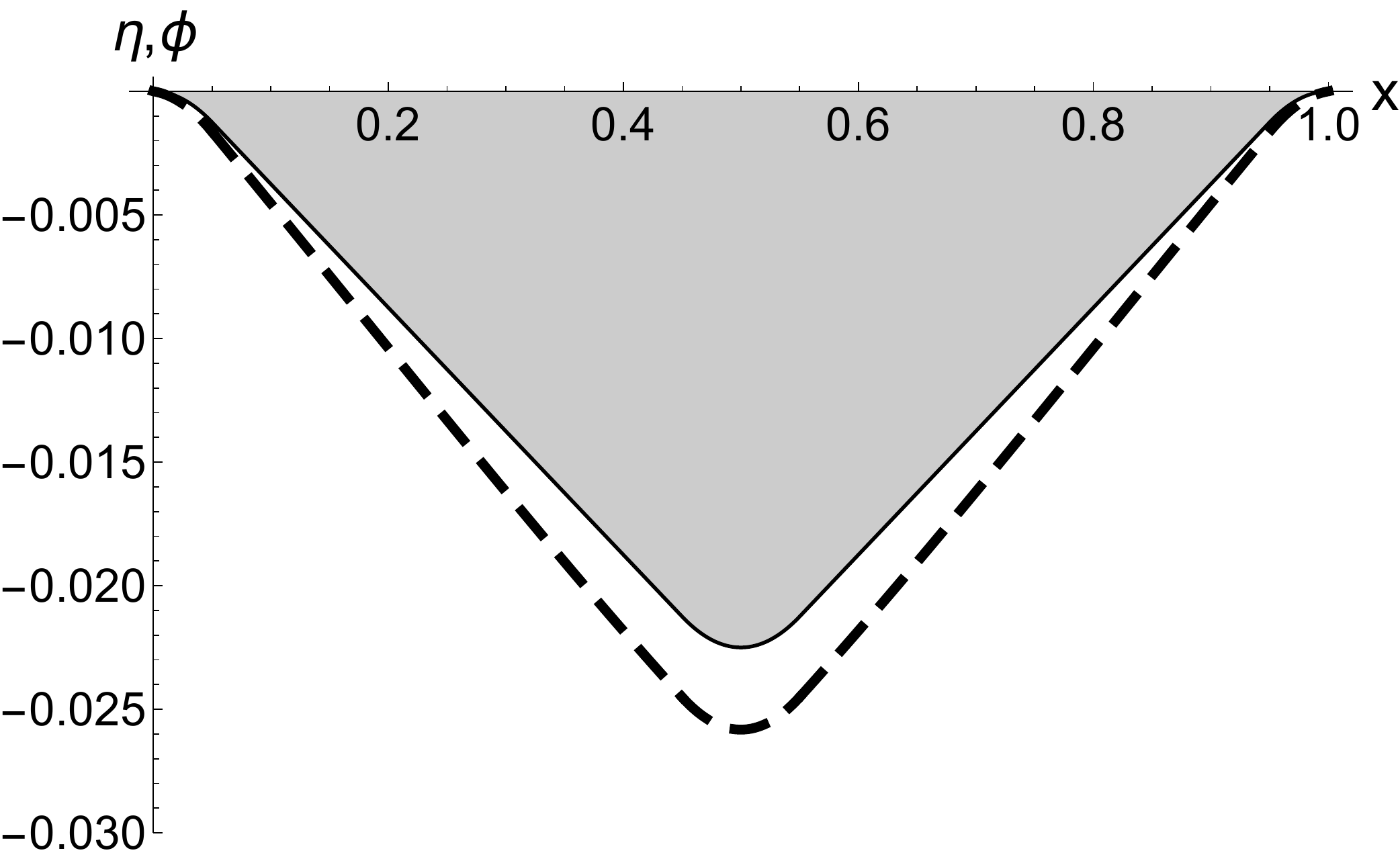}}
\subfigure[]{\includegraphics[width=0.4\textwidth,height=0.15\textwidth, angle =0]{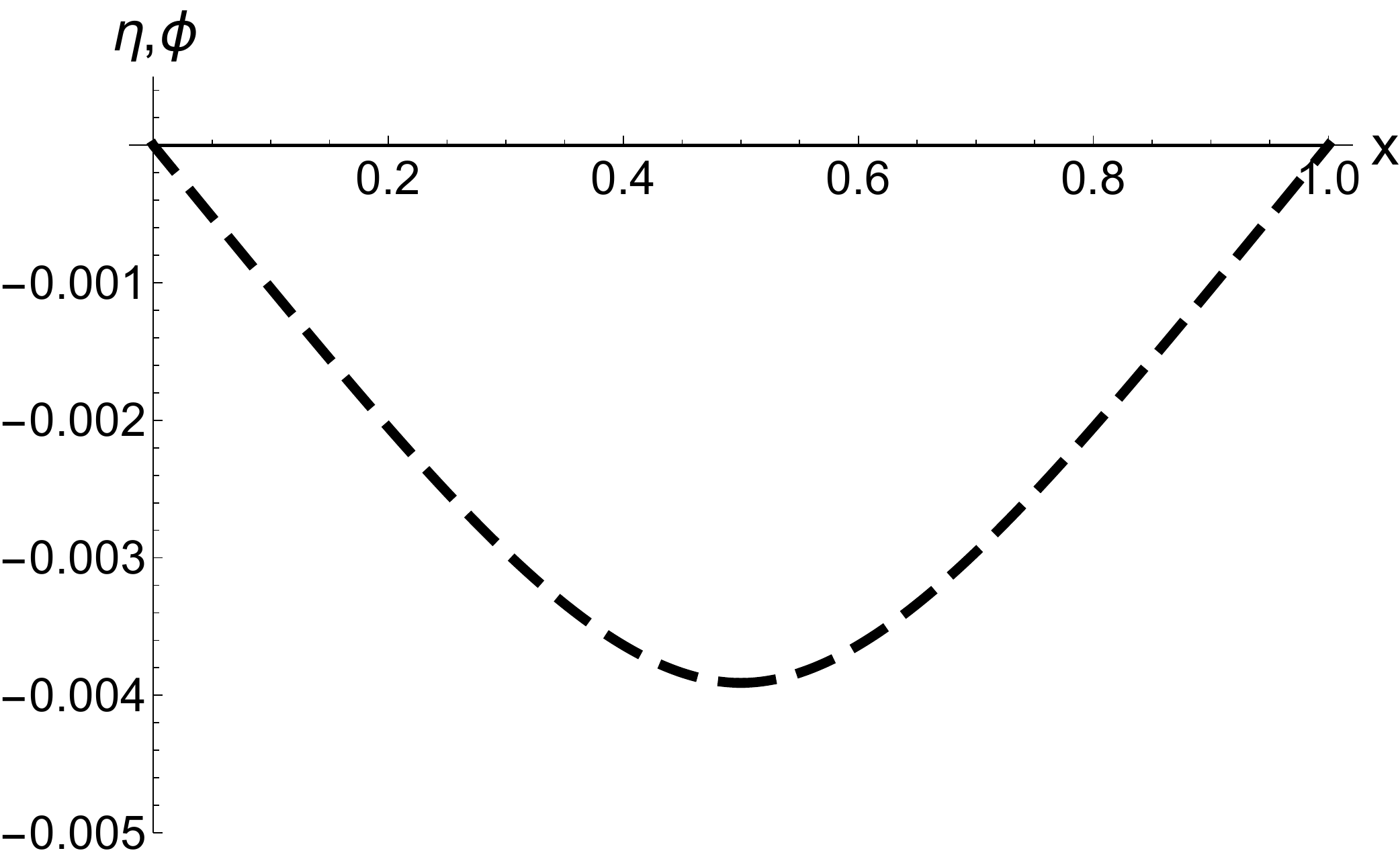}}
\caption{The exact breather $\phi(t,x)$ in the signum-Gordon model (shadowed region) and the approximated oscillon solution $\eta(t,x)$ (dashed line) in the perturbed model with $l=1$. The figures correspond to (a) $t=0.15$, (b) $t=0.25$, (c) $t=0.45$, (d) $t=0.5$.}\label{F11}
\end{figure*}

In Fig.\ref{F11} we present four time snapshots which compare the evolution of the analytical approximate solutions for
 $l=1$ in the perturbed model and the exact solution in the non-perturbed model. For $t$ being close to zero, 
the breather $\phi$ and the oscillon $\eta$ remain close to each other. The difference between them 
becomes visible only for larger $t$, approximately $t> 0.25$.  Note, that the partial solutions $\eta^-_{L_3}(t,x)$ and $\eta^-_{R_3}(t,x)$ do not vanish at $t=\frac{1}{2}$, Fig.\ref{F11}(d).   

The non-vanishing of the approximate solution at $t=\frac{1}{2}$ suggests that the period of the oscillon in the
 perturbed model is higher that the period of the exact signum-Gordon breather. As  the approximate solution is quite close 
to zero at $t=l/2$ we expect that its period is slightly larger than this period in the signum-Gordon model.
 We will come back to this feature in the next subsection.

\subsubsection{Solution for $t\in[\frac{l}{2},t^*]$}

The instant of time $t^*$ is defined by the condition $\eta(t^*,x=\frac{l}{2})=0$ {\it i.e.} it corresponds to the
 instant of time in which the solution starts to change its sign at the centre of the oscillon {\it i.e.} at $x=\frac{l}{2}$.
 Such behaviour of the field is observed in numerical simulations that will be the subject of the next section. 
In this subsection we will estimate the value of $t^*$. Although we are not able to construct the analytical solution
 for one complete oscillation, we can use the value of the time $t^*$ to estimate the period of the oscillon as $T=2t^*$.

In order to compute $t^*$ one needs to find a solution $\eta(t,x)$ in the region containing the point $x=\frac{l}{2}$. 
Since $\eta(\frac{l}{2},x)\ll \eta(\frac{l}{4},x)$ then in the interval $[\frac{l}{2},t^*]$ we have $-\eta\ll 1$. 
In such a case the absolute value of the perturbation $-\eta$ in the field equation is much smaller than absolute value
 of the term ${\rm sgn}(\eta)$. This justifies the approximation of the field equation by the signum-Gordon equation. 
The evolution of the field $\eta(t,x)$ in the interval $[\frac{l}{2},t^*)$ can be approximated by the solution of the
non-homogeneous wave equation $(\partial^2_t-\partial^2_x)\eta=1$ which satisfies the initial conditions 
\begin{eqnarray}
{\textstyle\eta(\frac{l}{2},x)}=\left\{\begin{array}{lcc}
f(x)&{\rm for}&0\le x\le \frac{l}{2} \\
f(l-x)&{\rm for}&\frac{l}{2} \le x\le l
\end{array}\right.
\qquad {\rm where}\qquad f(x):={\textstyle \eta_{L_3}(\frac{l}{2},x)}
\end{eqnarray}
and
\begin{eqnarray}
{\textstyle\partial_t\eta(t,x)}|_{\frac{l}{2}}=\left\{\begin{array}{lcc}
g(x)&{\rm for}&0\le x\le \frac{l}{2} \\
g(l-x)&{\rm for}&\frac{l}{2} \le x\le l
\end{array}\right.
\qquad {\rm where}\qquad
g(x):=\left.\partial_t\eta_{L_3}(t,x)\right |_{\frac{l}{2}}.
\end{eqnarray}

Let us denote by $\eta_{L_4}(t,x)$ the partial solution on the segment $x\in[t-\frac{l}{2},-t+l]$ and by $\eta_{R_4}(t,x)$ the
 partial solution on $x\in[t,-t+\frac{3l}{2}]$. Note that $\eta_{R_4}(t,x)=\eta_{L_4}(t,l-x)$. The solution $\eta_{L_4}(t,x)$ is given by the expression
\be
\eta_{L_4}(t,x)=\frac{l^2}{8}+\frac{t}{2}(t-l)+\frac{1}{2}\left[{\textstyle f(x+t-\frac{l}{2})
+f(x-t+\frac{l}{2})}\right]+\frac{1}{2}\int_{x-t+\frac{l}{2}}^{x+t-\frac{l}{2}}dw\,g(w).\label{etaL4}
\ee
The function $g(w)$ is quite complicated so the analytical integration of the integral in \eqref{etaL4}
is not possible in this case. Fortunately, for $l$ not too large (comparing it with $l=1$), the initial velocity 
at $t=\frac{l}{2}$ is very close to the expression  $-v(x)$, where $v(x)$ is given by \eqref{rightvelocity}. 
Thus we see that $g(x)\approx x$ and so \eqref{etaL4} becomes
\be
\eta_{L_4}(t,x)=\left(t-\frac{l}{2}\right)\left(x+\frac{t}{2}-\frac{l}{4}\right)+\frac{1}{2}\left[{\textstyle f(x+t-\frac{l}{2})+f(x-t+\frac{l}{2})}\right].
\ee

The partial solution $\eta_D(t,x)$ emerges at the segment $x\in[-t+l,t]$ and it matches the solutions 
$\eta_{L_4/R_4}(t,x)$ at the light cones. After some further computations we find that
\be
\eta_D(t,x)=\frac{1}{2}\left[{\textstyle f(-x-t+\frac{3l}{2})+f(x-t+\frac{l}{2})}\right]-\frac{1}{8}[3l^2-4lt+4x(x-l)].
\ee
The other partial solutions that match $\eta_{L_4/R_4}(t,x)$ at $x=t-\frac{l}{2}$ and
 $x=-t+\frac{3l}{2}$ are not needed for our purpose so we do not present them here. 

Using the expressions presented above we note that the condition $\eta_D(t^*,\frac{l}{2})=0$ takes the form
\be
f(l-t^*)+\frac{1}{4}l(2t^*-l)=0.\label{eqt*}
\ee

\begin{figure*}[h] 
\centering
\subfigure[]{\includegraphics[width=0.45\textwidth,height=0.3\textwidth, angle =0]{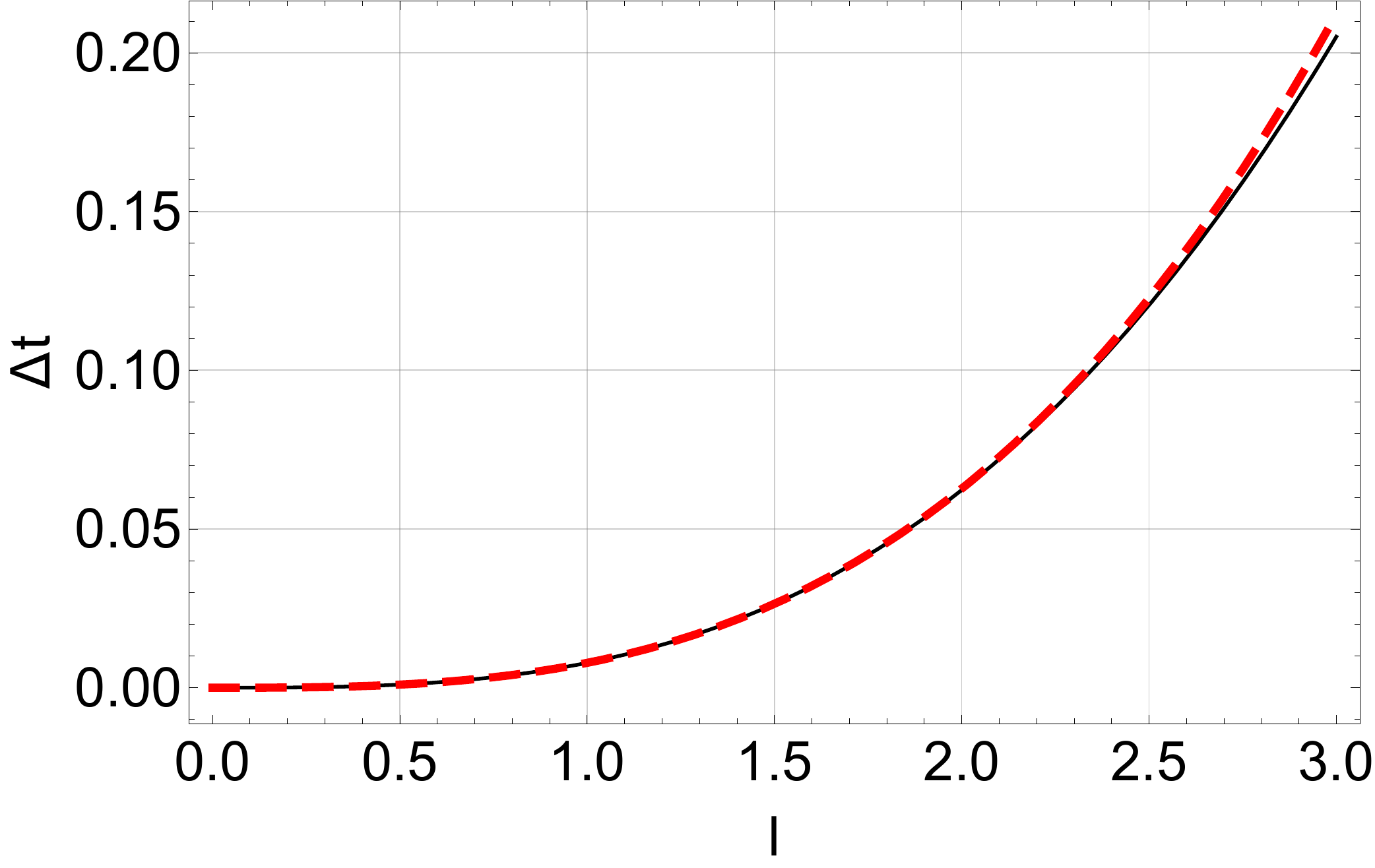}}
\subfigure[]{\includegraphics[width=0.45\textwidth,height=0.3\textwidth, angle =0]{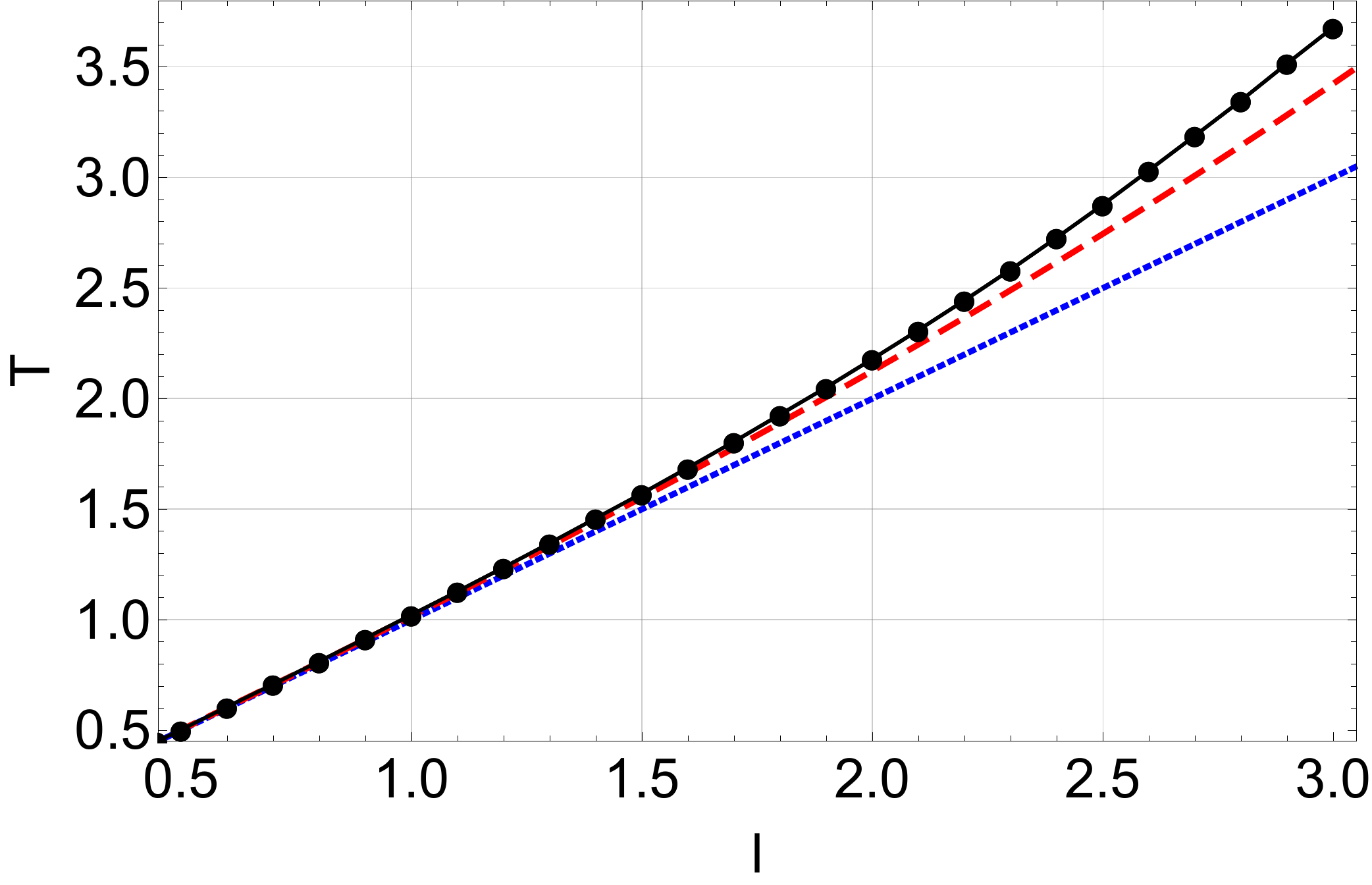}}
\caption{(a) The expression $\Delta t$, where the solid line is a numerical solution of \eqref{eqt*} whereas 
the dashed line represents the expression \eqref{deltat}. (b) The period of the oscillon $T(l)$. The solid line with dots stands for result obtained from numerical simulation. The dashed line represents analytical expression \eqref{periodoscilation}. The dotted straight line stands for the period of the exact breather.}\label{Fig-period}
\end{figure*}

Unfortunately, we cannot solve \eqref{eqt*} exactly due to its complexity. However, it is still possible to obtain 
its numerical and approximate analytical solutions.  In order to get an analytical expression for $t^*$ we expand the left hand side
 of \eqref{eqt*} around $t^*=\frac{l}{2}$ up to the linear term and solve the resultant linear equation 
\be
b_0+b_1(t^*-{\textstyle \frac{l}{2}})+{\cal O}((t^*-{\textstyle \frac{l}{2}})^2)=0.\label{lin-eqt*}
\ee
The solution of \eqref{lin-eqt*} gives us
\be
t^*=\frac{l}{2}+\Delta t,
\ee
where
\be
\Delta t:=-\frac{b_0}{b_1}=\frac{3}{4}\,\frac{2048[\cosh(\frac{l}{4})-1]+512l\sinh(\frac{l}{4})+3l^3-64l^2}{96l\cosh(\frac{l}{4})-768\sinh(\frac{l}{4})-l^3+48l}>0.\label{deltat}
\ee
In Fig.\ref{Fig-period}(a) we present the plot of the solutions of equation \eqref{eqt*} as functions of $l$. The solid line represents the
numerical solution of this equation, whereas the dashed line corresponds to the expression \eqref{deltat}. It turns out that
 our approximation of the solution by means of a power series expansion is quite good. The error of this approximation
takes the value of $3.35\,\%$ for $l=3$,  $0.73\, \%$ for $l=2$ and only $0.048\, \%$ for $l=1$. 

The period of oscillation can be approximated by 
\be
T=2t^*=l+2\Delta t.\label{periodoscilation}
\ee

In figure Fig.\ref{Fig-period}(b) we plot the period of the oscillation  \eqref{periodoscilation}.
 This period is represented by the dashed curve. The dotted straight line represents the period of the
exact breather in the signum-Gordon model. One can see from this plot that the period of the oscillon grows faster 
than linearly with the parameter $l$.
 Moreover, we have also compared this expression with the period obtained from the numerical simulations. 
This is represented by the solid curve which interpolates between the dots. We see that our analytical approximation
 is quite good for $l$ not being significantly larger than $l=2$.  We also note that the analytical computation
 gives the period which is shorter then the period obtained by the numerical calculations.

Naturally, one can ask how close or far away is the approximate solution from the true solution?
 In order to try to answer this question and get some ideas about correctness of our approximate solutions we have 
solved the equation \eqref{eqpert1} numerically  and in the next section we present and discuss some of our results.

\subsection{Numerical solutions}

In this section we discuss our numerical studies of the solutions in the perturbed model. In this discussion we
 concentrate our attention on initial configurations which differ from each other by values of the scale parameter $l$ or of the parameter $\epsilon$.

\subsubsection{Oscillons}
The first class of initial data is given by the configuration which is appropriate to obtain an exact breather in 
the signum-Gordon model {\it i.e.} this is exactly the configuration  \eqref{ic2}, which has been already studied
 in the previous section in the context of approximate analytical solutions. The initial configuration is of the form
\begin{eqnarray}
\eta(0,x)=0,\qquad \partial_t\eta(t,x)|_{t=0}=\epsilon\, v_l(x).\label{inieps}
\end{eqnarray}
In this section we put $\epsilon=1$ and $v_l(x)$ is nontrivial on the segment $x\in[0,l]$. More general 
configurations (with $\epsilon\neq 1$) are discussed in the next section. 
\begin{figure*}[h] 
\centering
\subfigure[]{\includegraphics[width=0.3\textwidth,height=0.15\textwidth, angle =0]{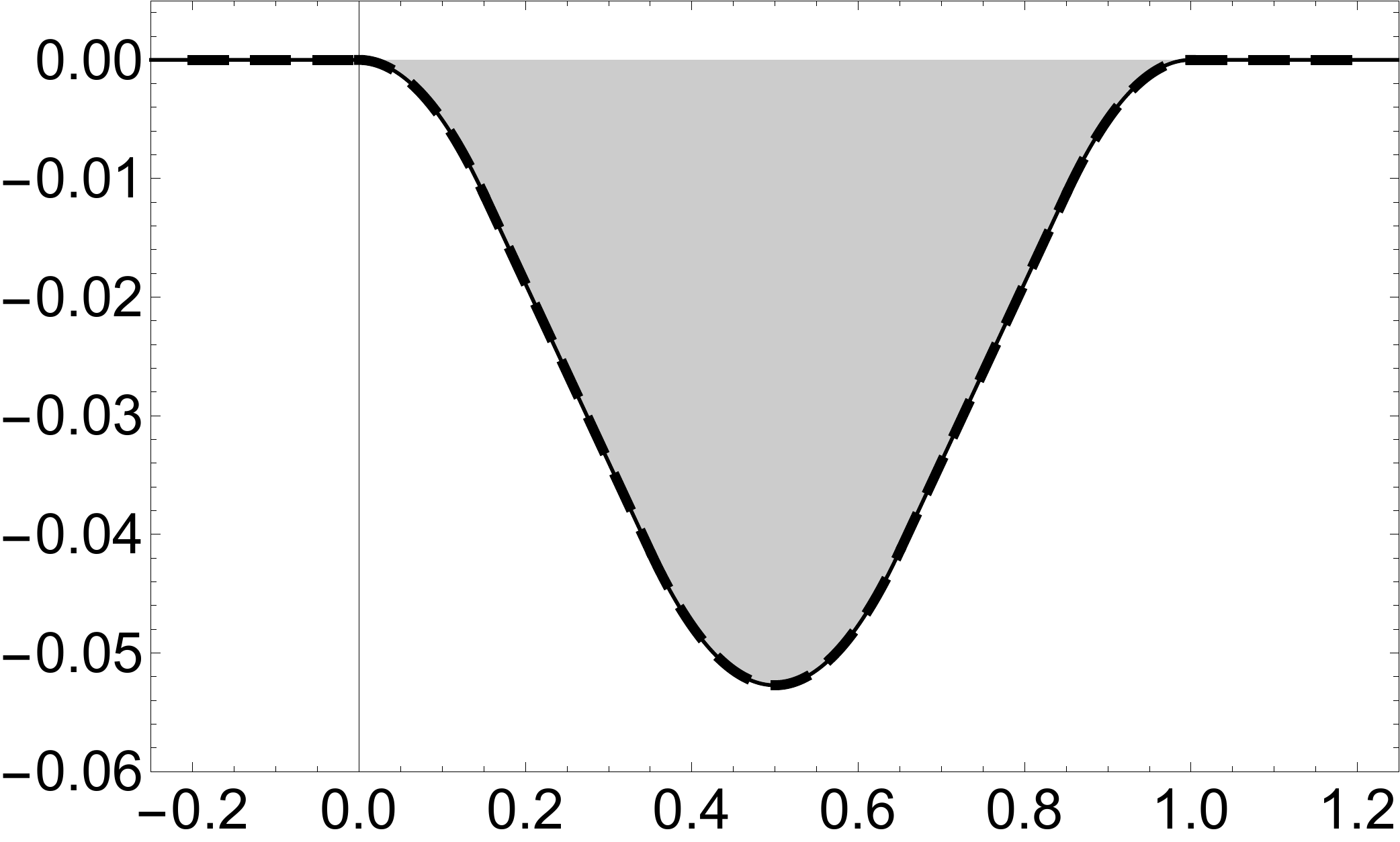}}
\subfigure[]{\includegraphics[width=0.3\textwidth,height=0.15\textwidth, angle =0]{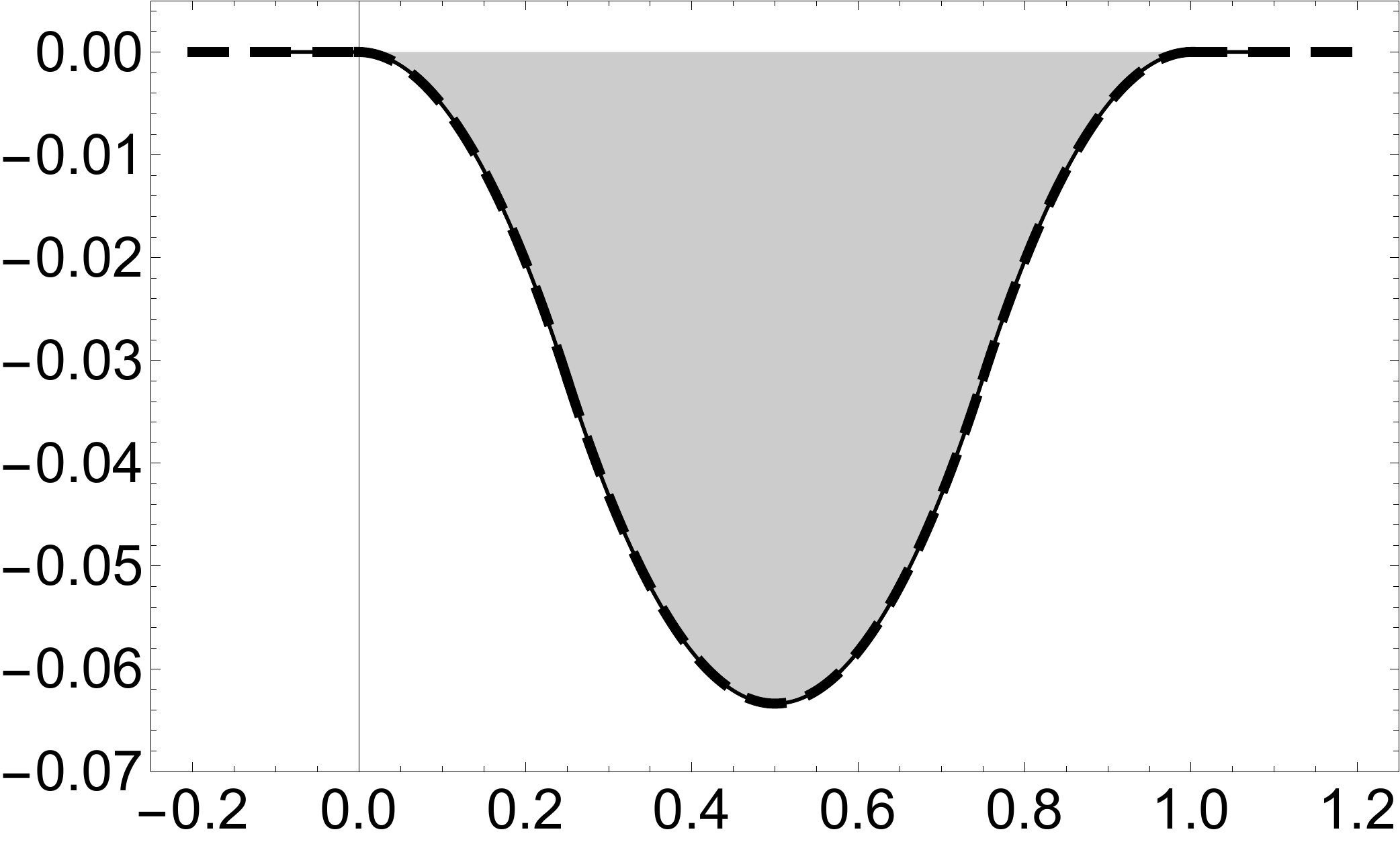}}
\subfigure[]{\includegraphics[width=0.3\textwidth,height=0.15\textwidth, angle =0]{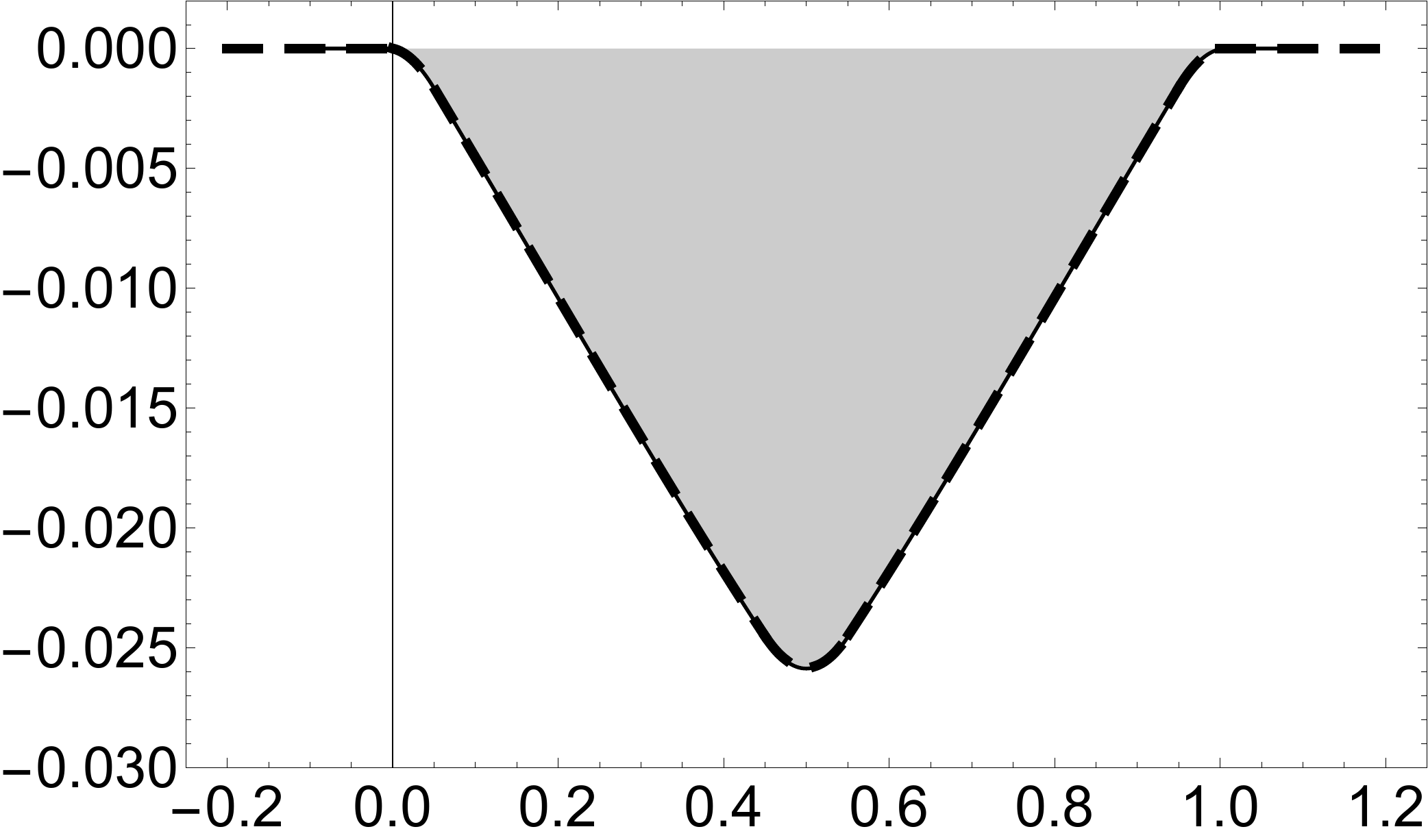}}
\subfigure[]{\includegraphics[width=0.3\textwidth,height=0.15\textwidth, angle =0]{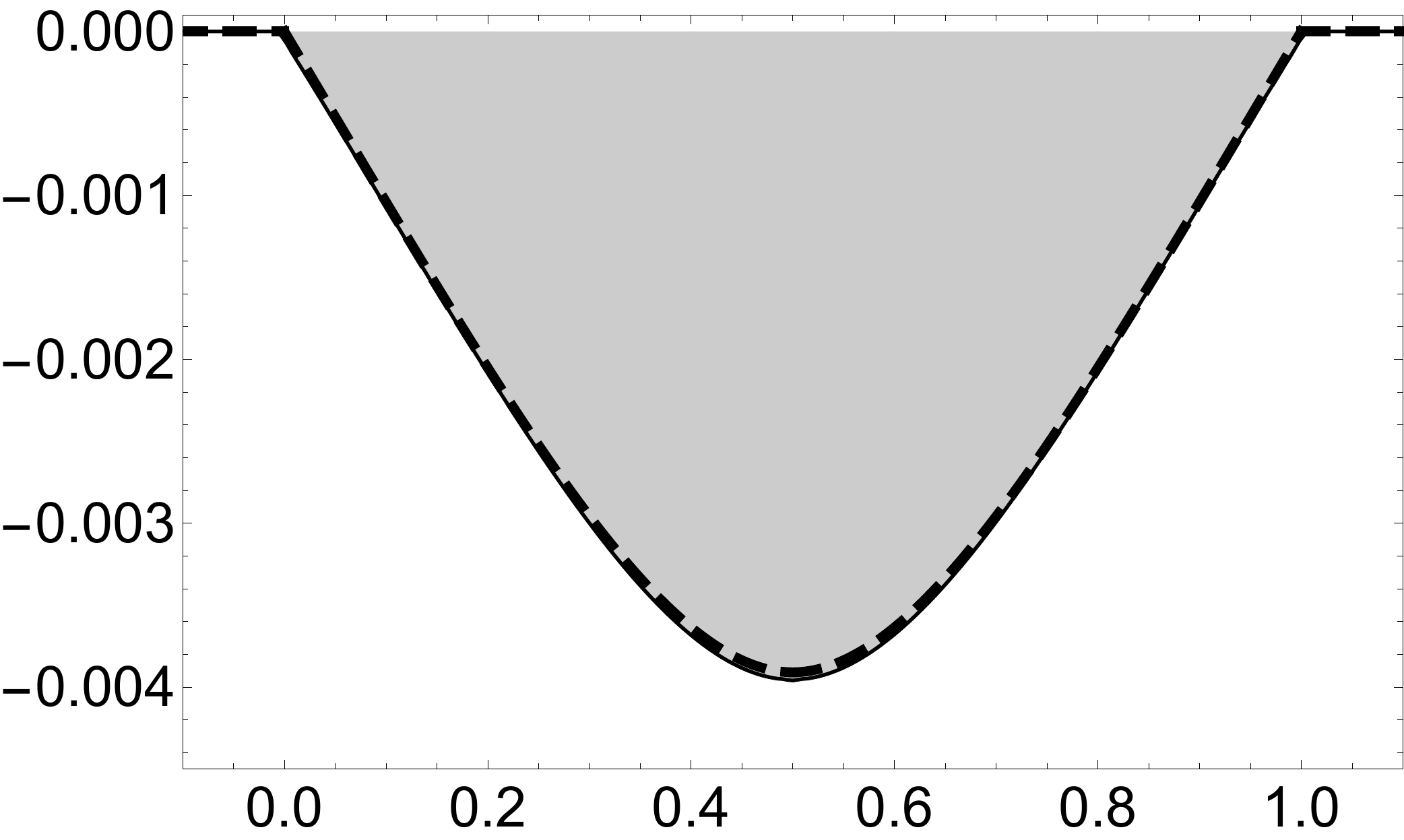}}
\subfigure[]{\includegraphics[width=0.3\textwidth,height=0.15\textwidth, angle =0]{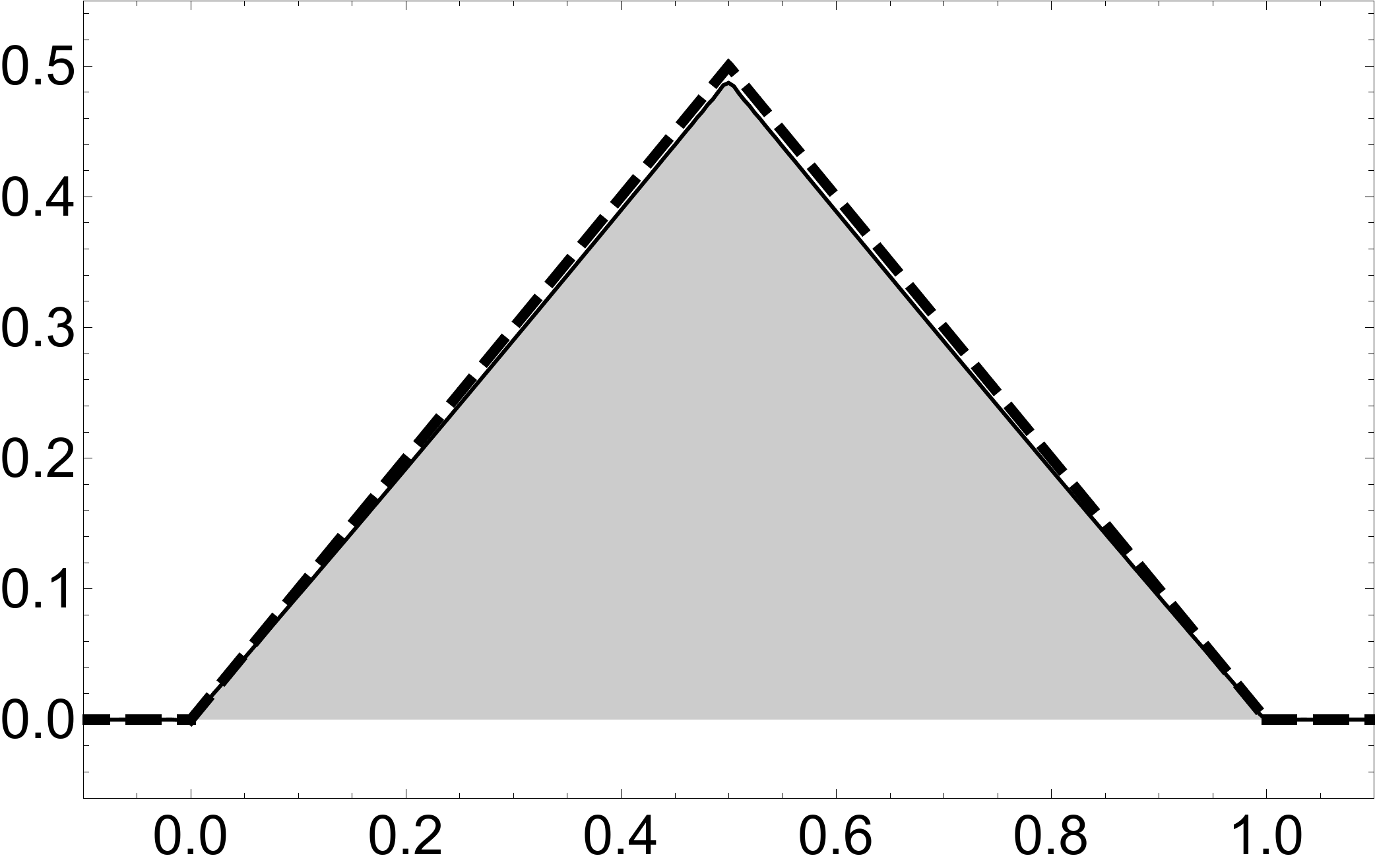}}
\subfigure[]{\includegraphics[width=0.3\textwidth,height=0.15\textwidth, angle =0]{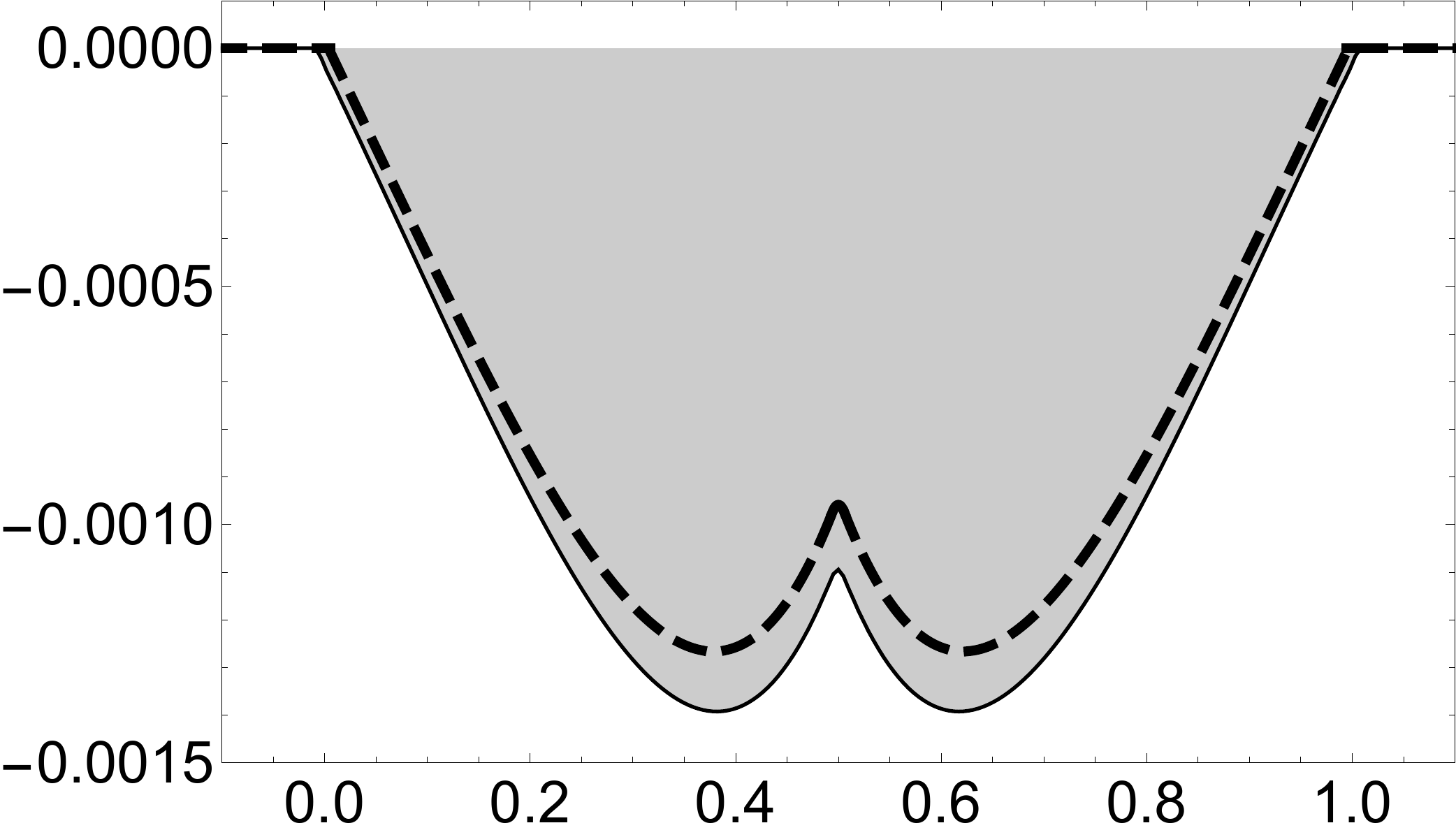}}
\caption{The numerical solutions $\eta(t,x)$ (shadowed region under the solid curve) and the approximated oscillon solution (dashed curve) in the perturbed model \eqref{eqpert1} for $\epsilon=1$ and $l=1$ at  (a) $t=0.15$, (b) $t=0.25$, (c) $t=0.45$, (d) $t=0.5$. (e) The derivative $\partial_t\eta$ at  $t=0.5$. (f) The solution at $t=0.5059$.}\label{F12}
\end{figure*}
In other words, here we present the results of evolving the exact signum-Gordon breather in the
 modified model (\ref{L-gen+}). As the exact breather of the signum-Gordon model with the size $l$ has amplitude
 $A=l^2/16$ one may expect that for not too large values of $l$ the true solution of the modified theory would not differ too much 
from the original signum-Gordon breather. These expectations are based on the fact that the analyzed modification is
 a subleading term for small amplitude solutions. On the other hand we know that even a small perturbation of
 the signum-Gordon model could strongly influence the stability of the topologically unprotected solutions.

\begin{figure*}[h] 
\centering
\subfigure[]{\includegraphics[width=0.3\textwidth,height=0.15\textwidth, angle =0]{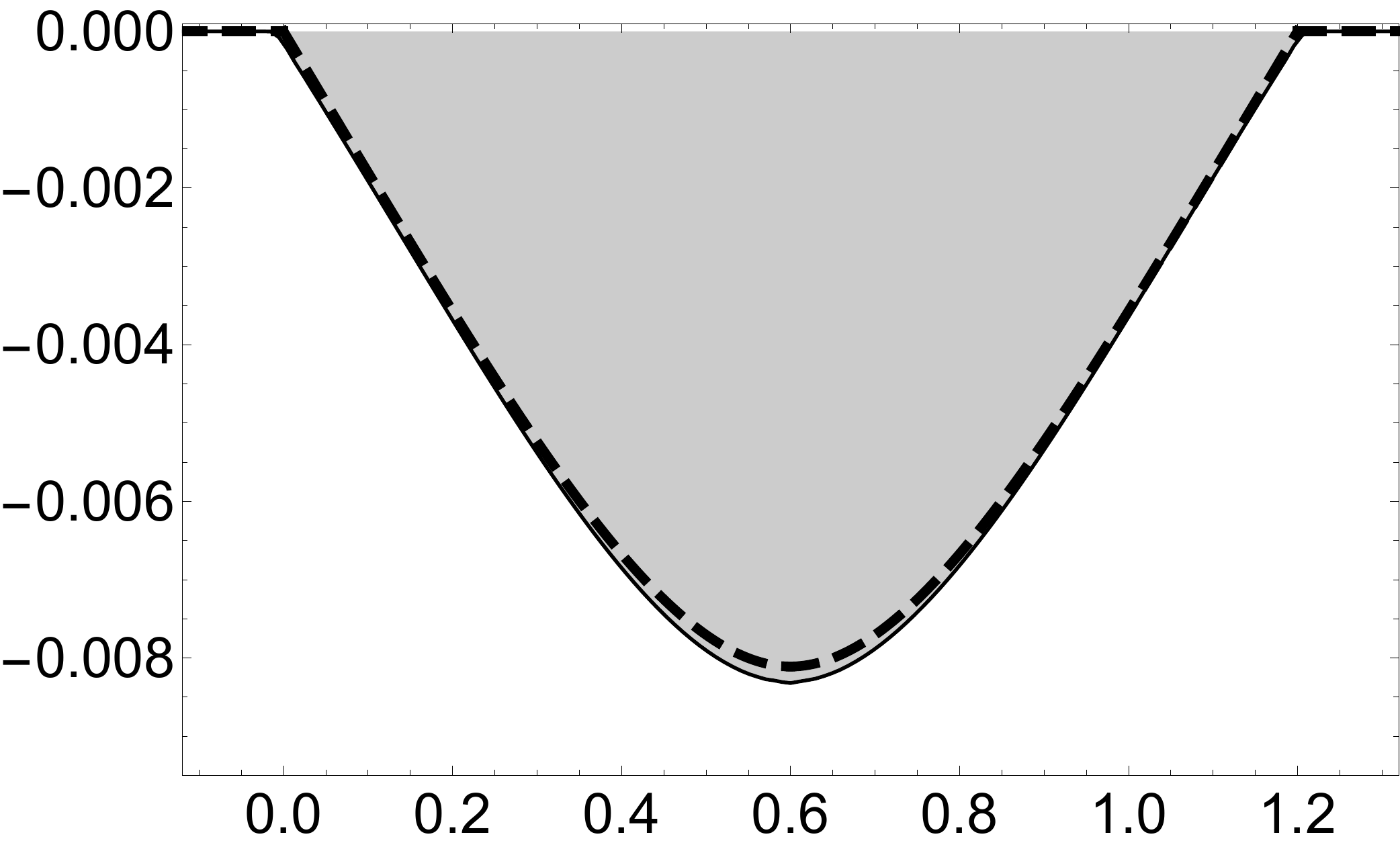}}
\subfigure[]{\includegraphics[width=0.3\textwidth,height=0.15\textwidth, angle =0]{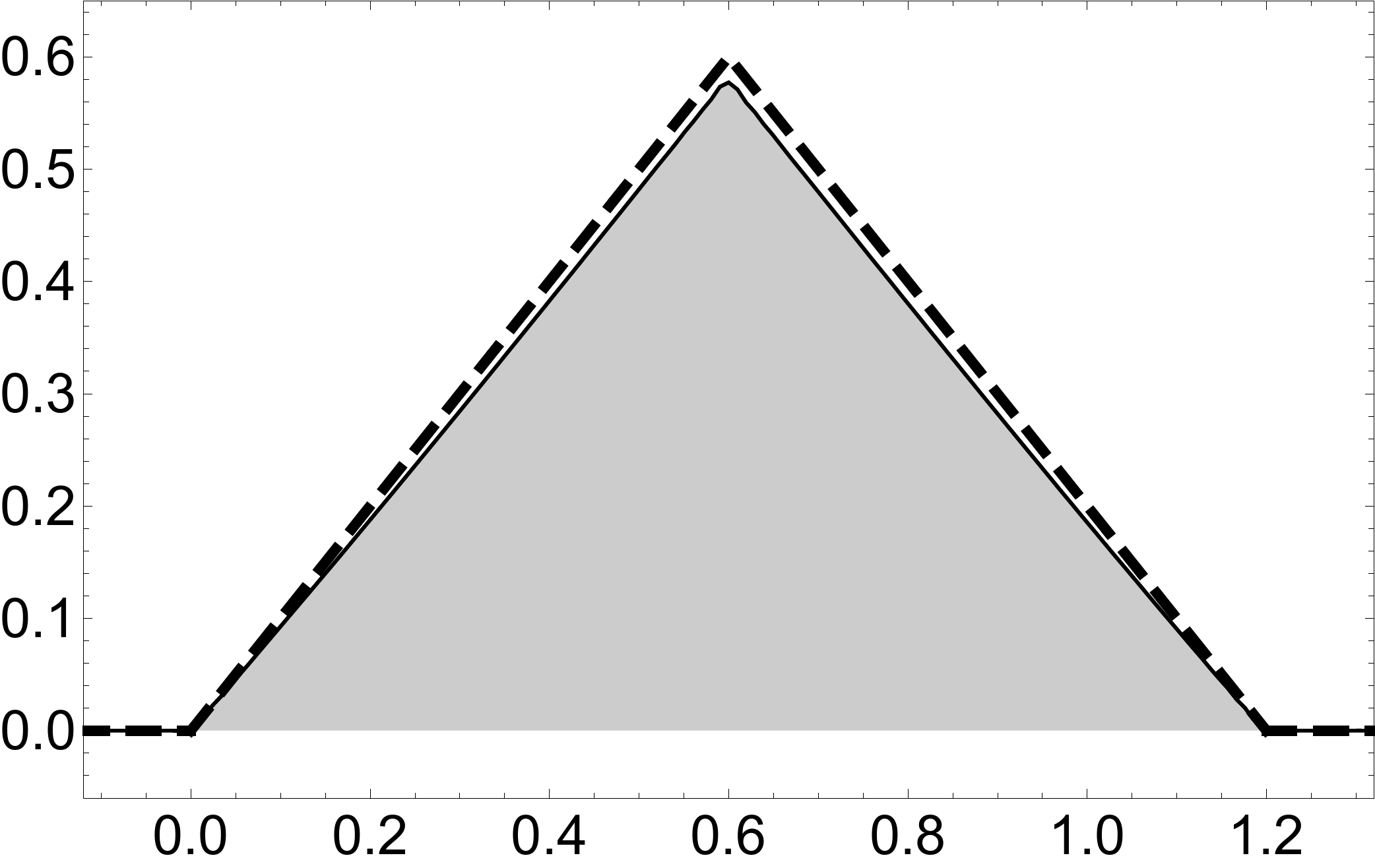}}
\subfigure[]{\includegraphics[width=0.3\textwidth,height=0.15\textwidth, angle =0]{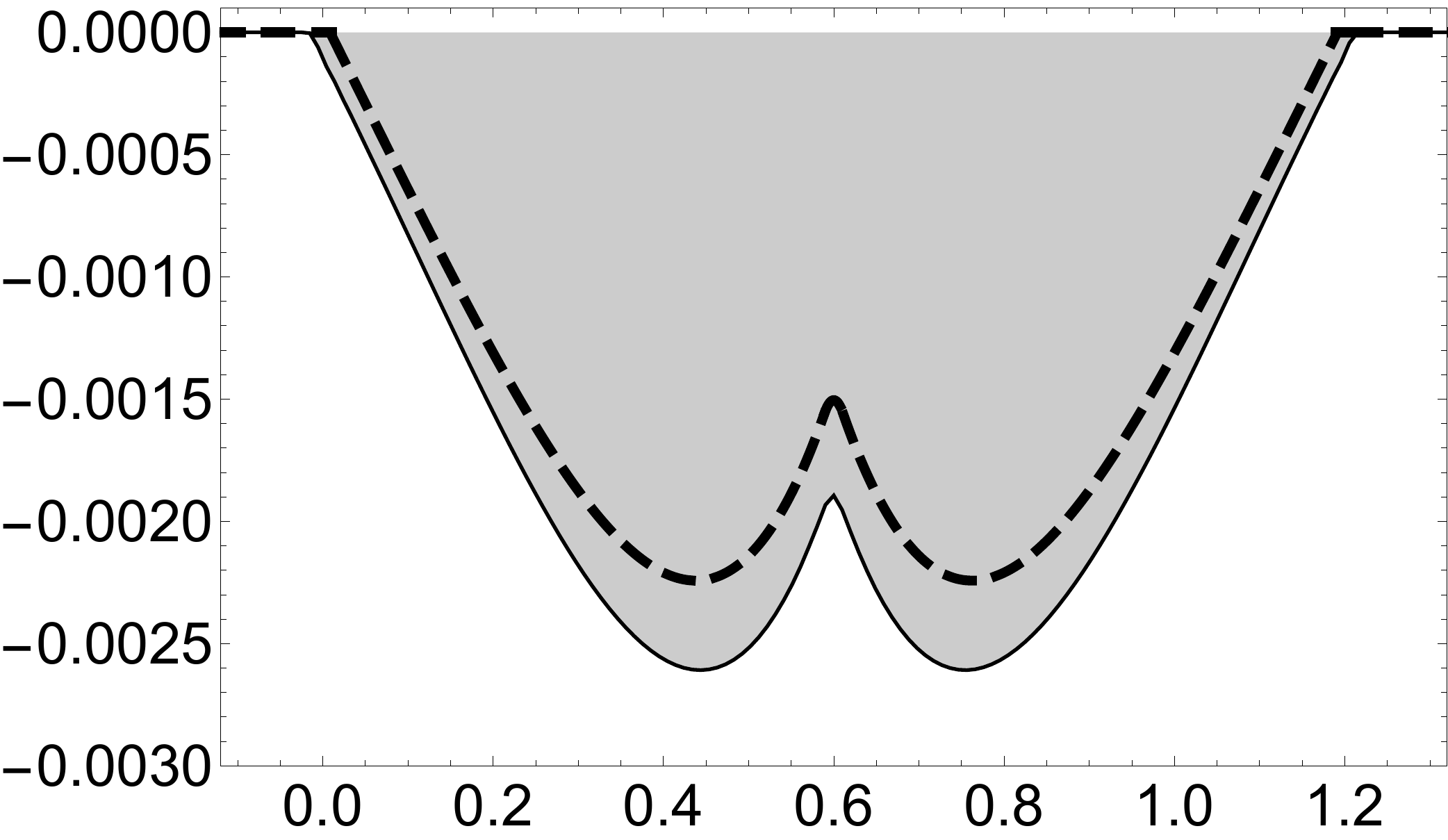}}
\subfigure[]{\includegraphics[width=0.3\textwidth,height=0.15\textwidth, angle =0]{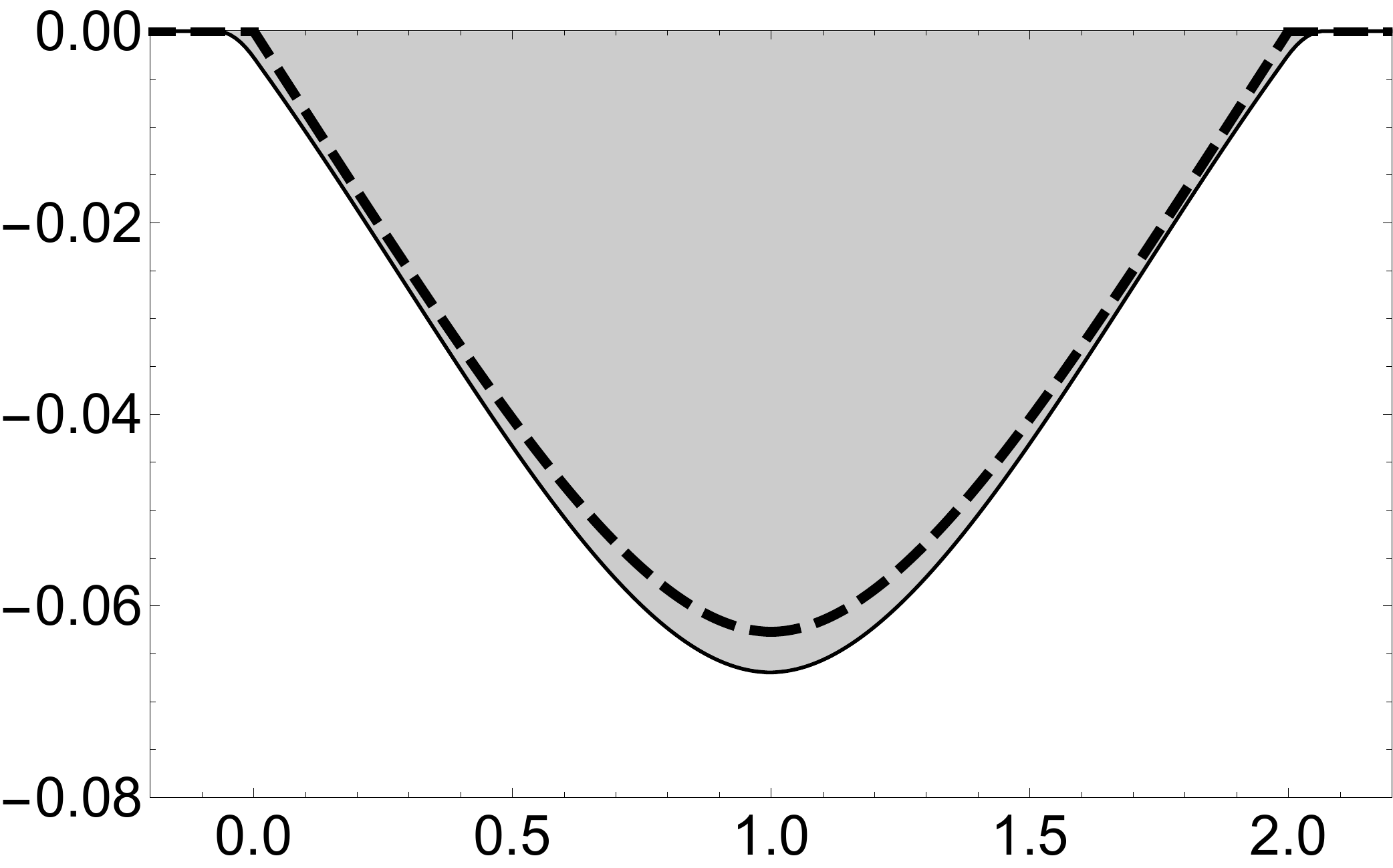}}
\subfigure[]{\includegraphics[width=0.3\textwidth,height=0.15\textwidth, angle =0]{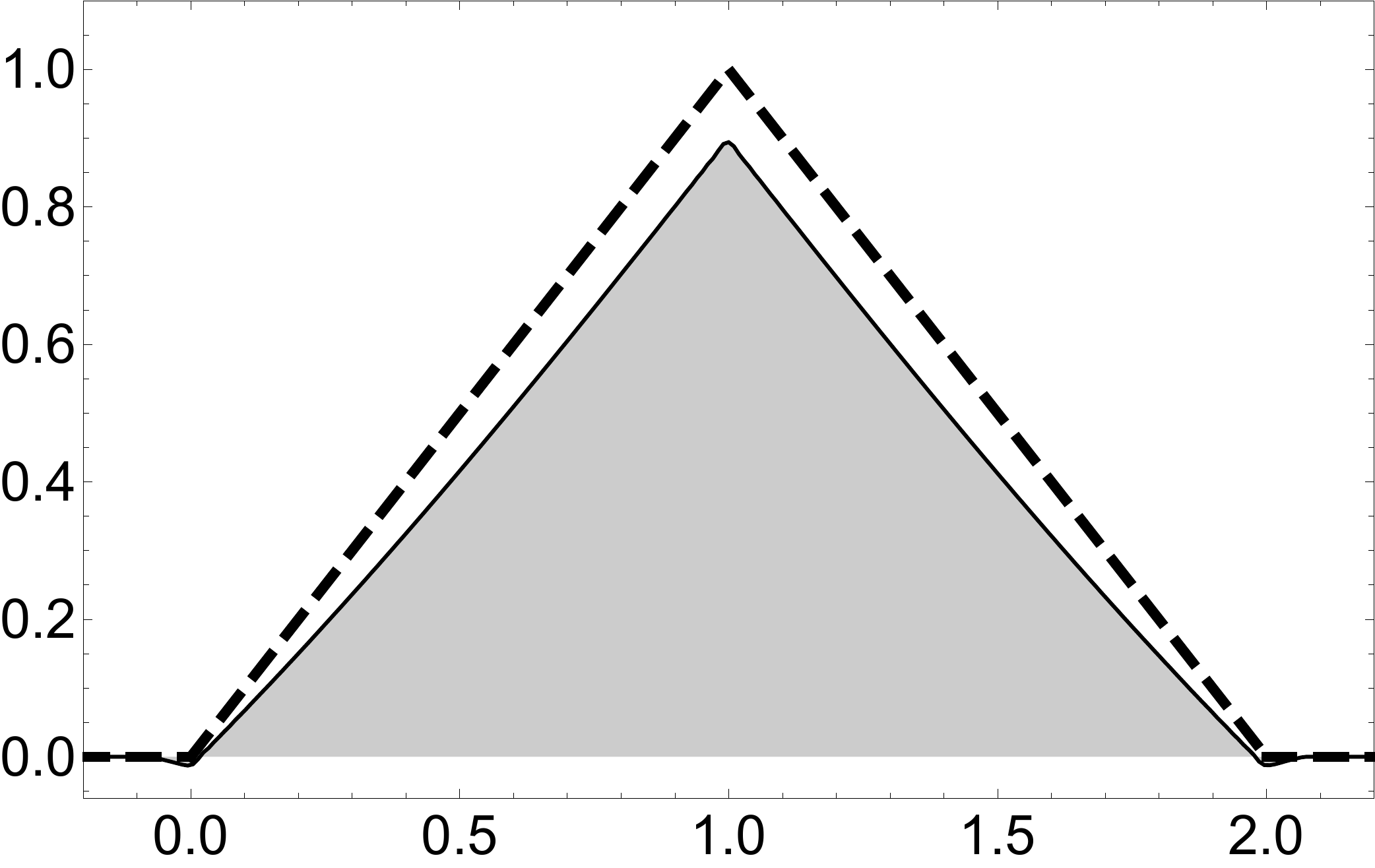}}
\subfigure[]{\includegraphics[width=0.3\textwidth,height=0.15\textwidth, angle =0]{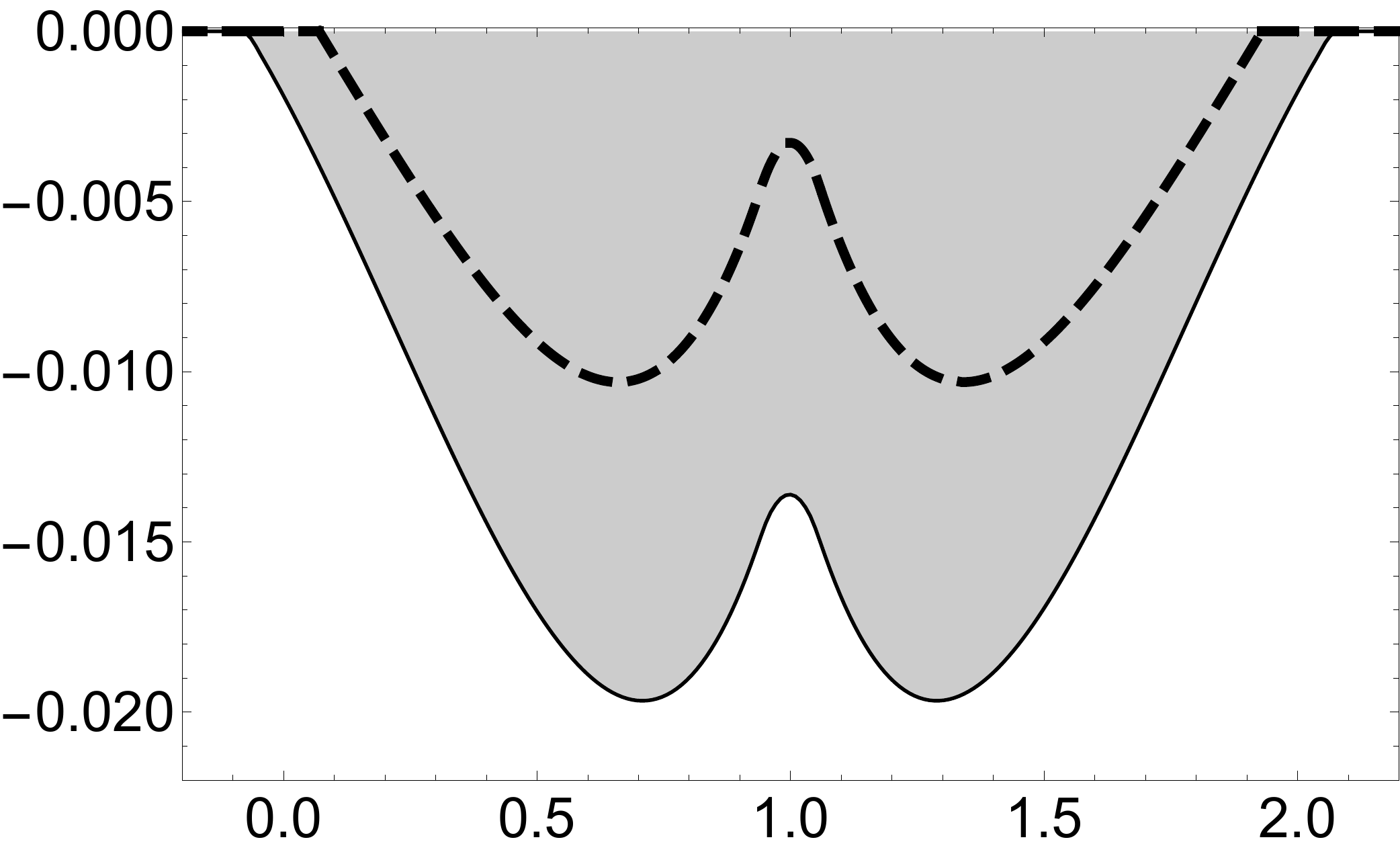}}
\caption{The numerical solutions (shadowed region under the solid curve) and the approximated oscillon solution (dashed curve) in the perturbed model \eqref{eqpert1} for $\epsilon=1$ and $l=1.2$ (a,b,c) and $l=2.0$ (d,e,f) where (a,d) show the field $\eta(t,x)$ at $t=\frac{l}{2}$, (c) at $t=0.611$, (f) at $t=1.059$.  (b, e) The derivative $\partial_t\eta$ at  $t=\frac{l}{2}$.}\label{lbigger}
\end{figure*}

In Fig.\ref{F12} we present our results obtained for the initial data \eqref{inieps} and compare them with the analytical
 approximate solution for $\epsilon=1$ and $l=1$. Our choice of the initial data leads to the exact breather solution
 when $\eta$ is  the signum-Gordon field. The presence of an extra term $-\eta$ in the equation of motion \eqref{L-gen+},
 results in the discrepancy between the oscillon solution and the exact breather.
 The dashed curve represents the analytical approximate solution. We conclude that our analytical approximate solution 
provides a surprisingly good approximation to the numerically derived data for $0\le t \le \frac{l}{2}$. 
The difference between these two curves is very small and visible only when the field is close to its change of sign,
 {\it i.e.} when the amplitude of the field is very small, as in the figure Fig.\ref{F12}(f). The difference between
 the numerical and the analytical approximate solutions increases for oscillons with higher amplitudes {\it i.e.} for $l>1$.
 In Fig.\ref{lbigger} we present the cases of $l=1.2$ and $l=2.0$. As expected, the approximate analytical solution for $l=2.0$ 
and $t>1.0$ begins to differ quantitatively from the numerical one. However, the analytical approximation 
still provides a very good qualitative picture of the evolution of the true solution. 

\begin{figure*}[h]
\centering
\subfigure[]{\includegraphics[width=0.45\textwidth,height=0.2\textwidth, angle =0]{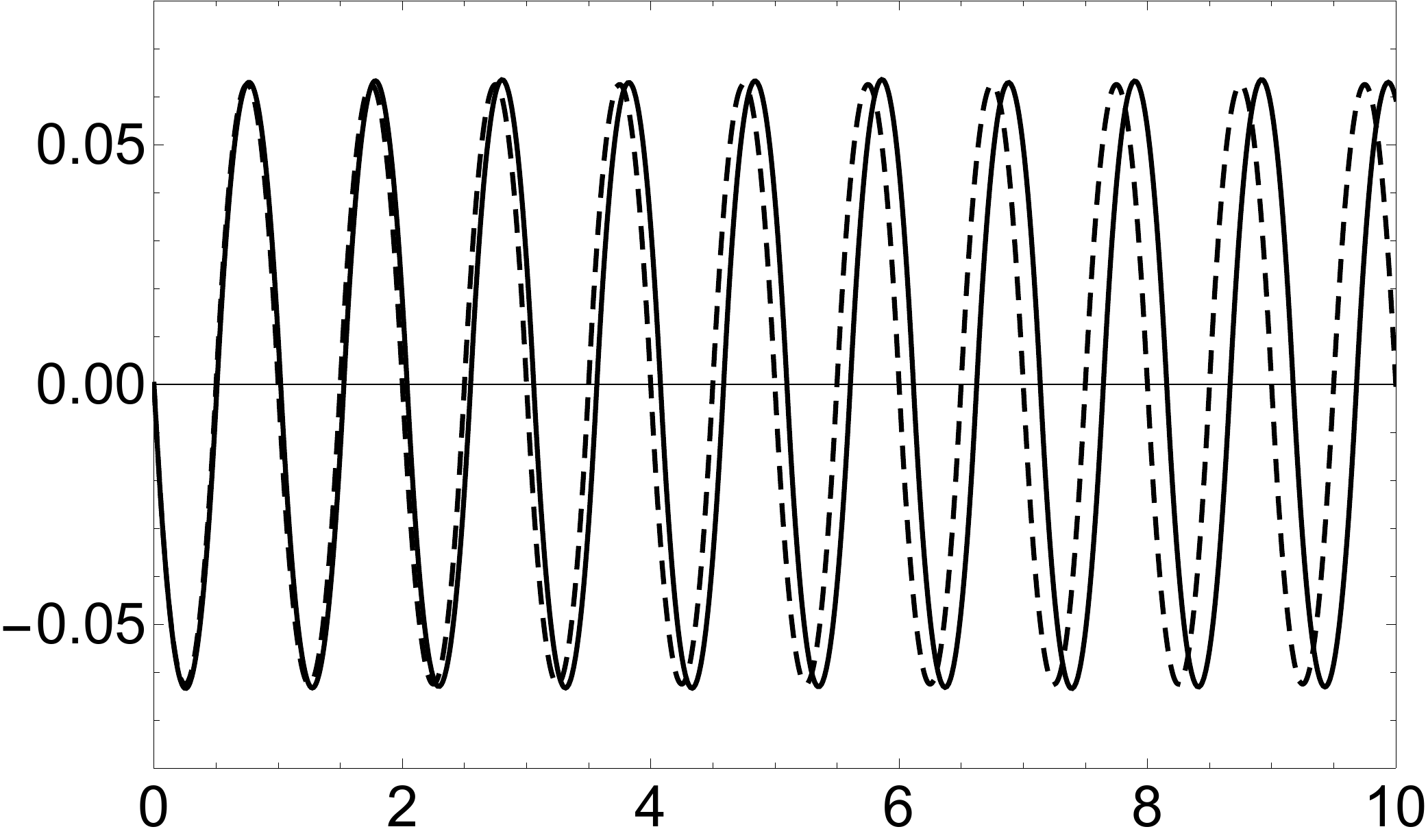}}
\subfigure[]{\includegraphics[width=0.45\textwidth,height=0.2\textwidth, angle =0]{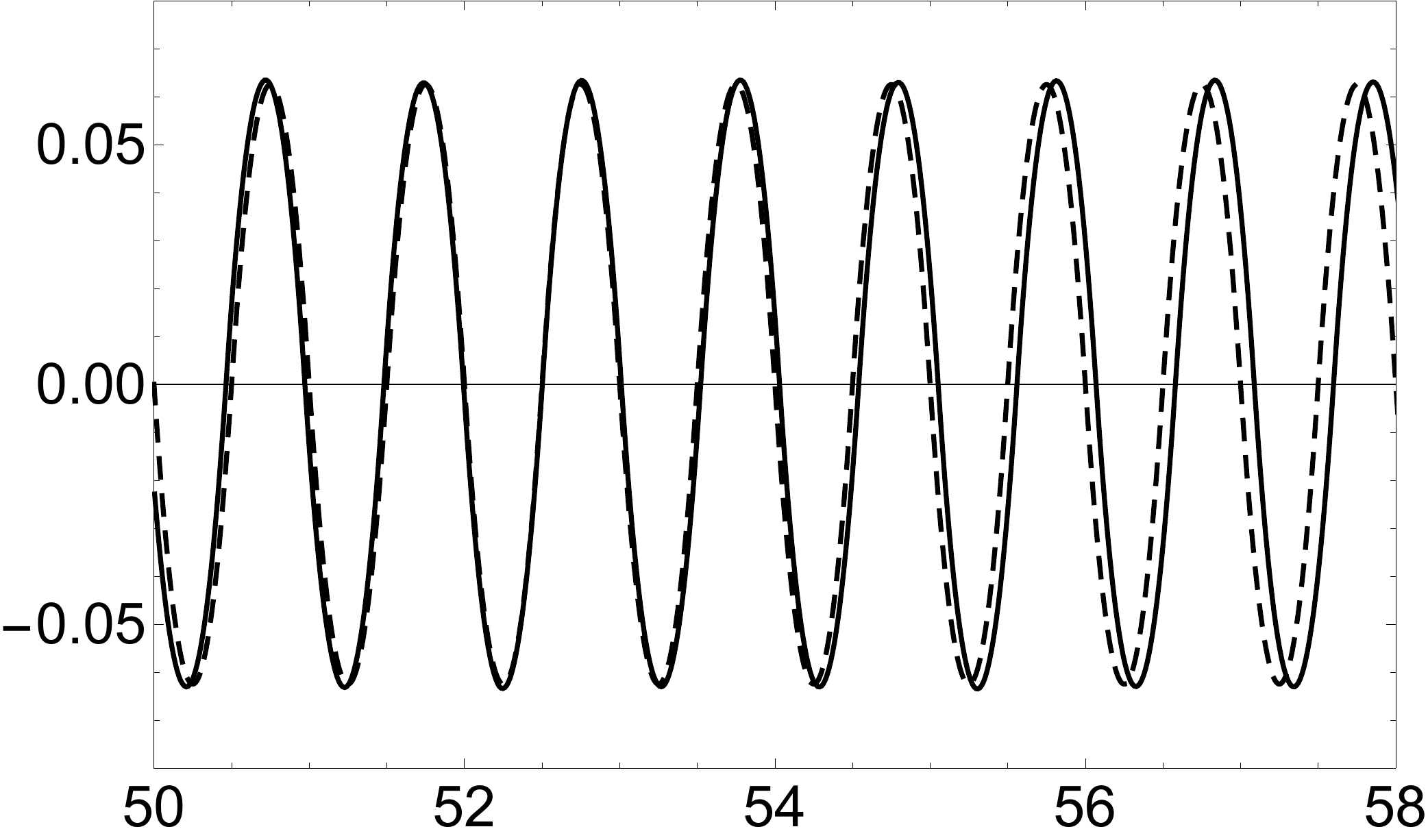}}
\caption{Numerical solution $\eta(t,\frac{1}{2})$ (solid line) in the perturbed model with $l=1$ and $\epsilon=1$. The exact breather solution $\phi(t,\frac{1}{2})$ (dashed line) is given by dashed line. The figures correspond to (a) $t\in[0,10]$ and (b) $t\in[50, 58]$.}\label{F14}
\end{figure*}

\begin{figure*}[h]
\centering
\subfigure{\includegraphics[width=0.45\textwidth,height=0.2\textwidth, angle =0]{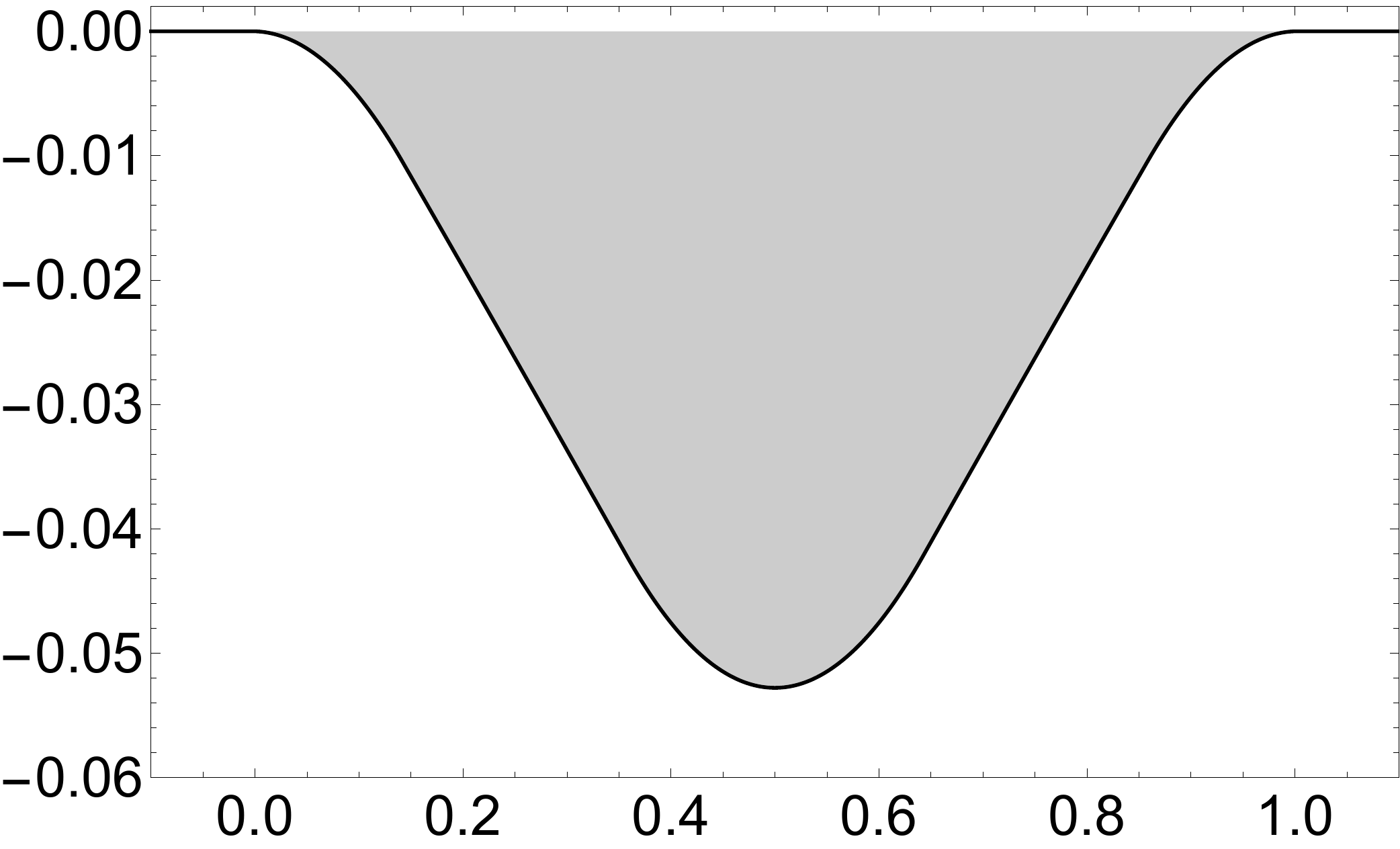}}
\caption{The numerical solution $\eta(t,x)$ in the perturbed model with $l=1$ and $\epsilon=1$ at $t=10.344+0.15=10\times {\rm period}+0.15$. The solution is very similar to the solution shown in Fig.\ref{F12}(a)}\label{tenperiods}
\end{figure*}

As expected from the previous analytical considerations the numerical oscillon solution oscillates 
slower than the exact breather in the signum-Gordon model.  
In figure Fig.\ref{F14} we present plots representing the trajectory of the centre $x=\frac{l}{2}$ of the oscillon 
with $l=1$, which form a very regular oscillating curve. This means that the exact small amplitude breathers are very 
weakly deformed by the new term in the Lagrangian. The long living solutions continue to exist and there is no visible emitted 
radiation within our numerical accuracy. The numerical solution is amazingly periodic in the sense that
 the oscillations are very regular. Plots a) and b) show that even after sixty periods of oscillations the two are very 
similar to what they were at the very beginning of the simulations. The solution at $t=10.494$ is presented in 
 Fig.\ref{tenperiods}. The plot represents the field shown in Fig.\ref{F12}(a) after $N=10$ oscillations. 


\begin{figure*}[h] 
\centering
\subfigure[]{\includegraphics[width=0.3\textwidth,height=0.15\textwidth, angle =0]{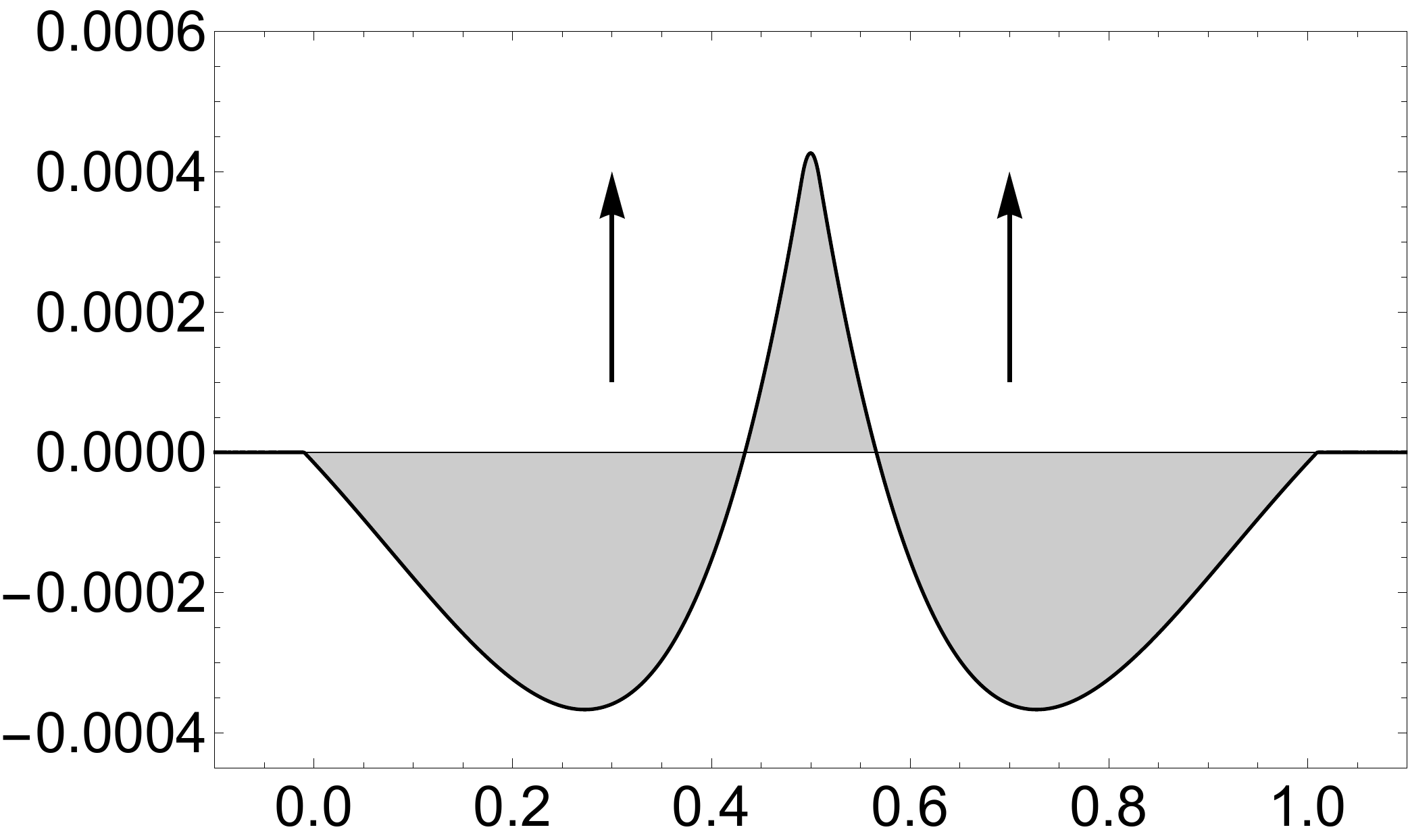}}
\subfigure[]{\includegraphics[width=0.3\textwidth,height=0.15\textwidth, angle =0]{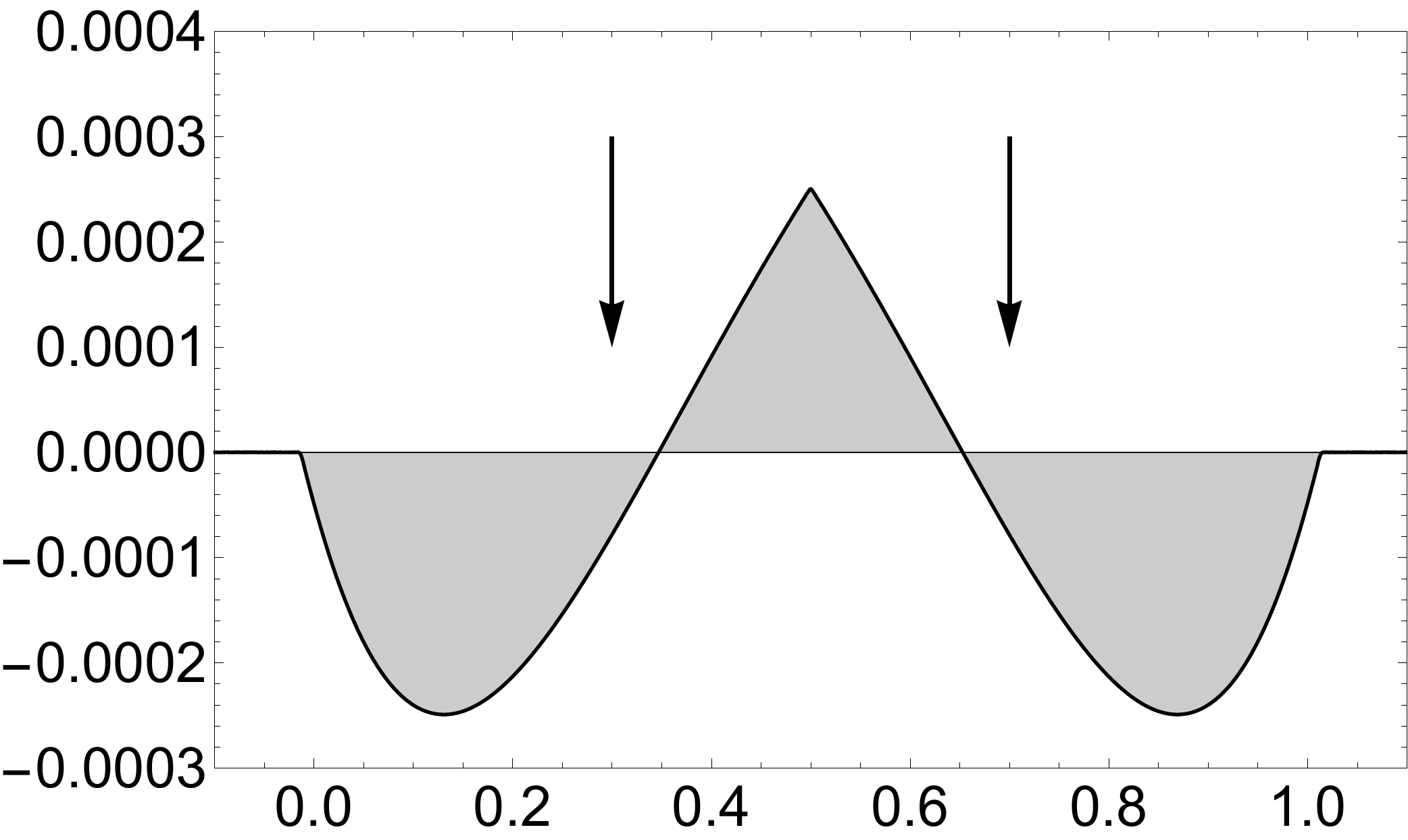}}
\subfigure[]{\includegraphics[width=0.3\textwidth,height=0.15\textwidth, angle =0]{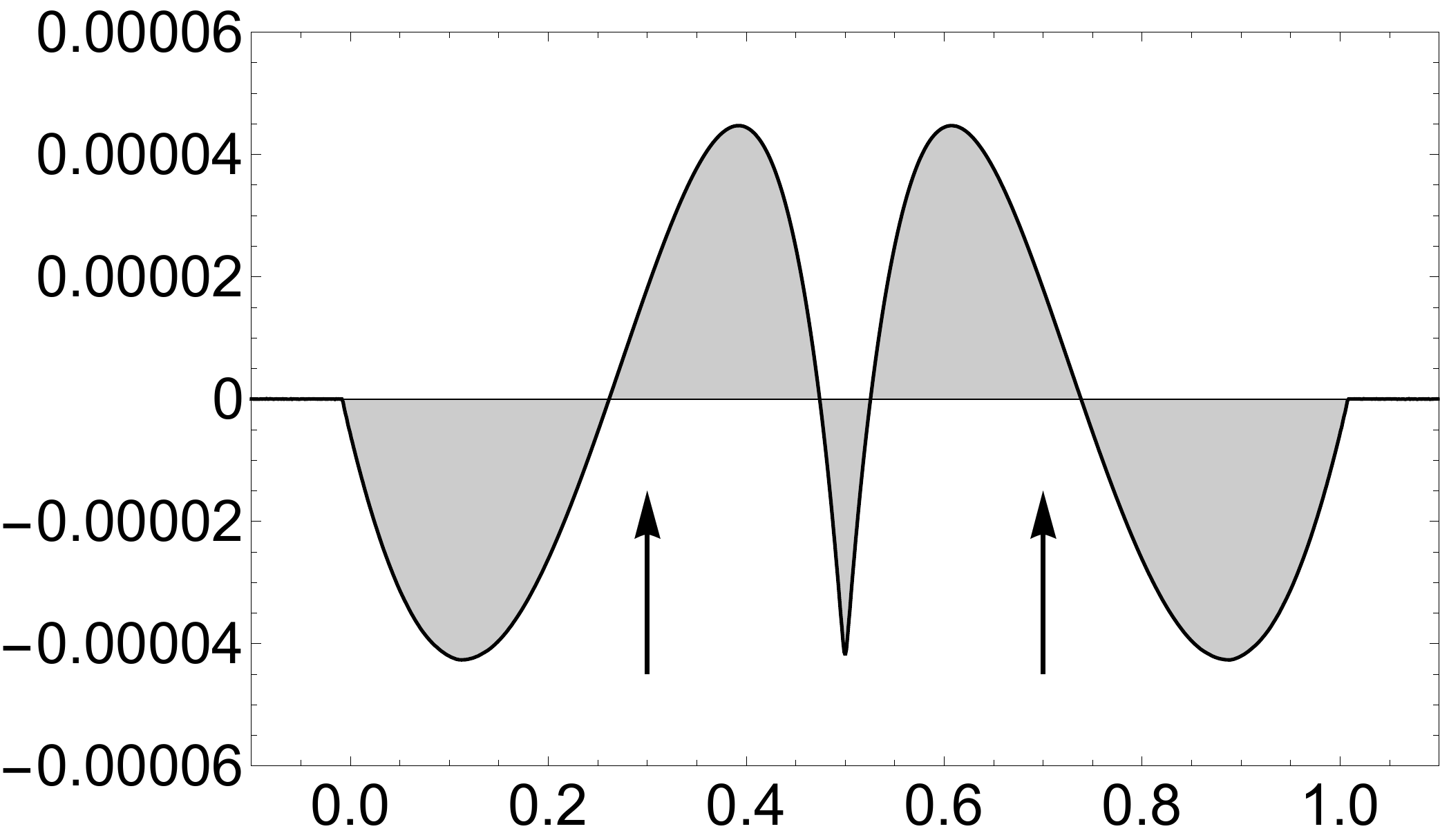}}
\subfigure[]{\includegraphics[width=0.3\textwidth,height=0.15\textwidth, angle =0]{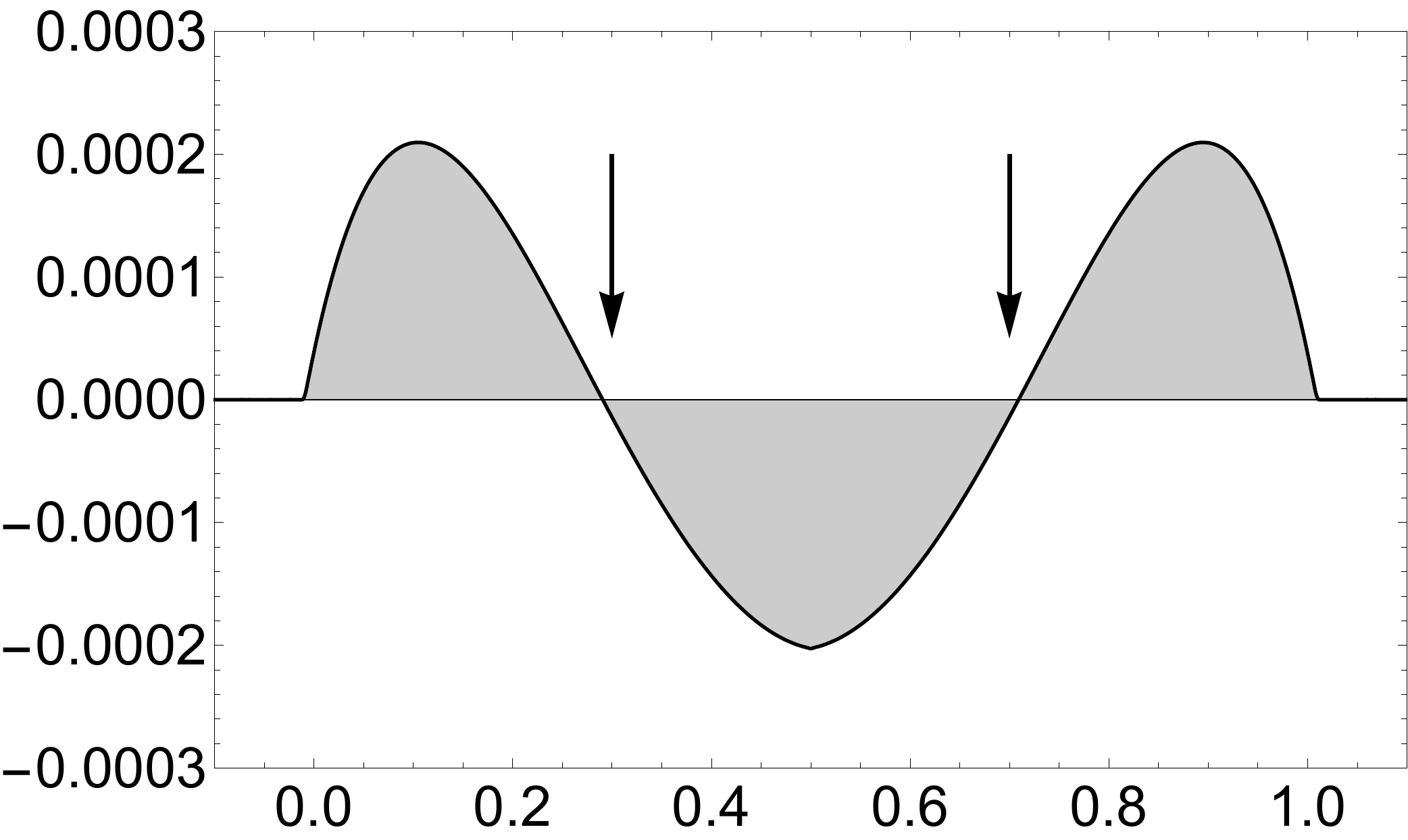}}
\subfigure[]{\includegraphics[width=0.3\textwidth,height=0.15\textwidth, angle =0]{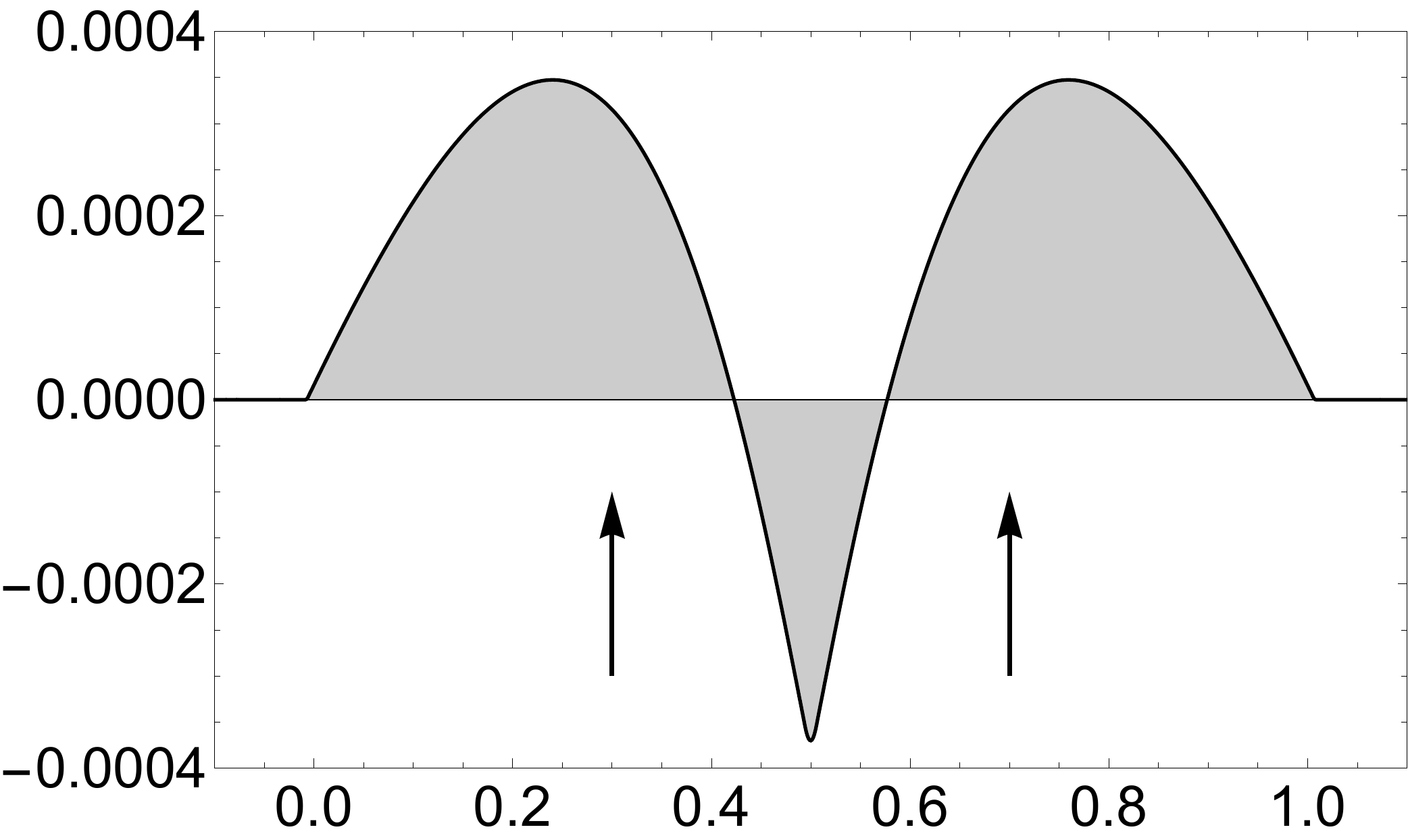}}
\subfigure[]{\includegraphics[width=0.3\textwidth,height=0.15\textwidth, angle =0]{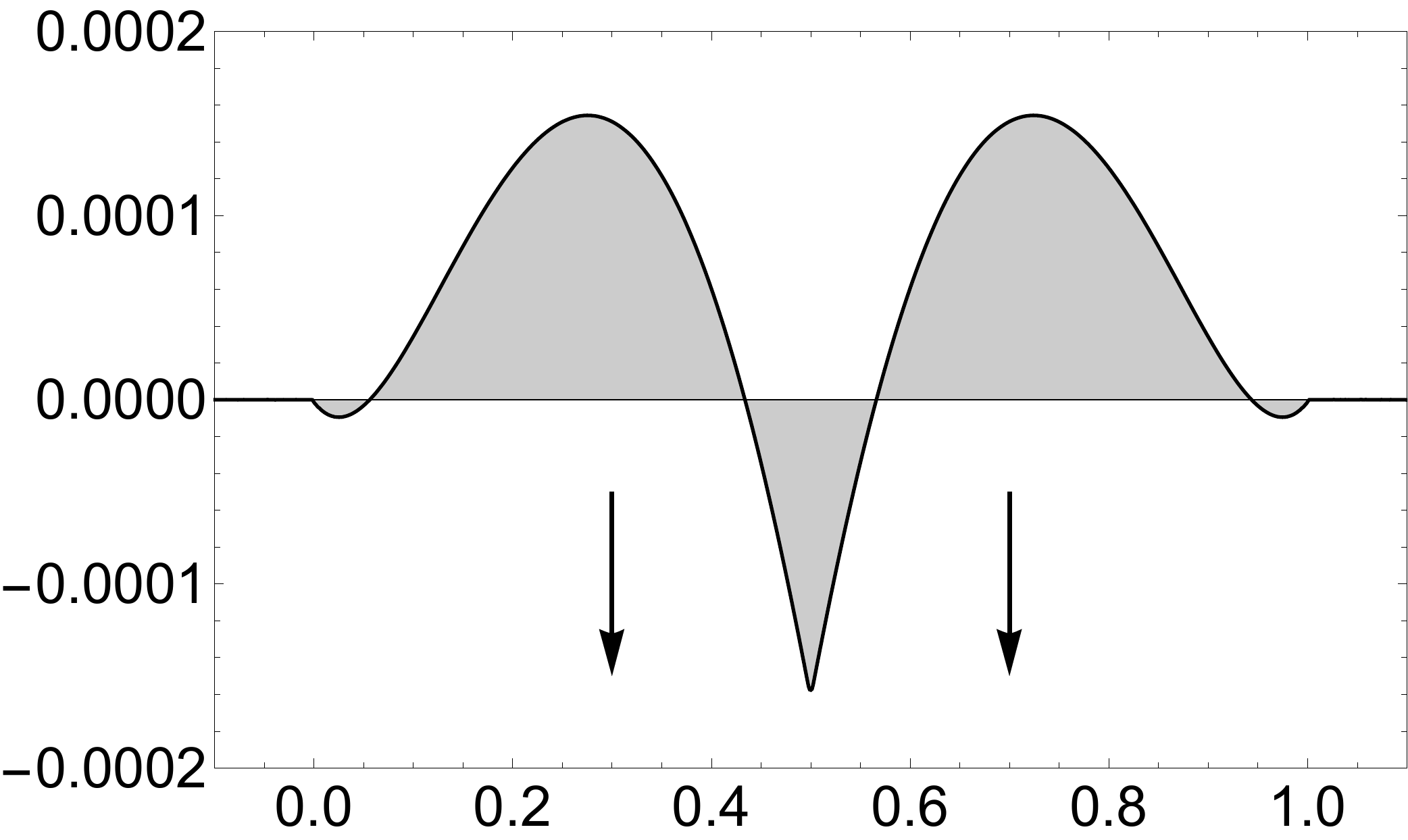}}
\subfigure[]{\includegraphics[width=0.3\textwidth,height=0.15\textwidth, angle =0]{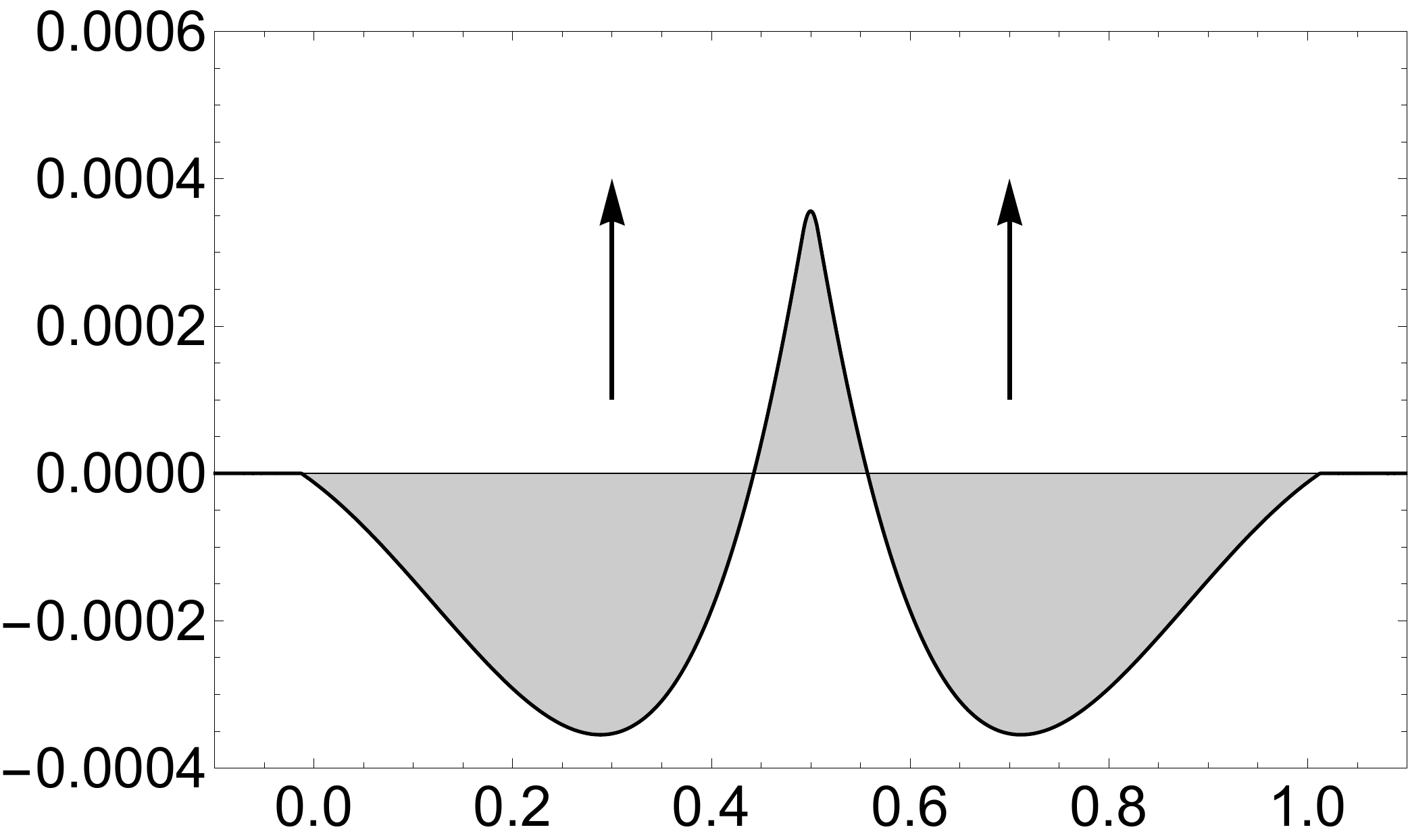}}
\subfigure[]{\includegraphics[width=0.3\textwidth,height=0.15\textwidth, angle =0]{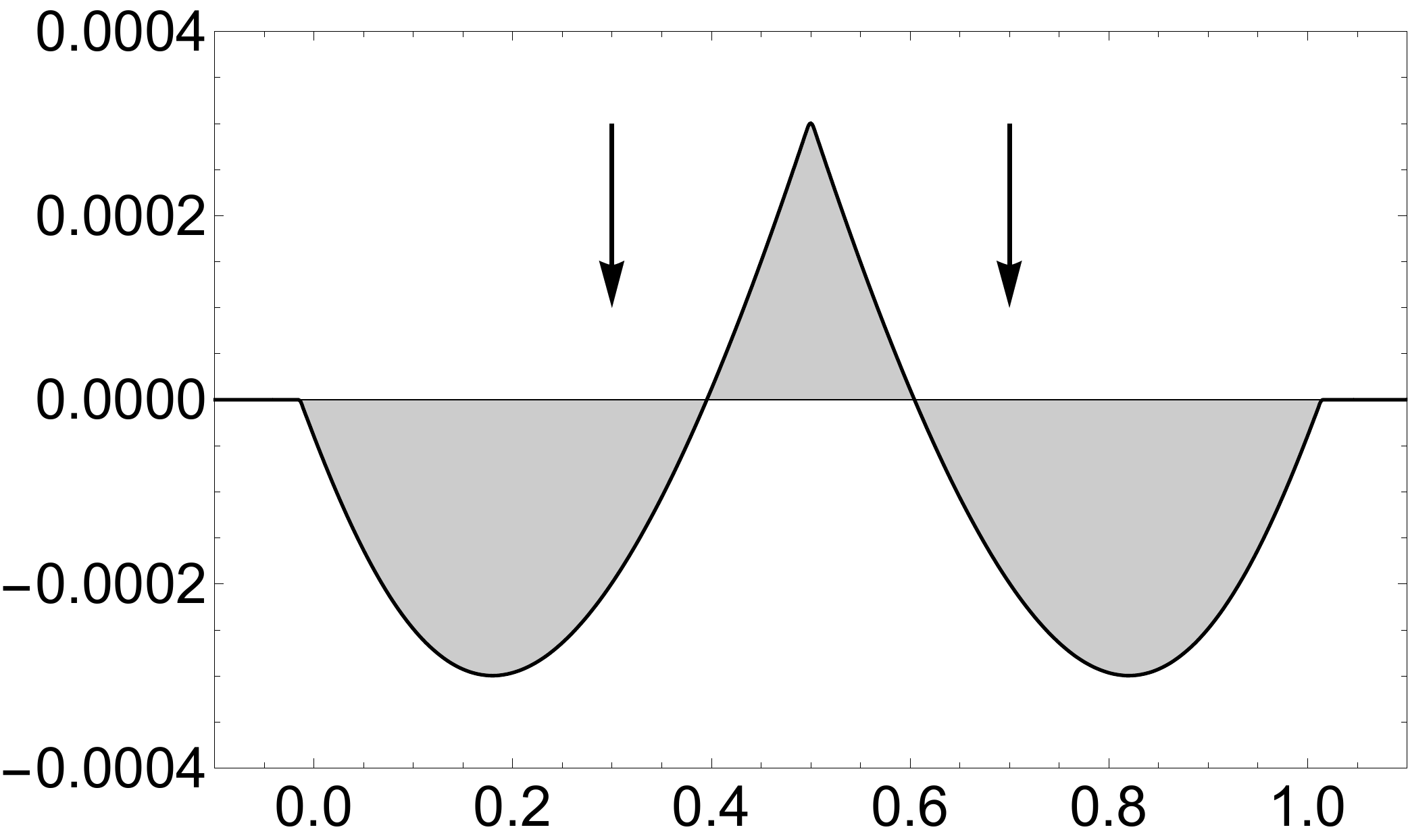}}
\subfigure[]{\includegraphics[width=0.3\textwidth,height=0.15\textwidth, angle =0]{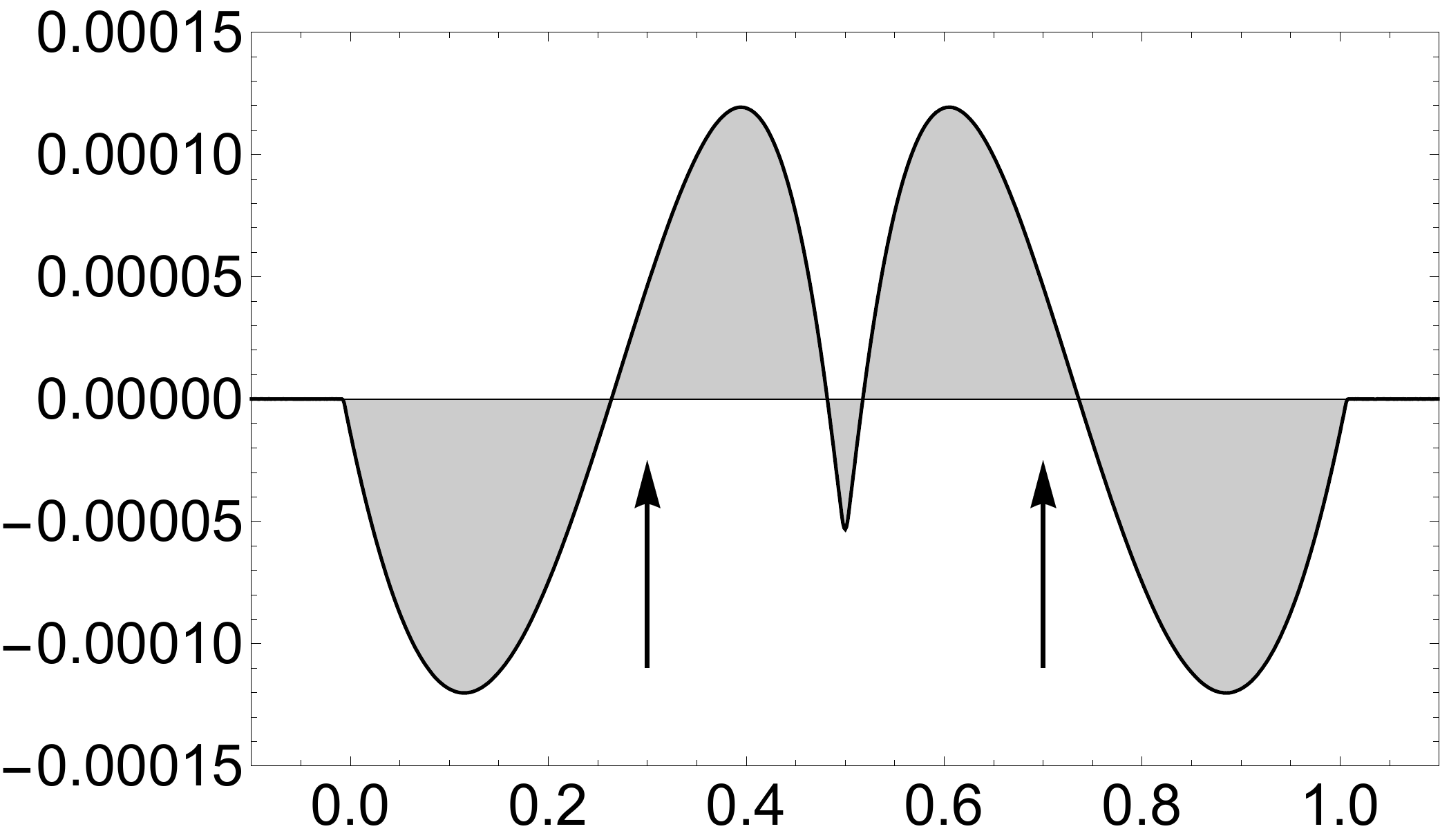}}
\subfigure[]{\includegraphics[width=0.3\textwidth,height=0.15\textwidth, angle =0]{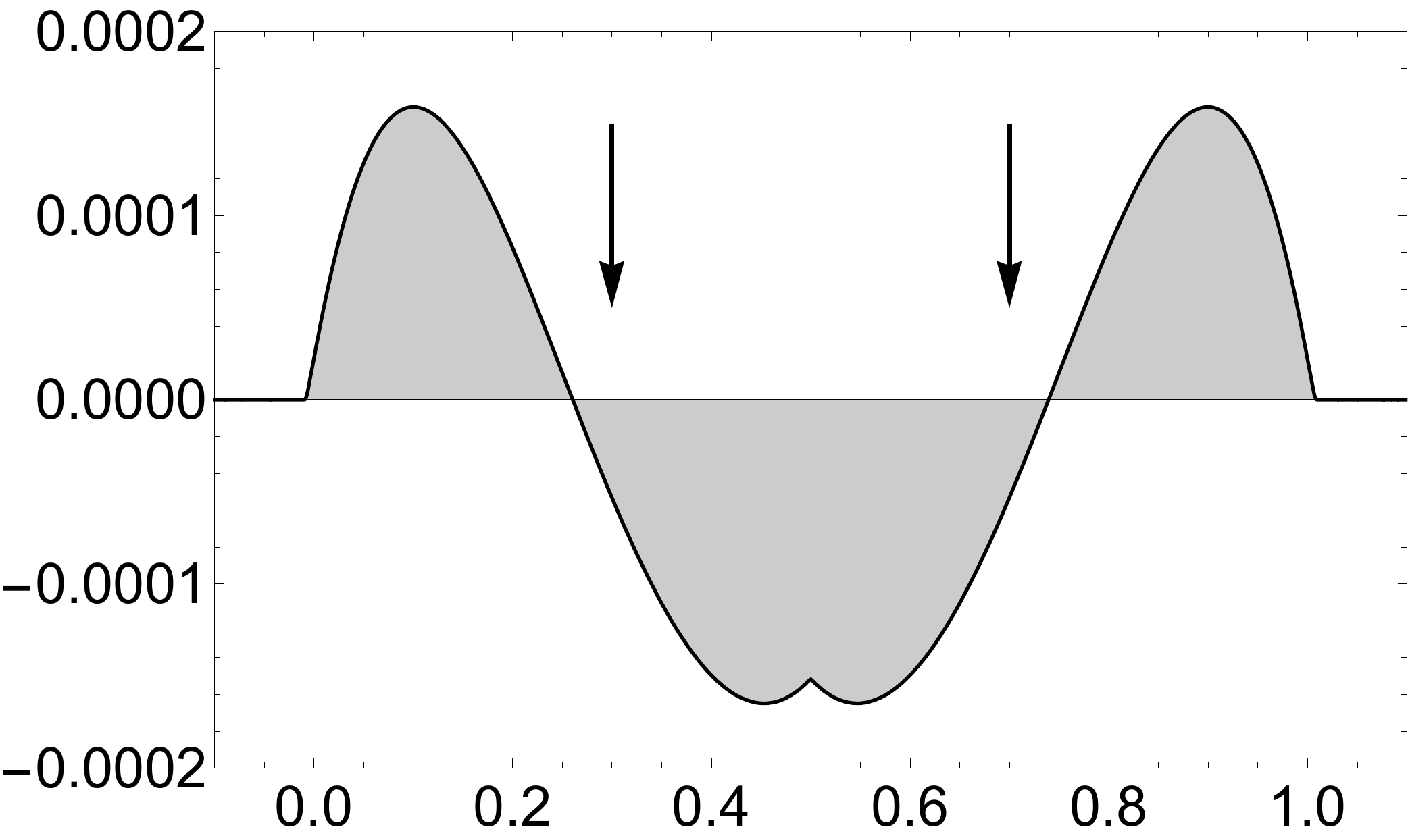}}
\subfigure[]{\includegraphics[width=0.3\textwidth,height=0.15\textwidth, angle =0]{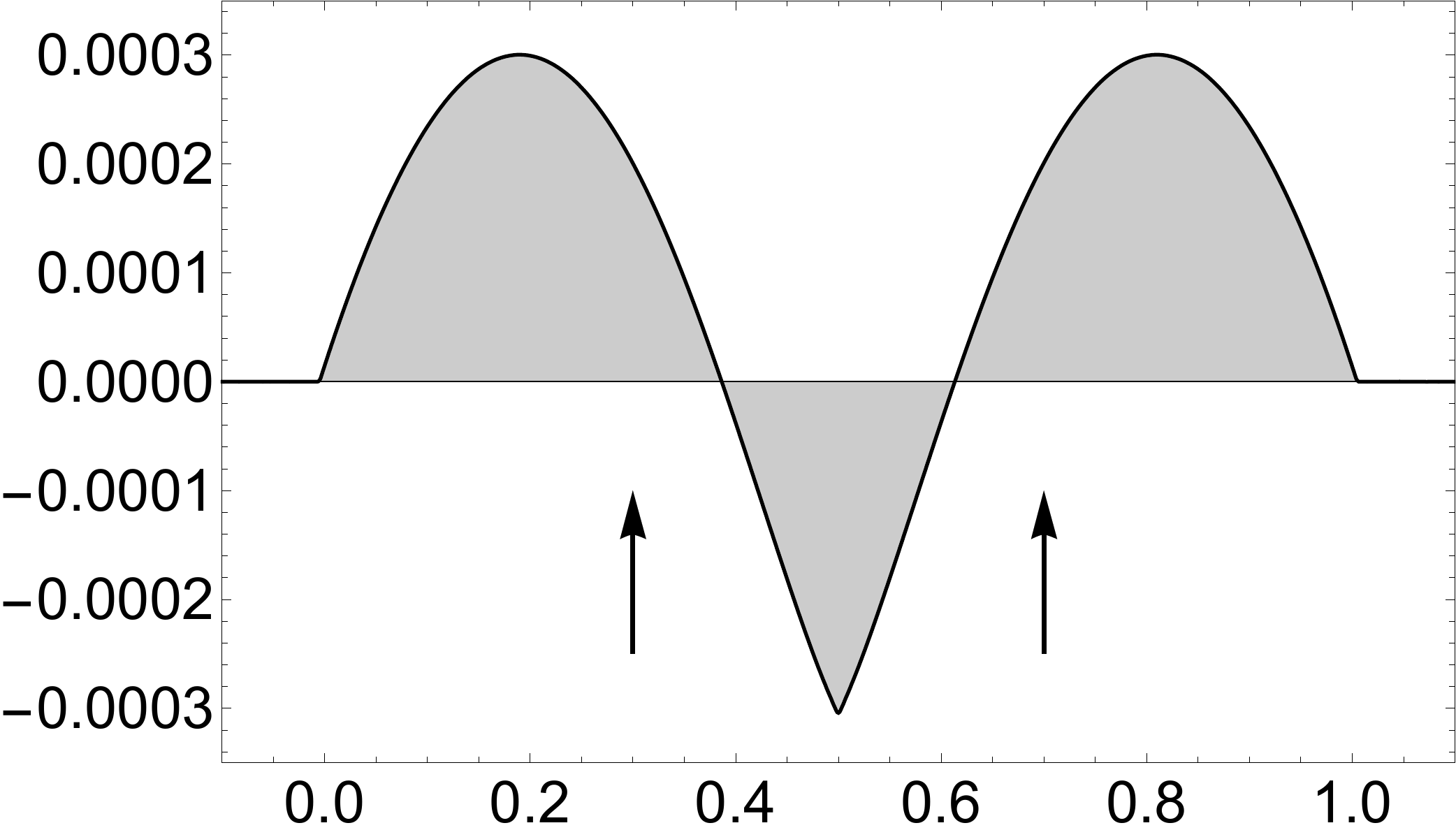}}
\subfigure[]{\includegraphics[width=0.3\textwidth,height=0.15\textwidth, angle =0]{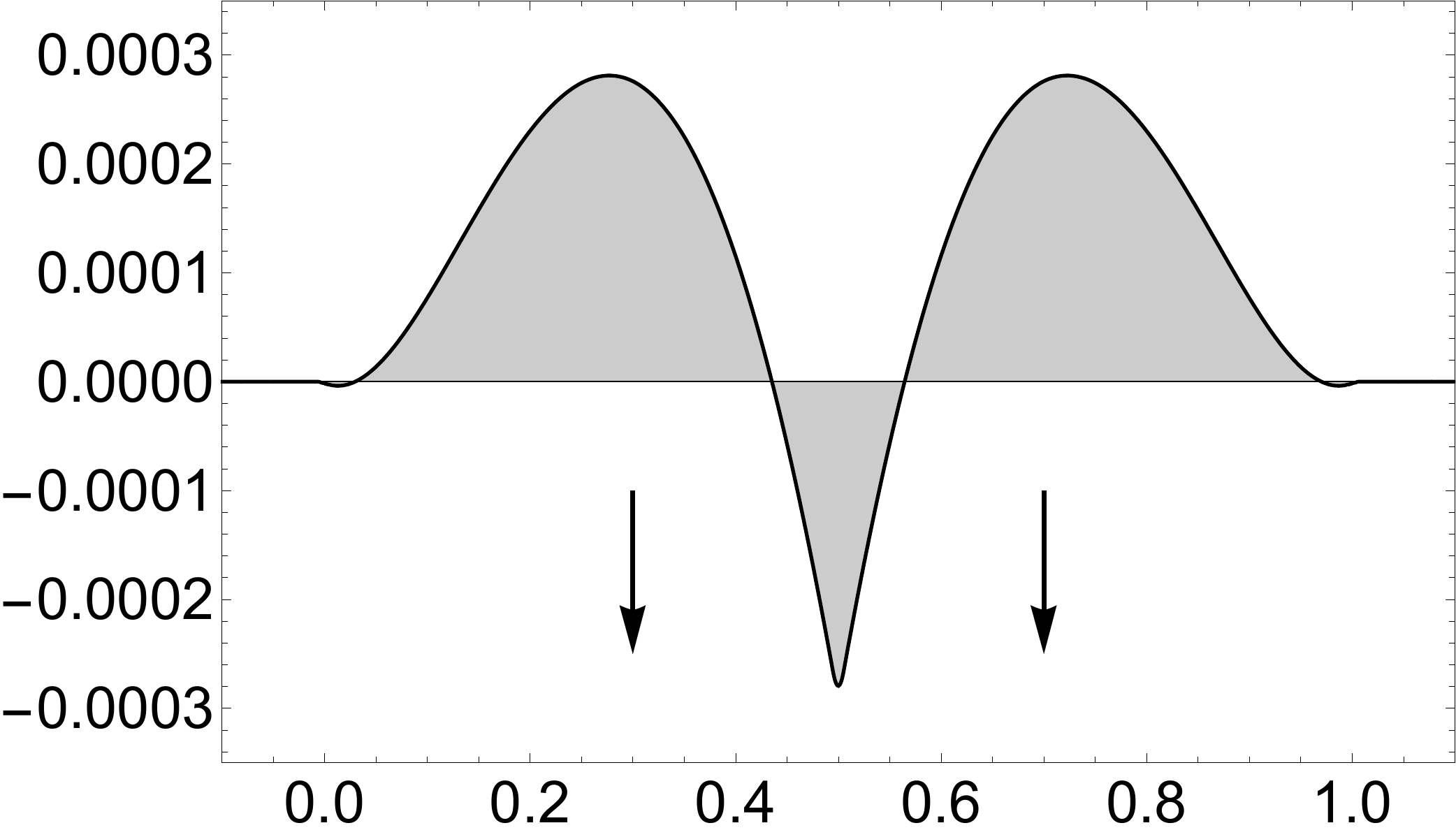}}
\caption{Numerical solutions  in the perturbed model for $\epsilon=1.0$ and $l=1$ at instants of time $t_n\approx n \,t_1$, where $n=1,2,\cdots,12$. (a) $t_1=0.5090$, (b) $t_2=1.0197$, (c) $t_3=1.5291$, (d) $t_4=2.0387$, (e) $t_5=2.5492$, (f) $t_5=3.0579$, (g) $t_7=3.5672$, (h) $t_8=4.0781$, (i) $t_9=4.5872$, (j), $t_{10}=5.0970$, (k), $t_{11}=5.6072 $, (l), $t_{12}=6.1159$. The arrows indicate direction of oscillation.}\label{F13}
\end{figure*}

When comparing this numerical solution to the exact breather we note the existence of some subtle differences.
 They are especially visible when the field changes its sign.
In Fig.\ref{F13} we show a sequence of snapshots of the field that correspond with the moments of time in which the
 oscillon solution changes its sign. The arrows indicate directions of the oscillation. Unlike for the signum-Gordon
 breather the change of the sign does not occur simultaneously in the full support of the solution. 
The evolution of the field in Fig.\ref{F13}(c) is qualitatively different from that shown in Fig.\ref{F13}(a).
 However, one can see that configurations of the field that differ by the period equal to three periods of the oscillations
 are more alike (although not identical). This can be checked by comparing subfigures that form pairs (a,g), (b,h) and (c,i).
 Similarly, a qualitatively similar behaviour of the field is visible in graphs that constitute the pairs (d,j), (e,k) and
 (f,l). Obviously, there are some small differences between corresponding plots. This is to be expected because 
the initial configuration of the field is only proper for the breather in the signum-Gordon model.
 A more carreful analysis shows that the oscillon expands and shrinks very little when oscillating.

\begin{figure*}[h!]
\centering
\subfigure[]{\includegraphics[width=0.45\textwidth,height=0.25\textwidth, angle =0]{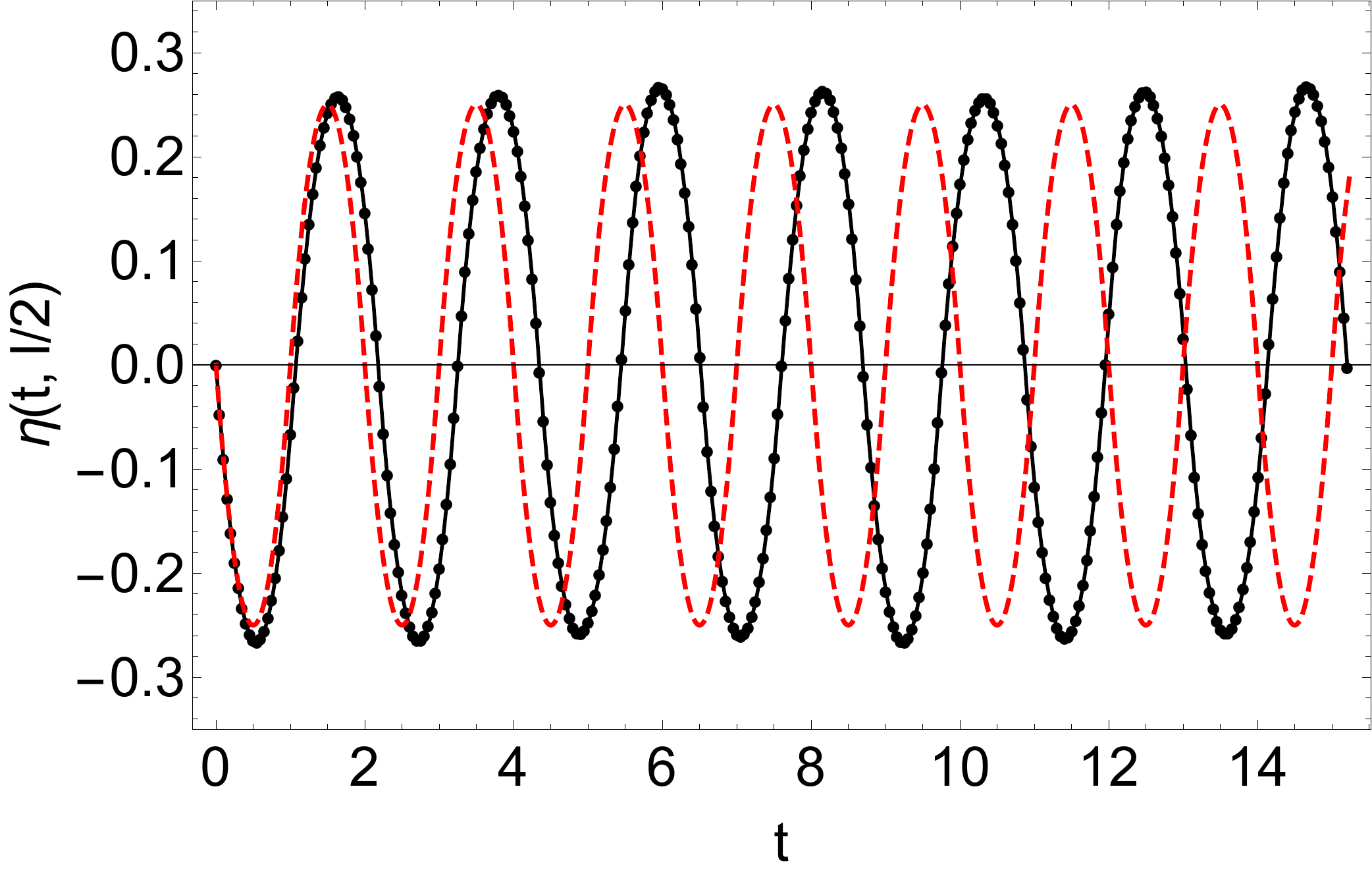}}
\subfigure[]{\includegraphics[width=0.45\textwidth,height=0.25\textwidth, angle =0]{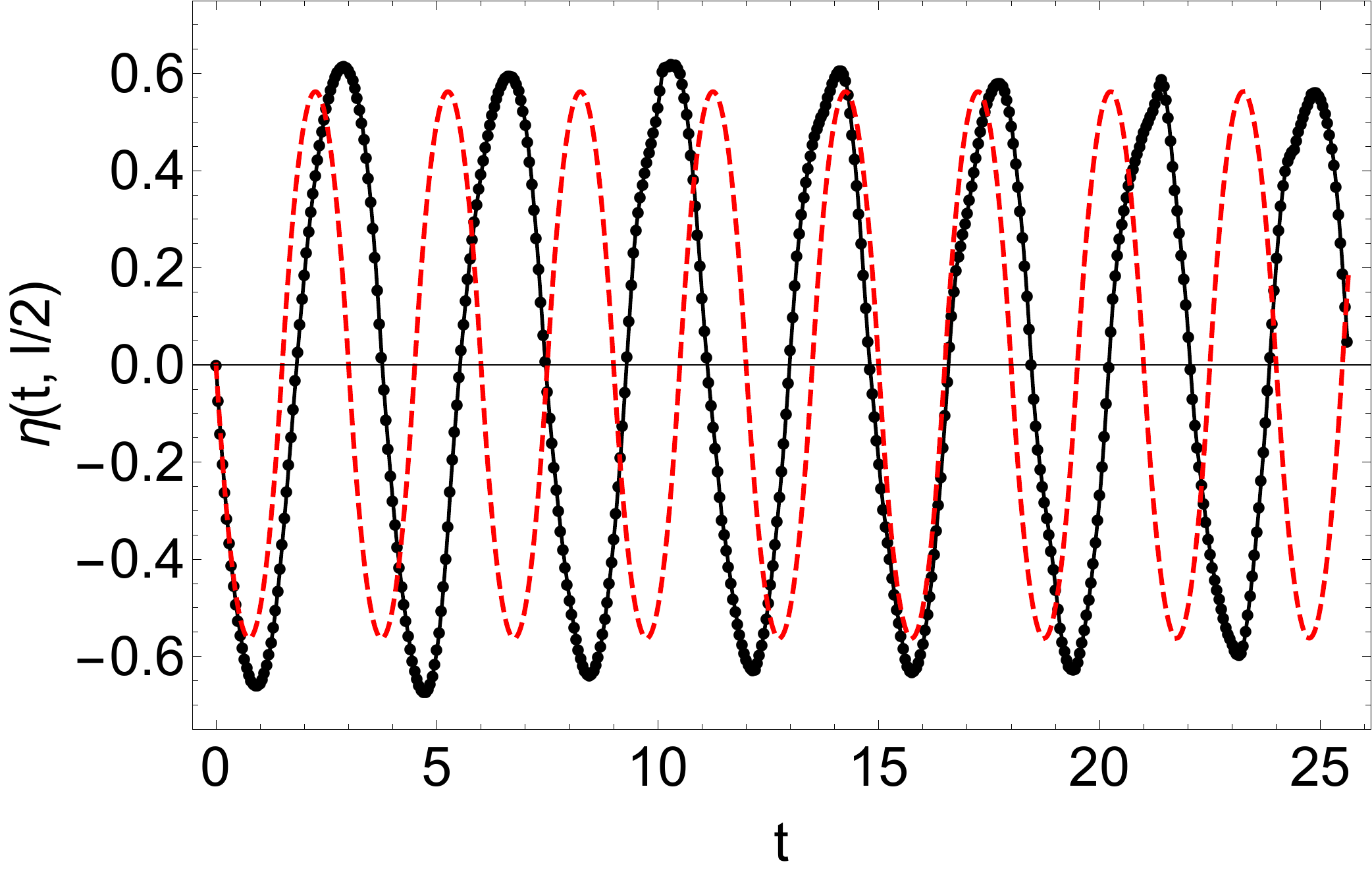}}
\caption{ Numerical solution $\eta(t,\frac{l}{2})$ (black line) in the perturbed model. The exact breather solution $\phi(t,\frac{l}{2})$ is given by dashed red line. The figures correspond to (a) $l=2.0$, (b) $l=3.0$.} \label{F22}
\end{figure*}

We have studied oscillons of this type with the size $l \in [0.5, 3]$. The observed
oscillations of the central point $x=\frac{l}{2}$) are described by the black dotted curve in Fig.\ref{Fig-period}(b).
We have calculated the corresponding period of the oscillations and we have found that, for the solution with $l=1$,
 the Fourier analysis of the numerical 
data gave us the value of $T$ as being approximately equal to $T=1.0224$.
Overall, it is clear that our analytical approximation works quite well - especially for $l<1.2$. Once the amplitude of the initial signum-Gordon breather is chosen to be larger, the nonlinear effects begin to modify the solution in a more significant way, which requires higher orders of approximation. This is manifested by a deformation of the sinus-like oscillation curve for higher $l$ - see Fig. \ref{F22}. 
 
If the amplitude of the initial configuration is large enough, and so the system possesses sufficiently large amount of energy, a new phenomenon does occur. This comes from the  fact that our modification of the signum-Gordon model introduces a new vacuum at $\eta=2$ 
(or infinitely many vacua in the unfolded variable $\eta$). Therefore, topological compact solitons (kink and anti-kink) 
can exist

\be
\eta_{\rm kink} (x) =  \left\{
\begin{array}{ll}
0 & x \leq 0 \\
1-\cos x & x \in [0, \pi] \\
2 & x \geq \pi
\end{array}
\right.,\qquad
\eta_{\rm anti-kink} (x) =  \left\{
\begin{array}{ll}
2 & x \leq 0 \\
1+\cos x & x \in [0, \pi] \\
0 & x \geq \pi
\end{array}
\right..
 \label{kink}
\ee
Note that these kink and anti-kink solutions also arise in the folded target space variable $\bar \eta$.

Due to this, the signum-Gordon breathers with sufficient large amplitude, if taken as initial states of the modified model, do nontrivially feel the second vacuum and, in a consequence, a collection of kink-antikink states can be created. This is exactly what we have observed in our simulations. In the next subsection we say a few words about such states. 

\subsubsection{Kink-antikink pairs}
Following the above observation we have studied the process of generation of 
kink-antikink pairs when the initial configurations (which produce the exact breather in the unmodified signum-Gordon model) have sufficiently large energy.  Assuming that such a pair has no extra kinetic energy and putting the solution \eqref{kink} into the expression for the energy we get
\be
E_{\rm pair}=2\int_{0}^{\pi}dx\left[\frac{1}{2}(\partial_x\eta)^2+V(\eta)\right]=\pi.
\ee
On the other hand, the energy of the initial configuration strongly depends on $l$ and is given by $E_{l}=\frac{l^3}{24}$ (which is of course the energy of the signum-Gordon breather of the size $l$, see \eqref{breatherenergy}). So we see that the value of the parameter $l$ must not be less then 
\be
l_{\rm min}=\sqrt[3]{24\pi}\approx4.225.\label{minsize}
\ee

\begin{figure*}[h!]
\centering
\subfigure[]{\includegraphics[width=0.3\textwidth,height=0.15\textwidth, angle =0]{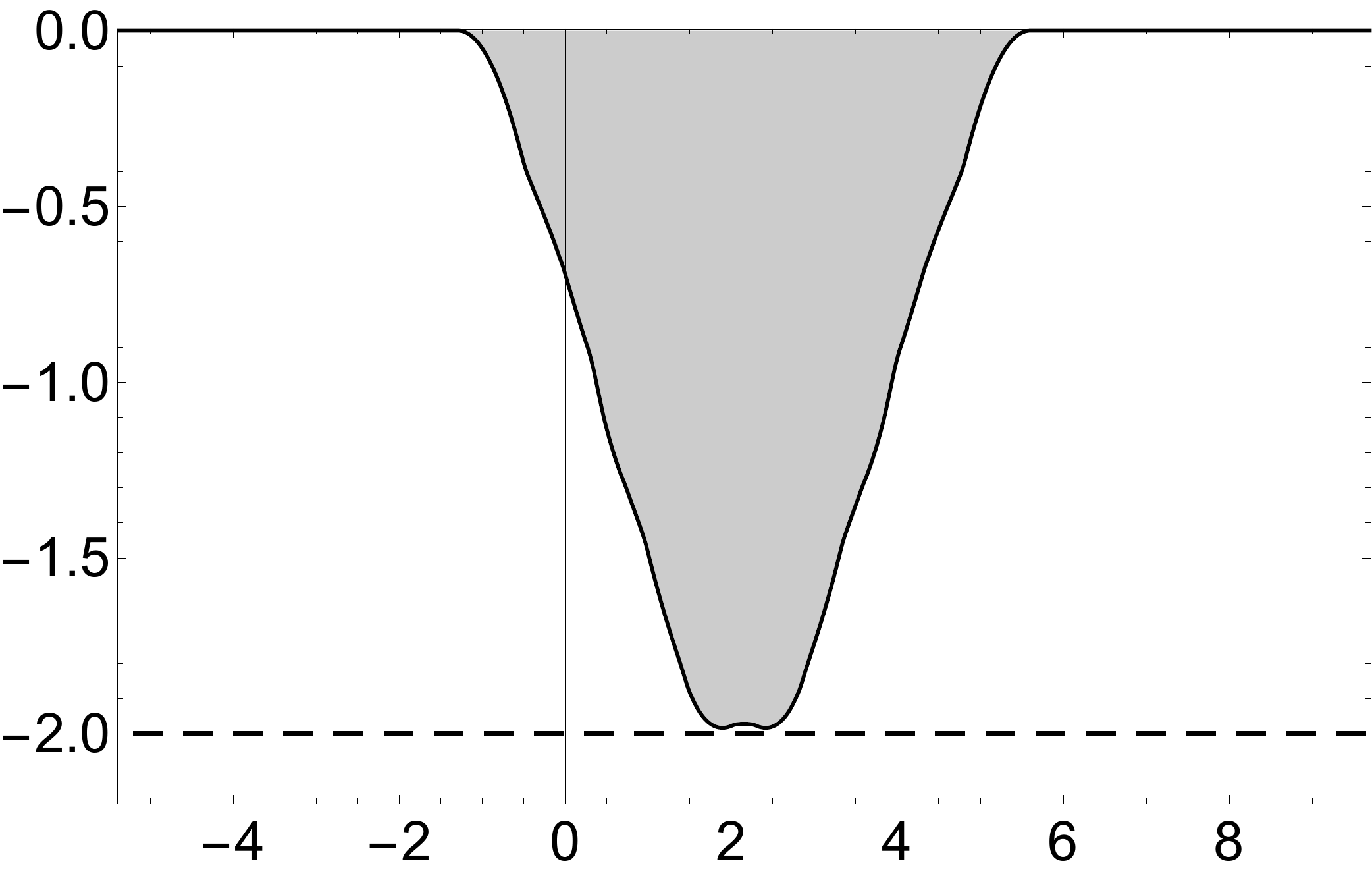}}
\subfigure[]{\includegraphics[width=0.3\textwidth,height=0.15\textwidth, angle =0]{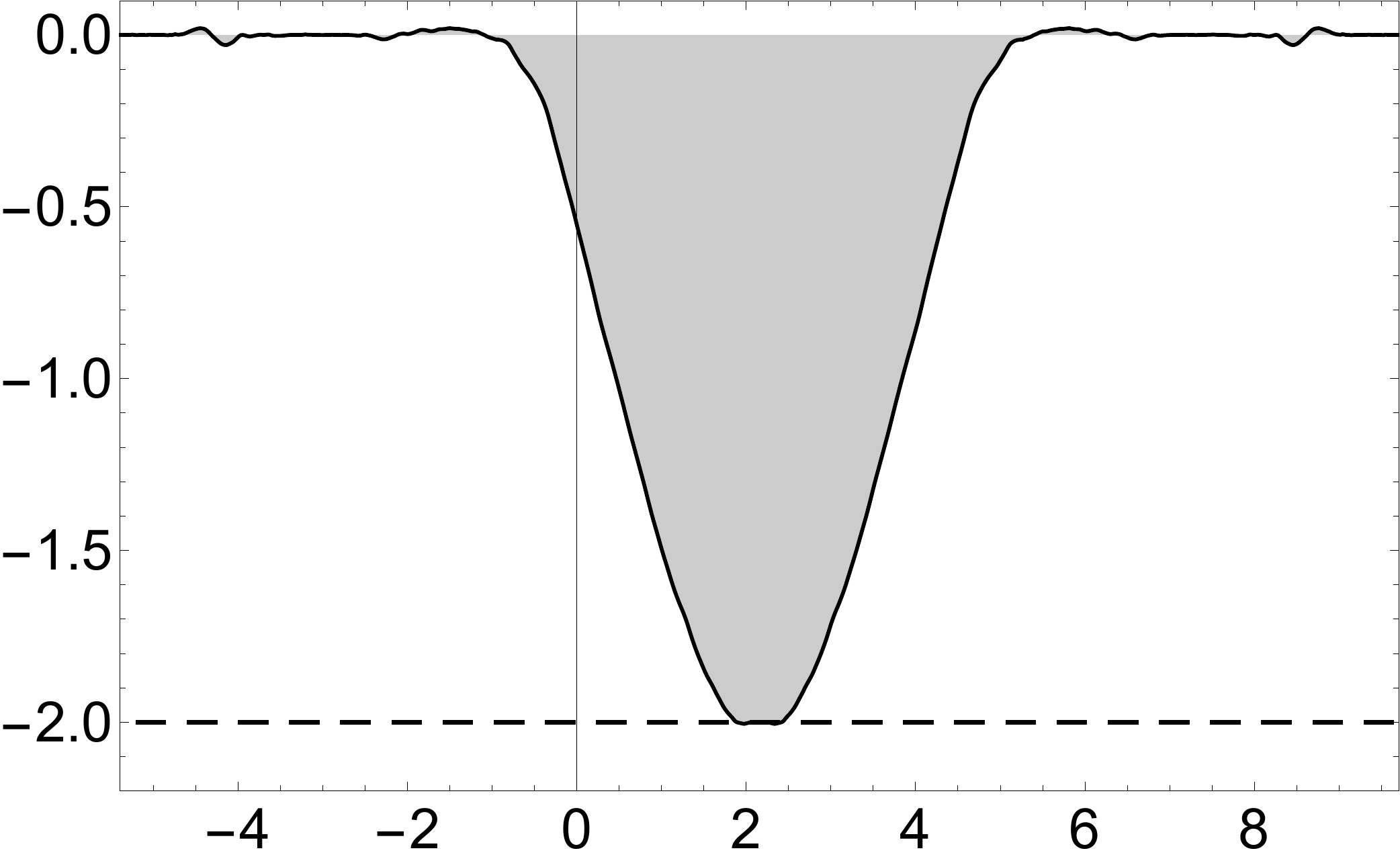}}
\subfigure[]{\includegraphics[width=0.3\textwidth,height=0.15\textwidth, angle =0]{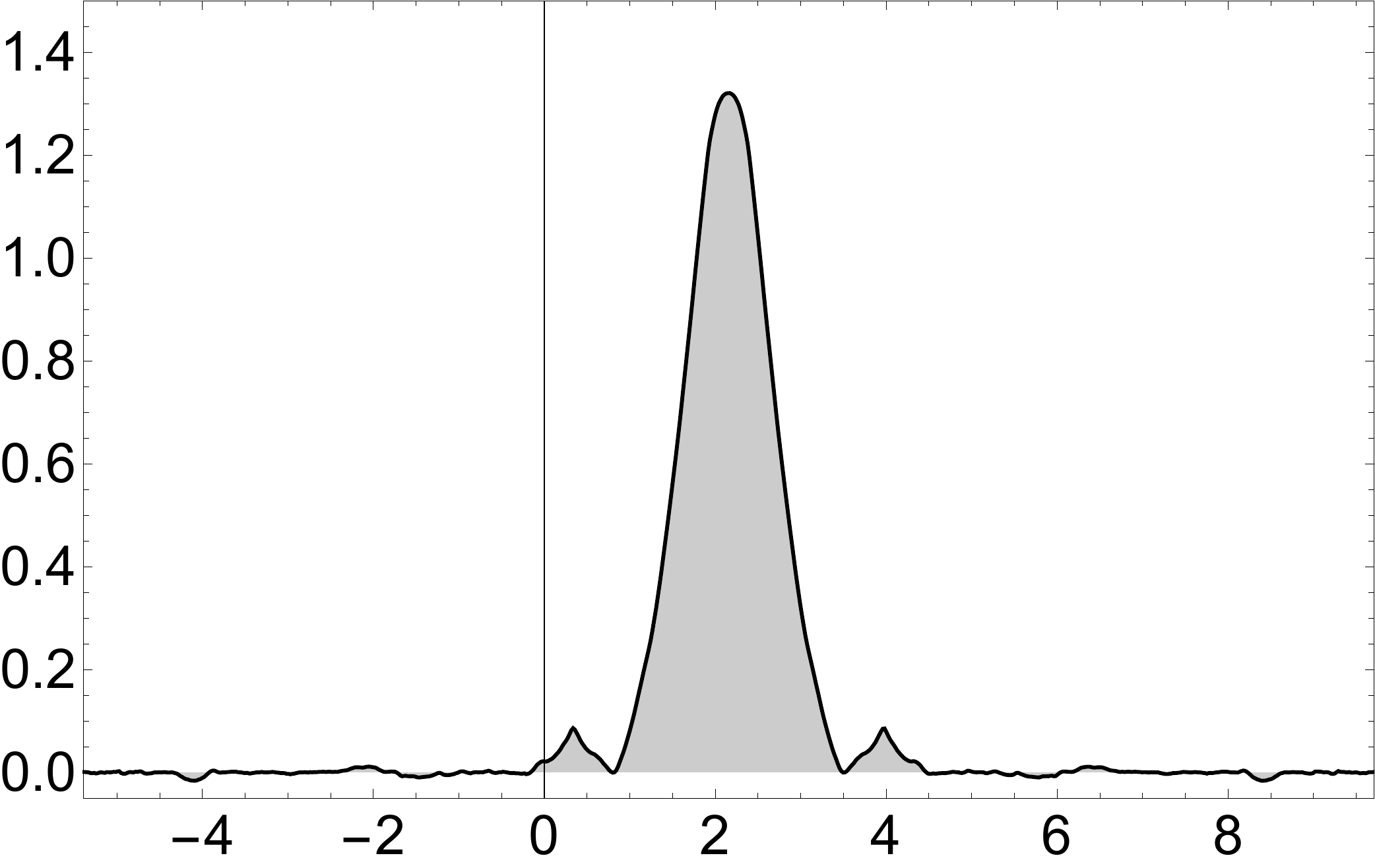}}
\caption{The metastable bounded state for $l=4.3146$ at  (a) $t=7.595$, (b) $t=17.274$, (c) $t=32.609$.}\label{F25}
\end{figure*}

\begin{figure*}[h!]
\centering
\subfigure[]{\includegraphics[width=0.3\textwidth,height=0.15\textwidth, angle =0]{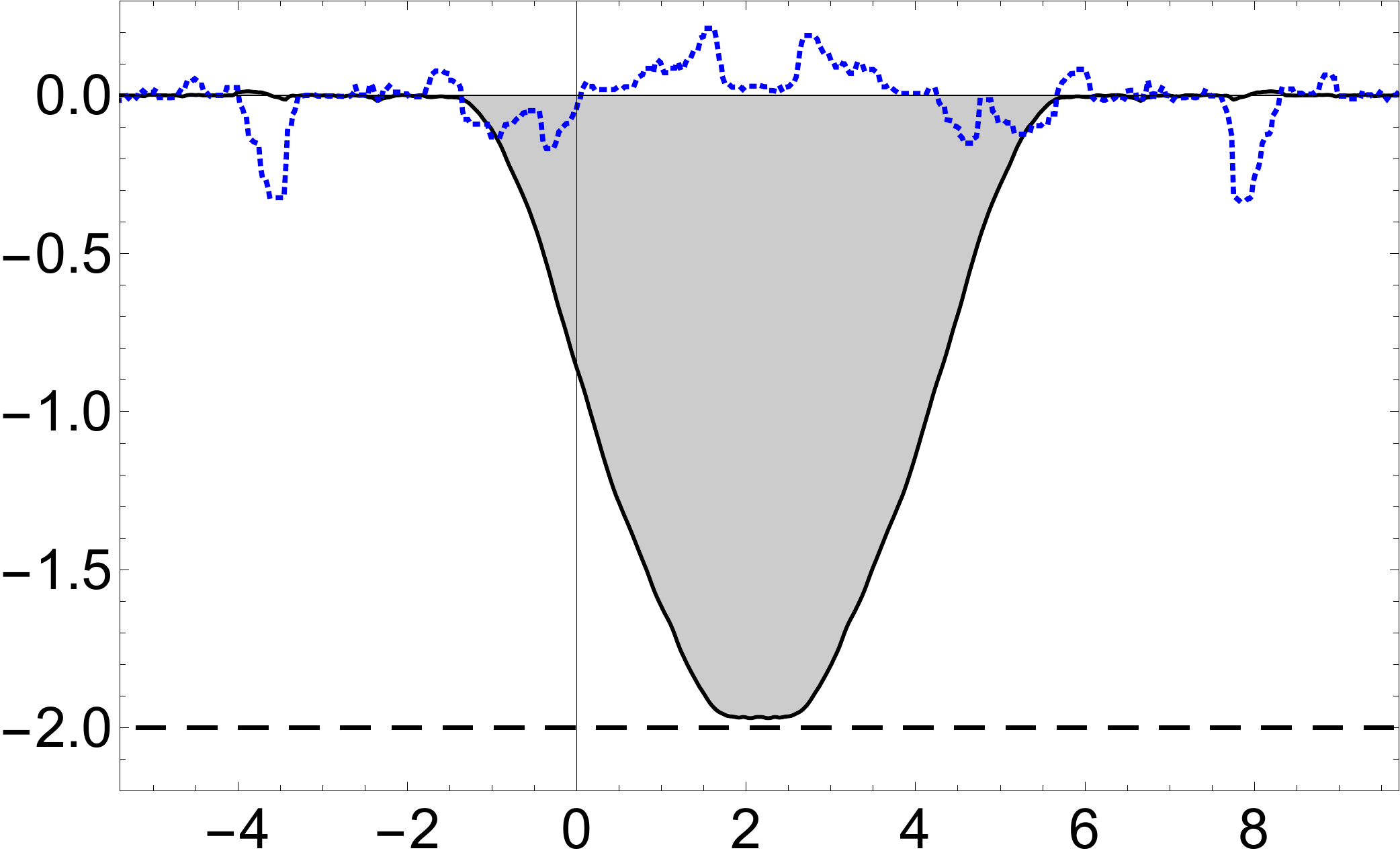}}
\subfigure[]{\includegraphics[width=0.3\textwidth,height=0.15\textwidth, angle =0]{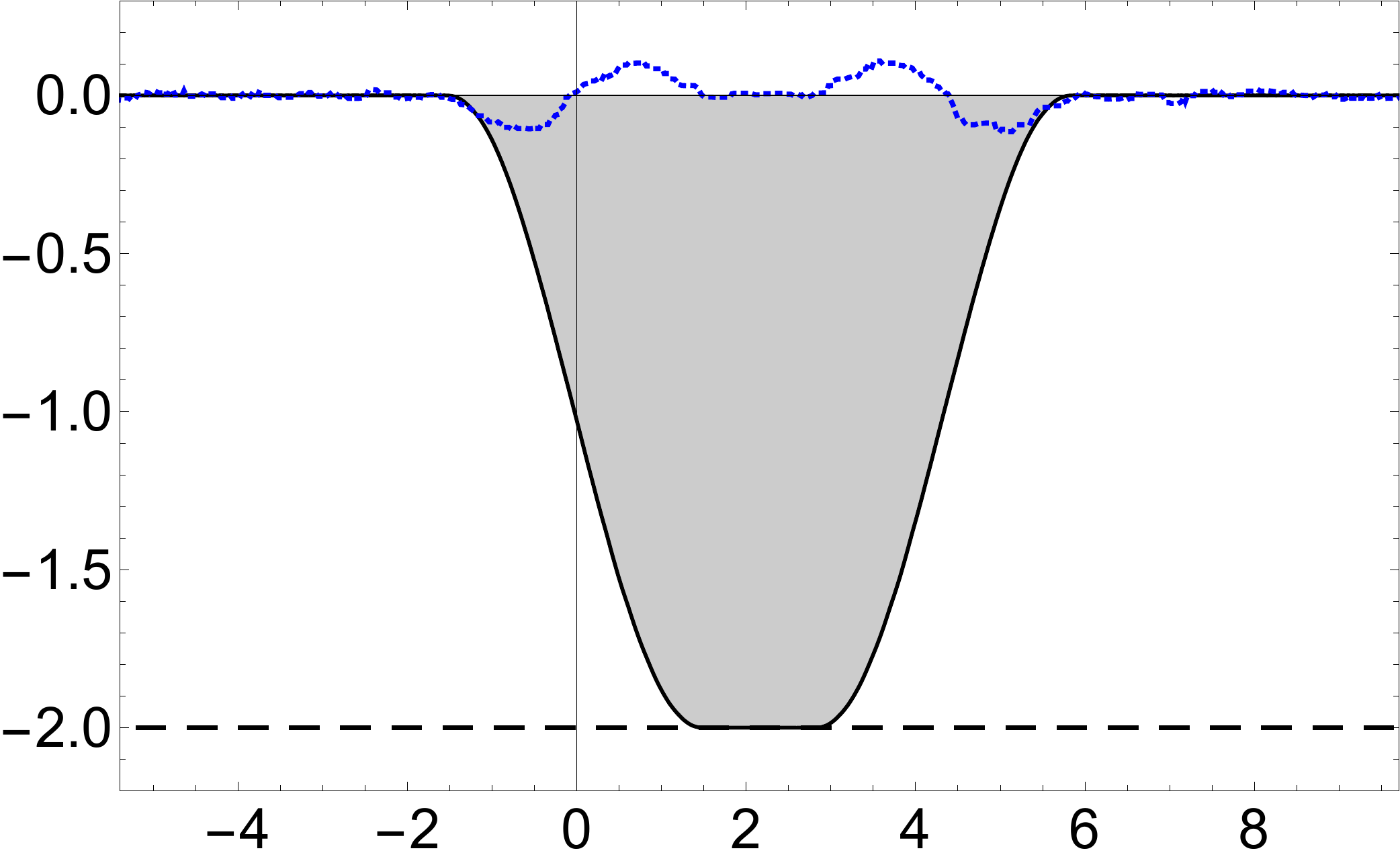}}
\subfigure[]{\includegraphics[width=0.3\textwidth,height=0.15\textwidth, angle =0]{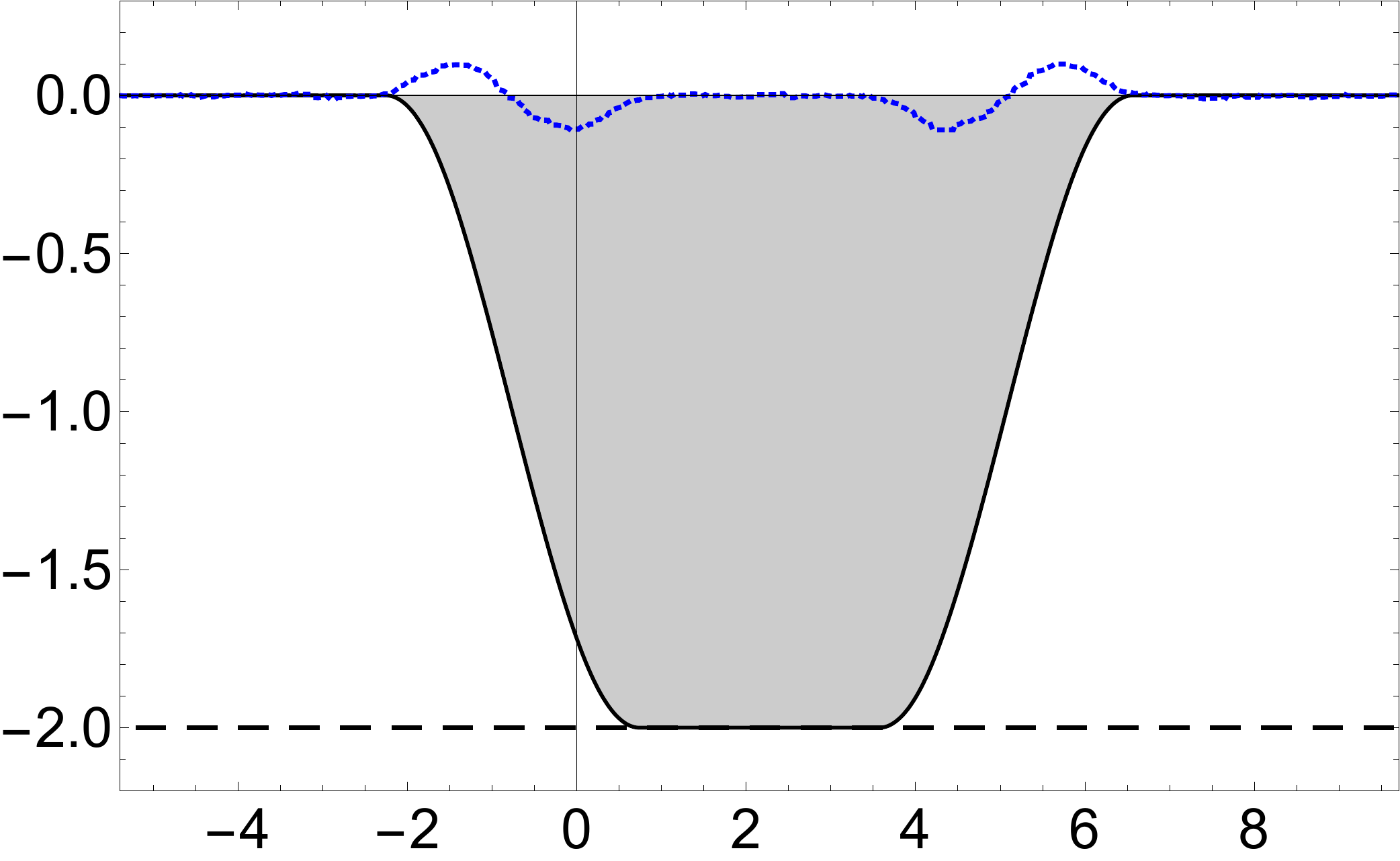}}
\caption{The stable bound state of two kinks for $l=4.3156$ at  (a) $t=19.030$, (b) $t=73.534$, (c) $t=237.208$. The dotted line represents derivative of the field with respect to time $\partial_t\eta$.}\label{F26}
\end{figure*}

In fact, the true critical value of the parameter $l$ is always larger than this minimal value $l_{\rm min}$ established by \eqref{minsize}. There is always some radiation emitted during the very initial phase of the evolution. 
For this reason the energy of the initial configuration must be a somewhat larger than $E=\pi$.  In consequence, $l=l_{\rm crit}>l_{\rm min}$. In Fig.\ref{F25} we plot a kink and antikink state which exists for a short period of time and then decays into an oscillon and some radiation. This configuration was obtained for $l=4.3146$. Such a pair had been created with very little kinetic energy and therefore the constituents did not have enough energy to escape before they formed an oscillon and stabilised with the energy below the creation threshold. If we increase $l$, {\it i.e.} add more energy, the created kink and antikink pair stays forever (in our simulation $t_{\max}\approx 237$). Such a configuration is plotted in Fig.\ref{F26} and it was obtained for $l=4.3156$.
This time the solitons have sufficient energy to escape from each other to a finite distance, which due to their compacton nature, guarantees that now they interact weekly via emitting small oscillons. The initial kink-antikink state loses some energy in the form of radiation and the system evolves to a configuration plotted in Fig.\ref{F26}(c). An interesting fact about this `final' state is that kinks are not static (although their centres do not move). We have also studied the time derivative of the field $\partial_t\eta$ and this has shown us that the kinks wobble - shrinking and  expanding periodically around their own centers. This means that the extra energy has excited some of their internal modes. 

\begin{figure*}[h!]
\centering
\subfigure{\includegraphics[width=0.5\textwidth,height=0.2\textwidth, angle =0]{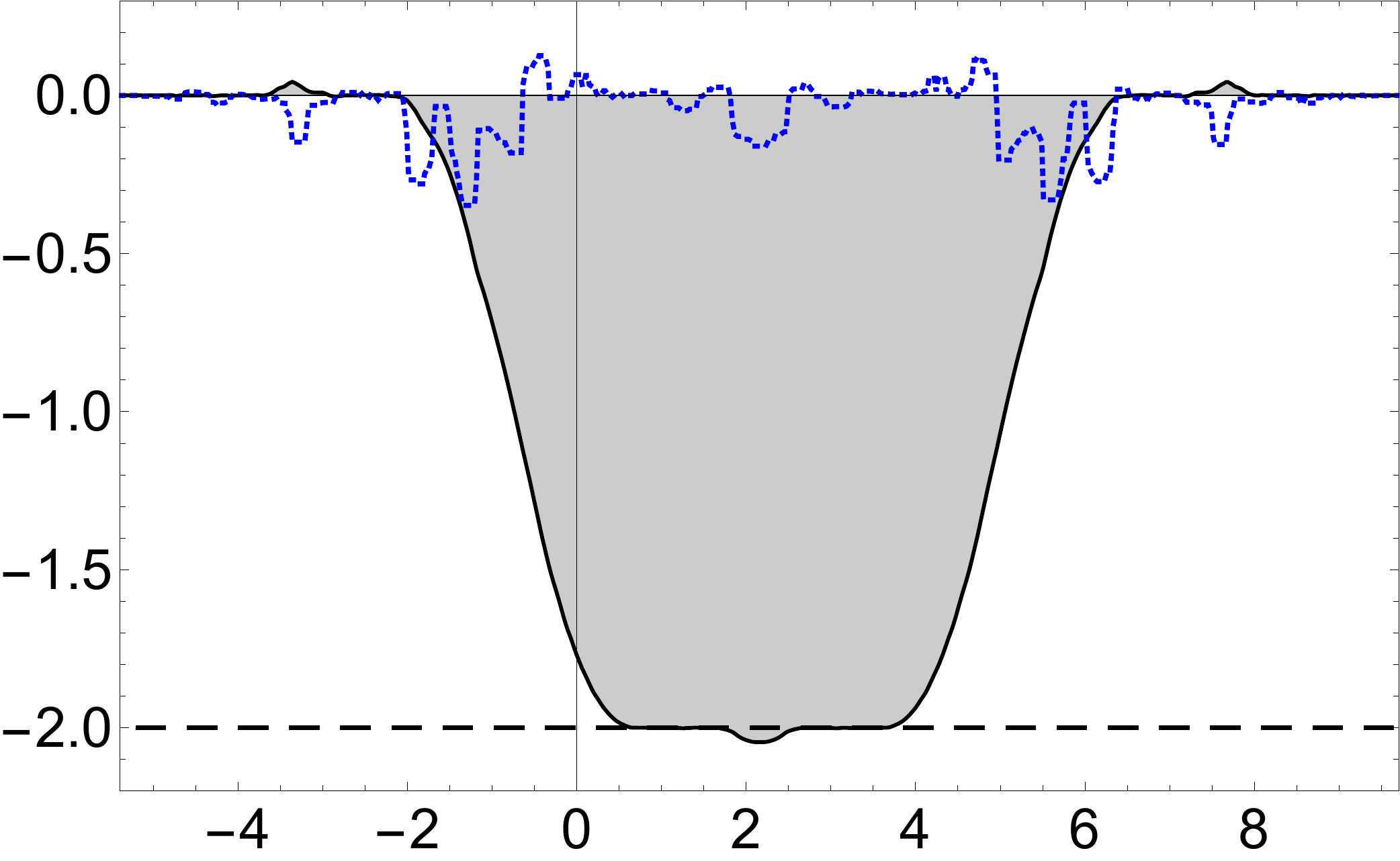}}
\caption{Two kinks and oscillon for $l=4.3246$ at $t=14.272$. The dotted line represents derivative of the field with respect to time $\partial_t\eta$.}\label{F27}
\end{figure*}

We have also looked at simulations which started with even a little more energy.  Of course, we have also got the kink-antikink pair. This time the surplus of energy has been transformed into {\it i)} the kinetic energy of the kinks (they move apart), {\it ii)} the appearance of the oscillon at the centre, and {\it iii)} some radiation which escapes from the region where the kinks are localized. Such a configuration  was obtained for $l=4.3246$. In Fig.\ref{F27} we present a snapshot of the field taken at $t=14.272$.

\begin{figure*}[h!]
\centering
\subfigure[]{\includegraphics[width=0.45\textwidth,height=0.2\textwidth, angle =0]{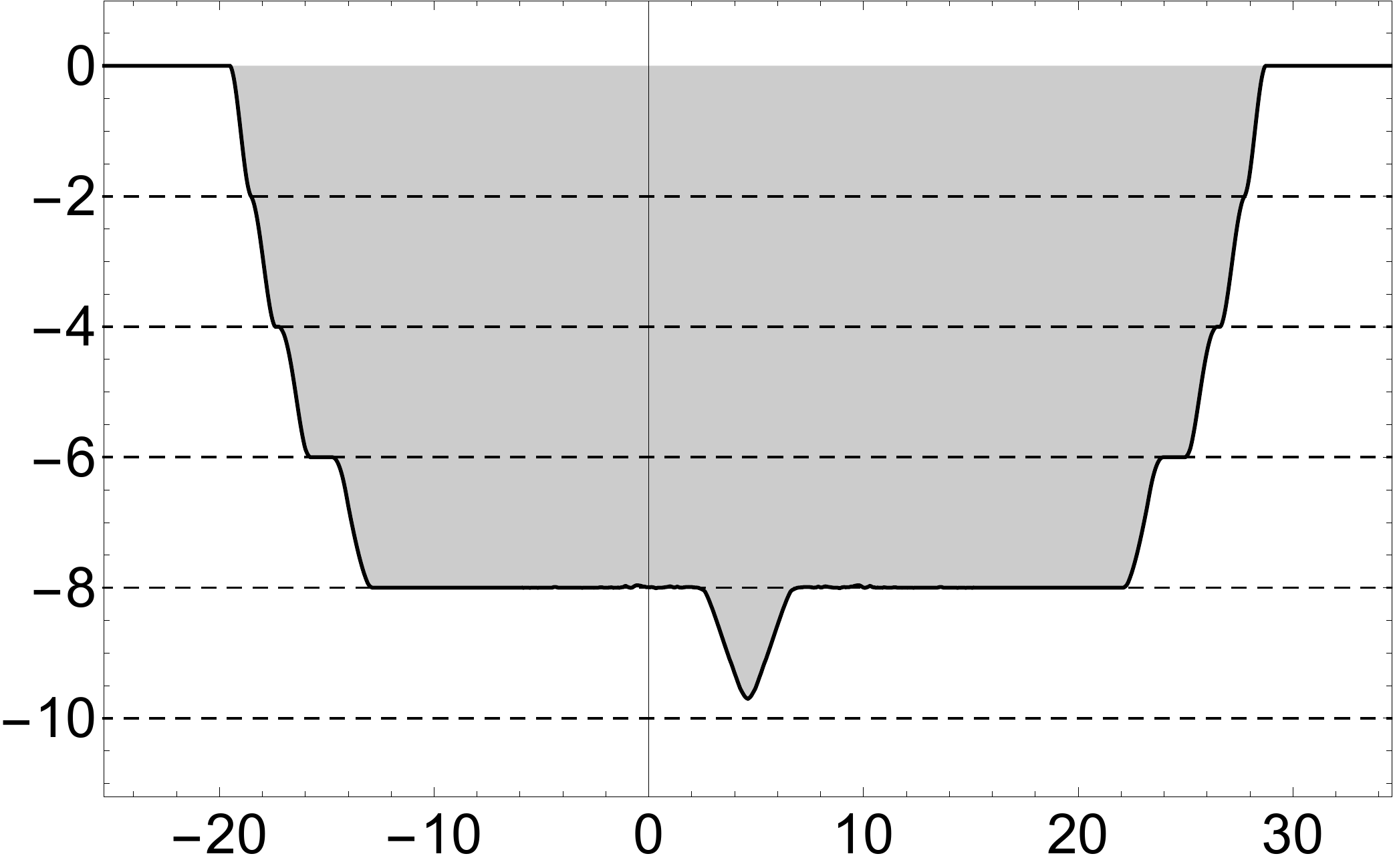}}
\subfigure[]{\includegraphics[width=0.45\textwidth,height=0.2\textwidth, angle =0]{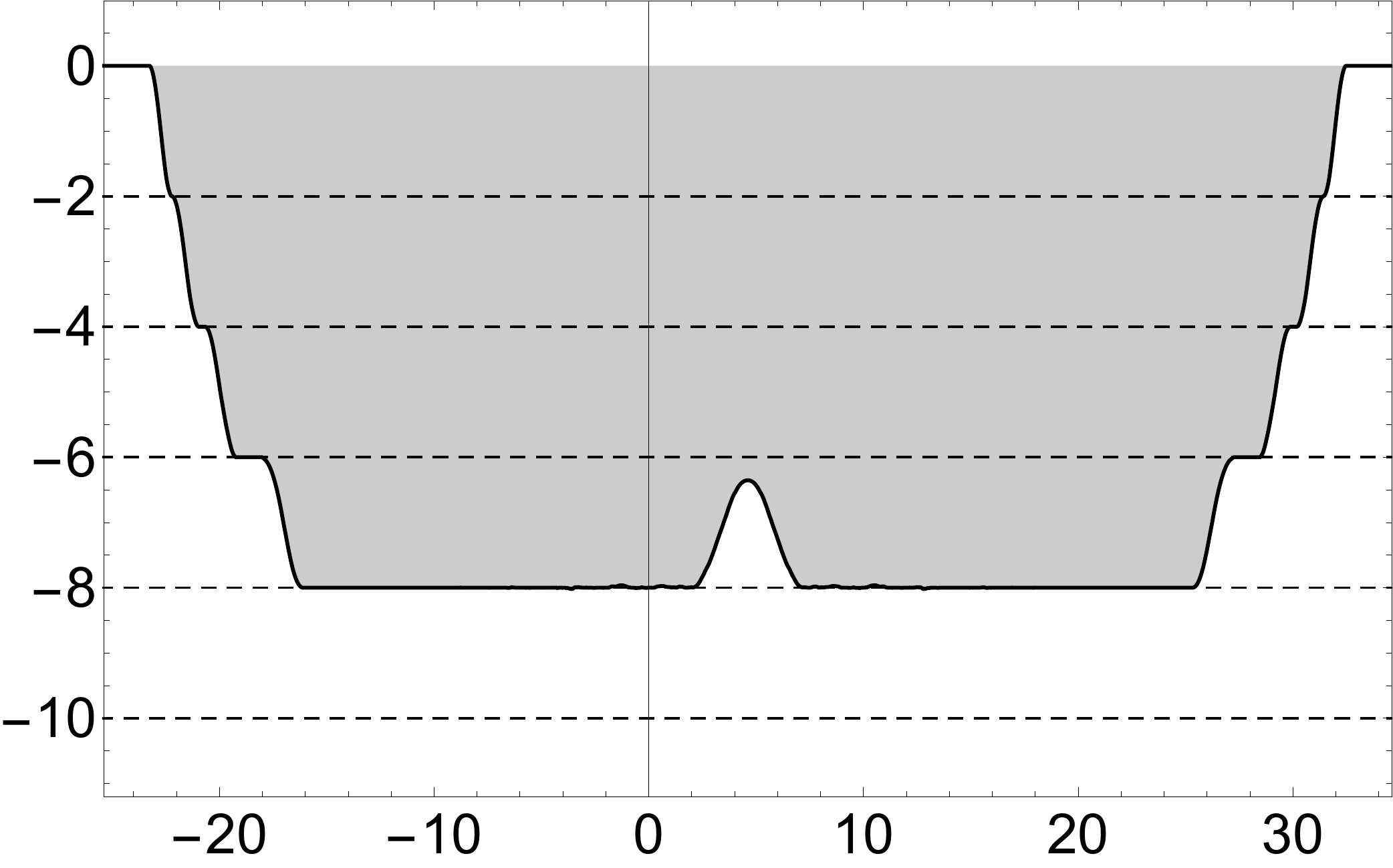}}
\subfigure[]{\includegraphics[width=0.45\textwidth,height=0.15\textwidth, angle =0]{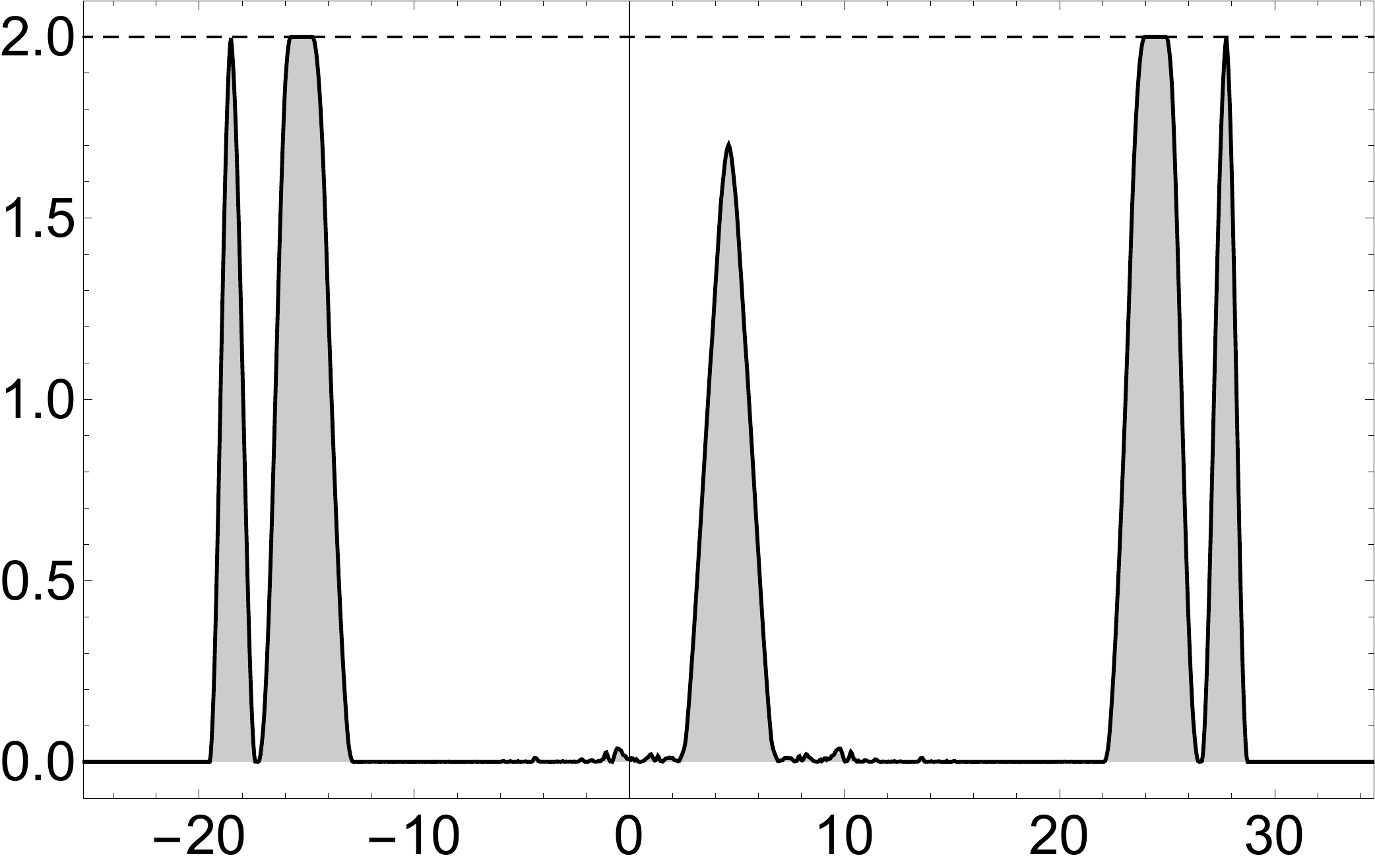}}
\subfigure[]{\includegraphics[width=0.45\textwidth,height=0.15\textwidth, angle =0]{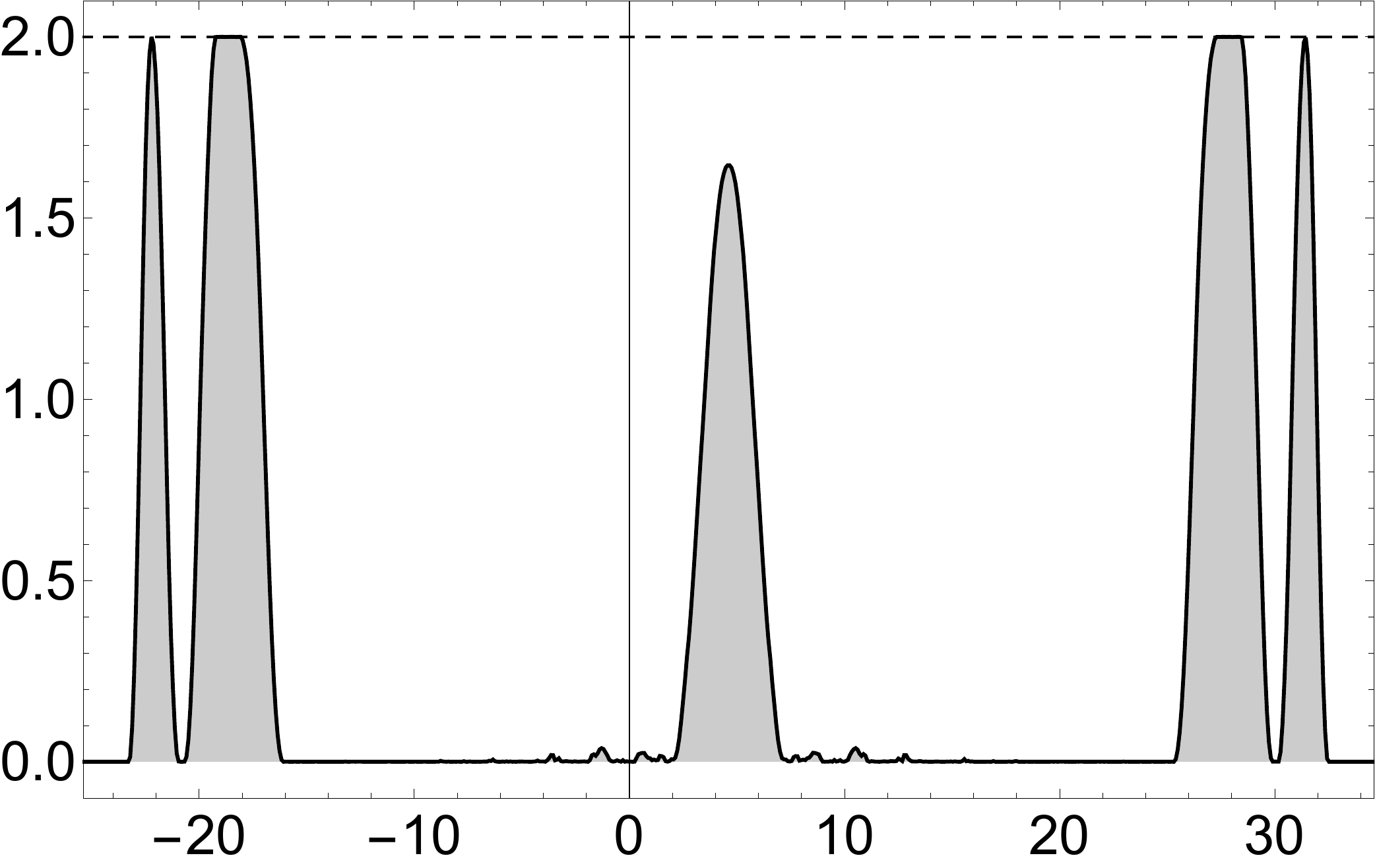}}
\caption{Four pairs of kink-antikink states and the oscillon in the centre for $l=9.2346$ at  (a,c) $t=23.879$, (b,d) $t=27.829$. Figures (c,d) represent folded (original) field $\bar\eta(t,x)$.}\label{F28}
\end{figure*}

When the initial configuration has the energy  which is significantly larger then this energy can be used to create many pairs of kinks and antikinks. A configuration containing four such pairs is presented in Fig. \ref{F28}. In this figure we also present the plots of the auxiliary unfolded field $\eta$ and of the original (folded) field $\bar\eta$. The
observed  process has proceeded in stages.
 The pair of kinks created first had higher velocities than pairs that arose later. This is clearly visible from the Lorentz contraction 
of the kinks in motion. A part of energy was also transformed into the creation of an oscillon at the centre (at $\frac{l}{2}$). We also see a small amount of radiation in the vicinity of the oscillon (pictures in folded variable).

\subsubsection{Perturbed oscillons}

We have also looked at the solutions for $\epsilon\neq 1$ which represents the original signum-Gordon breather plus a perturbation on top of it, immersed into the modified model. In this section we report the results of our studies with initial configurations \eqref{inieps} started with the scale parameter $l$ fixed at $l=1$. This means that our initial configuration had energy $\epsilon^2/24$.

\begin{figure*}[h!]
\centering
\subfigure[]{\includegraphics[width=0.3\textwidth,height=0.15\textwidth, angle =0]{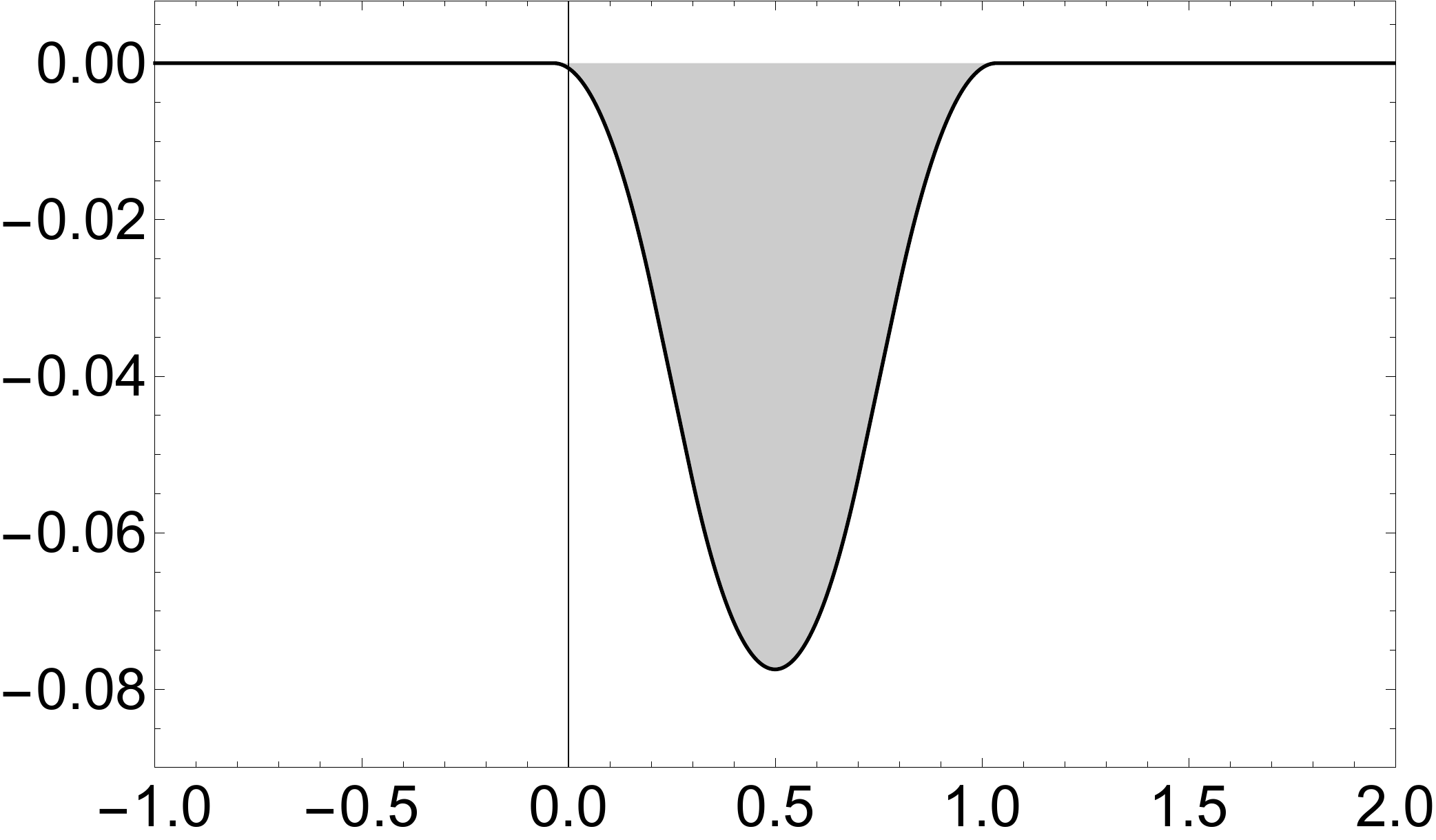}}
\subfigure[]{\includegraphics[width=0.3\textwidth,height=0.15\textwidth, angle =0]{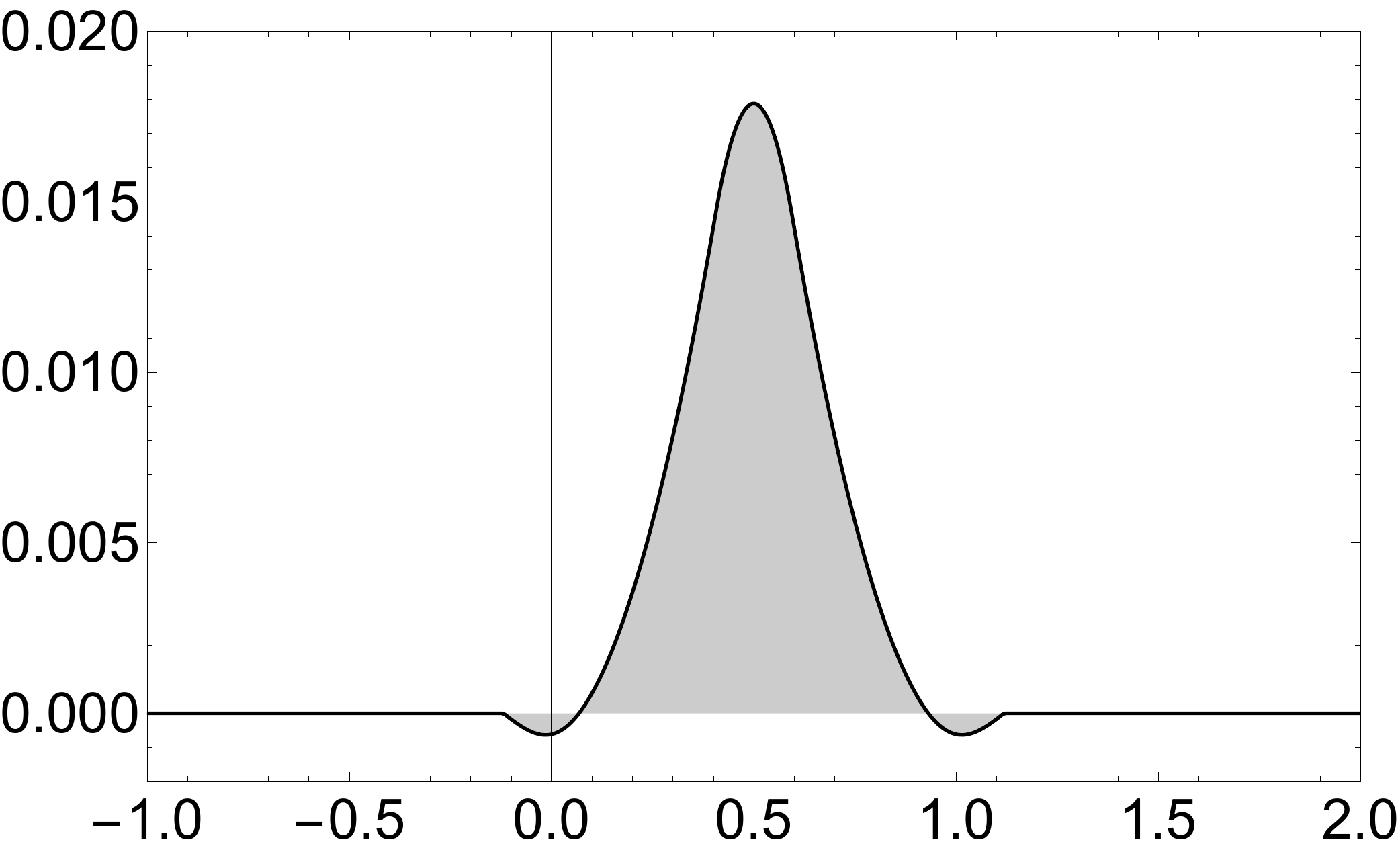}}
\subfigure[]{\includegraphics[width=0.3\textwidth,height=0.15\textwidth, angle =0]{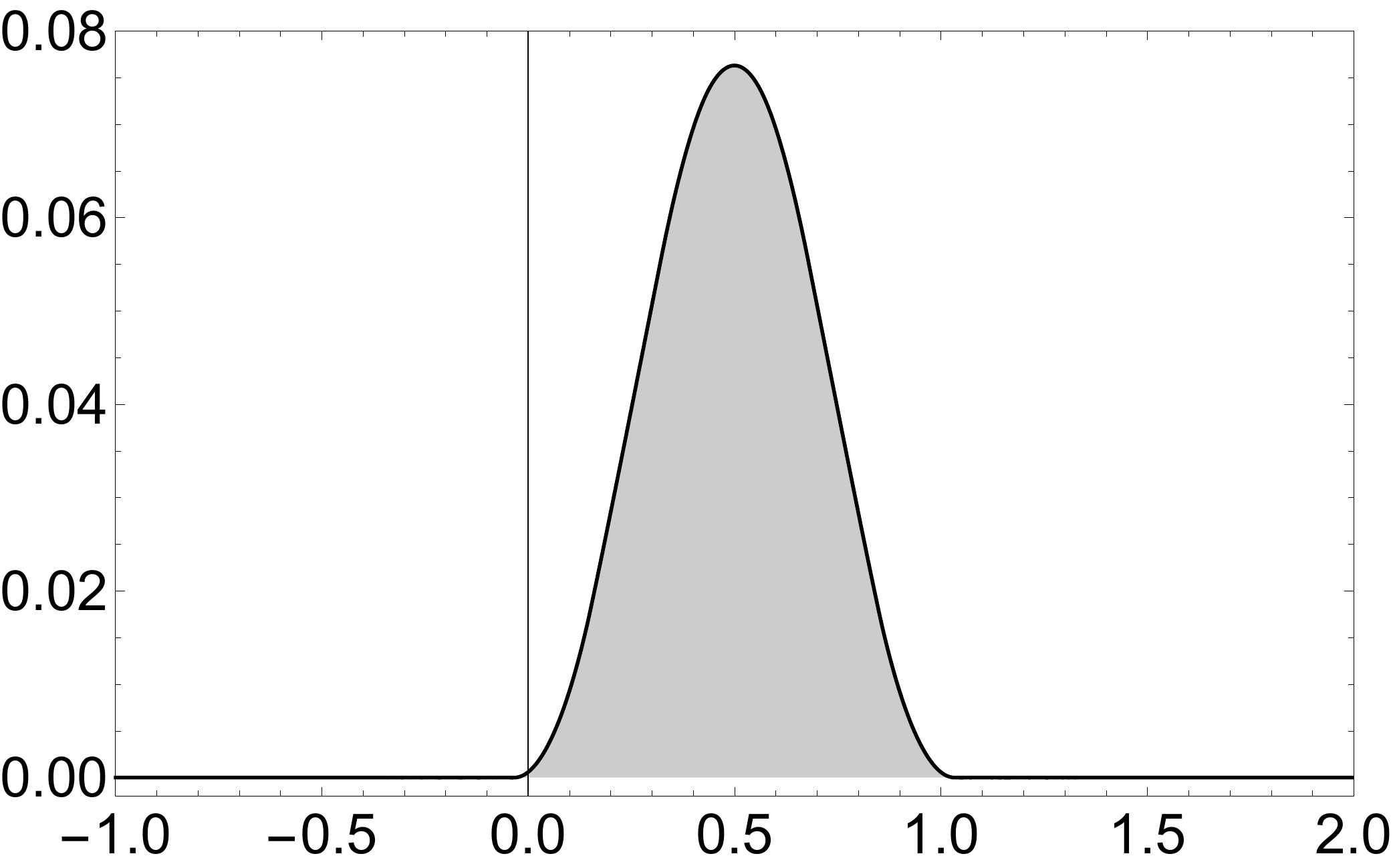}}
\subfigure[]{\includegraphics[width=0.3\textwidth,height=0.15\textwidth, angle =0]{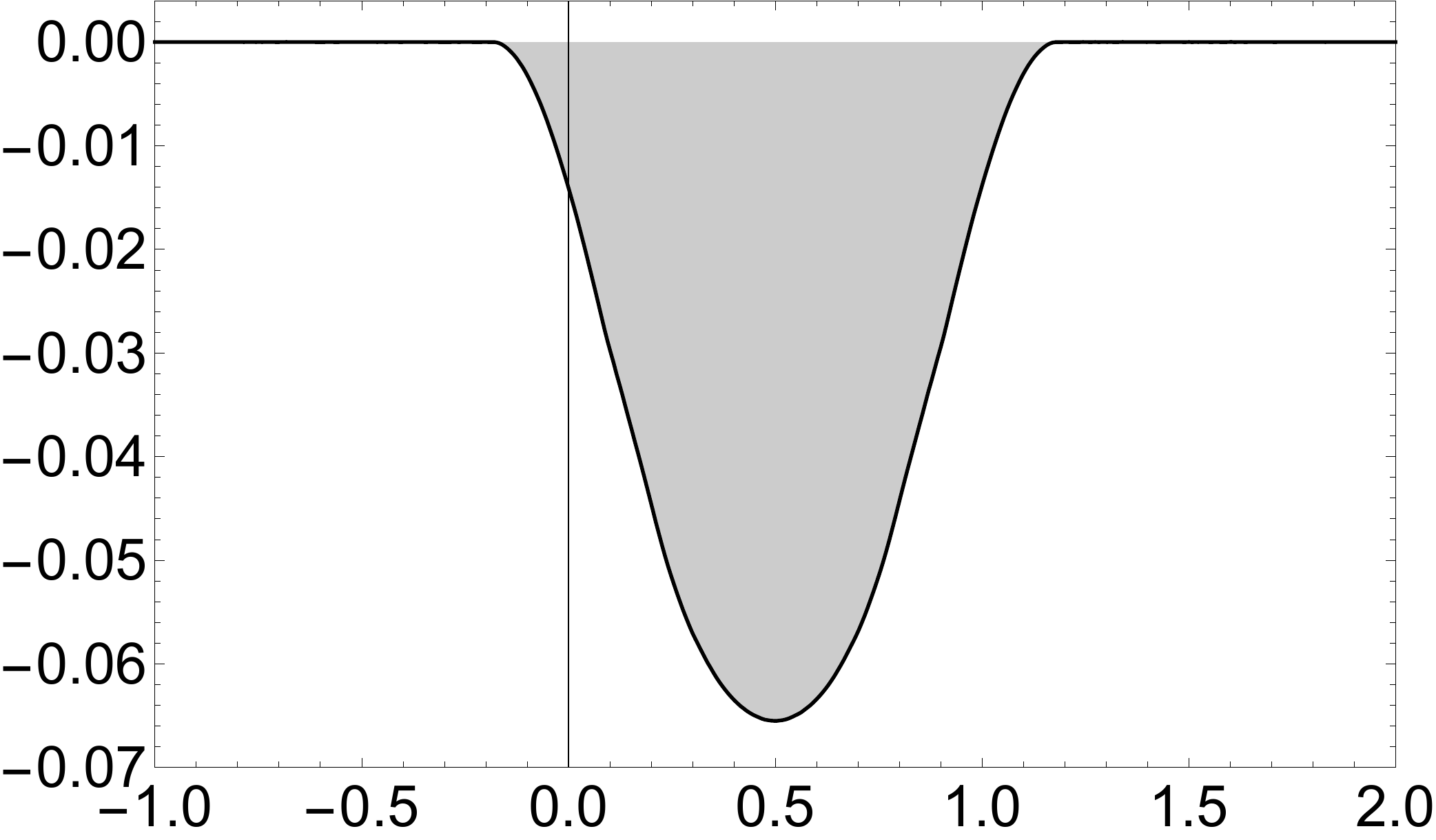}}
\subfigure[]{\includegraphics[width=0.3\textwidth,height=0.15\textwidth, angle =0]{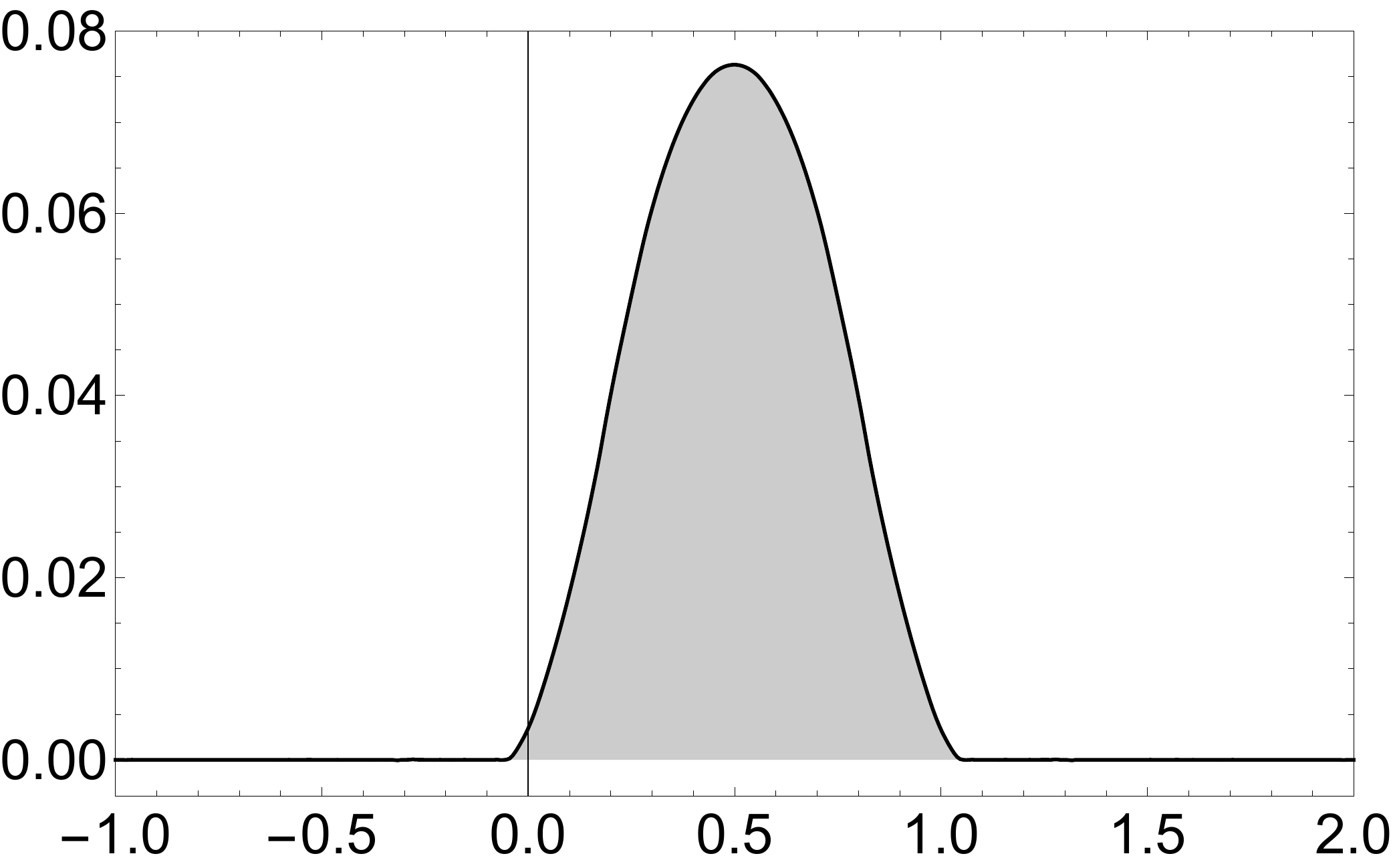}}
\subfigure[]{\includegraphics[width=0.3\textwidth,height=0.15\textwidth, angle =0]{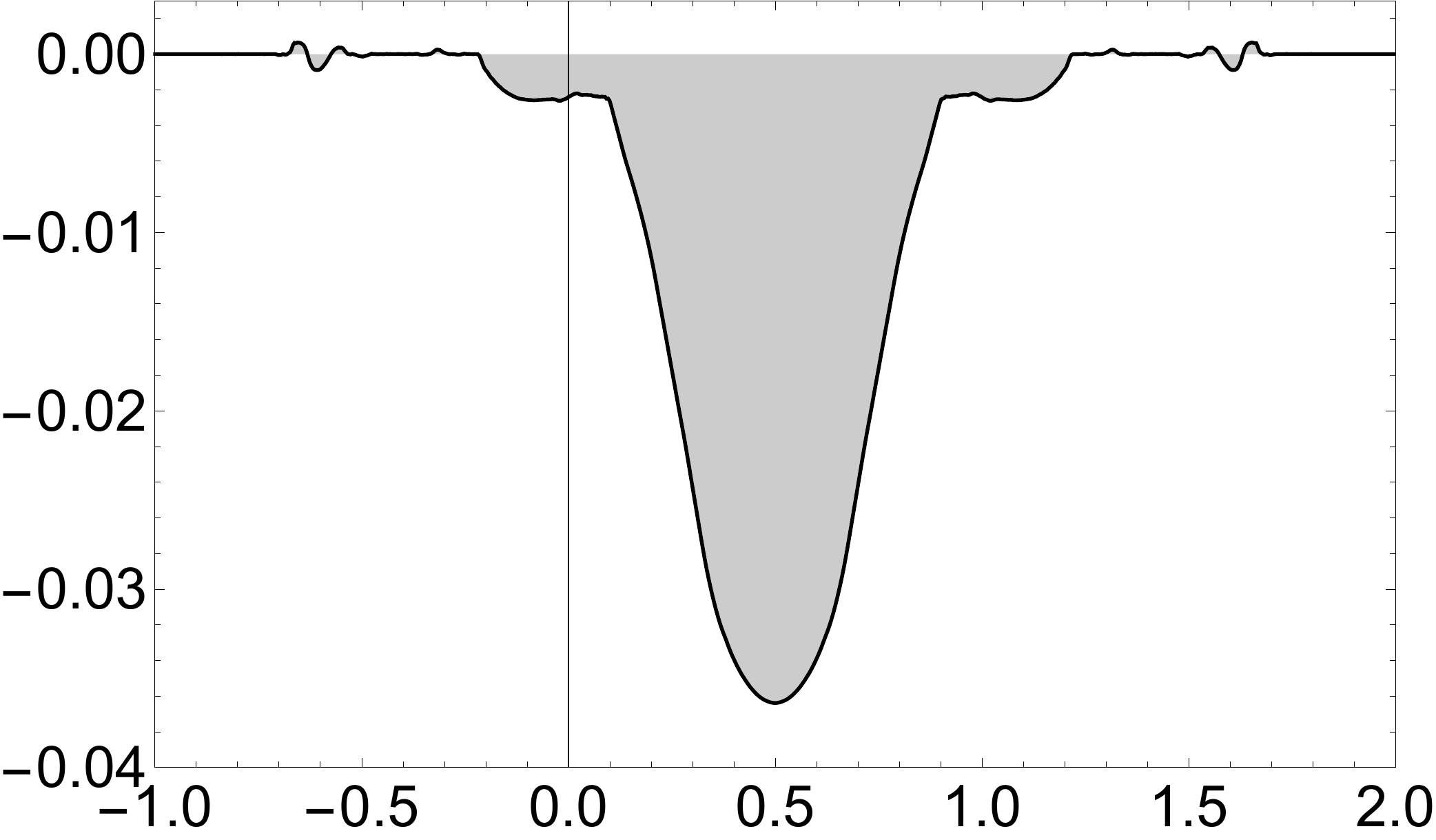}}
\caption{The perturbed signum-Gordon model for $\epsilon=1.2$ at (a) $t=0.2052$, (b) $t=1.1350$, (c) $t=5.4025$, (d) $t=26.9678$, (e) $t=27.5035$, (f) $t=32.8919$.}\label{PSG1.2}
\end{figure*}

We begun our studies by looking at small perturbations of the pure signum-Gordon breather. For this reason we have chosen  $\epsilon=1.2$, which corresponded to the case studied in the pure signum-Gordon model. In  Fig.\ref{PSG1.2} we present a few snapshots of the fields seen in our numerical simulations. Comparing these results with those obtained for the pure signum-Gordon model (see Fig.\ref{SG-e-1.2}) we conclude that the oscillations in the perturbed model are more regular than in the pure signum-Gordon one. This may appear to be a rather unexpected result. The radiation in the perturbed model becomes visible approximately at $t=30$ whereas in the pure signum-Gordon 
model it is already present at $t=13$. This suggests that the modification of the potential can enhance the stability of the solutions with respect to small perturbations.

\begin{figure*}[h]
\centering
\subfigure[]{\includegraphics[width=0.3\textwidth,height=0.15\textwidth, angle =0]{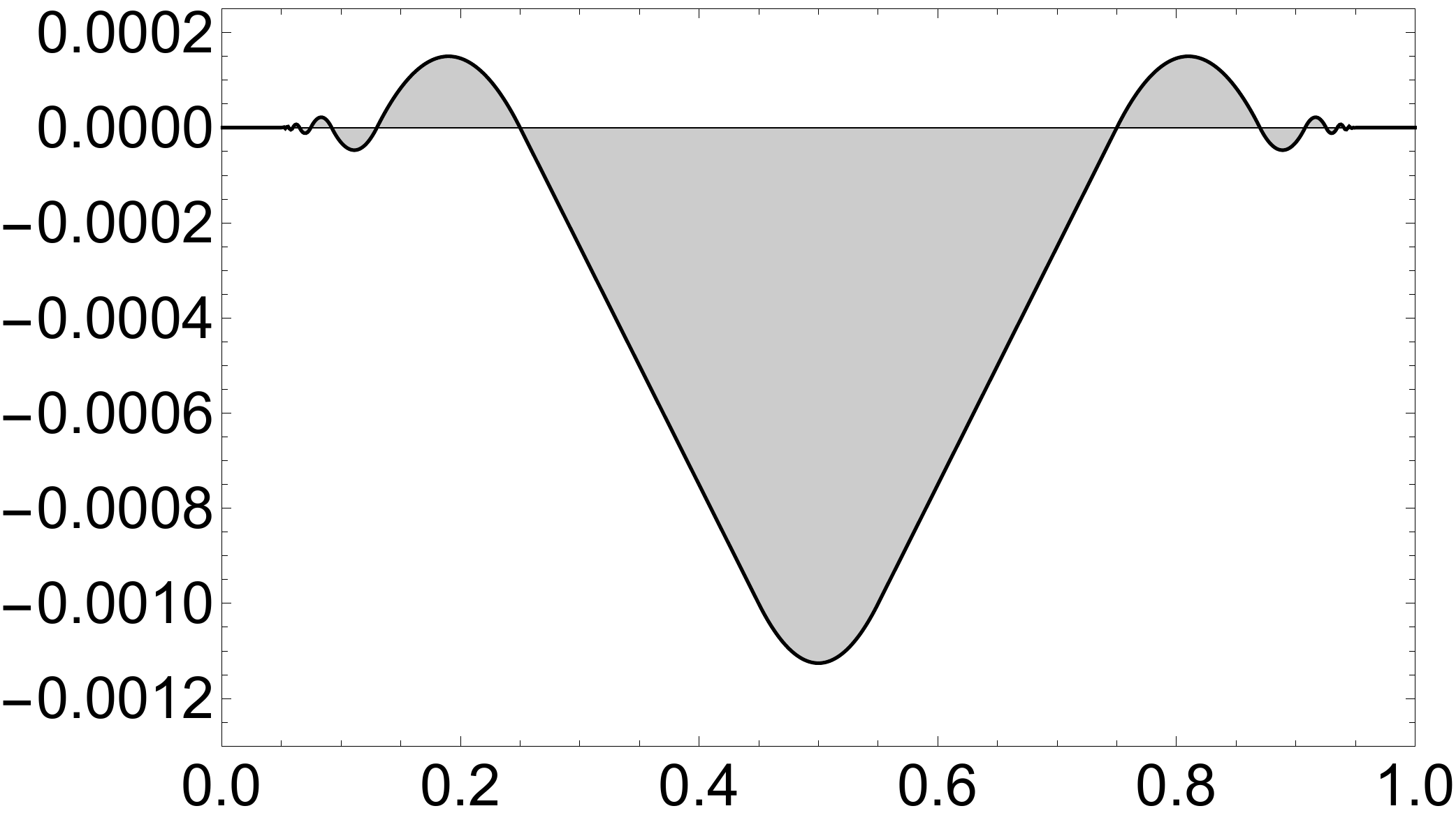}}
\subfigure[]{\includegraphics[width=0.3\textwidth,height=0.15\textwidth, angle =0]{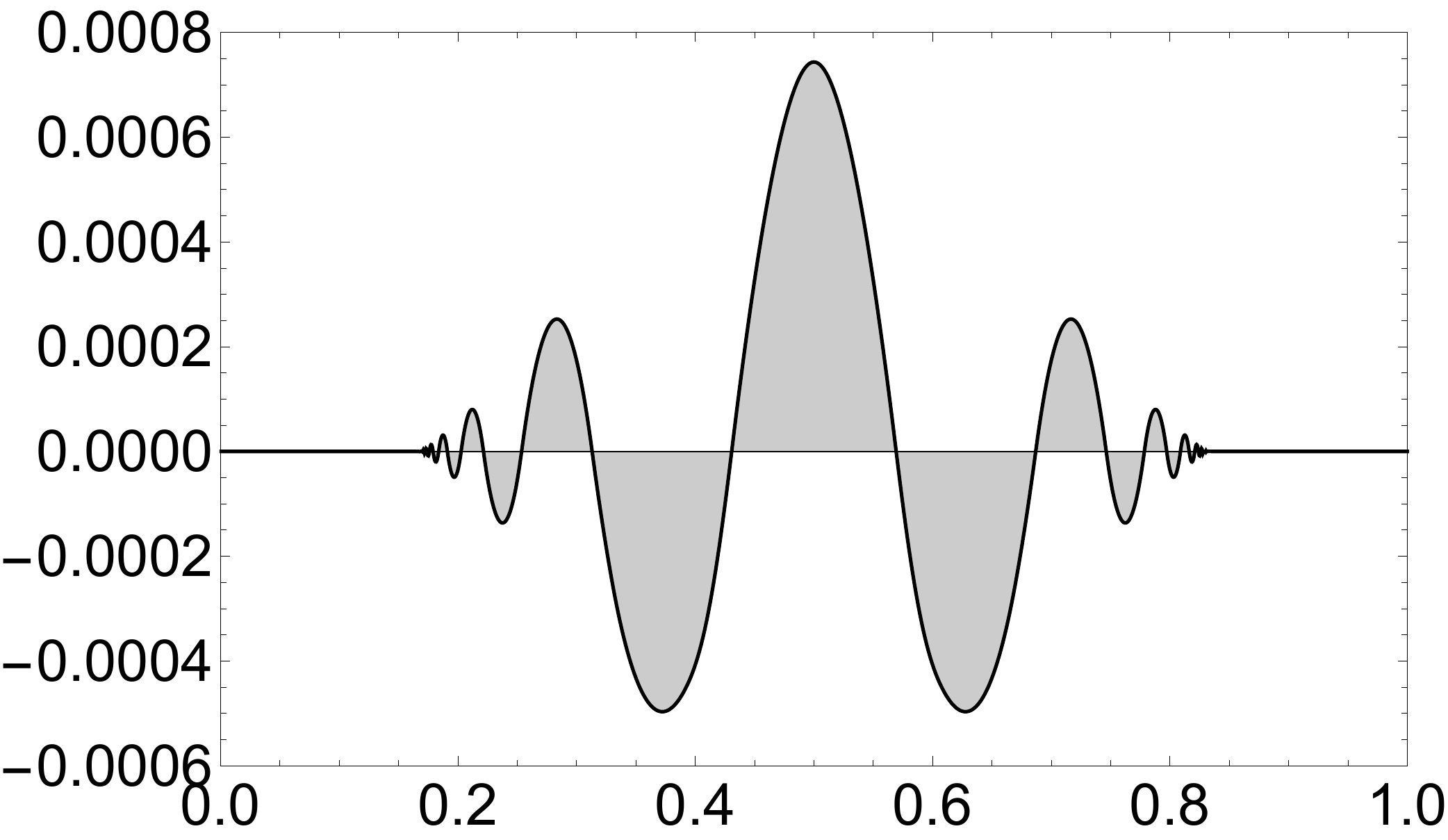}}
\subfigure[]{\includegraphics[width=0.3\textwidth,height=0.15\textwidth, angle =0]{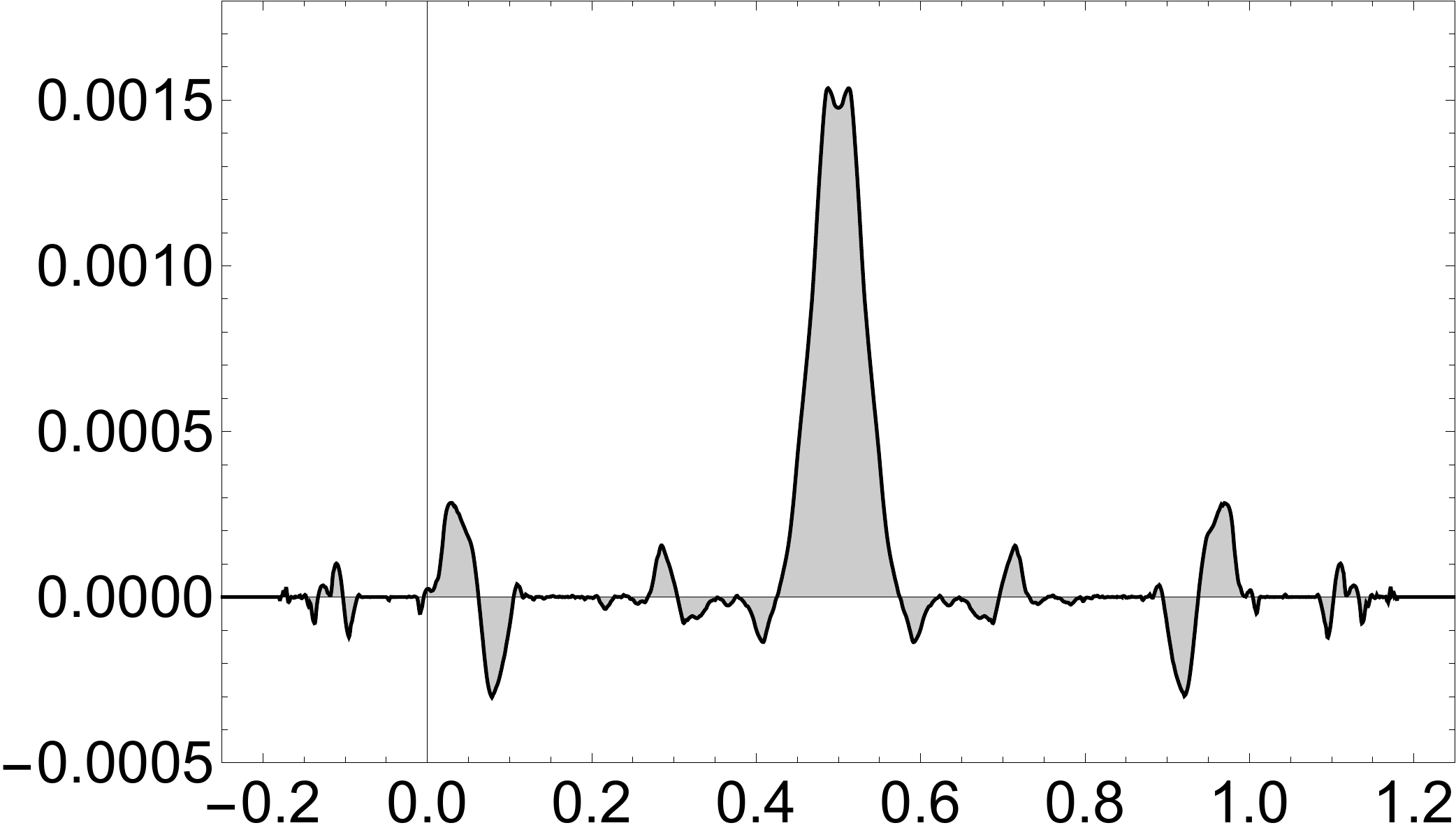}}
\caption{Numerical solutions  in the perturbed model for $\epsilon=0.1$ at  (a) $t=0.05$, (b) $t=0.17$, (c) $t=1.18$. }\label{F15}
\end{figure*}

Another observation we have made involves noting that for very small $\epsilon$ the field $\eta$ has a very small amplitude and so the term $-\eta$ in the field equation has no practical significance by comparison with the term ${\rm sgn}(\eta)$. As an example we have considered
the case of $\epsilon=0.1$. We have plotted the snapshots of the corresponding field in Fig.\ref{F15}  and we note that are very similar to the plots of the solution obtained for the signum-Gordon model which were presented in Fig.\ref{F9}. 


\begin{figure*}[h!]
\centering
\subfigure[]{\includegraphics[width=0.3\textwidth,height=0.15\textwidth, angle =0]{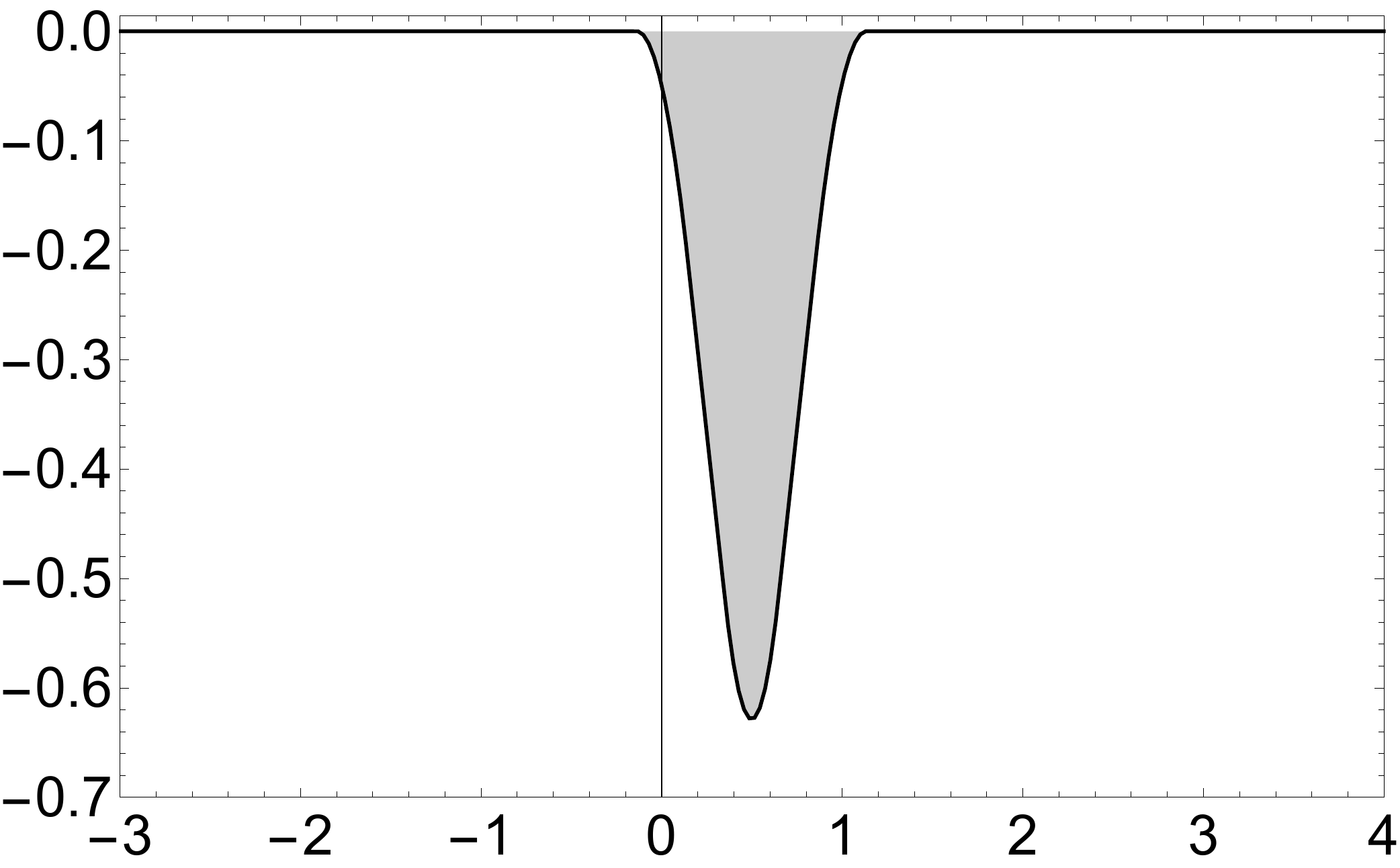}}
\subfigure[]{\includegraphics[width=0.3\textwidth,height=0.15\textwidth, angle =0]{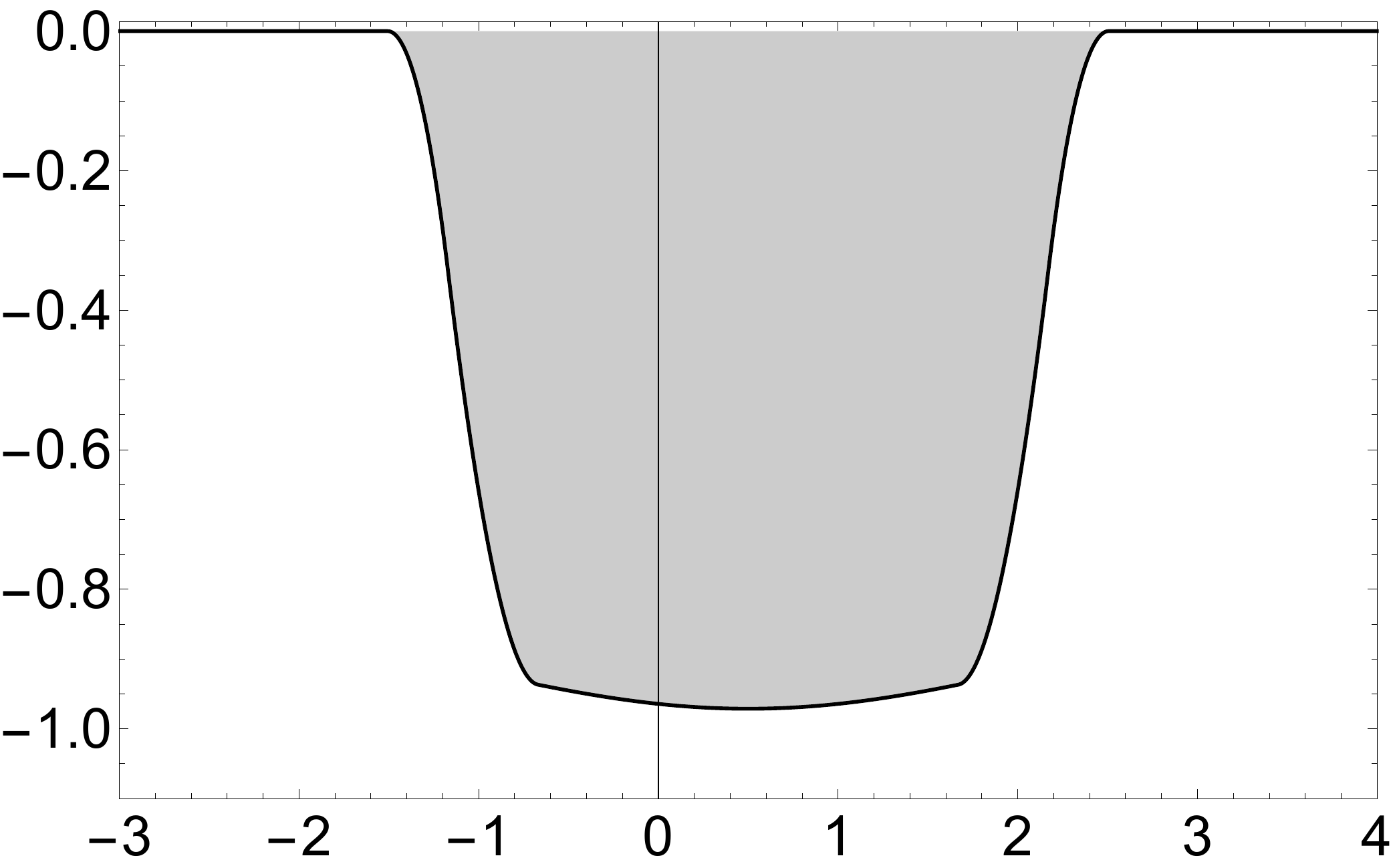}}
\subfigure[]{\includegraphics[width=0.3\textwidth,height=0.15\textwidth, angle =0]{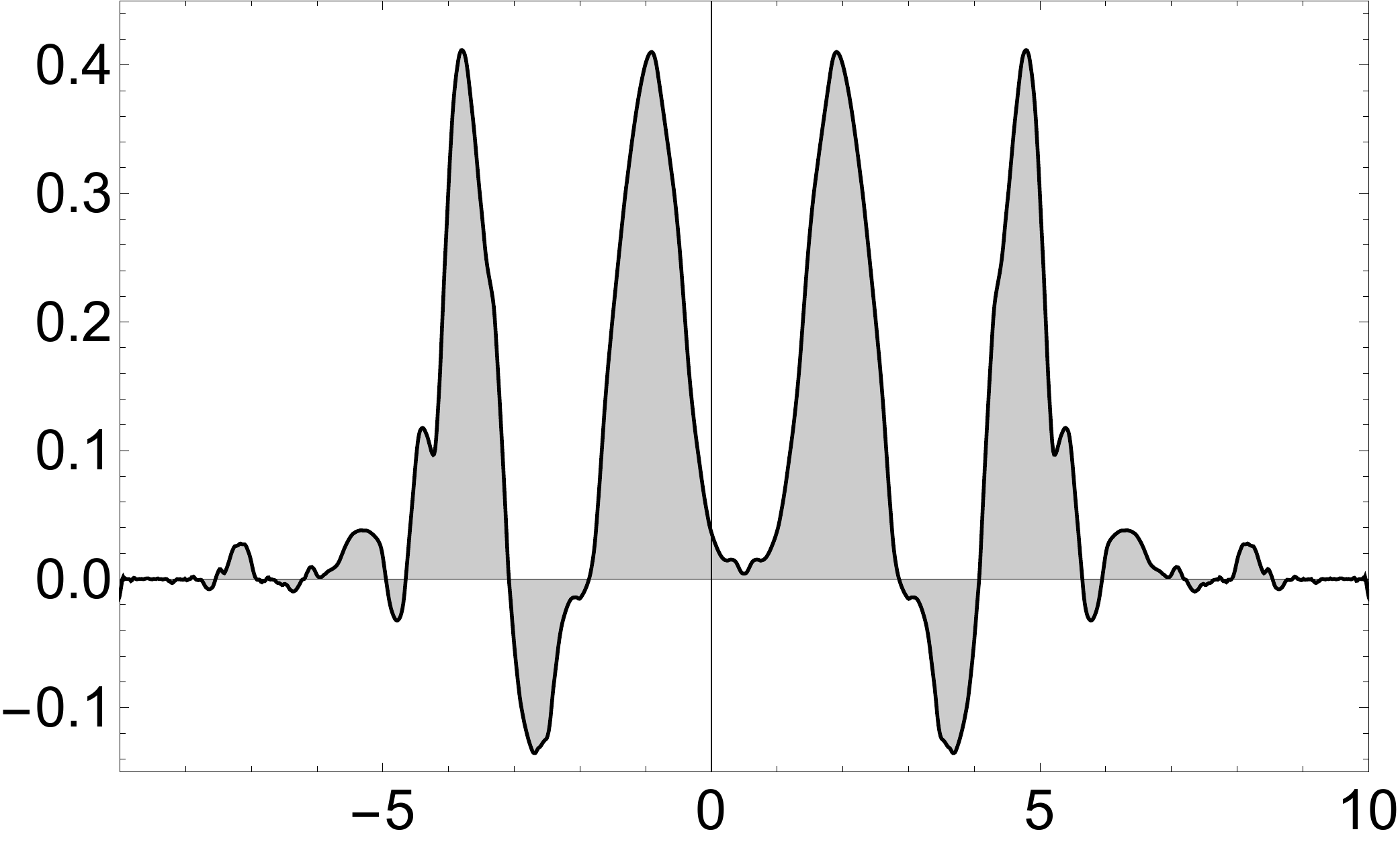}}
\caption{Numerical solutions  in the perturbed model for $\epsilon=10.0$ at  (a) $t=0.15$, (b) $t=1.67$, (c) $t=20.79$}\label{F16}
\end{figure*}

Once again, we have noted that, if we increase the perturbation, the modification of the model begins to nontrivially influence its solutions. In Fig.\ref{F16} we present the snapshots of the field seen in the simulation corresponding to $\epsilon=10$. There are significant quantitative differences in behaviour of this solution from the solution of the signum-Gordon model presented in Fig.\ref{SG-e-10}. We observe that for such high values of the parameter $\epsilon$ the solutions expand; however, they do not radiate significantly. In 
contradistinction to the solutions with $\epsilon=0.1$ we do not see any small waves that escape from the oscillating region and propagate with constant velocity.  
 
\begin{figure*}[h!]
\centering
\subfigure[]{\includegraphics[width=0.45\textwidth,height=0.15\textwidth, angle =0]{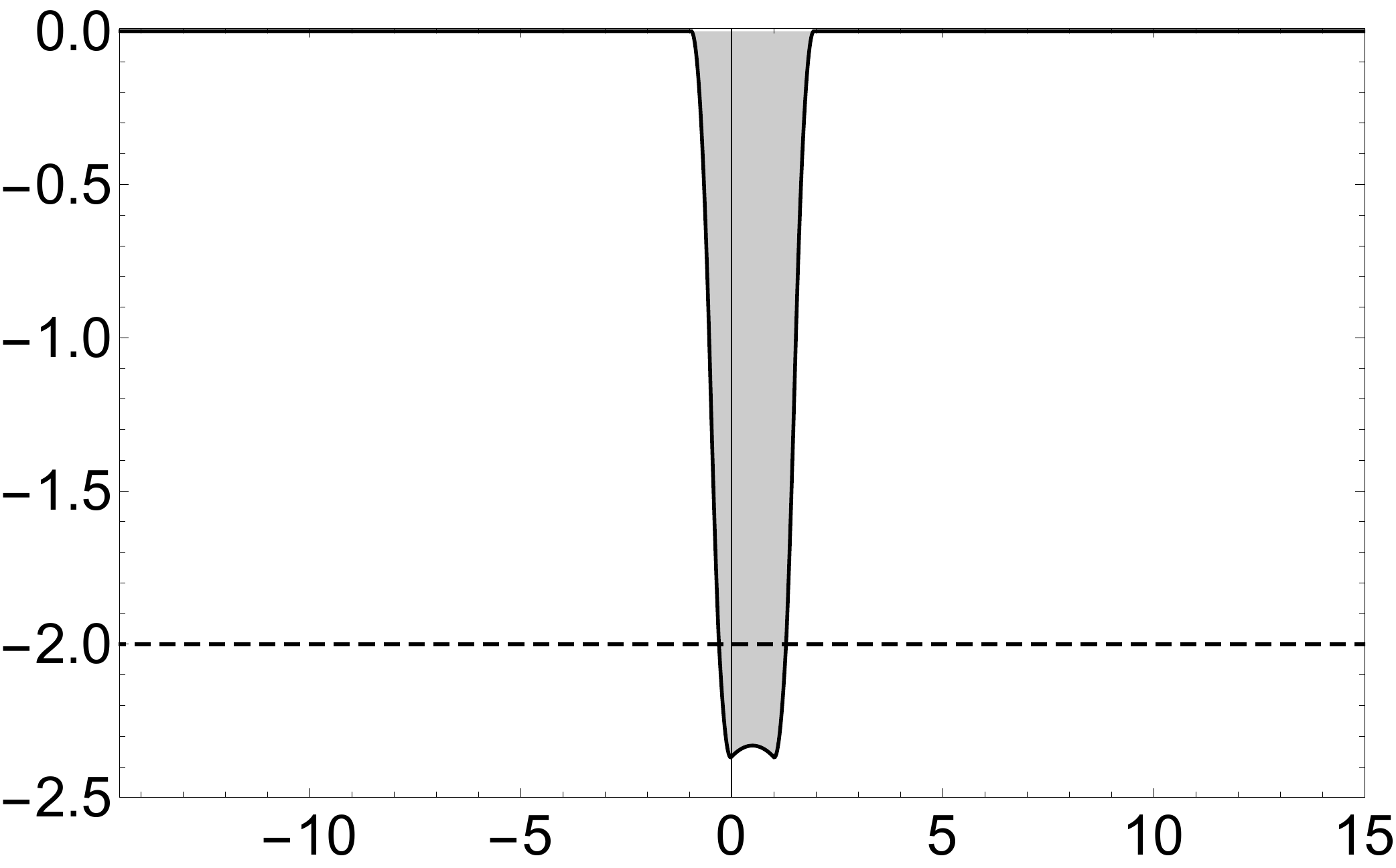}}
\subfigure[]{\includegraphics[width=0.45\textwidth,height=0.15\textwidth, angle =0]{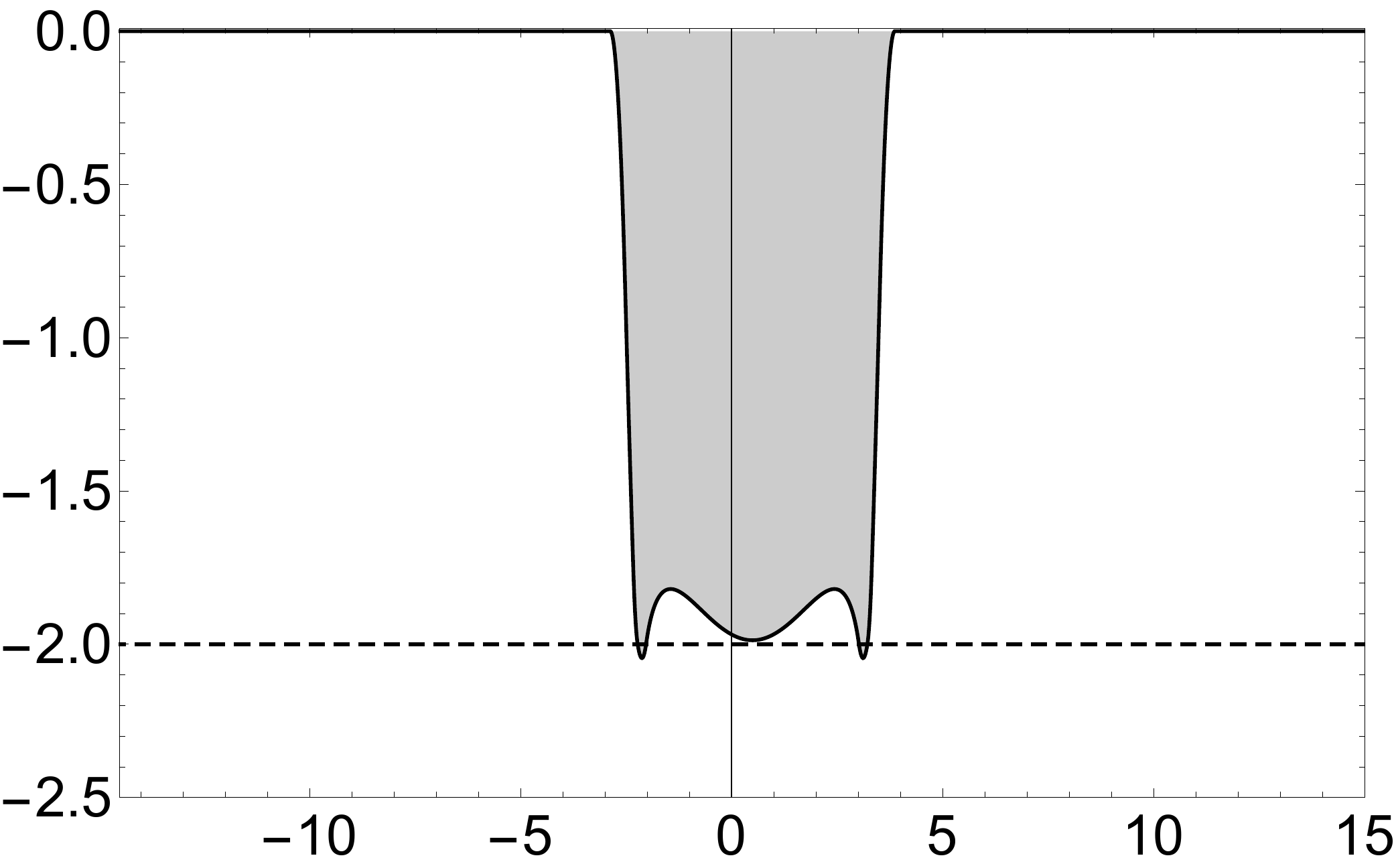}}
\subfigure[]{\includegraphics[width=0.45\textwidth,height=0.15\textwidth, angle =0]{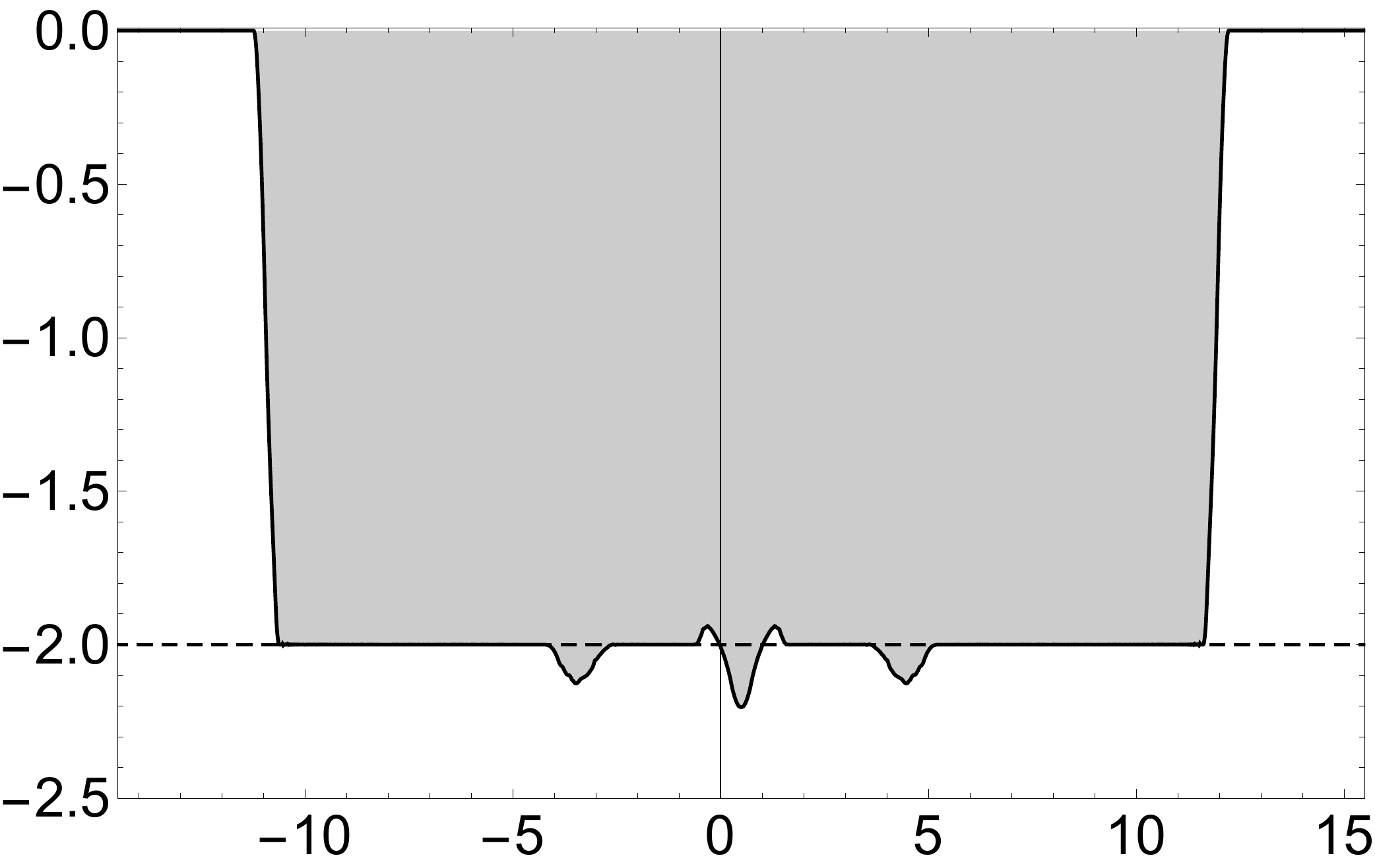}}
\subfigure[]{\includegraphics[width=0.45\textwidth,height=0.15\textwidth, angle =0]{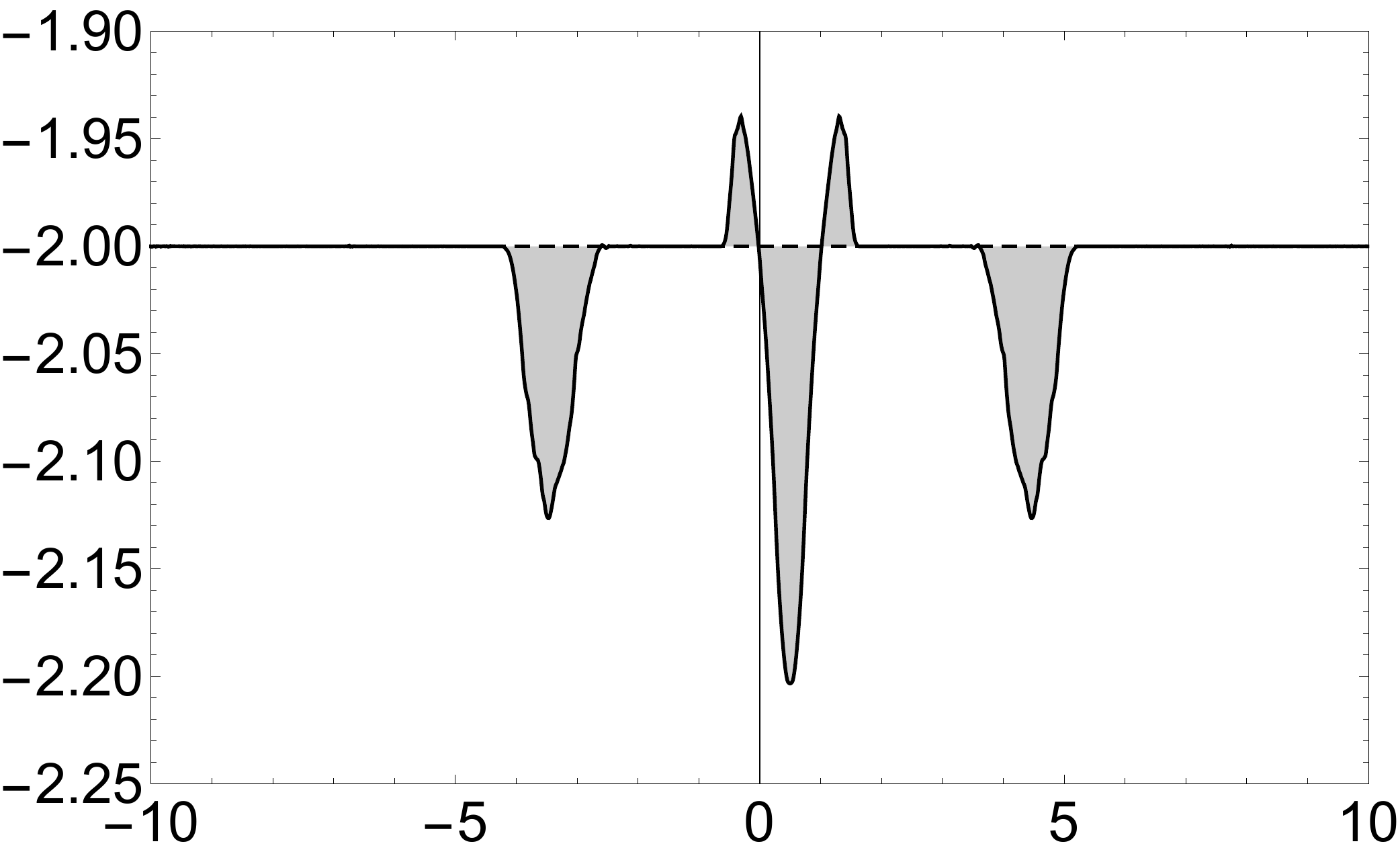}}
\caption{Numerical solutions  in the perturbed model for $\epsilon=20.0$ at  (a) $t=1.0030$, (b) $t=3.0180$, (c,d) $t=11.6490$. Dashed lines correspond to minima at $\eta=-2$.}\label{F17}
\end{figure*}

As before, for sufficiently large $\epsilon$  the second vacuum is visible and again, pairs of kink-antikink states get created. In figure Fig.\ref{F17} we present four snapshots of the field $\eta$ for $\epsilon=20.0$. We note that this time the kink and the antikink
that have been created propagate in opposite directions. Clearly, a part of the energy of the initial configuration has been transformed into oscillations of the field around $\eta=-2$.
Qualitatively, this process of kink-antikink pair creation is very similar to the process formerly considered for large $l$ initial configurations.

\section{Solutions of the modified signum-Gordon model on $\mathbb{R}_+$}

We have also looked at oscillon solutions in the radial modified signum-Gordon model. Again, there are two main new features. The first one is due to the backward scattering of left hand moving small oscillons and waves at the origin. In this case, they can travel back to the initial oscillon and interact with it. The second feature is related to the existence of ball shaped oscillons - inherited from the signum-Gordon model on $\mathbb{R}_+$.

\begin{figure*}[h!]
\centering
\subfigure[]{\includegraphics[width=0.3\textwidth,height=0.15\textwidth, angle =0]{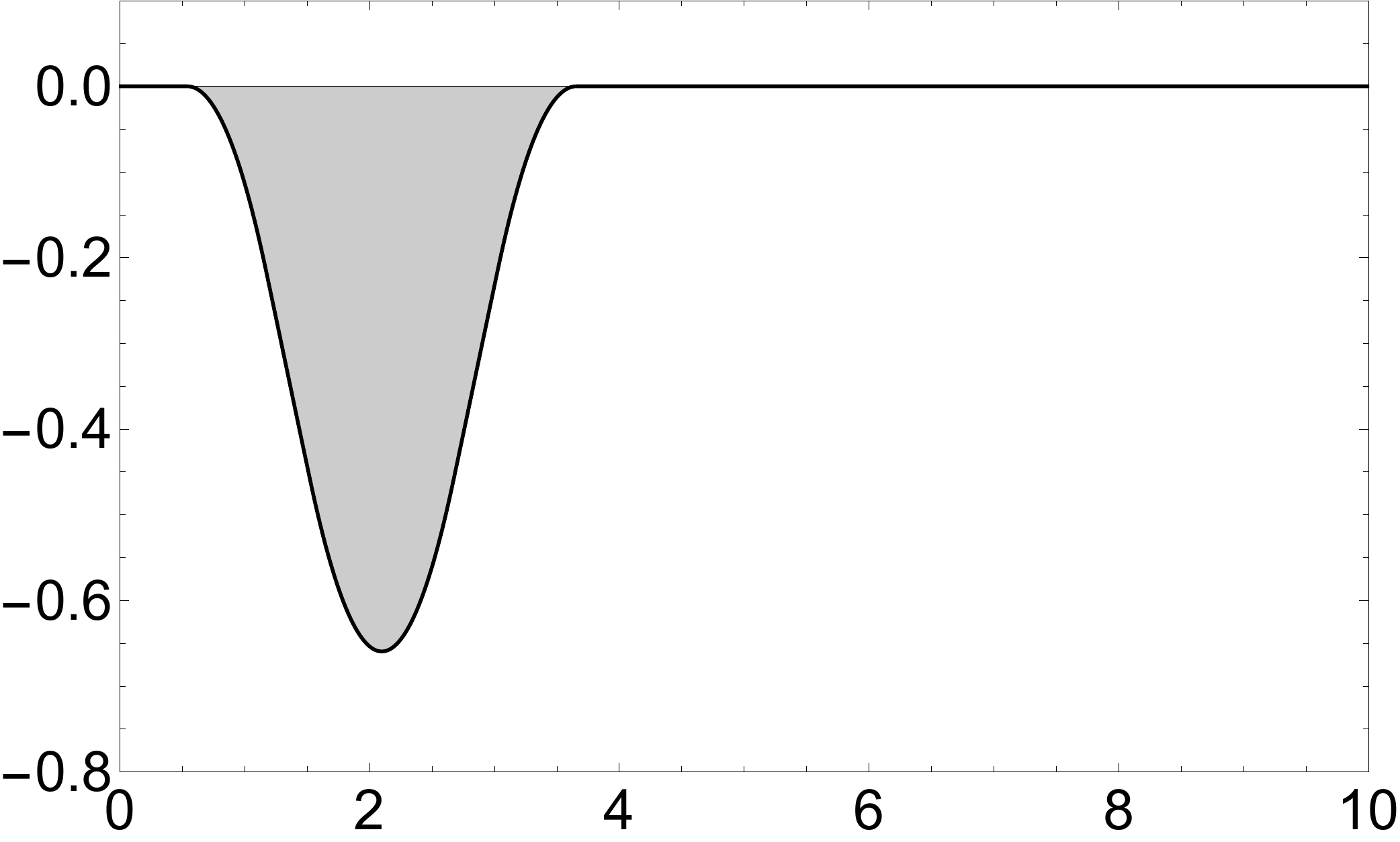}}
\subfigure[]{\includegraphics[width=0.3\textwidth,height=0.15\textwidth, angle =0]{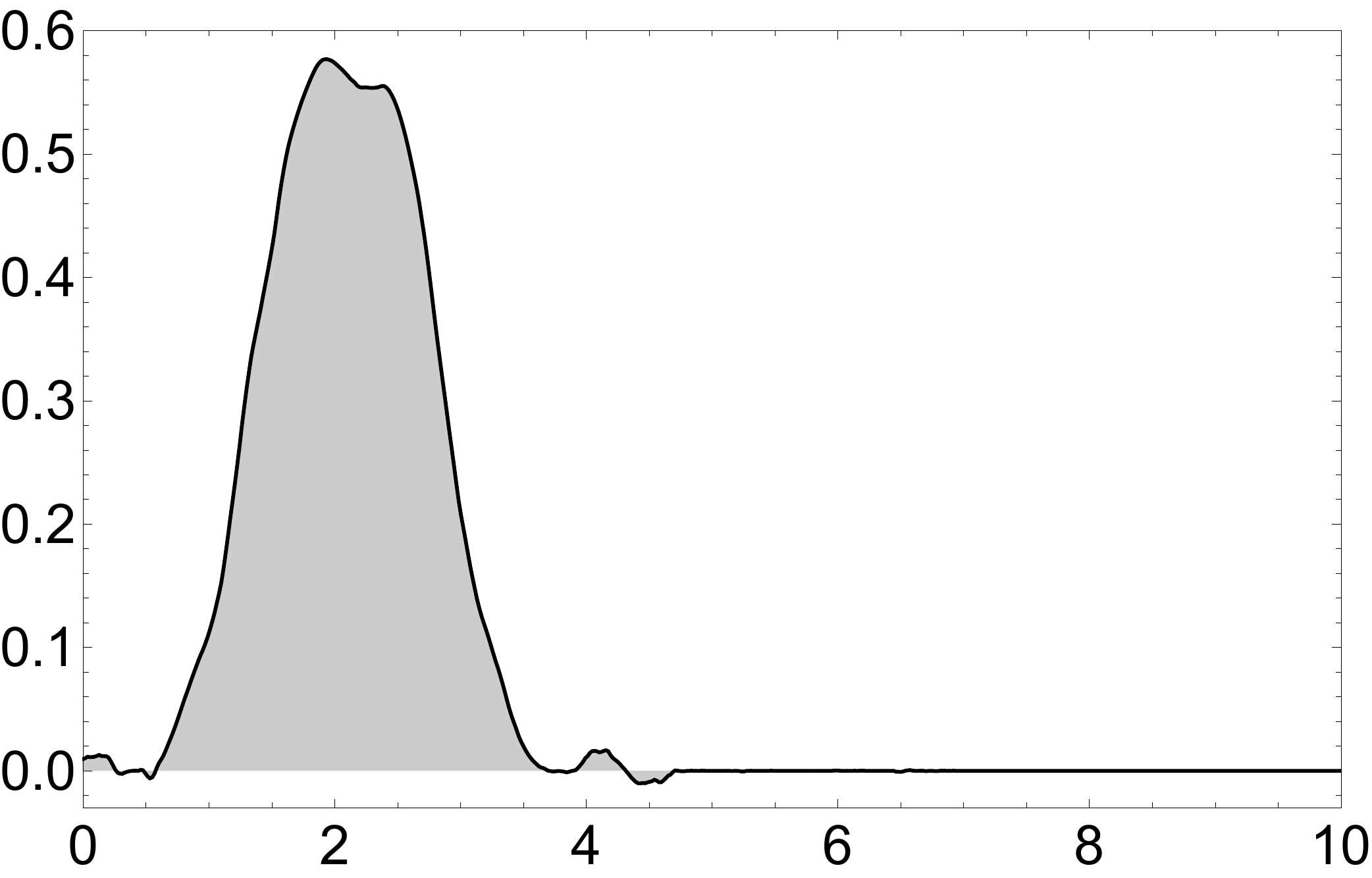}}
\subfigure[]{\includegraphics[width=0.3\textwidth,height=0.15\textwidth, angle =0]{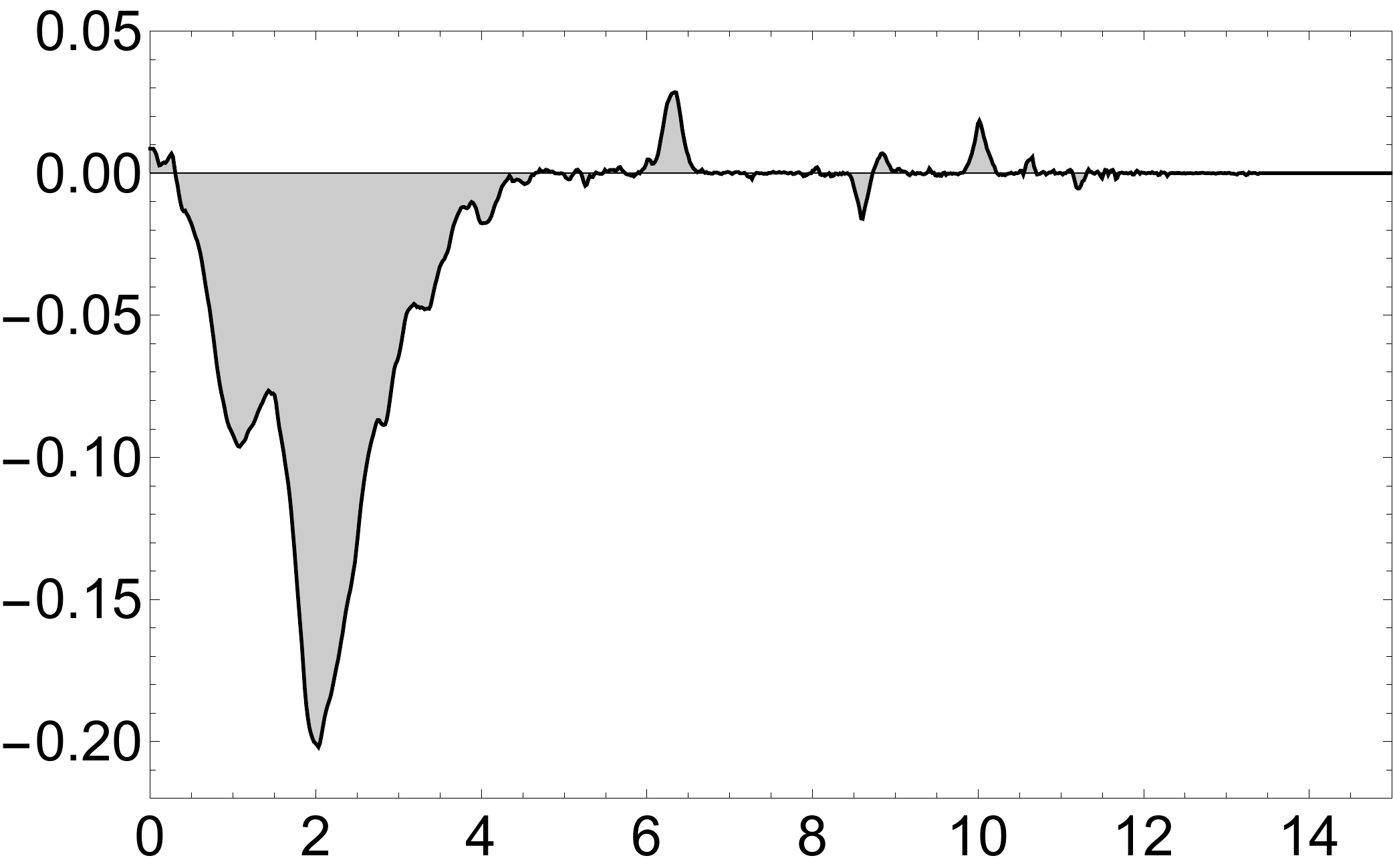}}
\caption{The shell-type  oscillon at ${\mathbb R}_+$ for $l=3.0$ at  (a) $t=0.931$, (b) $t=13.799$, (c) $t=25.765$.}\label{F29}
\end{figure*}

In Fig.\ref{F29} we present the evolution of the shell oscillating structure obtained for $l=3.0$. During the evolution the structure became more and more irregular. We have observed that the radiation emitted from the oscillon reflected at the centre $r=0$ and then returned to the oscillon. Of course, the part of the radiation that was emitted towards spatial infinity has never returned.

We know that by gradually increasing the value of the parameter $l$ we can reach the threshold of production of  kink-antikink pairs. 
One of the more intriguing facts about the solution containing bound states of kinks ($l=4.3156$) in the model on ${\mathbb R}$ is that only ``half'' of such a configuration would appear in the model on ${\mathbb R}_+$. So, such a configuration would describe a ball-shape solution whose border is provided by a wobbling kink.

\begin{figure*}[h!]
\centering
\subfigure[]{\includegraphics[width=0.3\textwidth,height=0.15\textwidth, angle =0]{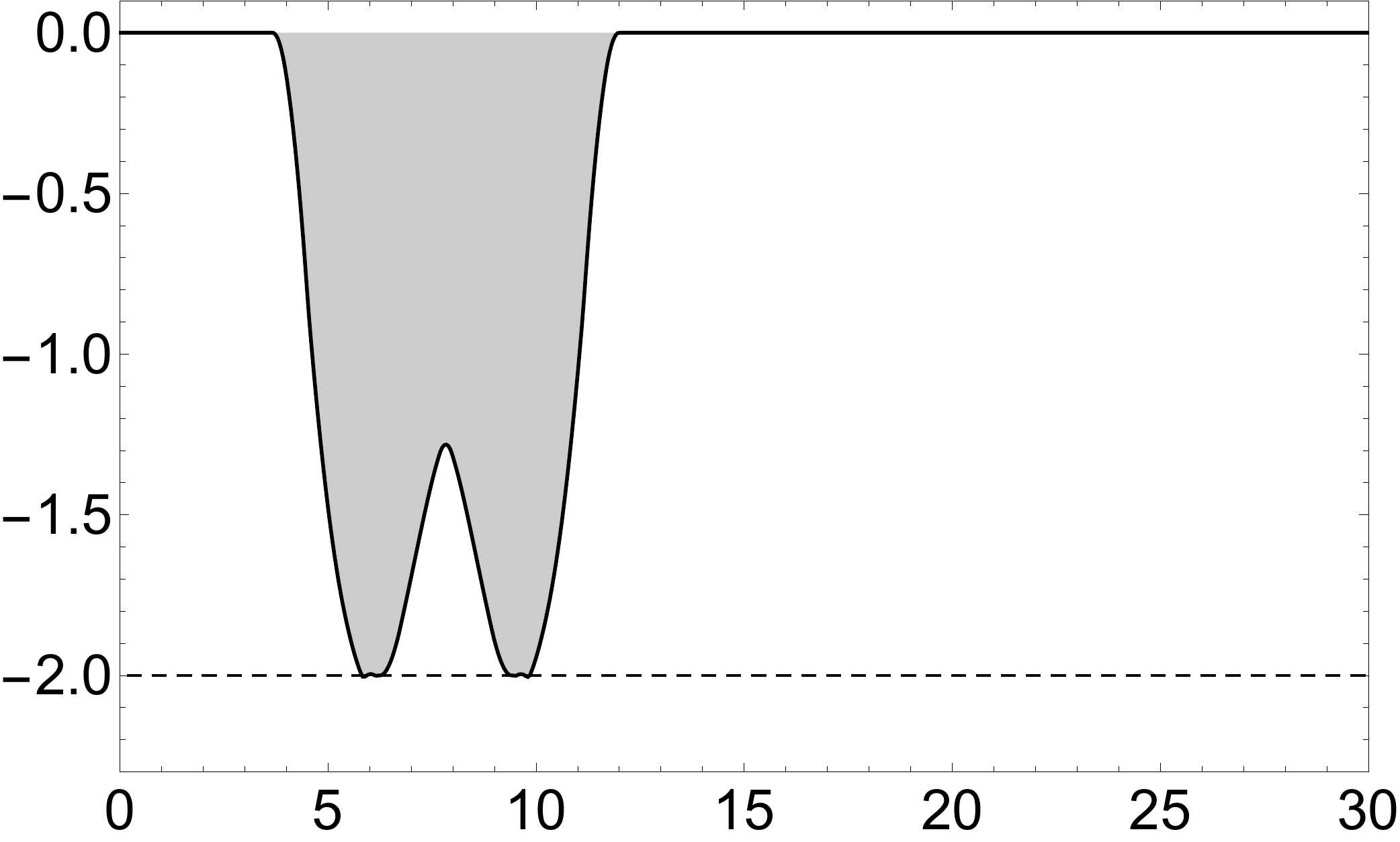}}
\subfigure[]{\includegraphics[width=0.3\textwidth,height=0.15\textwidth, angle =0]{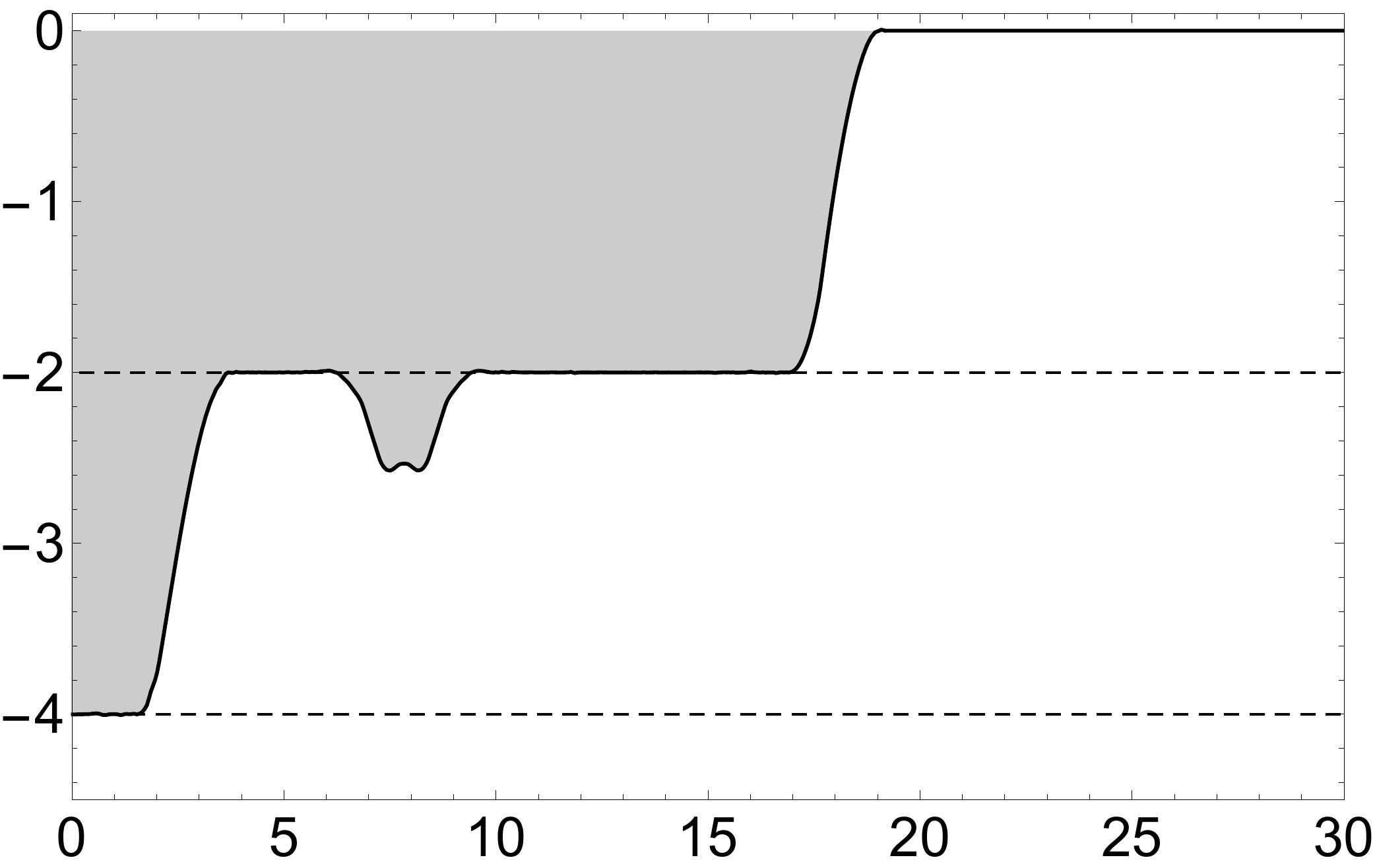}}
\subfigure[]{\includegraphics[width=0.3\textwidth,height=0.15\textwidth, angle =0]{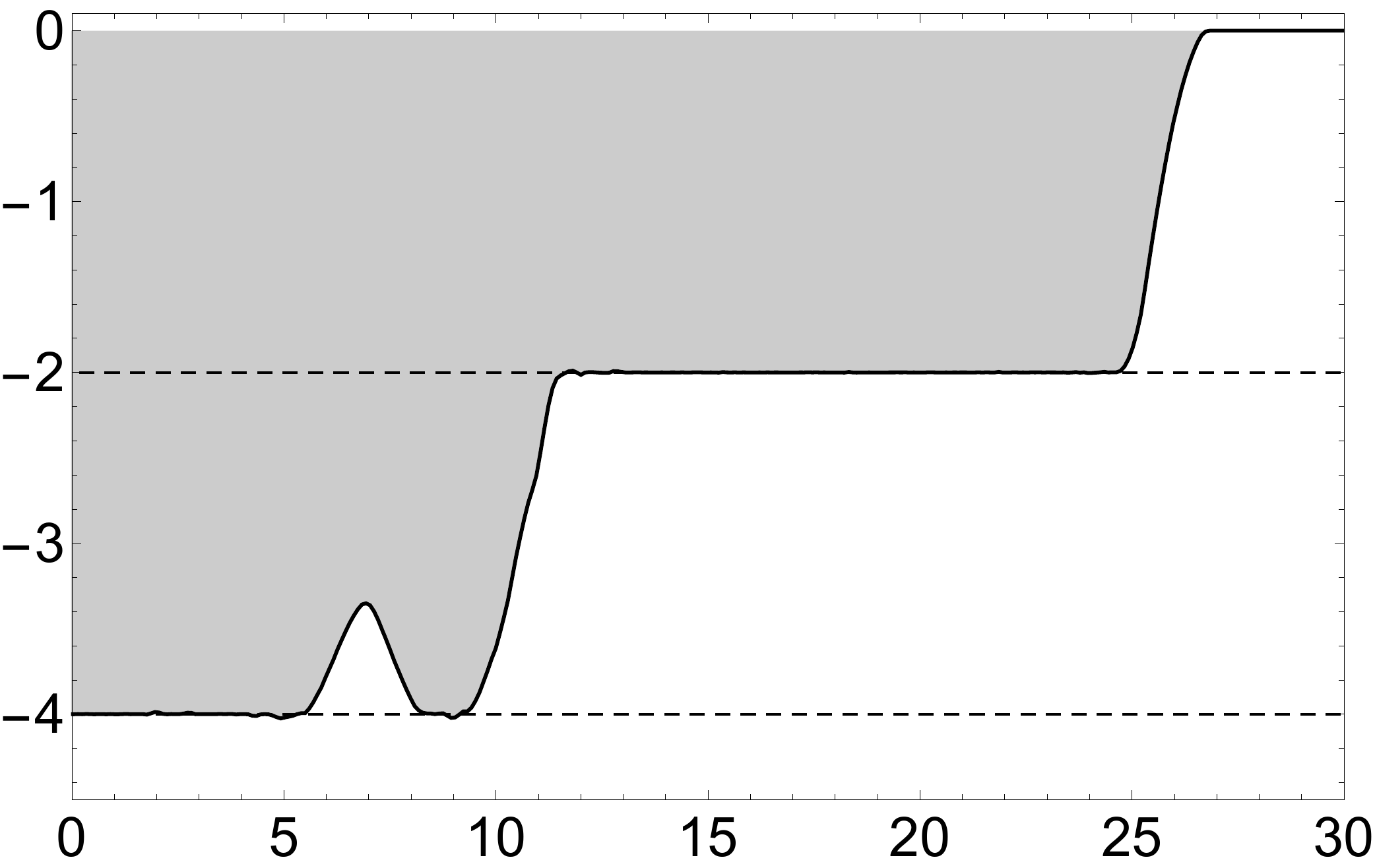}}
\subfigure[]{\includegraphics[width=0.3\textwidth,height=0.15\textwidth, angle =0]{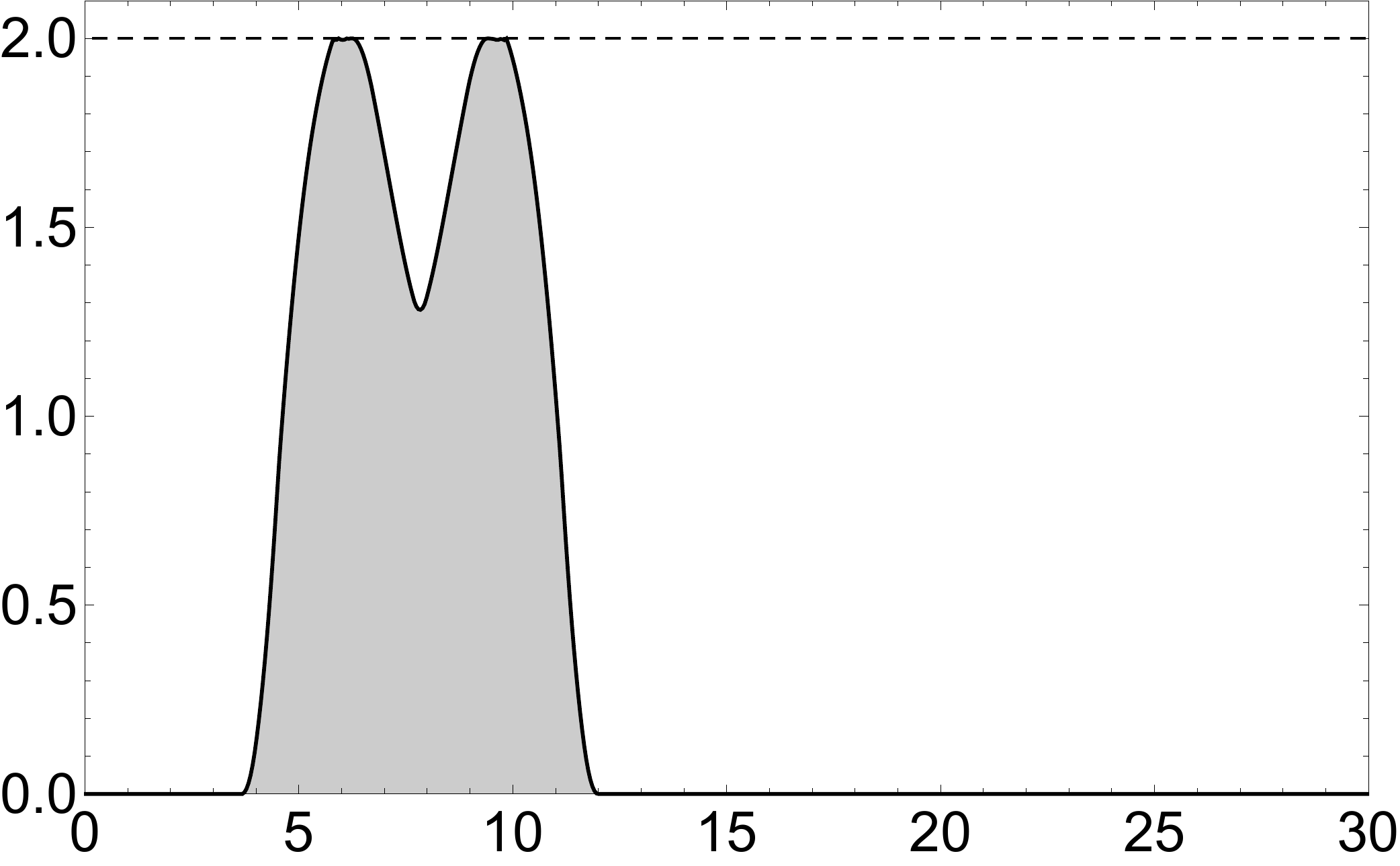}}
\subfigure[]{\includegraphics[width=0.3\textwidth,height=0.15\textwidth, angle =0]{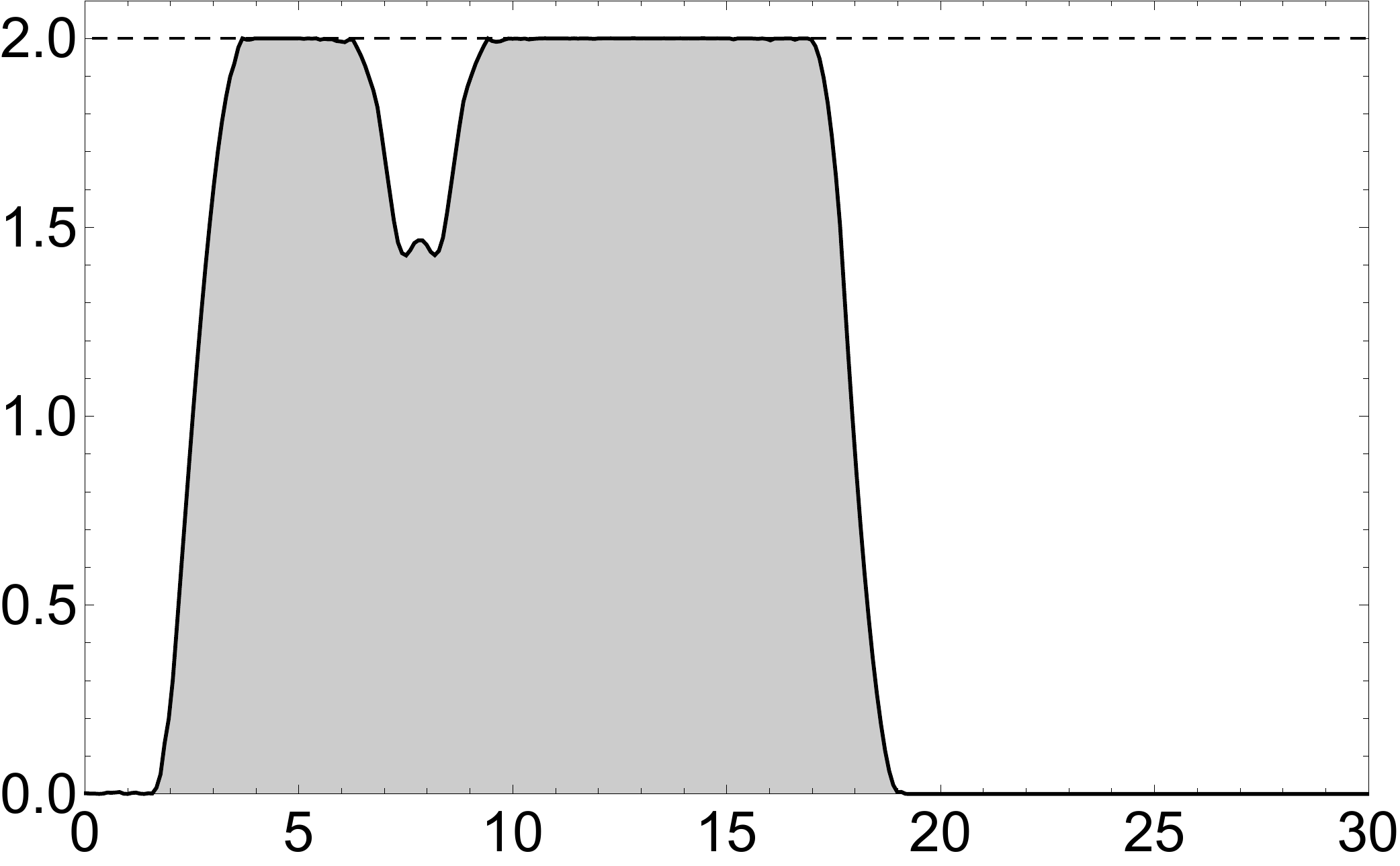}}
\subfigure[]{\includegraphics[width=0.3\textwidth,height=0.15\textwidth, angle =0]{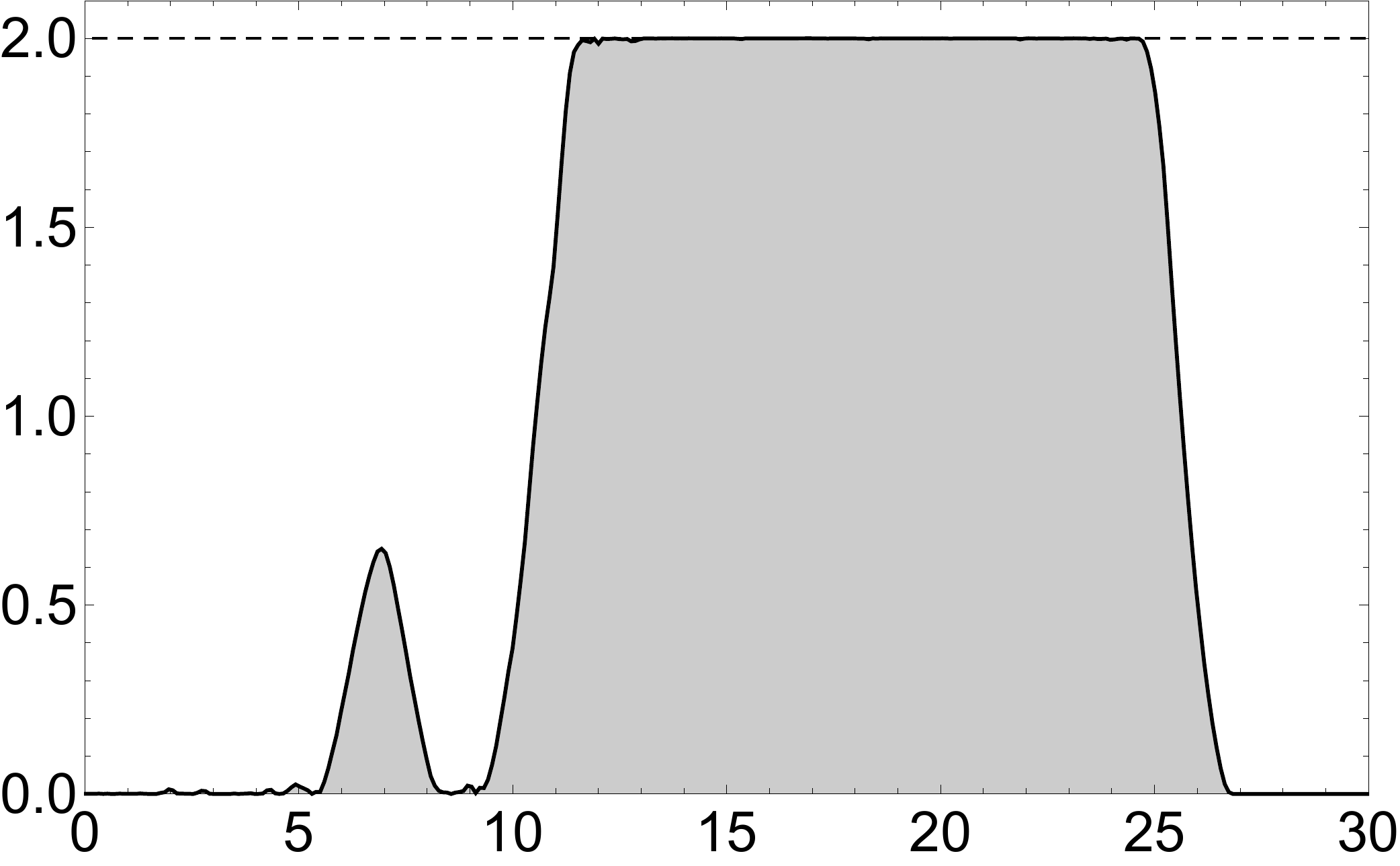}}
\caption{The shell-type  structure containing kinks at ${\mathbb R}_+$ for $l=5.224$ at  (a,d) $t=$, (b,e) $t=$, (c,f) $t=$. Figures (d,e,f) represent folded (original) field $\bar\eta(t,x)$.}\label{F30}
\end{figure*}

Finally, in Fig.\ref{F30} we present the results seen in our evolution of a shell type configuration that soon thereafter consisted of a kink-antikink pair. After the pair was created both objects started to move apart. Then the kink reflected at $r=0$ and started to move in the same direction as the antikink. This, effectively, resulted in the expansion of the shell. Comparing pictures Fig.\ref{F30}(e) with Fig.\ref{F30}(f) we observe a very interesting phenomenon. The oscillon which existed at the second minimum $\bar\eta=2$ has collided with the kink.   After the passage of the kink the oscillon `had jumped' to the first minimum.

\section{Conclusions}
The main aim of this work has been the study of the fate of the signum-Gordon breathers in a modified model, whose modification was motivated by 
very recent results obtained in some analytical investigations of the BPS structures of the Skyrme model. 

\vspace*{0.2cm} 

To accomplish this program we started with a careful study of the stability of perturbed breathers in the signum-Gordon model. We have found that such breathers, although not stabilized by topological considerations and with a continuous spectrum of energies, are remarkably stable under small perturbations. Our studies have shown that for initial configurations involving a breather with a small perturbation, the perturbed breather oscillates for quite long time (as long as 25 oscillations) as one object without emission of any detectable radiation. Afterwards, it starts to emit small packages of energy (small oscillons). 
  To get a better understanding of this process we have performed two long time simulations of the field configuration with $\epsilon=1.0$ and $l=1.0$ and then other ones with $\epsilon=1.3$ and $l=1.0$. The first one involved an unperturbed (exact) breather, the second one a slightly perturbed breather.

The main reason for the numerical study of the exact breather was to check our numerical procedures; to see and to assess  whether and to what extent the small inherent numerical errors could destroy the breather by inducing radiation. We have found that the numerical solution started to differ  visibly from the exact one only for times larger than $t=50$. We have performed the simulation up to $t=130$ and have found that by then the energy of the solution decreased from the initial value $E=\frac{1}{24}\approx 0.0416$ to the value $E=0.0381$. Of course this was a purely numerical artefact. In the case of the simulation of configurations with $\epsilon=1.3$ ({\it i.e.} of the perturbed breather) the initial energy $E=\frac{\epsilon^2}{24}=0.0704$ decreased to the value $E=0.0462$ at $t=130$ and to the value $E=0.0418$ at $t=200$ which is very close to the value of the energy for the exact breather. We have concluded that, in spite of small numerical errors (smaller than expected), the slightly perturbed breathers have a tendency to emit the surplus of their energy and tend to a breather solution. Our numerical calculations involved the use of 4th order Runge-Kutta method of simulating the time evaluation of the fields and working in double precision. We were satisfied with our results and we feel we can trust them; getting smaller numerical errors would require more sophisticated numerical techniques and clearly lies beyond the scope of this paper.

We have also found that if the initial state represented a strongly deformed breather then the evolution was more involved. For large amplitude perturbations (large $\epsilon$) we observed a decay into a collection of (still perturbed) breathers. Again, one could study 
the pattern of the formation of such substructures:  their number, mutual interactions and their stability. For a significant amplitude suppression ($\epsilon < \frac{1}{2}$) we reach a regime where the boundary of the compacton is approached by an infinite number of oscillations of a self-similar solution. This destabilises the solution rather strongly and the energy is emitted very quickly from the region of the initial oscillon. 
  
  \vspace*{0.2cm} 

Having established the stability of the signum-Gordon breathers we have moved to the modified signum-Gordon model. In this case we first used the initial condition of the signum-Gordon breather. This is equivalent to an insertion of a signum-Gordon breather into the modified theory. As the modification breaks the scaling invariance of the signum-Gordon model we had to consider initial breather states with different supports $l$ to get some meaningful results. 

Thus, first of all, we have found that small amplitude breathers ($l<2.0$) in the modified model still behave as breathers (or very long living oscillons) {\it i.e.}, they are oscillating solutions (with a compact support as in the original signum-Gordon model) without any detectable radiation. Hence, not too large breathers undoubtedly survive in the modified model. The new term in the Lagrangian increases the period of the oscillations and modifies the local form of the oscillating solutions, which now develops a richer structure with a non-simultaneous changing sign transitions, {\it i.e.} different parts of the solution change its sign at slightly different times. Furthermore, using an approximate description of the field we have found the local structure of the solution in an analytical way which agrees remarkably well with the true numerical results during the first period of oscillation. This also allowed us to obtain an analytical description of the relation between the period and size of the oscillon $T=T(l)$. 
\\
We have studied the evolution of the initial $l=1$ breather with a small perturbation in the modified theory. Surprisingly, we
have found that the modification increases the stability of the breathers against small perturbations. Initially perturbed breather oscillates as one unit longer than in the original signum-Gordon model and the emission of energy, again via small oscillons, takes place less frequently and possesses a more regular pattern.

Secondly, we have found that the modification can introduce a second vacuum in the theory and so may lead to the existence of topological compact solitons - absent in the original signum-Gordon model. This shows that for large enough initial configurations a pair of kink-antikink states can be created. Such a phenomenon has been analyzed in many $(1+1)$ dimensional scalar field theories, however, with usual infinitely extended solitons \cite{tom} rather than compactons. We studied such a creation mechanism in two ways: with large enough $l$ and $\epsilon=1$ (initial conditions for the signum-Gordon bretahers) and with $l=1$ and large enough $\epsilon$ (a large perturbation of the unit support signum-Gordon breather). In both cases a creation of such pairs of compact kink-antikinks has been observed. Usually, constituents of the pair (kink and antikink) have some kinetic energy which allows them to escape from each other leaving  an oscillon at the origin. The process can repeat itself until the central oscillon has too little energy to create such a pair. In addition some small radiation waves (probably in the form of small oscillons) are also emitted. For very special initial conditions it is possible to find a non-moving kink-antikink pair with a constant distance between the constituents of the pair. Such a solution can be described as a wobbler which shrinks and expands periodically. 

  \vspace*{0.2cm} 
  
In the final part of our work we have considered the modified theory on a semi-infinite line which corresponds to the radial coordinate $r$. Due to the compact nature of solitons in the modified signum-Gordon model, the transition from $\mathbb{R}$ to $\mathbb{R}_+$ does not change anything as long as the (breather) oscillon does not reach the origin. This corresponds to the existence of oscillon shell structures in the first BPS Skyrme submodel with the properties analogous to the properties of oscillons in the modified signum-Gordon theory. Moreover, the ball type oscillons are also possible. They are just a partial oscillons located at the origin ({\it i.e.} one of the structures is located at the origin). 
\\
From the  dynamical point of view the only difference in the semi-infinite line case, resides in the fact that the left moving perturbations emitted by an oscillon come back to it after a reflection at the origin. This leads to slightly more involved relaxation pattern as emitted small oscillons and waves can travel many times between the oscillon and the origin.  

  \vspace*{0.2cm} 
  
All this shows that the first BPS Skyrme submodel has solutions which are described by small amplitude oscillons of both shell and ball types. They are stable (at least to within our numerical accuracy) within our ansatz where only the profile function of Skyrmions depends on radial and temporal coordinates while the $\mathbb{S}^2$ part of the Skyrme field is "frozen" in a given rational map. Whether they remain to be stable in the full first BPS submodel without any restrictions on the form of the solution is an open problem. In this case, the dynamics does not follow the modified signum-Gordon and the complex field can have a nontrivial  impact on the existence and features of the oscillons. Another question is the fate and a possible role of this oscillon in the full Skyrme model.  

\section*{Acknowledgements}
WJZ work was supported in part by the Leverhulme Trust Emeritus Fellowship. The authors thank Luiz A. Ferreira for helpful discussions.


\begin{thebibliography}{100}
\bibitem{skyrme} T.H.R. Skyrme, {\em Proc. Roy. Soc. Lon.} {\bf 260} (1961)
127; {\em Nucl. Phys.} {\bf 31} (1962) 556; {\em J. Math. Phys.} {\bf
12} (1971) 1735

\bb{ASW-BPS} C. Adam, J. Sanchez-Guillen, A. Wereszczynski, Phys. Lett. {\bf B691} (2010) 105;
C. Adam, J. Sanchez-Guillen, A. Wereszczynski, Phys. Rev. {\bf D82} (2010) 085015
  
  \bb{mar} E. Bonenfant, L. Marleau, Phys. Rev. D {\bf 82} (2010) 054023; 
E. Bonenfant, L. Harbour, L. Marleau, Phys. Rev. D {\bf 85} (2012) 114045; M.-O. Beaudoin, L. Marleau, Nucl. Phys. B {\bf 883} (2014) 328
  \bb{Sut-BPS} P. Sutcliffe,
JHEP {\bf 1008} (2010) 019;
JHEP {\bf 1104} (2011) 045
  \bb{Leeds-BPS} M. Gillard, D. Harland, M. Speight, Nucl. Phys. {\bf B895} (2015) 272; M. Gillard, D. Harland, E. Kirk, B. Maybee, M. Speight, Nucl.Phys. {\bf B917} (2017) 286
  \bb{bjarke} S. B. Gudnason, Phys.Rev. {\bf D93} (2016) 065048; S. B. Gudnason, M. Nitta, Phys. Rev. {\bf D94} (2016) 065018; S. B. Gudnason, B. Zhang, N. Ma, Phys. Rev. {\bf D94} (2016) 125004
  
  \bibitem{nappi} G.S. Adkins, C.R. Nappi, E. Witten, {\em Nucl. Phys. 
B} {\bf 228} (1983) 552; G.S. Adkins, C.R. Nappi, {\em Nucl. Phys. B} {\bf 233} (1984) 109
  
  \bb{vib} C.J. Halcrow, Nucl. Phys. {\bf B904} (2016) 106; C.J. Halcrow, C. King, N.S. Manton, Phys. Rev. {\bf C95} (2017), 031303
  
  \bb{rat-map}
R. Battye, P. Sutcliffe, Phys. Rev. Lett {\bf 79} (1997) 363; C. Houghton, N. Manton, P. Sutcliffe, Nucl. Phys. B{\bf 510} (1998) 507 
  
  \bb{stefano} S. Baldino, S. Bolognesi, S. B. Gudnason, D. Koksal, Phys.Rev. {\bf D96} (2017) 034008 
  
  \bb{roper-old} C. Hajduk, B. Schwesinger, Phys. Lett. {\bf B140} (1984) 172; U. B. Kaulfuss, U.-G. Meissner, Phys. Lett. {\bf B154} (1985) 193; A. Hayashi, G. Holzwarth, Phys. Lett. {\bf B140} (1984) 175; L. C. Biedenharn, Y. Dothan, M. Tarlini, Phys. Rev. {\bf D31} (1985) 649; I. Zahed, U.-G. Meissner, U. B. Kaulfuss, Nucl. Phys. {\bf A426} (1984) 525; J. D. Breit, C. R. Nappi, Phys. Rev. Lett. {\bf 53} 1984) 889; W.T. Lin, B. Piette, Phys.Rev. {\bf D77} (2008) 125028; M. Heusler, S. Droz, N. Straumann, Phys. Lett. {\bf B271} (1991) 61; C. Adam, C. Naya, J. Sanchez-Guillen, A. Wereszczynski, Phys.Lett. {\bf B726} (2013) 892
  \bb{bizon} P. Bizon, T. Chmaj, A. Rostworowski, Phys.Rev. {\bf D75} (2007) 121702
 \bb{roper-BPS}  T. Ioannidou, A. Lukacs, J. Math. Phys. {\bf 57} (2016) 022901; C. Adam, M. Haberichter, T. Romanczukiewicz, A. Wereszczynski, Phys. Rev. {\bf D94} (2016) 096013
 \bb{roper-new} C. Adam, M. Haberichter, T. Romanczukiewicz, A. Wereszczynski, arXiv:1710.00837
  
  \bb{iso-rot} D. Harland, J. Jaykka, Y. Shnir, M. Speight, J. Phys. {\bf A46} (2013) 225402; R. Rajaraman, H. M. Sommermann, J. Wambach, H. W. Wyld, Phys. Rev. {\bf D33} (1986) 287; A. Halavanau, Y. Shnir, Phys. Rev. {\bf D88} (2013) 085028; R. Battye, M. Haberichter, Phys. Rev. {\bf D87} (2013) 105003; R. Battye, M. Haberichter, Phys. Rev. {\bf D88} (2013) 125016
  
  \bb{scatter} D. Foster, N. S. Manton, Nucl.Phys. {\bf B899} (2015) 513; D, Foster, S. Krusch, Nucl.Phys. {\bf B897} (2015) 697
  
\bb{new-BPS1} C. Adam, J. Sanchez-Guillen, A.
  Wereszczynski, Phys. Lett. {\bf B769} (2017) 362
  
    \bb{new-BPS2} C. Adam, D. Foster, S. Krusch, A.
  Wereszczynski, arXiv:1709.06583

 
\bb{arodzklimas} H. Arodz, P. Klimas, Acta Phys. Pol. {\bf B36} (2005) 787 

\bb{arodz1} H. Arodz, P. Klimas, T. Tyranowski, Acta Phys. Polon. {\bf B36} (2005) 3861
\bb{arodz2} H. Arodz, P. Klimas, T. Tyranowski, Phys. Rev. {\bf D77} (2008) 047701
\bb{arodz3} H. Arodz, Acta Phys. Polon. {\bf B38} (2007) 3099
\bb{Klimas} P. Klimas, J. Phys. {\bf A41} (2008) 095403 


\bb{zak osc} B. Piette, W. J. Zakrzewski, Nonlinearity {\bf 11} (1998) 1103
\bb{hopf in Sk} D. Foster, J.Phys. A50 (2017) 405401 

\bb{tom} P. Dorey, K. Mersh, T. Romanczukiewicz, Y. Shnir, Phys. Rev. Lett. {\bf 107} (2011) 091602;  T. Romanczukiewicz, Y. Shnir, Phys.Rev.Lett. {\bf 105} (2010) 081601; T. Romanczukiewicz, Y. Shnir, arXiv:1706.09234 




\end{thebibliography}
\end{document}